\documentclass[twoside,12pt]{article}
\usepackage{epsfig}

\def\Journal#1#2#3#4{{#1} {#2} (#4) #3 }

\def\NPA{{\em Nucl. Phys.} A}

\def\PLB{{\em Phys. Lett.} B}

\def\PRL{\em Phys. Rev. Lett.}

\def\PREP{\em Phys. Rep.}

\def\PRD{{\em Phys. Rev.} D}
\def\PRC{{\em Phys. Rev.} C}

\def\om{\omega}

\newcommand{\be}{\begin{equation}}
\newcommand{\ee}{\end{equation}}
\newcommand{\bea}{\begin{eqnarray}}
\newcommand{\eea}{\end{eqnarray}}

\newcommand{\lsim}{\stackrel{\scriptstyle <}{\phantom{}_{\sim}}}
\newcommand{\gsim}{\stackrel{\scriptstyle >}{\phantom{}_{\sim}}}
\begin{document}

\title{ \vspace{1cm} Structure formation during phase transitions in strongly interacting matter}
\author{D. N.\ Voskresensky,$^{1,2}$
\\
$^1$BLTP, Joint Institute for Nuclear Research, RU-141980 Dubna, Russia\\
$^2$National Research Nuclear
    University (MEPhI),  Kashirskoe shosse 31,\\
    115409 Moscow, Russia\\}
\maketitle
\begin{abstract} A broad range of problems associated with  phase transitions in  systems characterized by the strong interaction between particles and with formation of structures is reviewed.
A general phenomenological mean-field model is constructed describing phase transitions of the first and the second order to the homogeneous,  $k_0=0$, and inhomogeneous, $\vec{k}_0\neq 0$ , states, the latter may occur even in case, when the interaction is translation-invariant.
Due to fluctuations, the phase transition to the  state, $\vec{k}_0\neq 0$, becomes the transition of the first order. Various specific features   of the phase transitions to the  state $\vec{k}_0\neq 0$ are considered such as the anisotropic spectrum of excitations, a possibility of the formation of various structures including running and standing waves, three-axis structures, the chiral waves, pasta mixed phases, etc. Next, a formal transition to hydrodynamical variables is performed. Then focus is made on description of the dynamics of the order parameter at the phase transitions to the states with $\vec{k}_0= 0$ and $\vec{k}_0\neq 0$.
In case of the phase transition to the inhomogeneous state the dynamics has specific features.
Next the non-ideal hydrodynamical description of the phase transitions of the liquid--gas type in nuclear systems is performed. The ordinary Ginzburg--Landau model proves to be  not applicable for description of an initial inertial stage of the seeds. Surface tension and viscosity are the driving forces of the phase transitions. Quasi-periodic  structures are developed during the transitions.
Next, the specific example of the pion condensation phase transition to the $\vec{k}_0\neq 0$ state in dense, cold or warm nuclear matter  is considered and then the nuclear system at high temperature and small baryon chemical potential  is studied, when baryons  become completely blurred and light bosons, e.g., pions, may condense either in $\vec{k}_0= 0$ or $\vec{k}_0\neq 0$ states.
Then, for the scalar collective modes the phenomena of the  Pomeranchuk instability and the Bose condensation in $\vec{k}_0= 0$ or $\vec{k}_0\neq 0$ states are studied and a possibility of a metastable dilute nuclear state is discussed.
Next, possibility of the condensation of  Bose excitations in the $\vec{k}_0\neq 0$ state in the moving media is  considered. Then Bose-Einstein condensation of pions with dynamically fixed number of particles is studied.  Finally,   specific  purely non-equilibrium effects are demonstrated on an example of the sudden breaking up of the box filled by nucleons.
\end{abstract}
%\eject
\tableofcontents
\section{Introduction}
It is commonly believed  that the  strong interaction theory on fundamental level  deals with quarks and gluons  described
by the QCD Lagrangian and the Lagrange equations of motion. The specific property of the QCD interactions  is the presence of the so called asymptotic freedom at short distances (associated with  vanishing of the running QCD coupling, $\alpha_s (q)$, at high momenta) and the confinement at large distances,
  $d\gsim r_\Lambda\sim (0.2 -0.4)$ fm, related to small momenta.
  %, that results in the unlimited growth of $\alpha_s (q)$ for $q\to 0$.  The latter property does not allow for existence of  quarks and gluons in the free state. However in a strongly pressed or heated  baryon matter,  quarks may undergo the deconfinement phase transition.

In the laboratory the quark-gluon interactions are studied in hadron-hadron, hadron-nucleus and nucleus-nucleus collisions at high collision energies. Even at highest  energies in nucleus-nucleus collisions at the Large
Hadron Collider (LHC) in  TeV region of the momenta of colliding particles,    $\alpha_s (q)\gsim 10^{-1}$ and the region of the asymptotic freedom  is not yet reached. In dense baryon matter a perturbative QCD description is expected to be valid only for baryon densities  $n\gsim  40n_0$, where $n_0\simeq 0.16$fm$^{-3}\simeq 0.5 m^3_\pi$  is the nuclear saturation density, $m_\pi\simeq 140$ MeV is the pion mass, $\hbar =c=1$. Nevertheless most of the researchers believe that the deconfinement regime has been  reached in the violent heavy-ion collisions at the Super Proton Synchrotron (SPS) at CERN,  the Relativistic
Heavy-Ion Collider (RHIC) at Brookhaven, and at the LHC at CERN energies, when typical temperature of excited matter is $T\gsim m_\pi$ and the particle density may vary from $n\ll n_0$ to $n$ signficantly exceeding $n_0$. However the state of the matter is not weakly interacting quark-gluon plasma, as it was  believed during a long time \cite{Shuryak:1980tp},  but the  strongly interacting quark-gluon plasma, being not  badly   described by the relativistic non-ideal hydrodynamics at very small but finite value of the shear viscosity, cf. \cite{Shuryak2005,Romatschke:2007mq}. Being formed in the nucleus-nucleus collision, the nuclear fireball is expanded into vacuum, distance between quarks increases, they are hadronized at certain  expansion stage and then   the system breaks up. The last stage is already well described in terms of the purely hadronic degrees  of freedom, cf. \cite{Karsch}. At lower collision energies,  $\lsim $ several GeV $/A$ (per nucleon),   one deals with the hadron matter when the baryon density is  $n\lsim 3n_0$ and $T<m_\pi$, cf. \cite{Reisdorf2007,Voskresensky:1989sn,Voskresensky:1993ud,Maslov:2019dep}.

Calculations show that in the heaviest compact stars the baryon density in the center may reach up to $(6-8) n_0$,  e.g. cf. \cite{{Klahn:2006ir}}. Again typical distances between particles are such that asymptotic freedom regime is not realized. Various possibilities are considered: purely hadronic neutron stars, hybrid stars, where a part of the interior consists of the strongly interacting quark-gluon plasma,  quark nuclearites (strange-quark objects of sizes of meteorites) and smaller size quark nuggets   consisting of strange quark matter, cf. \cite{Glendenning2000,Witten:1984rs,DeRujula:1984axn}. New branch of compact object astrophysics is related with the gravitational wave detections by LIGO and VIRGO Collaborations from mergers of compact stars. The Neutron
Star Interior Composition Explorer (NICER) observations of rotation-powered millisecond pulsars get  information about the mass-radius relation of compact stars and the equation of state of the
dense matter in their cores, cf. \cite{Bogdanov:2021yip}. Future gravitational wave measurements may help to constrain the onset densities of the strong phase transitions \cite{Blacker:2020nlq}.
 The future  laboratory experiments at Nuclotron-based
Ion Collider Facility  (NICA) at Dubna and Antiproton and Ion Research (FAIR) at Darmstadt will complement actively continuing astronomical observations, cf.  \cite{Senger2022}.

In case of the quark-gluon plasma one deals with quarks and gluons as the building blocks of matter, whereas in  the description of hadronic matter  with baryons and mesons as the building blocks glued together. In  condensed matter physics the building blocks of matter are  the  atoms and electrons. Physicists are interested in how these basic building blocks,  clued  together by quantum forces,   form
new  possible states of the matter.  For example,
they can form crystalline solids, liquid crystals, amorphous media, liquids, gases, magnets, superfluids  and superconductors, etc.
The  success of the condensed matter and  the particle physics in the previous and this  centuries is associated with usage of  the principle of spontaneous symmetry breaking \cite{Anderson1977}. For example, a crystalline solid breaks translation symmetry, although the interaction among the  building blocks is translation invariant.  The pattern of the symmetry breaking (below the critical point of the transition) is described by   order parameters  heaving non-vanishing expectation values  in the ordered state, and anomalous dispersion characterized by strong long-range correlations. In  superconductors  the  gauge
symmetry is broken. The effective field theory,  is generally called the Landau theory in  case of the real scalar order parameter and
the Ginzburg--Landau theory in case of the complex order parameter. The latter was  developed to describe superconductivity in metallic superconductors and the Ginzburg-Pitaevskii theory was constructed in application to neutral superfluids, as $^4$He, cf.  \cite{LL5,LP1981,Ginzburg1976,Tilly-Tilly}.  More complex  phases may exist in superfluid $^3$He, cold gases, liquid crystals, solids, anomalous superconductors  and other substances, cf. \cite{Leggett,Nozieres,GP-FL,PitString,Chandrasekhar,Voskresensky:2019zcp}. These phenomenological models  give  universal  description of  various phase transitions.  In discovered new topological insulators, the bulk of two-dimensional samples is insulating, and the electric current is carried only along the edge of the sample  \cite{Klitzing1980}. The flow of this unidirectional current avoids dissipation and gives rise to a quantized Hall effect. These states are possible due to a
combination of spin orbit interactions and time reversal symmetry \cite{Hasan2010}. A somewhat similar possibility of existence of a  superconducting surface pion current in the piece of the dense nuclear matter in the ground state was found in \cite{Voskresensky:1980vt}.

\begin{figure}\centering
\includegraphics[width=8.8cm,clip]{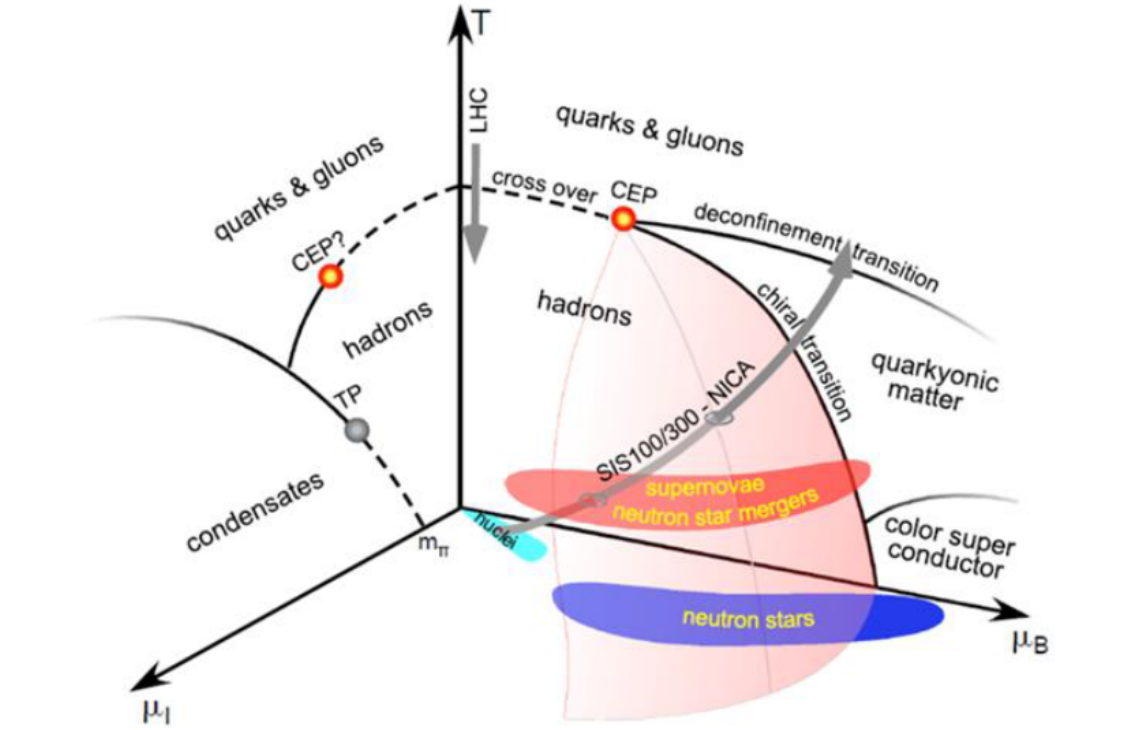}
\caption{Sketch of the phase diagram for nuclear matter taken from \cite{NUPECC}.
 }\label{QCDphaseDiagr}
\end{figure}

Even ordinary water at different conditions may exist in 12 crystalline, 3 glass states, as well as in liquid and  vapor phases.
The   phase diagram of strongly interacting matter also  contains many possible phases with  phase transitions between them \cite{Glendenning2000,Fischer2019,Philipsen2021,Guenther2022,Glozman2022}.
A part  of  various  possibilities is shown in  Fig. \ref{QCDphaseDiagr} from \cite{NUPECC} as the temperature dependence on the baryon-  and isospin- chemical potentials. One bounds the critical endpoint (CEP) of a possible deconfinement phase transition at finite density by $\mu_B \gsim 2.5 T$  and $T \lsim m_\pi$, where $\mu_B$ is the baryon chemical potential.  Within the Schwinger--Dyson approach from the temperature dependent quark propagator one extracts the order parameter for the chiral transition, the quark condensate value, and order parameters for the deconfinement
transition, the dressed Polyakov loop and the dual scalar
quark dressing \cite{Fischer2009,Fischer2011}.

At small baryon-chemical potentials and high temperatures QCD predicts a smooth crossover transition from hadronic matter to the quark-gluon plasma at a pseudo-critical temperature. Fit of the particle production at top RHIC and LHC energies done with the help of the hadron resonance model  \cite{Stachel} yields the value for the pseudo-critical temperature $T_c\simeq (155-160)$ MeV. Chemical nonequilibrium approach \cite{Letessier,Petran} yields a smaller value $T_c$. The value of the critical temperature predicted by the Lattice QCD calculations  \cite{Ding2019} for a chiral phase transition is still smaller, $T_{\rm chir}<135$ MeV.
A primordial hadron-quark particle-antiparticle soup \cite{NUPECC} (or a hadron-quark porridge, as dubbed in   \cite{Voskresensky:2004ux}),   existed  in the early Universe  during microseconds after the Big Bang and is produced in the laboratory at  high energy heavy-ion collision experiments at RHIC and at LHC.
One \cite{Rohrhofer2019,Rohrhofer2020} suggests that chiral quarks are bound by color-electric flux tubes in  a regime, which  has been dubbed a “stringy liquid”. When temperature increases
beyond $T_{\rm s} \sim 3T_{c}$, the color-electric interactions between
quarks get screened and the symmetry reduces to the expected chiral symmetry corresponding to a quark gluon plasma.

At a  hot quasi-equilibrium stage  very specific phenomena of the baryon blurring \cite{Dyugaev:1993mn,Voskresensky:2004ux,Voskresensky:2008ur} and,  might be, a hot Bose condensation  for effectively light bosons  are possible, since the strong coupling effects dominate over thermal disordering, cf. \cite{Voskresensky:2004ux,Voskresensky:2008ur}. For pions the hot Bose condensation may appear either in the s-wave state or  in p-wave state at finite pion momentum, $k_0\neq 0$, owing to a strong p-wave pion-nucleon interaction. In the latter case the condensate matter may form a glass-like state.

On the other hand, as it was shown in \cite{Voskresensky1994} and in a number of subsequent works, cf. \cite{Kolomeitsev1995,Voskresensky:1995tx,KKV1996,
Voskresensky1996,BFR2014,Kolomeitsev2018,Nazarova2019,Blaschke:2020afk,
Kolomeitsev:2019bju}, the Bose-Einstein  condensation  of pions  can occur  at a small baryon chemical potential in a nonequilibrium stage characterized by the dynamically fixed pion number provided an initial nonequilibrium state was overpopulated.
Similarly, the Bose-Einstein  condensation of gluons may arise \cite{Blaizot2012,Xu:2014ega,Blaizot2017,Greiner2019,Peshier2019}.  

Dynamics of the first-order phase transition has many specific features \cite{Skokov:2008zp}.
At a low temperature $T\lsim (15-20)$MeV, in the interval of the nucleon density $0.3n_0\lsim n\lsim 0.7 n_0$, there may occur the first-order gas--liquid phase transition, which signatures were already observed in low-energy heavy-ion collisions, cf. \cite{RMS,RMS1,SVB,Chomaz:2003dz,Margueron:2002wk,
Ducoin2006,Gulminelli2007,INDRA,Maslov:2019dep}.
 References \cite{Kolomeitsev:2016zid,Kolomeitsev:2017foi} suggested a  possibility of the condensation of the scalar quanta,  as the result of the Pomeranchuk instability arising in some region of densities at $n<n_0$ in approximately isospin-symmetric matter. This phenomenon may also result in appearance of a metastable nuclear state at $n\ll n_0$. The clusterization  and  $\alpha$ Bose condensation have been also studied, cf. \cite{Ropke:2012qv,Ropke:2017dur,RVKB2018,Hui} and references there.

Many effective models  predict a first-order phase transition with a critical end-point (CEP), cf.
\cite{Fukushima2011}. Most dense and not too hot matter is expected to be formed in the heavy-ion collisions  at NICA energies, as it is shown in Fig. \ref{QCDphaseDiagr}. The hadron-quark phase transition may have similar signatures, as the liquid--gas one \cite{Skokov:2008zp,Skokov:2009yu,Skokov:2010dd,Steinheimer:2013gla,
Steinheimer:2016bet}.
Passing of the spinodal instability region, the matter forms structures, which effects might be  easier to observe than those associated with passing of the vicinity of the CEP, due to the slowing down effects in the latter case. Spinodal instabilities are also relevant for consideration of the pasta phase in neutron stars, cf. \cite{Fang2016,Fang2017}.

Many other  phases of QCD and hadron matter have
been proposed, such as quarkyonic matter, which can be considered as a Fermi gas of quarks with confined thermal excitations   \cite{McLerran2007},  and a parity doubled baryon matter as a candidate for a chiral spin-symmetric regime of cold and dense QCD,
which can be naturally embedded into the quarkyonic matter. The existence of a first-order phase transition ending in a critical point, as indicated in Fig.  \ref{QCDphaseDiagr}, is
still under debate. For example, following the concept of the quark-hadron continuity, it is conjectured  \cite{Baym2018} that quark degrees of freedom may emerge gradually with increasing density and a partial restoration of the chiral symmetry.
%%%

The theory of normal Fermi liquids was built up by
Landau and Migdal in Ref.~\cite{FL1956,FL1956a,Mjump,Mjump1,FL1956b}, see in textbooks~\cite{LP1981,Nozieres,GP-FL}.
The Fermi liquid approach to the description of nuclear systems
was developed by Migdal in ~\cite{Mjump2,M67}, see also \cite{M67a}. In the Fermi liquid
theory the low-lying excitations are treated explicitly whereas the short-range correlations are described with the help of  several
phenomenological Landau parameters. Pomeranchuk has shown in
Ref.~\cite{Pomeranchuk} that Fermi liquids are stable only, if certain
inequalities on the values of the Landau parameters are fulfilled.
%%%
The Fermi-liquid
approach with the explicit separation of the in-medium pion exchange was formulated  by A. B. Migdal for description of the cold nuclear matter, cf.  \cite{Migdal78}, and then it was  generalized for equilibrium systems at finite temperatures, cf. ~\cite{MSTV90,Voskresensky:1982vd,Dyugaev:1982gf}, and  for nonequilibrium systems, cf.  ~\cite{Voskresensky:1987hm,Voskresensky:1988pxw,Voskresensky:1989sn,
MSTV90,Voskresensky:1993ud,Kolomeitsev2011}. It was shown that in nuclear matter soft modes with pion quantum numbers and the momenta $k\neq 0$  become efficiently occupied by pion-like excitations at non-zero temperature forming so called liquid phase of the pion condensate for $n> n_{c}^{(1)} \sim (0.5-0.8)n_0$, cf. \cite{Dyugaev:1982gf,Voskresensky:1993ud}.  With increase of the  baryon density these modes may become unstable to formation of the liquid-crystal-like or solid-like pion condensate. This pion condensation in a warm and dense nuclear matter predicted to occur for $n> n_c^\pi \sim (2-3)n_0$ is the example of the phase transition to the inhomogeneous state $k_0\neq 0$, cf. \cite{Migdal78,Ericson:1988gk,MSTV90}.  In some models the pion condensation in the isospin symmetric matter may arise also in the s-wave state \cite{Voskresensky2022}, whereas in neutron star matter $\pi^{\pm,0}$ may condense only in states with $k_0\neq 0$, cf. \cite{Migdal78,MSTV90}, due to the presence of the repulsive Weinberg-Tomazawa pion-nucleon interaction term.

 Many different structures can be formed in the process of formation of the inhomogeneous condensate state $k_0\neq 0$, cf. \cite{Voskresensky:1984rd,MSTV90,Voskresensky:1993ux,Muto:1992gb,Dautry1979}.  The most energetically profitable proved to be   the alternating-layer structure \cite{Takatsuka:1978ku,Takatsuka:1993pv}, in the dynamics a polycrystal is  probably formed \cite{Voskresensky:1993ux}.

 Antikaon $K^-$ and $\bar{K}^0$ condensations may arise  as in the s-wave state, cf.  \cite{Glendenning2001}, as in the p-wave one  \cite{Kolomeitsev:1995xz,Kolomeitsev:2002pg,Kolomeitsev:1996bh}.
 If the effective $\rho$ meson mass decreases with increasing density,
 the s-wave non-abelian $\rho^-$ meson condensation in interiors of the neutron stars may occur \cite{Voskresensky:1997ub,KolVosk2005,Kolomeitsev:2017gli}.

Being studied in the mean-field approximation the critical temperature of the pion condensation phase  transition to the inhomogeneous state $k_0\neq 0$ is rather high (typically $\gsim \epsilon_{{\rm F},N}$, where $\epsilon_{{\rm F},N}$ is the nucleon Fermi energy), cf. \cite{Voskresensky:1978cb,Baym:1980jx}. Even in case of the second-order phase transition at $k_0=0$ fluctuations may significantly  modify the phase diagram of the strongly interacting matter, cf. \cite{Ginzburg1976,Larkin2005}. In the mean-field theory, three phases meet at what  is known as a Lifshitz point. One of the phases that meet at the tricritical point is an inhomogeneous phase. In the context of the Cooper pairing phases in dilute isospin-asymmetric nucleon matter discussion of the Lifshitz three-critical point can be found in \cite{Stein}. Fluctuations smear  the Lifshitz point \cite{PST2019}.  Strong color superconducting fluctuations in dense nuclear matter are expected to occur even significantly above CEP, cf. \cite{Voskresensky2003arx,Voskresensky:2004jp}. In heavy-ion collisions they might be manifested at $T\lsim (1.5-2)T_c$,  provided the critical temperature of color superconductivity, $T_c$, is sufficiently large (e.g., if $T_c\gsim (50-70)$MeV). Reference \cite{Nishimura:2022kqa} supported these estimates of \cite{Voskresensky2003arx,Voskresensky:2004jp}.
Di-lepton and photon yields in heavy-ion collisions can be affected by pre-critical fluctuations \cite{Nishimura:2022kqa,Kerbikov}.
Note here also the geometrical effect on  fluctuations near the critical point, cf. \cite{Voskresensky2003arx,Voskresensky:2004jp}. Effectively one deals with the 3d geometry,  if the coherence  length $l\sim 1/\sqrt{|T-T_{c}|}$ (esimated here in the mean-field approximation) fulfills inequalities $l<R_{\parallel}, l<R_{\perp}$; one deals with    2d geometry, if $R_{\perp}>l>R_{\parallel}$; and
with the 1d geometry,  if $R_{\parallel}>l>R_{\perp}$, where $R_{\parallel}$ and $R_{\perp}$ are typical sizes of the system in parallel and perpendicular directions. For $l>R_{\parallel}, l>R_{\perp}$ the geometry is so called 0-dimensional. These effects  reflect , e.g., in the temperature dependence of the specific heat (associated with the variance of the energy), $C_V\sim 1/\sqrt{|T-T_{c}|}$ for 3d and
$C_V\sim 1/(T-T_{c})^2$ for d=0. So, in principle, such a geometrical effects can be manifested in not-central heavy-ion collisions.

Note here that even if in the mean-field consideration  the  phase transition to the state $k_0=0$ is of the second-order, inclusion of electromagnetic fluctuations  results in a weakly first-order phase transition in superconductors, cf. \cite{Halperin1974}. Fluctuations near the critical point of the phase transition to the state $k_0\neq 0$ prove to be so strong that they necessarily lead to the change of the order of the phase transition from the second-order to the first-order, cf. \cite{Brazovskii1975,Dyugaev:1975dk,Voskresensky:1981zd,Voskresensky:1982vd,Dyugaev:1982gf,
Dyugaev:1982ZHETF,Kolehmainen:1982jn,Dyg1,Schulz:1984cb,Voskresensky:1984rd,Voskresensky:1989sn,MSTV90,
Voskresensky:1993ud,Tatsumi2016,Pisarski2021}.

The matter  with a large baryon chemical potential is reached in compact stars such as neutron stars and hypothetical hybrid and strange stars, cf.  \cite{Glendenning2001,Blaschke:2004cc,Ivanov:2005be,Okamoto:2013tja,
Ayriyan:2017nby,Maslov:2018ghi,Thi2022} and references therein.
Thereby, figure \ref{QCDphaseDiagr} indicates also the regions of cosmic nuclear matter realized in neutron stars, neutron
star mergers, and hypothetical hybrid stars and strange stars, and nuclearites  and smaller size strange quark nuggets, located at large baryon chemical potentials, low temperatures, and finite isospin chemical potentials. The idea of nuclearites and strange stars \cite{Witten:1984rs,DeRujula:1984axn}  was preceded by the idea of pion condensate superheavy nuclei \cite{Migdal:1971cu,Migdal:1974yx} and nuclei--stars \cite{Voskresensky:1977mz,MSTV90} glued by the inhomogeneous $k_0\neq 0$ charged pion condensate and electrons. Recent paper \cite{Shahrbaf} studied  consequences for neutron star  phenomenology that
would follow from the existence of a possible stable sexaquark ($S$) state with the quark content $uuddss$. The authors showed that the hypothesis does not contradict to the   current astrophysical  constraints.
If in the nature existed very heavy (with the mass $m_{\rm DM}\gg m_N$, where $m_N$ is the nucleon mass)  stable charged dark matter (DM) particles, there would also exist nuclearites and nuclei--stars consisting of ordinary approximately isospin-symmetric nuclear matter at $n\simeq n_0$ stabilized by heavy charged dark matter particles \cite{Gani:2018mey}.

Nucleons in nuclear matter  form Cooper pairs but with  small gaps, typically  $T\lsim $ several MeV, cf. \cite{Sedrakian2019,Kolomeitsev2011,Voskresensky:2019zcp}. Paired neutrons form neutron superfluid phases, paired protons form superconducting phases. In presence of not a strong deviation of proton and neutron Fermi seas the nucleons may form $np$  pairs  and the gap may form a one-dimensional  wave (Fulde--Ferell--Larkin--Ovchinnikov state \cite{Fulde1964,Larkin1965,Sedrakian2009}).

At densities above $(2-3)n_0$ Fermi seas of the hyperons \cite{Maslov:2015msa} and $\Delta$ isobars \cite{Drago:2014oja,Kolomeitsev:2016ptu,Raduta2020,Haratyunyan,Marquez2022} can be occupied. It typically occurs by the third-order phase transition (in the sense of Ehrenfest at the third-order phase transition a thermodynamic potential  turns out to have a discontinuity in the third derivative in $n$ or $T$ in critical point, whereas first and second derivatives are continues.

Hybrid stars and strange stars should be color superconductors at $T<T_c$,  cf. \cite{Barrois1977,BalinLove1984}. Value of the pairing gap and the critical temperature depend on the phase under consideration. Many phases of the color superconducting matter were suggested, cf. \cite{Rajagopal2001,ARRW,Alford2001,Alford2008}.   Estimates for the gaps vary from $T\sim 1$MeV for spin-color-locking phase till  hundred MeV or more for color-flavor-locked phase. Description of color superconductors by the generalized  Ginzburg--Landau   model was suggested in \cite{Iida:2000ha} and it was applied to description of fluctuations in \cite{Kitazawa2002,Voskresensky2003arx,Voskresensky:2004jp,Kitazawa2005,
Nishimura:2022kqa,Kerbikov}. An inhomogeneous diquark condensation may  develop  when the gap energy
is comparable with $m_s^2/\mu_q$, where $m_s\simeq (90-150)$ MeV is the strange quark mass and $\mu_q =\mu_B/3$ is the quark chemical potential. Such an inhomogeneous color superconducting  state   gives rise to a crystal structure, cf. \cite{Alford2001,Sedrakian2009,Anglani}.

 The phase structure at high  baryon densities and moderate temperatures can be  considerably 
modified by  the presence of the inhomogeneous  QCD phases. Pion and kaon condensates may appear not only in the hadron matter but also in the quark matter, cf. \cite{Mannarelli}.  Basing on the consideration of the $\sigma-\pi^0$ condensation \cite{Dautry1979} in nucleon matter characterized by the chiral wave with $k_0\neq 0$, many models were suggested, cf. \cite{Nakano:2004cd,Carignano2014,Carignano2015,TGLee2015,Tatsumi2016,Gronli:2022cri}. Reference \cite{Canfora:2020uwf} derived inhomogeneous configurations of pion fields characterized by a
non-vanishing topological charge that can be identified with baryons. Quark and pion condensates at finite isospin density studied in chiral perturbation theory may result in existence of pion stars \cite{Adhikari:2020ufo}.

Another interesting feature of the first-order phase transition in isospin-asymmetric matter characterized by  more than one conserved charge is the possibility of the existence of so-called pasta phases  \cite{G92}.  In case of the  compact stars the  conserving charges are the baryon and electric charges.  The interplay   between Coulomb and surface tension leads to the possibility of the appearance  of structures of different geometry as droplets, rods and slabs and configurations of a more whimsical form, cf. \cite{Ravenhall1983,Lorenz1993,Watanabe2000}. Important role is played by the Debye screening effects, cf. \cite{Voskresensky:2001jq,VYT2003}. Only with taking into account of finite size effects one is able to construct a correct description of the pasta phases \cite{Voskresensky:2001jq,VYT2003,Maruyama:2005vb,MTVTEC,Maruyama2008,Okamoto:2013tja,
Ayriyan:2017nby,Maslov:2018ghi}. Nuclear pasta may exist in the inner crusts of neutron stars at $n\lsim 0.7 n_0$, cf. \cite{Ravenhall1983,Lorenz1993,Maruyama:2005vb,Thi2022}. In  dense interiors there may  exist regions of the hadron-quark \cite{Voskresensky:2001jq,VYT2003,Maruyama2008} and pion and kaon \cite{MTVTEC} pasta phases. In heavy-ion collisions the conserving quantities are the baryon number and isospin. Specifics of the first-order phase transitions in collisions of  asymmetric nuclei has been discussed in  \cite{Chomaz:2003dz,Margueron:2002wk,
Ducoin2006,Gulminelli2007,INDRA,Maslov:2019dep}. It
would be interesting to seek  possible manifestations  of
 the formation of the pasta structures in  heavy-ion
reactions \cite{Maslov:2019dep}.

Another part of problems is associated with the behavior of the quark-gluon and hadron media in external fields.  First estimates \cite{Voskresensky:1980nk} of
 the value of the strength of the magnetic field reached  in peripheral heavy-ion collisions at energies $\lsim $ few GeV$/ A$    yielded  $H\sim (10^{17}-10^{18})$ G, that was later supported by  numerical calculations  \cite{Skokov:2009qp}. Presence of strong magnetic fields and the rotation (with frequencies $\omega \lsim 10^{22}$Hz in case of  heavy-ion collisions)  influences the superfluids. Strong magnetic fields $\gsim 10^{15}$G exist  at the surface of magnetars. Fields $H\sim 10^{17}$G  influence the pasta phase in the inner core, cf. \cite{Fang2016,Fang2017}.  In interiors of neutron stars the magnetic field may reach even  higher values, $H\sim 10^{18}$G, cf. \cite{Voskresensky:1980nk,MSTV90}. Superconductors and superfluids specifically react on the magnetic field and the rotation, respectively. Abrikosov mixed phases  can be formed. Formes of the vortices in case of inhomogeneous ($k_0\neq 0$) and homogeneous    ($k_0=0$) condensates prove to be different, cf. \cite{Voskresensky:1980nk,MSTV90}.

 Most of the problems touched above were studied for the equilibrium systems. Models describing inhomogeneous condensates are much less developed than those considering homogeneous condensates. Dynamics  of the formation of various phases is still  less studied.
 To partially cover this gap,  the present review discusses besides effects of the condensation at $k_0 = 0$, the condensation at $k_0\neq 0$ at equilibrium and nonequilibrium conditions.

 The paper is organized as follows. Section \ref{General-sect} discusses a general phenomenological model for description of phase transitions.
 First we introduce a phenomenological description of dissipative dynamics of phase transitions at $k_0=0$. Then  Green function
 of the Bose excitations in matter is introduced and its expansion in frequencies and momenta is performed. Next, the spectrum  of over-condensate excitations is found and its peculiarities in case of the running wave condensate with $k_0\neq 0$  are discussed. Then various static
 configurations of the  condensate field at $k_0\neq 0$, one-axis and three-axis structures, chiral waves etc,  are  described. The specific role  of fluctuations
 at the phase transition to the $k_0\neq 0$ state is discussed. Then transition from the equation for the order parameter to hydrodynamical variables is performed as in case of the phase transition to the state $k_0=0$ as for $k_0\neq 0$. In Section \ref{First-order-transitions} the phase transitions in slowly evolving  systems
are studied. The congruent first-order phase transition of the gas--liquid type and   the non-congruent transition  to the pasta phase are considered. The dynamics of the first-order phase transition to the state $k_0=0$ is described.  It depends on whether the initial seed of the stable phase is formed in    the metastable region or within the spinodal region.
Besides spherical droplets, dynamics of strongly non-spherical configurations is considered.
Then focus is made on the description of dynamics of configurations at the first-order phase transition to the state $k_0\neq 0$. Dynamical descriptions  of the transitions between  inhomogeneous phases and between homogeneous and inhomogeneous phases are performed. Formation of  polycrystal and  glass phases are discussed.
In Section \ref{hydro} the hydrodynamical description  of the first-order and the second-order phase transitions of the liquid-gas type in terms of the local density-temperature variables is performed at assumption of a small over-criticality. From non-ideal hydrodynamical equations, the equation for the order parameter is derived. The initial (inertial) stage of the evolution of  seeds  differs from that described by the ordinary Ginzburg--Landau theory whereas  the long-time tail is described by the Ginzburg--Landau equation. The dynamics of the growth of seeds formed in  metastable phase is studied. At a low viscosity quasi-periodic  oscillations occur. Then, the dynamics of fluctuations in the spinodal region is considered and again during time evolution, quasi-periodic configurations appear.
In Section \ref{Pion-section} we focus on the explicit example  of the pion condensate phase transition to the state $k_0\neq 0$ in  cold and warm nuclear matter. The liquid/amorphous phase of the condensate appears already for $n>n_{c}^{(1)}\sim (0.5-0.7)n_0$ and the solid-like phase of the pion condensate may arise for $n>n_{c}^\pi\sim (1.5-3)n_0$.
 Peculiarities of  the dynamical description of the pion condensate phase transition are reviewed. Section \ref{Blurring} studies issues of the fermion blurring and hot Bose condensation  at conditions of a small or zero baryon chemical potential. For pions obeying a strong p-wave pion-baryon interaction, the hot Bose condensation
may occur either in the state with $k_0= 0$ or in the state $k_0\neq 0$.
Section \ref{ScalarSection}  describes non-pionic excitations in a Fermi liquid
and a possibility of the condensation of the scalar quanta in the region of the low densities, $n<n_0$, where there arises the Pomeranchuk instability.
Section \ref{MovingSection} discusses peculiarities of the condensation of Bose excitations in the state $k_0\neq 0$ in the uniformly moving  media.
Section \ref{NonequilibriumSection} demonstrates the role  of quantum and finite size effects in nonequilibrium distributions.  First,  the Bose--Einstein condensation of bosons in nonequilibrium systems is considered at the stage when inelastic processes are not efficient. The possibility of the Bose--Einstein condensation of pions in ultratelativistic heavy-ion collisions is discussed. Then,
 a toy   model of the breaking up of a box filled by nucleons is considered. Nucleon distributions prove to be oscillating and the  pion distributions follow a power law. Main results are discussed in the Conclusion.

\section{Phenomenological description of phase transitions and structure formation }\label{General-sect}

\subsection{Phenomenological description of dissipative dynamics at phase transitions}
Dissipative dynamics of the phase transitions is usually described by the phenomenological Ginzburg--Landau equation, cf. \cite{LL5,Onuki,Patashinsky},
\be
\partial_t \psi =-\hat{\Gamma} (\hat{\Delta})\frac{\delta \delta F}{\delta \psi}+\xi (t,\vec{r}),\quad
\hat{\Delta}=\partial_i^2\,, \quad i =1,2,3\,,\label{Leqorderparam}
 \ee
 where $\delta F$ is  the  part of the free-energy density dependent on    the order parameter $\psi$. The order parameters can be of the scalar, vector, tensor origin in dependence of the system under consideration. For description of the quantum liquids one uses the macroscopic wave function as the complex order parameter, cf. \cite{LP1981,Tilly-Tilly,Leggett}.
 In case of smooth distributions the relaxation rate $\hat{\Gamma} =\Gamma_0+\Gamma_1\hat{\Delta}+...$ is    expanded in the series of the operator $\hat{\Delta}$, $\Gamma_i$ are coefficients of the expansion. For $\Gamma_0 >0$, $\Gamma_1 =0$ one deals with long-wavelength behavior of the non-conserving order parameter $\psi$.
For $\Gamma_0 =0$, $\Gamma_1 >0$, the long-wavelength behavior of the  conserved order parameter  is well described by
the stochastic diffusion equation. The meaning of Eq. (\ref{Leqorderparam}) is that in a weakly nonequilibrium configuration  a small rate of the relaxation of the system to the equilibrium is proportional to the thermodynamic force describing deviation from the equilibrium. In the linear response theory the kinetic coefficients $\Gamma_i$ are related to correlators of the ﬂuctuations of thermodynamic quantities in the hydrodynamical limit, cf.  \cite{Kovtun2012,Jeon2015}. The stochastic noise term $\xi (t,\vec{r})$  is assumed to be Gaussian and white. Its width
is determined by the fluctuation-dissipation relation
\be
\langle\xi (t,\vec{r})\xi (t',\vec{r}')\rangle =a \delta (t-t')\delta(\vec{r}-\vec{r}\,')\,,
 \ee
%$a=2T\Gamma$,
where  the value $a$,   dependent on the diffusion coefficient and the
temperature $T$, cf. \cite{LP1981}, as well as  coefficients $\Gamma_i$, are  supposed  to be  smoother  $t,\vec{r}$ variables compared to the order parameter $\psi (t,\vec{r})$.
Recent applications of diffusive dynamics of critical fluctuations to heavy ion collisions see, e.g., in \cite{Bluhm2019,Bluhm2020}.
Within the mean field consideration one may put the $\xi$ term zero.

\subsubsection{The phase transition to the state $k_0=0$. Mean field approximation}\label{MFhomSect}

In case of the phase transition to the state $k_0=0$, the $\psi$ dependent part of the free energy is usually also expanded in the order parameter and the operator $\hat{\Delta}$  up to a linear term. Expansion in $\psi$ is performed at least up to $\psi^4$ term. For consideration of the phase transitions in non-uniform state $k_0\neq 0$ expansion is done at least up to a quadratic term.
In the simplest case for description of quasi-uniform quasi-equilibrium configurations
one chooses the order-parameter dependent part of the free-energy density in the form \cite{LL5}
\be
\delta {\cal F} =\frac{1}{2} (\nabla \psi)^2+ \frac{1}{2} m^2 \psi^2+ \frac{1}{4} \Lambda \psi^4- h\psi\,,\label{freeenorderpar}
\ee
with temperature and density dependent coefficients $ m^2$,  $h$ and $\Lambda$. In the mean-field approximation the coefficients are expanded in the Taylor series near the critical point, e.g., in $T-T_c$. To describe the second-order phase transition in the simplest case one chooses  $m^2=\alpha_0 (T-T_c)$, i.e.  changing the sign in the critical point of the second-order phase transition, $\alpha_0 >0$ to get nontrivial solution for $\psi$ at $T<T_c$. The value  $\Lambda$ is chosen to be positive for the stability of the resulting state, $h=0$ in case of the second-order phase transition and $h\neq 0$ in case of the first-order phase transition.

To describe superfluidity and superconductivity, when the flux/current can be  nonzero, one should use the complex order parameter such as  the macroscopic wave function. In these cases one deals with the second-order phase transitions and one uses \cite{LP1981},
\be
\delta {\cal F} =\frac{1}{2} |\nabla \psi|^2+ \frac{1}{2} m^2 |\psi|^2+ \frac{1}{4} \Lambda |\psi|^4\,.\label{freeenorderpar1}
\ee

\subsubsection{Fluctuations of the order parameter, case $k_0=0$}\label{FlhomSect}
Taking into account fluctuations of the order parameter in the vicinity of  the critical point of the second-order phase transition results in a non-analyticity  of the thermodynamical potentials and in divergences  in the critical point of their second derivatives, such as the specific heat and magnetic susceptibility. Therefore Taylor expansion of the thermodynamical potential in $T-T_c$,  being employed  within the mean-field approximation, fails in the  fluctuation region near the critical point, cf. \cite{Ginzburg1976}.

In classical systems and also at not too small temperatures in quantum
systems, quantum fluctuations are suppressed compared to thermal fluctuations.   Fluctuation theory of phase transitions is a well developed field, cf. \cite{Larkin2005}. The width of the fluctuation region near the critical point is estimated from the so called Ginzburg criterion: the probability of the fluctuation $W\sim e^{-\delta F(V_{\rm fl})/T}$ becomes $\sim 1$ for $\delta F(V_{\rm fl})\sim T_{\rm fl}$
where $\delta F(V_{\rm fl})$ is the work necessary to prepare the fluctuation in the minimal volume  $V_{\rm fl}\sim l_0^3$ for fluctuation, where $l_0$ is the typical length of the change of the order parameter. In case of the phase transition to the state $k_0=0$ one may estimate $l_0\sim 1/|m|$, where $m$ is the coefficient in (\ref{freeenorderpar}), (\ref{freeenorderpar1}). As it was mentioned, dealing with the second-order phase transition  in the mean field approximation one usually takes $m^2=\alpha_0 (T-T_{c})$ and $\alpha_0, \Lambda$ are positive constants. Then
the coherence  length behaves as $l_0\sim 1/\sqrt{|T-T_{c}|}$ and one easily estimates the value  $T_{\rm fl}$.

Cooper pairing fluctuations
are associated with the formation and breaking of excitations of
Cooper pairs out of the condensate.
In case of clean metallic superconductors  the region $|T_c-T_{\rm fl}|$ proves to be very narrow
and thereby the mean field approximation works very well. In case of the superfluid with a strong interaction between particles, as $^4$He, fluctuations prove to be strong up to $T=0$.
The width of the energy region near the critical temperature, where fluctuations essentially contribute, is characterized by the Ginzburg number,  $Gi= |T_{\rm fl}-T_c| /T_c$. $Gi$ proves to be tiny for clean  conventional  superconductors but might be $\sim 1$ for strongly interacting systems.

To take into account fluctuations more carefully one either uses the methods of scale invariance and $\epsilon$-expansion or, as in case of the description of superfluid $^4$He,  in (\ref{freeenorderpar1}) one presents coefficients $m^2 =\alpha_0 (T-T_c)|T-T_c|^{\alpha -1}$ and $\Lambda =\beta_0 |T-T_c|^{\beta}$ with algebraic values $\alpha$ and ${\beta}$, cf. \cite{Ginzburg1976}.
In the latter case one may find the coefficients $\alpha$ and $\beta$ from the condition that  the Ginzburg number should not depend on the quantity $T_{\rm fl}-T_c$ and from the experimental fact that the specific heat in this case may diverge not stronger than logarithmically, cf. \cite{V88}. Thus one finds $\alpha =4/3$ and $\beta =2/3$, these values are in agreement with  the data, see \cite{Ginzburg1976}.

 In interiors of most massive compact stars a color superconducting matter  may exist.
Temperatures are not as high and mean-field consideration is relevant for many purposes.
To produce a color superconducting matter in the laboratory
one needs to cook a dense but not too hot Cooper-paired quark matter.   For that   most relevant are probably the NICA and FAIR facilities.
Although the temperatures at relevant densities permitted  in heavy-ion collisions are most likely larger than the critical temperature $T_c$ of the color superconductivity, one may rise the question about a possible manifestation of the precursor phenomena of the color superconducting phase transition, if the value $T_c$ is rather high (for $T_c \gsim (50 - 70)$ MeV). Reference \cite{Kitazawa2002} considered such a possibility within the Nambu--Jona-Lasinio model and estimated $|T_{\rm fl}-T_c|\lsim (0.1-0.2)T_c$.
References \cite{Voskresensky2003arx,Voskresensky:2004jp} employed estimate \cite{Larkin2005} $Gi\sim  A(T_c/\mu_q)^4$, with $A \sim 500$ for the case of color superconductors. To compare, for clean metals \cite{Larkin2005} one has  $A \sim 100, \mu_q \to \mu_e$, where $\mu_q$ and $\mu_e$ are  the quark and electron chemical potentials. Thus in case under consideration  $Gi\sim 1$, if $T_c$ is rather high,
$T_c \sim ( 1/3 - 1/5 )\mu_q$, and we expect a broad region of temperatures, where fluctuation effects might be important. Thus Refs. \cite{Voskresensky2003arx,Voskresensky:2004jp}
 estimated $T_{\rm fl,<} \lsim 0.5T_c$ for $T<T_c$ and $T_{\rm fl,>} \sim 2T_c$ for $T>T_c$ and Ref. \cite{Nishimura:2022kqa} also estimated $|T_{\rm fl}-T_c|\sim 0.5T_c$. These estimates are rather optimistic to seek at least some signatures of the
color superconducting fluctuations of the di-quark gap in the course of the heavy ion collisions.
The color superconducting  fluctuations might be also relevant for an initial stage
of the hybrid star evolution, if $T_c$ is $\gsim 50$ MeV. They may affect the neutrino radiation of the most massive hot hybrid stars  and the heat transport at an
initial stage of their evolution.

 \subsubsection{Effective Lagrangian, equation of motion, phase transition to the state $k_0\neq 0$.}

Let us first consider a complex classical condensate field.
In case of the pair  interaction including the retardation effects, the equation of motion for the order parameter describing a second-order phase transition, like that occurring in $^4$He, can be presented as follows \cite{Voskresensky:1993ux}:
\be
\hat{D}^{-1}(\partial_t,\nabla)\psi (t,\vec{r})-\psi(t,\vec{r}) \int |\psi (t+t',\vec{r}+\vec{r}^{\,\prime})|^2 U(t', \vec{r}^{\,\prime})dt' d\vec{r}^{\,\prime}=0
%-h(\psi(t,\vec{r})/\psi^*(t,\vec{r}))^{1/2}
\,,\label{efpot}
 \ee
 where $D(\omega,k)$ is the  boson Green function related to the $\psi$ field in the medium, $\hat{\omega}= i\partial_t$, $\hat{\vec{k}}= -i\nabla$,
 cf. the Gross--Pitaevskii equation that follows for the static local potential $U$, cf. \cite{LP1981,PitString}.

Expanding $\psi (t+t',\vec{r}+\vec{r}^{\,\prime})$ in the series in $t',\vec{r}^{\,\prime}$ one presents Eq. (\ref{efpot})  as
  \be
\hat{D}^{-1}(\partial_t,\nabla)\psi (t,\vec{r})-\psi(t,\vec{r})\hat{\Lambda}|\psi (t,\vec{r})|^2=0\,, \label{genGGP}
 \ee
 with $\Lambda =\Lambda_0 +\Lambda_{12} \partial_t +\Lambda_{21} \Delta +...,$ where
 $\Lambda_0>0$ for stability of the ground state and $\Lambda_{12}$, $\Lambda_{21}$... are some  complex constants. The  equation describing weakly non-ideal Bose gas placed  in an external static field \cite{LP1981} follows from Eq. (\ref{genGGP}) as
 \be
 i\partial_t \psi =-\left(\frac{1}{2m}\Delta +\mu\right)\psi +\psi \int |\psi (t,\vec{r}^{\,\prime})|^2 U(\vec{r}-\vec{r}^{\,\prime})d \vec{r}^{\,\prime}\,.
  \ee
For $\mu <0$ and $U>0$  it describes the phase transition to the homogeneous condensate state, $k=0$, since the kinetic $k^2$ term enters the free energy with the positive sign.

Pions undergo a strong pion-nucleon attraction in the p-wave that may cause pion condensation in sufficiently dense nucleon matter \cite{Migdal78,MSTV90}. To describe the pion condensate phase transition to the state $k\neq 0$ in a dense ($n>n_c$)  equilibrium nuclear matter one may employ the following Fourier representation of the Fourier transform of the
Landau free energy, cf. \cite{Dyugaev:1982ZHETF},
\be
\delta F [\phi]=\sum_{\vec{k}} [\omega_0^2 -\alpha_{4} (\vec{k}^{\,2}-k_0^2)^2]\vec{\phi}_{\vec{k}}
\vec{\phi}_{-\vec{k}}+\Lambda_1 (\sum_{\vec{k}}\vec{\phi}_{\vec{k}}
\vec{\phi}_{-\vec{k}})^2 +\sum_{\vec{k}_i}\Lambda_2 (\vec{k}_i)\vec{\phi}_{\vec{k}_1}
\vec{\phi}_{\vec{k}_2}\vec{\phi}_{\vec{k}_3}\vec{\phi}_{\vec{k}_4}
 \ee
 with positive $\Lambda_1$ and $\Lambda_2$, $\vec{\phi}=(\phi_1,\phi_2,\phi_3)$ is the isospin-vector pion field.
Differences between various crystalline and liquid structures are mainly determined by the term $\propto \Lambda_2$, which is usually rather small, $\Lambda_2\ll \Lambda_1$. The phase transition occurs for  $\omega_0^2 =-D^{-1}(\omega_c,k_0)<0$, $\omega_c$ is the critical frequency and $k_0\neq 0$ is the value of the momentum of the condensate corresponding to the minimum of $-D^{-1}(\omega_c,k)$, $\vec{\phi}=(\phi_1,\phi_2,\phi_3)$ is the isospin-vector pion field.

Further to be specific let us consider the case of the two-component (complex) classical field interacting with the  medium consisting of heavy fermions. The part of the effective Lagrangian density depending on the condensate field, the fermion density $n$ and the temperature $T$, can be expanded in the condensate field and time-space gradients as
\begin{eqnarray}\label{Omoper}
&\langle {\cal L}\rangle =\left[\psi^* \Re \hat{D}^{-1}(\hat{\omega} ,\hat{\vec{k}},n, T)\psi -\frac{1}{2}|\psi|^2\Re\hat{\Lambda} (\hat{\omega} ,\hat{\vec{k}},n, T) |\psi|^2 \right]
\\
&+h\psi\left(\frac{\psi^* e^{2im\pi}}{\psi}\right)^{1/2}+h\psi^*\left(\frac{\psi e^{2im\pi}}{\psi^*}\right)^{1/2}\,.\nonumber
\end{eqnarray}
Here and further we use  notations $\Re Z=\mbox{Re}Z$, $\Im Z=\mbox{Im}Z$. Averaging is assumed to be done over all degrees of freedom besides the given boson field described by the complex $\psi$ variable.

The last two terms in (\ref{Omoper}), with $h>0$, permit to describe first order phase transitions from metastable to stable  homogeneous, $k_0=0$,  or inhomogeneous, $k_0\neq 0$, states.
We have chosen a simplified form of the $|\psi|^4$ term.
The multiplier $e^{2im\pi}=1$ is introduced to recover two signs of the square root for $m=0,1$; and $-1$ is treated as $e^{i\pi}$, so $\psi\sqrt{\frac{\psi^* e^{2im\pi}}{\psi}}$ for $\psi =-1$ yields $-\sqrt{e^{-2i\pi}e^{2im\pi}}=\pm 1$ for $m=0$ and $m=1$ respectively, i.e.,    the same as for $\psi =1$. As we shall see, the continues solutions will correspond to $m=0$ within the volume of the stable phase, for  $\vec{r}\in V_{<}$, and to $m=1$ within the volume of the metastable  phase, for  $\vec{r}\notin V_{<}$.
For the real field $\psi$, e.g., for $\pi^0$ or $\sigma$ meson fields,
\be
\langle {\cal L}\rangle =\left[\psi \Re \hat{D}^{-1}(\hat{\omega} ,\hat{\vec{k}},n, T)\psi -\frac{1}{2}\psi^2\Re\hat{\Lambda} (\hat{\omega} ,\hat{\vec{k}},n, T) \psi^2 \right]+2h\psi\,.\label{Omoperreal}
\ee
To get this expression from $(\ref{Omoper})$ for positive $\psi$ describing the stable phase we should use $m=0$ and for negative $\psi$ describing the metastable phase, $m=-1$.

Equation of motion for the condensate field related to $(\ref{Omoper})$ can be presented as \cite{Migdal78}:
\be
 \hat{D}^{-1}(\hat{\omega} ,\hat{\vec{k}},n, T)\psi -\psi\hat{\Lambda} (\hat{\omega} ,\hat{\vec{k}},n, T) |\psi|^2  +he^{im\pi}(\psi/\psi^*)^{1/2}=0\,.\label{eqMotPhi}
\ee
Here
\be
D^{-1} (\omega, \vec{k}) = \omega^2 -m^2 -\vec{k}^{\,2}-\Sigma^R (\omega, \vec{k}, n,T)\,,\label{eqMotPhi1}
\ee
$\Sigma^R (\omega, \vec{k}, n,T)$ is the boson retarded polarization operator. In case of  the real field $\psi$ at $k_0=0$, the last term simply yields $h$. In example of the pions interacting with the baryons $n$ is the baryon density, $\langle {\cal L}\rangle$ is effective pion Lagrangian density, $\Sigma^R (\omega, \vec{k}, n,T)$ is the retarded  polarization operator of particles heaving the pion quantum numbers, cf. \cite{Migdal78,MSTV90}.

In case of the infinite matter for $n>n_c$, for $\hat{\Lambda}\simeq \Lambda_0$, for  $\omega_0^2 =-D^{-1}(\omega_c,k_0)<0$, equation of motion (\ref{eqMotPhi}) has the running-wave classical solution
\be
\psi =a\,\mbox{exp}[-i\omega_c t+i\vec{k}_0 \vec{r}]\,.
 \ee
 Here $\omega_c$ and $\vec{k}_0$ are the condensate frequency and momentum, $a$ is the amplitude of the condensate field.

 Local charge neutrality in the infinite system suggests that
 $$n_p +e\left(\partial \delta {\cal F}/\partial\omega\right)_{\omega_c}  =0\,,$$ where $n_p$ is the non-condensate contribution to the charged density, e.g., the
 positive charge density of the baryon sub-system plus negative charge density of the lepton sub-system in case of the neutron star. This condition determines the critical frequency of the condensate $\omega_c$. The condition of the absence of the current in the equilibrium state,
 $$\left(\partial\partial \delta {\cal F}/\partial \vec{k}\right)_{\vec{k}_0}=0\,,$$
 determines the value of the condensate momentum $k_0$.

\subsection{Description of   Bose excitations}
 \subsubsection{The $\omega, k$ expansion of Green function}
Let us study small variations from the equilibrium state. Then $D^{-1} (\omega, \vec{k})$ can be expanded in the series of $k^2-k_0^2$ and $\omega -\omega_c$, cf \cite{Voskresensky:1980vt,Voskresensky:1984rd,MSTV90}:
\be
D^{-1} (\omega, \vec{k})\simeq -\omega_0^2 (\omega_c,k_0)+\alpha_0 (\omega -\omega_c)
+\alpha_1 (\omega -\omega_c)^2 +\alpha_2 (k_0^2-k^2)(\omega-\omega_c)+\alpha_3 (k_0^2 -k^2) +\alpha_4 (k_0^2 -k^2)^2+...\,,\label{Dexp}
\ee
\begin{eqnarray}
&\omega_0^2 (\omega_c,k_0)=m^2 +k_0^2+\Re\Sigma^R (\omega_c,k_0)-\omega_c^2\,,\quad \alpha_0 =\left(2\omega -\frac{\partial\Sigma^R}{\partial\omega}\right)_{\omega_c, k_0}\,,\quad
\alpha_1 =1 -\left(\frac{\partial^2\Sigma^R}{2\partial\omega^2}\right)_{\omega_c, k_0}\,,\nonumber\\
&\alpha_2 =\left(\frac{\partial^2\Sigma^R}{\partial k^2\partial\omega}\right)_{\omega_c, k_0}\,,\quad
\alpha_3 =1+\left(\frac{\partial\Sigma^R}{\partial k^2}\right)_{\omega_c, k_0}\,,\quad
\alpha_4 =-\frac{\gamma}{2}=-\frac{1}{2}\left(\frac{\partial^2\Sigma^R}{\partial (k^2)^2}\right)_{\omega_c, k_0}\,.
\end{eqnarray}
In general case $\alpha_i =\alpha_{i1}+i\alpha_{i2}$, where $\alpha_{i1}$ and $\alpha_{i2}$ are real quantities, $\Im\Sigma^R(\omega_c,k_0)=0$.
In case of the phase transition to inhomogeneous state, $k_0\neq 0$, from the condition of the absence of the current in the ground state follows that  $\alpha_3 =0$ and $\alpha_{4}<0$ for stability of the state.
 In case of the condensation in homogeneous state, $k_0=0$, we have $\alpha_3 > 0$,  $\alpha_{4}=0$.

Similar expansion exists for $\Lambda$. Simplifying the consideration we use
\be
{\Lambda}  =\Lambda_0  +
\Lambda_1 (\omega-\omega_c) +...\,,\label{LambdaExp}
\ee
where $\Lambda_0 =\Lambda (\omega_c, \vec{k}_0)>0$ to provide stability of the condensate state.

Expansion (\ref{Dexp}) holds also for the case of the  non-relativistic bosons, $D^{-1}=\omega -k^2/(2m)-\Sigma^R$, but now with coefficients given by \cite{Voskresensky:1993ux},
\begin{eqnarray}
&\alpha_0 =1 -\left(\frac{\partial\Sigma^R}{\partial\omega}\right)_{\omega_c, k_0}+\frac{\omega_c}{m}\simeq 1 -\left(\frac{\partial\Sigma^R}{\partial\omega}\right)_{\omega_c, k_0}\,,\quad
\alpha_1 =\frac{1}{2m} -\left(\frac{\partial^2\Sigma^R}{2\partial\omega^2}\right)_{\omega_c, k_0}\simeq -\left(\frac{\partial^2\Sigma^R}{2\partial\omega^2}\right)_{\omega_c, k_0}\,,\nonumber\\
&\alpha_2 =\left(\frac{\partial^2\Sigma^R}{\partial k^2\partial \omega}\right)_{\omega_c, k_0}\,,\quad
\alpha_3 =\frac{1}{2m}+\left(\frac{\partial\Sigma^R}{\partial k^2}\right)_{\omega_c, k_0}\simeq \left(\frac{\partial\Sigma^R}{\partial k^2}\right)_{\omega_c, k_0}\,,\quad
\alpha_4 =-\frac{\gamma}{2}=\left(\frac{\partial^2\Sigma^R}{\partial (k^2)^2}\right)_{\omega_c, k_0}\,.
\end{eqnarray}

The existence of gapless phases in “Bose-Luttinger liquids”, which in some respects can be regarded as bosonic versions of Fermi liquids, becomes now  a hot topic \cite{Lake2021}.
The model considered in \cite{Lake2021} is the limit case of (\ref{Dexp}) at $\alpha_0=\alpha_2=\alpha_3=0$.
Similarly,  the condensates with $k_0\neq 0$ arise, when the spectra of the sigma mesons  and pions have a “moat”, where the minimum of the energy lies over a sphere of nonzero radius in momentum.
A moat spectrum can arise in a large region of the phase diagram, as in QCD  and in different 1+1 dimensional models \cite{Basar2008,Basar2008a,Basar2008b,Carignano2015,Pisarski2021,Pisarski2021a,Pisarski2021b}.
The spectrum of excitations is degenerate along a sphere of radius $k_0$ in momentum space, which is refered to as a “Bose surface” in \cite{Lake2021}.

\begin{figure}\centering
\includegraphics[width=4.8cm,clip]{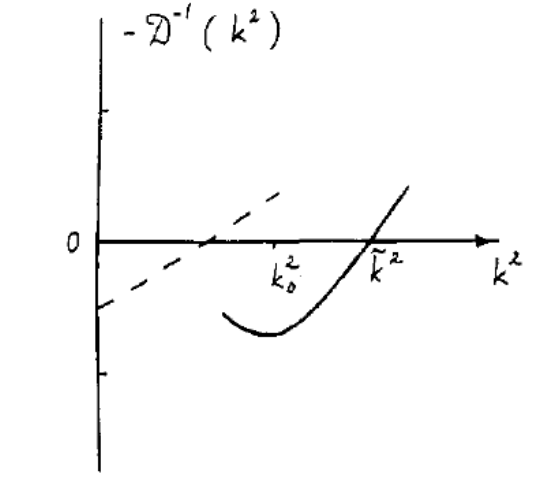}
\caption{Dependence $-D^{-1}(k^2)=-\Re D^{-1}({\omega_c} ,{\vec{k}^{\,2}},n,T)$ for  $n>n_c$ for the case of the phase transition to inhomogeneous state $k_0\neq 0$ (solid line) and to homogeneous state $k_0= 0$ (dashed line).
 }\label{InversedGreenF}
\end{figure}

For the case $\alpha_3 = 0$ the function $\Re D^{-1}({\omega_c} ,{\vec{k}}^{\,2})$ has the form shown in Fig. \ref{InversedGreenF} and we deal with the phase transition  to the inhomogeneous state $k_0\neq 0$.
Also, simplifying consideration let us put for a while $h=0$, then we deal with   the  second-order phase transition.
Interaction of the condensate with the own electromagnetic field is introduced with the help of the minimal coupling.
The part of the effective Lagrangian density depending on the order parameter related to the equation of motion (\ref{eqMotPhi}) with $D^{-1}$ presented in the form (\ref{Dexp}) it is convenient to present in a symmetric form
\begin{eqnarray}
&\delta {\cal L}[\psi] = [-\omega_0^2 +\alpha_4 k_0^4]|\psi|^4+ \frac{1}{2}(\alpha_{01}+\alpha_{21} k_0^2)[\psi^*(i\partial_t -V)\psi +c.c. ]-
\alpha_{11} (i\partial_t +V)\psi^* (i\partial_t -V)\psi\nonumber\\
&-2\alpha_4 k_0^2 (\nabla +ie\vec{A})\psi^*(\nabla-ie\vec{A})\psi
+\frac{1}{4}\alpha_{21}[(\nabla +ie\vec{A})^2\psi^*(i\partial_t -V)\psi+
\psi^*(\nabla -ie\vec{A})^2(i\partial_t -V)\psi +c.c.]\nonumber\\
&-\alpha_{41}[(\nabla +ie\vec{A})^2\psi^*(\nabla-ie\vec{A})^2\psi
+ (\nabla +ie\vec{A})^3\psi^*(\nabla-ie\vec{A})\psi +(\nabla -ie\vec{A})\psi(\nabla+ie\vec{A})^3\psi^*]\nonumber\\
&-\frac{1}{2}\Re\hat{\Lambda} |\psi|^4 +he^{im\pi}\psi (\psi^*/\psi)^{1/2}+e^{im\pi}h\psi^*(\psi/\psi^*)^{1/2} -\frac{1}{4}F_{\mu\nu}F^{\mu\nu}\,.
\end{eqnarray}
Interaction with electromagnetic field $A_\mu$ is included with the help of the gauge replacement, $F_{\mu\nu}=\partial_\mu A_\nu -\partial_\nu A_\mu$, $e$ is the electron charge, $V=eA_0$, $\mu,\nu =0,1,2,3$.
This expression was extensively used in the problem of the pion condensation in nuclear matter \cite{Anisimov:1979kt,Voskresensky:1980nk,MSTV90} and in some condensed matter problems \cite{Voskresensky:1984rd,Voskresensky:1993ux}. For static case a similar extension of the ordinary Ginzburg--Landau expression was suggested to describe so called superdiamagnets, a class of materials with strong diamagnetism but differing from conventional superconductors \cite{Ginzburg1979}.

\subsubsection{Spectrum of excitations}

We further assume $\alpha_3 =0$, $\alpha_1, -\alpha_4 >0$, and also $\alpha_0$ and $\alpha_2$ to be real that is satisfied for the case of the charged pion propogation in neutron star matter but not fulfilled for the case of isospin-symmetric matter. Besides that let us assume that $|\alpha_i|\sim 1$, $|\omega_0^2|\ll 1$ in units $m =1$. Also, consider excitations  dropping small  $\propto h$ terms.

To be specific let us bear in mind  the case of charged pion condensate in neutron star matter. Then one should add to the Lagrangian density $\delta {\cal L}[\psi]$ the contribution of other relevant fields:  of the nucleons and electrons, and, in case of  sufficiently high density, also of  muons and other baryons, and other mean  meson fields. Note that in case of the pion condensation in nuclear matter $k_0\sim p_{{\rm F},N}$, where $p_{{\rm F},N}$ is the nucleon Fermi momentum, $p_{{\rm F},N}(n_0)\simeq 1.9 m_\pi$, $m_\pi$ is the pion mass, and one expects $|\omega_0|\lsim m_\pi$. Simplifying consideration with demonstration aim let us consider only interaction of the classical charged  field with the positively charged particles (protons) assuming their  density, $n_p$, to be constant inside the system. The proton charge is compensated in the medium by the $\pi^-$ condensate charge.
Then the  spectrum of excitations is found from the equations of motion for the classical field $\phi$ and electromagnetic fields $V, \vec{A}$, which follow from the variation of the action in the fields.
We choose solution in the form
\be
\psi =a(1+\rho')\mbox{exp}[-i\omega_c t+i\vec{k}_0 \vec{r}+i\chi]\,.\label{pioncondfieldexc}
 \ee
  Performing the variable replacements $\vec{A}\to \vec{A}-\nabla\chi/e$, $V\to V+\partial_t \chi$, $e^2/(4\pi)=1/137$ we see that the Goldstone variable $\chi$ is absorbed by these transformations.
The linearized equations of motion for the fields $\rho'\ll 1$, $V'=V-\omega_c$ and $\vec{A}'=\vec{A}$ for $|V'|\ll \omega_c$, $|e \vec{A}'|\ll m_\pi$ render
\begin{eqnarray}
&2\omega_0^2\rho'-\alpha_1\ddot{\rho'}-2\alpha_2 \vec{k}_0\nabla \dot{\rho'}
+\alpha_4\Delta^2 \rho' -4\alpha_4 (\vec{k}_0\nabla )^2\rho' -\alpha_0 V' -\frac{\alpha_2}{2}\Delta V'\nonumber\\
&+i\alpha_0 \dot{\rho'}+i\alpha_2 \Delta \dot{\rho'}+4\alpha_4 i \vec{k}_0 \nabla\Delta \rho' -i\alpha_1 \dot{V'}-i\alpha_2 \vec{k}_0\nabla V'+i\alpha_2 \vec{k}_0 e\dot{\vec{A}'}-\alpha_4 \Delta \nabla e\vec{A}'\nonumber\\
&+\frac{\alpha_2}{2}\nabla e\dot{\vec{A}'}+2\alpha_4 \vec{k}_0\Delta e\vec{A}'+2\alpha_4 (\vec{k}_0\nabla)(\nabla e\vec{A}')+4\alpha_4 (\vec{k}_0\nabla)(\vec{k}_0 e\vec{A}')=0\,,
\end{eqnarray}
\be \partial_\mu F^{\mu\nu}=j^\nu\,,\quad j^\nu =(\rho, \vec{j})\,,\quad
\rho =en_p +\rho_\pi\,,\ee
\be
\rho_\pi = e\partial \delta {\cal L}/\partial V=[-\alpha_0-2\alpha_0 \rho'-\alpha_2 \Delta \rho'+2\alpha_1V'-2\alpha_2\vec{k}_0 e{\vec{A}'} ]ea^2\,,
\ee
\be
\vec{j} = \partial \delta {\cal L}/\partial \vec{A}=[\alpha_2 \nabla\dot{\rho'}+4\alpha_4 \vec{k}_0 \Delta\rho' +4\alpha_4 (\vec{k}_0 \nabla)\nabla \rho' -2\alpha_2\vec{k}_0 V'+8\alpha_4 (\vec{k}_0
e\vec{A}')\vec{k}_0 ]ea^2\,.
\ee
Here we dropped small contributions of the type $e^2a^2 (\ddot{ A_{\mu}'}-\Delta A_{\mu}')$ since such terms also appear in higher-order terms in  the expansion of $D^{-1}(\omega,k^2)$ in $\omega-\omega_c$ and $k^2-k_0^2$.  The condition of the local charge neutrality yields $\rho_p=\alpha_0 a^2$.

%%%%%%%%%%%%

The spectrum of excitations is found employing the Lorenz gauge $\partial_t V+\mbox{div}\vec{A}=0$.
It proves to be strongly anisotropic. We seek  fluctuating fields $\rho^{\prime},\chi,V^{\prime},\vec{A}^{\prime}$ in the form $\cos (\vec{q}\vec{r}-\omega t)$. Using expansion in small $|\vec{q}|$, for $\vec{A}\,'\perp \vec{k}_0$,  one finds
the branches \cite{Voskresensky:1980vt,MSTV90}:
\be ({\rm I}):\quad \omega^2 =\vec{q}^{\,2}\,;\ee
and
\be ({\rm IIa}):\quad \omega^2 =\frac{2|\omega_0^2|}{\alpha_1}-\frac{\alpha_0^2 e^2}{\Lambda}
-\frac{\alpha_1\alpha_0^2e^2}{2\Lambda|\omega_0^2|}\vec{q}^{\,2}-\left(\frac{\alpha_4}{\alpha_1}+
\frac{\alpha_1^2\alpha_0^2e^2}{4\Lambda\omega_0^4}\right)\vec{q}^{\,4}+...\,,\ee
for $\alpha_0\neq 0$, $\vec{k}_0\vec{q}=0$, $1\gg |\omega_0^2|/m^2_\pi\gg e^2/\Lambda$;
\be ({\rm IIb}):\quad \omega =\sqrt{\frac{2|\omega_0^2|}{\alpha_1}}-\frac{\alpha_0^2 e^2}{2\Lambda}\sqrt{\frac{\alpha_1}{|\omega_0^2|}}+\frac{\alpha_2}{\alpha_1}\vec{k}_0\vec{q}+...\,,\ee
for $\alpha_2\neq 0$ and $\vec{k}_0\vec{q}\neq 0$;
and
\be ({\rm IIIa}):\quad \omega^2 =\frac{\alpha_0^2 e^2}{\Lambda}+\vec{q}^{\,2}\,,\ee
 for $\alpha_0\neq 0$, $\vec{k}_0\vec{q}=0$; and
\be ({\rm IIIb}):\quad \omega =\frac{|\alpha_0| e}{\Lambda}-\frac{\alpha_2\alpha_0^2 e^2}{2\Lambda|\omega_0^2|}\vec{k}_0\vec{q}+...\,,\ee
for $\alpha_2\neq 0$, $\vec{k}_0\vec{q}\neq 0$.

For oscillations at $\vec{A}\,\vec{k}_0\neq 0$ the spectrum is as follows
\be ({\rm Ia}):\quad \omega^2 =-8\alpha_4 k_0^2 a^2 e^2 +\frac{4|\omega_0^2|\alpha_2^2}{\alpha_0^2}a^2 e^2 k_0^2
 +\vec{q}^{\,2}+...\,,\ee
for $\alpha_0\neq 0$ and $\vec{k}_0\vec{q}= 0$; and
\be ({\rm Ib}):\quad \omega =2\sqrt{-2\alpha_4} k_0 a e +\frac{|\omega_0^2|\alpha_2^2}{\sqrt{-2\alpha_4}\alpha_0^2}a e k_0
 +\frac{\alpha_2^2 (\alpha_0+4\alpha_2 k_0^2 )}{2\alpha_0^2}a^2 e^2 \vec{k}_0\vec{q}+...\,,\ee
for $\alpha_0\neq 0$ and $\vec{k}_0\vec{q}\neq 0$. The branches ${\rm II a,b}$ and ${\rm III a,b}$ are the same as for the case $\vec{A}\,\vec{k}_0= 0$. All excitations except branch ${\rm I}$ at $\vec{A}\,\vec{k}_0= 0$, are gapped. The former branch describes Goldstone excitations.
The spectrum is strongly anisotropic. For $\alpha_2\neq 0$ ``magnetic'' excitations have a minimal mass $2\sqrt{-2\alpha_4} k_0 a e$, ``electric''  excitations have a higher mass
$|\alpha_0| e/\sqrt{\Lambda}$, and ``meson'' excitations have still larger mass $|\omega_0|\sqrt{2/\alpha_1}$.
The system is superfluid and superconducting.

It is curious to find spectrum of over-condensate excitations for the case of the complex condensate field of the form (\ref{pioncondfieldexc}) but for $e^2=0$, cf. \cite{Voskresensky:1980vt}. This consideration generalizes to the case $k_0\neq 0$ the consideration of the excitations in He-II.
For $\vec{k}_0\vec{q}= 0$ one finds for $\omega, |\vec{q}|\ll m_\pi$  the branches
\be ({\rm Ia}):\quad \omega^2 =\frac{\alpha_0^2}{\alpha_1^2}-\frac{2\omega_0^2}{\alpha_1}
 -\frac{2\alpha_0\alpha_2}{\alpha_1^2} \vec{q}^{\,2}+...\,,\ee
\be ({\rm IIa}):\quad \omega^2 =
 +\frac{2\alpha_4 \omega_0^2}{\alpha_0^2-2\alpha_1 \omega_0^2} \vec{q}^{\,4}+...\,,\ee
whereas for $\vec{k}_0\vec{q}\neq 0$,
\be ({\rm Ib}):\quad \omega =\frac{\alpha_0}{\alpha_1}+\frac{2\alpha_2}{\alpha_1} \vec{k}_0\vec{q} -\frac{\alpha_2}{\alpha_1}\vec{q}^{\,2}- \frac{\alpha_4}{\alpha_0} (\vec{k}_0\vec{q})^2+...\,,\ee
\be ({\rm IIb}):\quad \omega =\frac{\sqrt{-8\alpha_4}}{\alpha_0}
|\omega_0\vec{k}_0\vec{q}|+...\,\ee
Thus there is the gappless non-Goldstone mode $\omega\propto \vec{q}^{\,2}$ and the gapped mode for $\vec{k}_0\vec{q}=0$ and there is the Goldstone  and the gapped anisotropic modes  for $\vec{k}_0\vec{q}\neq 0$.

Knowledge of the spectrum of over-condensate excitations is important for description of the transport processes, e.g., in $\pi$ condensate regions of the neutron stars.

\subsection{Static inhomogeneous configurations and finite size effects}

 There are many  configurations of the condensate field, which at the fixed volume $V_3$ correspond to the same volume energy $\delta F= -\frac{\omega_0^4}{2\Lambda_0}V_3 $ and different surface energies. For example, these solutions are: \\
 slabs
 \be
 \psi =ae^{i\vec{k}_0\vec{r}}\,,
  \ee
 rods
 \be
 \psi =ae^{i{k}_0\rho}\,,\label{infrod}
  \ee
 spherical drops
 \be
 \psi =ae^{i{k}_0{r}}\,,\label{infspher}
  \ee
 disordered phase
 \begin{eqnarray}
 &\psi \simeq \frac{1}{N}\left\{\sum_{1}^{N_1}a_i \mbox{exp}[ik_0\sqrt{(x-x_i)^2+(y-y_i)^2+(z-z_i)^2}\,\,]\right.\nonumber\\
 &\left. +
 \sum_{N_1+1}^{N}a_i \mbox{exp}[ik_0\sqrt{(x-x_i)^2+(y-y_i)^2}\,\,]+a_0 e^{ik_0x}
 \right\}\,,\quad \frac{1}{N}\sum_0^N a_i^2 =a^2\,.
 \end{eqnarray}
 The phase transition to the inhomogeneous state $k_0\neq 0$ is always of the first-order due to strong fluctuations with $\vec{k}\sim \vec{k}_0\neq 0$, cf. \cite{Brazovskii1975,Dyugaev:1975dk,Voskresensky:1981zd,Voskresensky:1982vd,
 Dyugaev:1982gf,Dyugaev:1982ZHETF,Dyg1}. If the system is initially in the metastable state, the transition to the stable phase occurs
 by the growth of   seeds of stable phase of an overcritical size, being formed in fluctuations. The system first reaches the state with different structures
 in different domains and only after a passage of a long time it  may reach the  equilibrium state with the most energetically profitable structure of the condensate field in the whole system \cite{MSTV90,Voskresensky:1993ux}. If  the   system in its evolution  crosses  the isothermal spinodal line, the fluctuations with different directions of $\vec{k}_0$ grow rapidly and  the system arrives at the glass-like state disordered at distances $r\gg 1/k_0$.

Consider inhomogeneous phase characterized by the wave vector $\vec{k}_0$ and frequency  $\omega_c$.
To describe surface effects one needs to consider deviations of $\vec{k}$  from $\vec{k}_0$.
Setting $\psi\propto e^{-i\omega_c t}$ in equation of motion (\ref{eqMotPhi}) one gets
\be
[-\omega_0^2 +\alpha_4 (k_0^2 +\Delta)^2]\psi -\psi \hat{\Lambda} (\Delta, \overline{|\psi|^4}, \overline{|\psi|^2})|\psi|^2 +he^{im\pi}(\psi/\psi^*)^{1/2}=0\,.\label{eqMphistat}
 \ee
 For the static field one should put $\omega_c =0$.
Here we use simplified presentation of (\ref{Dexp}) for $\alpha_1=\alpha_2=\alpha_3 =0$ and  put $\hat{\Lambda}=\Lambda_0$, which value however may depend on the structure of the condensate field.

Let us again consider the case $h=0$ corresponding to the description of the second-order phase transition.
The free-energy density associated with the static classical field $\psi$ is as follows,
\be
\delta{\cal F} =-\psi^*D^{-1}(0, \hat{k}^2) \psi
+\Lambda_0 |\psi|^4/2\,.\label{Fbraz}
\ee
Let us consider the case of the sharp boundary between the medium and the vacuum. Inside the medium (to be specific let it be region $x<0$) we have   $-D^{-1}(0, \hat{k}^2)=\omega_0^2 -\alpha_4 (k_0^2 +\Delta)^2$.  In the vacuum (for $x>0$) we have  $-D^{-1}(0, \hat{k}^2)=m^2-\hat{k}^2$.
The equation of motion  is as follows
\be
(\beta_1 -\beta_2 \hat{k}^2 +\beta_3 \hat{k}^4)\psi -\Lambda_0 |\psi|^2\psi =0\,,\label{phisolinhom}
 \ee
 with $D^{-1}(\omega_c, \hat{k}^2)=\beta_1 -\beta_2  \hat{k}^2+\beta_3 \hat{k}^4$ at $\beta_1 =-\omega_0^2+\alpha_4 k_0^4$, $\beta_2 =-2\alpha_4 k_0^2$, $\beta_3 =\alpha_4$ for $x<0$ and respectively $-m^2, -1, 0$ in the vacuum, for $x>0$.

In the vacuum (for $x>0$) the field is described by the differential equation of the second order in derivatives, whereas inside the medium by the 4-th order one. To derive necessary boundary conditions, cf. \cite{Voskresensky:1984rd}, one may consider single differential equation of the 4-th order for all $x$, with coefficients in front of  the third- and fourth- derivative terms  tending to zero  in the sharp boundary layer (of the  typical length $l_{\rm b.l.}\ll l_{<},l_{>}$, where $l_{<},l_{>}$ are typical length scales, respectively for $x<0$ and $x>0$, characterizing $\psi (x)$ obeying  the equation of motion (\ref{phisolinhom}) with constant coefficients.  One boundary condition is  the continuity of $\psi$ at $x=0$. Other conditions are derived by one, two and three integrations of the equation of motion for the smoothed boundary subsequently letting the boundary layer width tend to zero. This procedure yields, as the boundary conditions, the continuity of the quantities
 \be \beta_2\nabla\psi +\beta_3  \Delta\nabla\psi\,, \quad \beta_3\Delta \psi\,,\quad \beta_3 \nabla\psi\,, \quad \psi\,.\label{bcond}
 \ee
 In case of the Maxwell equations describing two dielectric media, two latter conditions (\ref{bcond}) coincide with the ordinary conditions of the continuity of the electric potential, $\Phi$, and the electric displacement, $\epsilon\nabla\Phi$, where $\epsilon$ is the dielectric constant.
 In case of the semi-infinite three-dimensional system with the flat $(x,y)$ boundary between the medium and the vacuum, provided  $l_< =l_0\sim 1/|\omega_0|\gg 1/k_0$ and $l_<\gg l_>\sim 1/m$,  one may use more simple conditions $\phi (x=0)=\phi^{\prime}_x (x=0)=0$.

\subsubsection{One-axis-like  structures}
There are  two types of one-dimensional modulations. One is of the Fulde--Ferrell  type \cite{Fulde1964}, characterized by modulations of the phase of a complex order parameter with constant amplitude,  the other is of the Larkin--Ovchinnikov  type \cite{Larkin1965}, where  the amplitude of the condensate field is modulated.

One-axis structures are of interest for description of A-smectic liquid crystals and the charged pion condensate.
With the field in the form of the running wave,
\be
\psi =a\chi (\vec{r})  e^{i\vec{k}_0\vec{r}}\,,\label{psirunning}
\ee
placed in Eq. (\ref{phisolinhom}) one obtains equation of motion for the $\chi$ amplitude
\be
\alpha_4\Delta^2\chi -4\alpha_4(\vec{k}_0\nabla)^2 -\omega_0^2 \chi +\omega_0^2 |\chi|^2 \chi=0\,,
\quad a^2 =(-\omega_0^2/\Lambda_0)\theta (-\omega_0^2)\,,\label{EqMrunning}
 \ee
 where  $\theta(x)$ is the step-function and we assumed that $\chi$ is a smooth function and we dropped $(\vec{k}_0\nabla)\Delta\chi $ term but retained $\Delta^2\chi$ term, being the main term in case $\vec{r}\perp\vec{k}_0$. Substituting
 (\ref{psirunning}) in (\ref{Fbraz}) one obtains
 $\delta{\cal F} =-(\omega_0^4/(2\Lambda_0))\theta (-\omega_0^2)$.

 After performing the transformation $x_i\to \tilde{x}_i=x_i 2k_{0i}\sqrt{\alpha_4/\omega_0^2}$, where $x_i = (x,y,z)$,
in dimensionless variables ($\nabla\to \tilde{\nabla}$), Eq. (\ref{EqMrunning}) renders
\be
-\alpha \tilde{\Delta}^2 \chi +(\vec{n}\tilde{\nabla})^2\chi +\chi -|\chi|^2\chi =0\,, \quad \vec{n}=\vec{k}_0/k_0\,,\quad
\alpha = \omega_0^2/(16k_0^4 \alpha_4)\,. \label{chidimfourth}
\ee
For a smooth function $\chi$, this is the  second-order derivative equation in the directions $\vec{r}\,\vec{k}_0\neq 0$ and  the fourth-order one for  $\vec{r}\perp\vec{k}_0$.

There are systems, which are ordered in the longitudinal
direction, but react as a liquid in the transverse direction \cite{Chaikin2010}.
Note that Eq. (\ref{chidimfourth}) does not coincide with the Landau--De'Gennes equation \cite{Chandrasekhar}, which is often used to describe A-phase of smectic liquid crystals. In the latter case the inter-plane distance $d$ is fixed. The molecules are  perpendicular to the layer planes and molecular centers are disordered. The density wave has the form
\be
\rho (z)=\bar{\rho} [1+2^{-1/2}|\psi|\cos (k_0 z -\phi)]\,,
 \ee
where $\bar{\rho}$ is the averaged density, $|\psi|$ is the amplitude, $k_0$ is the wave number,
$\phi$ is the phase. The order parameter  $\psi =|\psi|e^{i\phi}$ obeys the Landau--De'Gennes equation
\be
\frac{\partial_z^2\psi}{2M_V}+ \frac{(\partial_x^2+\partial_y^2)\psi}{2M_T}+\alpha_0 \psi -\Lambda |\psi|^2\psi =0\,,\label{LandauDeGennesEq}
 \ee
for $\alpha_0,\Lambda >0$, $M_V$ and $M_T$ play roles of the effective longitudinal and transverse masses, $M_V\neq M_T$. The solution for the  infinite system is given by
$|\psi|^2=\alpha_0/\Lambda >0$. The  free-energy density associated with $\psi$ is as follows
\be
{\cal F}=-\alpha_0 |\psi|^2+\frac{\Lambda |\psi|^4}{2} +\frac{|\partial_z\psi|^2}{2M_V}
+ \frac{(|\partial_x\psi|^2+|\partial_y\psi|^2)}{2M_T}\,.
\ee
Note also that Landau-De'Gennes model does not describe the phase transition from the isotropic uniform phase to the A smectic phase since $k_{0x}^2=k_{0y}^2\neq k_{0z}^2$ in this model. In order to  describe both mentioned phases one may use
Eq. (\ref{EqMrunning}), which follows from the free energy depending on $k_0^2$ rather than on
$k_{0z}^2$ and $k_{0x}^2=k_{0y}^2\neq k_{0z}^2$. In case of the model described by Eq. (\ref{EqMrunning}) the direction $\vec{k}_0$ arises by the spontaneous symmetry breaking.

 Let $-\omega_0^2>0$ in the half-plane $z<0$ and $\omega_0^2=m^2$ for $z>0$. Solutions of Eq. (\ref{EqMrunning})
are characterized by two length scales, $l_{\parallel}=(-2\alpha_4 k_0^2/|\omega_0^2|)^{1/2}$ in the directions $\vec{r}\,\vec{k}_0\neq 0$ and $l_{\perp}=(-2\alpha_4 /|\omega_0^2|)^{1/4}$ in the direction  $\vec{r}\,\vec{k}_0=0$.
The  solution for $\vec{k}_0\parallel z$ satisfying boundary conditions $\chi (z\to -\infty)\to 1$ and $\chi (z\to 0)\to 0$ is as follows,
\be
\psi \simeq a e^{i k_0 z}\mbox{tanh}[(z-z_0)/(\sqrt{2}\,l_{\parallel})]\,.\label{solparallel}
 \ee
To find solution for $\vec{k}_0\perp z$ one presents $\chi =1-\chi_1 -\chi_2-...$ with
$1\gg |\chi_1|\gg |\chi_2|...$. Thus one obtains \cite{Anisimov:1979kt,Voskresensky:1980nk,MSTV90},
\be
\psi \simeq a e^{i k_0 y}\left[1-C_1 e^{z/l_{\perp}}\cos (\frac{z}{l_{\perp}}+C_2)+O(\chi_1^2,\chi_2)
%-\frac{C_1^2}{10}e^{2z/l_{\perp}}\left(1+\sin^2(\frac{z}{l_{\perp}}+C_2)_...\right)
\right]\,.\label{solperp}
 \ee
Question on the boundary conditions is not trivial, as it has been mentioned,  since for $z<0$ we deal with the differential equation of the forth-order, whereas for $z>0$ with the differential equation of the second-order \cite{Voskresensky:1980nk,Voskresensky:1984rd}. For $l_{\perp}\propto 1/|\omega_0|^{1/2}\gg 1/k_0$ and one may apply simpler boundary conditions $\psi (0)=\psi_z'=0$
and one finds $C_1 =\sqrt{2}$, $C_2=\pi/4$. Solution (\ref{solperp}) yields the same volume part of the free energy as (\ref{solparallel}) but a smaller surface part, since
$l_{\perp}<l_{\parallel}$ (provided $|\omega_0^2|\ll m^2$). Thereby, choice of the direction $\vec{k}_0$ parallel to the medium boundary is energetically profitable. Generalization to the case of the slab $-z_0<z<z_0$, for $z_0 \gg l_{\perp},l_{\parallel}$, is given with the help of the replacement $z\to |z|$ in Eqs. (\ref{solparallel}), (\ref{solperp}). It is curious to notice that the relation
$l_{\perp}^2\sim l_{\parallel}/k_0$ indeed holds for A-smectic crystals \cite{Nelson1974}.

For the spherical system of the radius $R$ there is the following solution of Eq. (\ref{EqMrunning}), cf. \cite{Migdal:1974fk,Voskresensky:1993ux},
\be
\psi \simeq a e^{i k_0 z}\mbox{tanh}[(|z|-\sqrt{R^2-x^2-y^2})/(\sqrt{2}\,l_{\parallel})]\,\label{solparallelSph}
 \ee
 for $x^2+y^2<R^2$.
Also, this solution  holds for weakly deformed systems. Following this solution the condensate part of the surface energy decreases with increasing deformation and an elongation is energetically profitable.
The equilibrium deformation is found from competition of the  condensate and   non-condensate parts of the free energy. Solution (\ref{solparallelSph}) holds only for  weak deformations. Its value depends on the length scales $1/k_0$ and $l_{\parallel}$. With increasing deformation, there appears the dependence on the scale  $l_{\perp}$.
To demonstrate this dependence we may construct solution for the condensate field in the parallelepiped, cf. \cite{Voskresensky:1980nk,Voskresensky:1984rd},
\be
\psi =a e^{ik_0 y}\left[\mbox{tanh}\frac{y_0-|y|}{\sqrt{2}\,l_{\parallel}}-\frac{(1-i)
\mbox{sinh}((1+i)x_0/l_{\perp})\mbox{cosh}((1-i)x/l_{\perp})+{\rm c.c.}}{\mbox{sinh}(2x_0/l_{\perp})+\sin (2x_0/l_{\perp})}+{x\leftrightarrow z,\,} {x_0\leftrightarrow z_0}
\right]\,.
\ee
The gain in the condensate part of the free energy renders
\be
\delta F=-\int \frac{\Lambda_0 |\psi|^4}{2}d^3 x=-\frac{\omega_0^4}{2\Lambda_0}
\left[1-\frac{4l_{\parallel}}{3y_0}-4l_{\perp}\left(\frac{1}{x_0}
+\frac{1}{z_0}\right)\right]V_3\,,\quad V_3 =8x_0y_0z_0\,.
 \ee
Let us assume that the non-condensate part of the surface energy is negligibly small. Then optimal sizes of the system at fixed volume $V_3$ are given by \cite{Voskresensky:1980vt,MSTV90}:
\be
x_0=z_0=(3l_{\perp}/(l_{\parallel}))y_0\ll y_0\,.
 \ee
 Therefore in presence of the condensate, an initially spherical nucleus becomes elongated in the direction $\vec{k}_0$. Now optimal deformation exists, even if the non-condensate part of the surface energy is negligibly small.

The spherical solutions $\vec{k}_0\parallel \vec{r}$ render \cite{Voskresensky:1993ux}:
\be
\psi \simeq a e^{i k_0 r}\mbox{tanh}[(r-r_0)/(\sqrt{2}\,l_{\parallel})]\,,\label{solparallelsph}
 \ee
for $r_0\gg l_{\parallel}$.

Cylindric solutions  are
 \be
\psi \simeq a e^{i k_0 \rho}\mbox{tanh}[(\rho-\rho_0)/(\sqrt{2}\,l_{\parallel})]\,,\label{solparallelcyl}
 \ee
for $\rho_0\gg l_{\parallel}$, $\rho =\sqrt{x^2+y^2}$, and there are many other configurations of the same surface energy at different surface energies.

A polycrystal configuration constructed from  domains is as follows
\begin{eqnarray}
&\psi =a \sum_{i=1}^{N_1}e^{i k_0 |\vec{r}-\vec{r}_i|}\mbox{tanh}[(|\vec{r}-\vec{r}_i|-R_i)/(\sqrt{2}\,l_{\parallel})]
\theta(R_i -|\vec{r}-\vec{r}_i|)\nonumber\\
&+
a \sum_{N_1+1}^{N_2}e^{i k_0 z}
\mbox{tanh}[(|z|-\sqrt{R_i^2-x^2-y^2})/(\sqrt{2}\,l_{\parallel})]\theta (\sqrt{R_i^2-x^2-y^2}-|z|)\nonumber\\
&+\sum_{N_2+1}^{N_3}\{{\rm rod}\}+\sum_{N_3+1}^{N_4}\{{\rm slab}\}+...\,,\quad \sum_i N_i =N\,,
\label{polisol}
\end{eqnarray}
$R_i\gg l_{\parallel}$ is the domain radius, $\sum_i V_i \simeq V_{\rm tot}$ up to a surface term, where $V_{i}$ is the volume of the $i$-domain,  $V_{\rm tot}$ is the volume of the condensate system.

In case when by some reason one deals with the real condensate field, e.g., as for neutral pions,
equation of motion has the same form (\ref{phisolinhom}), but now for the real $\psi$.
Choosing  the field in the form of the standing wave,
\be
\psi =a_{\rm stan}\chi (\vec{r})\cos(\vec{k}_0\vec{r})\label{psistaying}
\ee
and substituting it  in (\ref{phisolinhom}) one finds $a^2_{\rm stan}=-4\omega_0^2/(3\Lambda_0)$.
Substituting this solution in
(\ref{Fbraz}) one obtains
 $\delta{\cal F} =-\omega_0^4/(3\Lambda_0)$ and instead of Eq. (\ref{polisol}) we get
 \begin{eqnarray}
&\psi =a_{\rm stan} \sum_{i=1}^{N_1}\Re\{e^{i k_0 |\vec{r}-\vec{r}_i|}\}\mbox{tanh}[(|\vec{r}-\vec{r}_i|-R_i)/(\sqrt{2}\,l_{\parallel})]
\theta(R_i -|\vec{r}-\vec{r}_i|)\nonumber\\
&+
a_{\rm stan} \sum_{N_1+1}^{N_2}\Re\{e^{i k_0 z}\}\mbox{tanh}[(|z|-\sqrt{R_i^2-x^2-y^2})/(\sqrt{2}\,l_{\parallel})]\theta (\sqrt{R_i^2-x^2-y^2}-|z|)\\
&+\sum_{N_2+1}^{N_3}\{{\rm two-dim.}\}+...\,,\quad \sum_i N_i =N\,.\nonumber
\end{eqnarray}
It is important to notice that in case of a complex field for $\Lambda =const$ the solution (\ref{psirunning}) produces a smaller free energy than (\ref{psistaying}).

\subsubsection{Three-axis structures}

Assume that the lattice structure has the form
\be
\psi =a \chi \cos (k_{0x}x)\cos (k_{0y}y)\cos (k_{0z}z)\label{coscoscos}
 \ee
with $k_0^2=k_{0x}^2+k_{0y}^2+k_{0z}^2$. Minimization of the free energy at $\chi =1$ yields  $a^2 =-(4/3)^3\omega_0^2/\Lambda_0>0$.  Then setting  (\ref{coscoscos}) in Eq.  (\ref{phisolinhom}) one derives equation for  $\chi$:
\be
\alpha_4 (\Delta^2 -4k_{0x}^2\partial_x^2 -4k_{0y}^2\partial_y^2-4k_{0z}^2\partial_z^2)\chi
-\omega_0^2 (\chi -\chi^3)=0\,.\label{chicoscos}
 \ee
The free-energy density becomes
\be
\delta {\cal F}=\delta {\cal F}_1 +\delta {\cal F}_2\,,
\ee
where
\be
\delta {\cal F}_1={\cal F}_0\left[\frac{\chi^2}{2}-\frac{\chi^4}{4}+\frac{2\alpha_4}{\omega_0^2}
(k_{0x}^2(\partial_x\chi)^2+k_{0y}^2(\partial_y\chi)^2+k_{0z}^2(\partial_z\chi)^2) \right]\,,
\ee
${\cal F}_0=-\frac{16}{27}\frac{\omega_0^4}{\Lambda_0}$, $\delta {\cal F}_2$ is associated with higher derivatives of $\chi$ and thereby
$|\delta {\cal F}_2|\ll
|\delta {\cal F}_1|$.
Note that although Eq. (\ref{chicoscos}) describes the three-dimensional lattice, it coincides  (up to higher derivative terms) with  Eq. (\ref{LandauDeGennesEq}) of the  Landau-De'Gennes model of the one-axis A-smectic liquid crystal.

Assuming lattice structure in the form
\be
\psi =a \chi [\cos (k_{0x}x)+\cos (k_{0y}y)+\cos (k_{0z}z)]\label{sumcos}
 \ee
one finds $a^2 =-(4/15)\omega_0^2/\Lambda_0>0$. Equation of motion for $\chi$ becomes
\be
\alpha_4 (\partial_x^4 + \partial_y^4+\partial_z^4  -4k_{0x}^2\partial_x^2 -4k_{0y}^2\partial_y^2-4k_{0z}^2\partial_z^2)\chi
-\omega_0^2 (\chi -\chi^3)=0\,.\label{chicospluscos}
 \ee
The free-energy density is given by
\be
\delta {\cal F}=\delta \tilde{\cal F}_1 +\delta \tilde{\cal F}_2\,,
\ee
where
\be
\delta \tilde{\cal F}_1=\tilde{\cal F}_0\left[\frac{\chi^2}{2}-\frac{\chi^4}{4}+\frac{2\alpha_4}{\omega_0^2}
(k_{0x}^2(\partial_x\chi)^2+k_{0y}^2(\partial_y\chi)^2+k_{0z}^2(\partial_z\chi)^2) \right]\,,
\ee
$\tilde{\cal F}_0=-\frac{4}{5}\frac{\omega_0^4}{\Lambda_0}<{\cal F}_0$, $\delta \tilde{\cal F}_2$ is associated with higher derivatives of $\chi$ and thereby
$|\delta \tilde{\cal F}_2|\ll
|\delta \tilde{\cal F}_1|$. Note that the value $k_0^2$ is determined by minimization of the free energy, whereas separate values $k_{0i}$ remain not fixed. This circumstance might be of interest for description of phase transitions in varios materials, e.g.,  like glassing transition, martensite structures in alloys, domain structures, etc. Although, as we see, solution (\ref{sumcos}) yields a smaller free energy than (\ref{coscoscos}) at the same value $\Lambda_0$, we pay attention that value $\Lambda_0$ may depend on the form of the solution.

Neglecting fourth-order derivative terms,  after transformation $x_i\to x_i 2k_{0i}\sqrt{\alpha_4/\omega_0^2}$, where $x_i = (x,y,z)$, Eqs. (\ref{chicoscos}), (\ref{chicospluscos}) require a simple form
\be
\tilde{\Delta} \chi +\chi -\chi^3 =0\,.\label{threehom}
 \ee
In these dimensionless variables equation for $\chi$ is the same as that describing the second-order phase transition to the homogeneous state $k_0=0$. In this sense, the consideration of the second-order phase transition to the three-dimensional lattice  states is similar to that for the phase transition to the state $k_0=0$. Thereby, in general there is a principal difference in description of the three-axis and the one-axis-like structures. Note that the Landau-De'Gennes model   used for description of A-phase of smectic liquid crystals employs simplified Eq.
(\ref{threehom}) rather than a more general Eq. (\ref{chidimfourth}).

Note again that the values of the coefficients $\Lambda_i$ depend on the structure of the condensate field. Thereby they are different for various one-axis, two-axis and three-axis structures.
Moreover instead of one real or complex order parameter there may exist many order parameters, e.g., the pion condensate order parameter is the vector in the isospin space. Some possible structures of the pion condensate were studied within the Thomas--Fermi approximation $k_0^2/4p^2_{{\rm F},N}\ll 1$, cf. \cite{Migdal78,MSTV90}. Examples of such structures are
\be
\vec{\phi}=(a\cos (k_0 x), -a\sin (k_0 x), 0)\,, \ee
\be \vec{\phi}=(2^{-1/2}a\cos (k_0 x), 2^{-1/2}a\cos (k_0 x), a\sin (k_0 x))\,,\ee
\be \vec{\phi}= (0,0,(2/3)^{1/2} a[\sin (k_0 x)+\sin (k_0 y)+\sin (k_0 z)])\,.
 \ee
 Then the isotopic and spatial structures of the pion condensate were considered in the so called alternating-layer-structure model, cf. \cite{Takatsuka:1978ku}.
It was demonstrated that the most energetically preferable structure likely corresponds to the standing $\pi^0$ wave with $\phi^0\propto \Re e^{ik_0 z}$
 and the running $\pi^\pm$ wave $\phi^{\pm}\propto e^{i\vec{k}_{0\perp} \vec{\rho}}$.
Simplifying consideration,  we further do not consider these complications supposing each $\Lambda_i$ to be constant corresponding to the given structure of the condensate field.

Ultracold atoms with strong dipole-dipole interactions
allow to study  many-body systems with long-range anisotropic interactions \cite{Baranov}.
Reference \cite{Maeda2013} demonstrated that such systems  enable one to realize in the laboratory analogs of meson condensation in nuclear matter, due to similarities of the electric and magnetic dipole interactions and the nuclear tensor force, with the same $r$-dependence of the interaction potential as for the $p$ wave $\pi NN$ interaction. Consideration of these systems goes beyond the framework of the present study.

\subsubsection{Chiral waves}
Within the sigma model  the pion and sigma mesons are unified in the Euclidean  4-vector: $\phi_i =(\sigma,\vec{\phi})$. Studies of pion  condensates with $k_0\neq 0$ within the sigma model were began in \cite{Campbell:1974qt,Campbell:1974qu,Baym:1975tm,Voskresensky:1978cb,
Voskresensky:1982vd}. The chiral spiral structures are also possible \cite{Dautry1979,Nakano:2004cd}, where the sigma-pion condensate $(\sigma, \vec{\phi})$
forms a structure, e.g., like
$$\phi_i=(\phi_0 \cos (k_0 z), 0,0, \phi_0 \sin (k_0 z))\,.$$ In the large $N_c$ limit to QCD the early studies suggested the emergence of the  chiral density wave \cite{Deryagin1992,Shuster2000}.
The phase structure at high  baryon densities and moderate temperatures can be
modified considerably also by the presence of the inhomogeneous  QCD phases,  cf. \cite{Nakano:2004cd,Carignano2014,Carignano2015,TGLee2015,Tatsumi2016}.
The chiral spiral structures were extensively discussed, where the sigma-pion condensate $(\sigma, \vec{\phi})$
forms a structure, e.g., like
$$(\phi_0 \cos (k_0 z), 0,0, \phi_0 \sin (k_0 z))\,,$$  the  quarkyonic
chiral spiral, cf.  \cite{McLerran2007,Kojo2009ha}, and  other systems with the moat-like  spectrum, cf.  \cite{Pisarski2021,Pisarski2021a,Pisarski2021b}.  Various inhomogeneous chiral condensed phases have been proposed, cf.  \cite{Carignano2015,Tatsumi2016} and references therein.   These features appear in mean-field calculations in the Nambu--Jona-Lasinio, the quark-meson  models \cite{Nakano:2004cd,Nickel2009wj} and the Dyson--Schwinger approach to the dense QCD  \cite{Muller2013}.

\subsection{Mean field versus fluctuations}
\subsubsection{Critical temperature in the mean field approximation}
In relativistic mean-field models \cite{Serot:1984ey,Glendenning2000,KolVosk2005} one assumes that the $T$ dependence enters only the fermion distribution functions.
In the Schwinger--Dyson approach to the fermion-boson system the   temperature  dependence enters  the particle Green functions. In the mean field approximation one takes into account the $T$ dependence of the fermion Green functions but ignores it in the boson Green functions.  In simplest case of the one fermion -- one boson system, like for nucleons and pions in isospin-symmetric matter, the parameter characterizing the temperature dependence of the fermion Green function is $(T/\epsilon_{\rm F})^2$, as it follows from the known Fermi integral expansions at low $T$, cf. \cite{LL5}. As the result $\langle|\psi|^2\rangle=-\omega_0^2(T)/\Lambda_0 (T)
\simeq -\omega_0^2(0)/\Lambda_0 (0)+C(T/\epsilon_{{\rm F},N})^2$, $C=const$, $\epsilon_{{\rm F},N}$ is the nucleon Fermi energy.
Thus in the mean-field (MF) approximation, cf. \cite{Voskresensky:1978cb}, $\langle|\psi|^2\rangle=0$ at
\be
T=T_c^{\rm MF}\sim  \epsilon_{{\rm F},N}(n-n_c)/n_c\,,
 \ee
provided $(n-n_c)/n_c\ll 1$.
In  mentioned case of the pion condensation also $\Delta$ isobars give essential contribution, and the typical temperature destroying the $\Delta$-particle -- nucleon hole contribution is $T\sim m_\Delta -m_N\simeq 2.1 m_\pi$. It proves to be that  with taking into account of $\Delta$-particle -- nucleon hole diagrams the value $T_c^{\rm MF}$ increases,
cf. \cite{Voskresensky:1978cb,Voskresensky:1982vd,MSTV90}.

\subsubsection{Fluctuations in case of the phase transition to the state $k_0\neq 0$}\label{flnonzero-sect}
As we have discussed in Section \ref{FlhomSect}, fluctuations of the order parameter become strong in the vicinity of the critical point of the phase transition to the state $k_0=0$. In case of the phase transition to the state $k_0\neq 0$  fluctuations  play even more important role than in case $k_0= 0$ because of their large phase space volume. As the result, with taking into account of fluctuations the phase transition to the state $k_0\neq 0$ proves to be of the first order, even if in the mean field approximation it was of the second order \cite{Brazovskii1975,Dyugaev:1975dk,Voskresensky:1981zd,Voskresensky:1982vd,Dyugaev:1982ZHETF,
Dyg1,MSTV90,Voskresensky:1989sn,Voskresensky:1993ud}.
A liquid/amorphous  phase of the pion condensate characterized by strong correlations with $k\sim k_0\neq 0$
may arise  in nuclear matter already at $n\gsim n_c^{(1)}\simeq (0.5-0.8)n_0$, cf. \cite{Dyg1,MSTV90,Voskresensky:1989sn,Voskresensky:1993ud}, i.e.,  at  smaller densities compared to the critical density $n_c^\pi$ for the pion condensation of the liquid-crystal-like or crystal-like  type.

The correlation function of two fluctuating complex fields reads, cf. \cite{Voskresensky:1981zd,Voskresensky:1984rd},
\be
N(\vec{r})=\langle\psi (\tau_1,\vec{r}_1) \psi^* (\tau_2,\vec{r}_2)\rangle|_{\tau_2=\tau_1-0}
=-T\sum_{n=-\infty}^{\infty} \int \frac{d^3 k}{(2\pi)^3}{D}_{\rm M}(E_n,\vec{k}^2)e^{-E_n0+i\vec{k}\vec{r}}\,,\quad \vec{r}= \vec{r}_1 -\vec{r}_2\,,\label{Matsum}
 \ee
where ${D}_{\rm M}(E_n,\vec{k})$ is the Matsubara boson Green function determined for discrete frequencies $E_n =2\pi nT$, $n=0,\pm 1,...$.

For non-relativistic systems the high temperature limit $|E_1|\gg |\omega_0^2|$ is usually realized. In this case one can put $n=0$ in the sum (\ref{Matsum}) and for $k_0\neq 0$, $\omega_0^2>0$, $\omega_0^2\ll m^2$ one obtains \cite{Voskresensky:1984rd}
\be
N(r)=N(0)e^{-r/(\sqrt{2}l_{\parallel})}\frac{\sin k_0 r}{k_0r}\,,\quad N(0)\simeq k_0T/[(4\pi)(-\alpha_4)^{1/2}|\omega_0|]\,.\label{NflCl}
 \ee
After the replacement $\omega_0^2\to -2 \omega_0^2$  the same result holds for $\omega_0^2<0$.
$N(r)$ is characterized by two length scales: it oscillates on a short scale $l_k\sim 1/k_0$ and smoothly falls off on a coherence length scale $l_{\parallel}$,  and $l_{\parallel}\gg l_k$ at least in the vicinity of the critical point when $|\omega_0^2|\ll m_b^2$. In case of the pion condensation both classical and quantum limits are relevant, see below Section \ref{PionFlSect}.

A good approximation to take into account fluctuation contribution
in the boson Green function $D$ is  inclusion of the diagram
\be\label{ladprtad1}
\includegraphics[width=2cm,clip=true ]{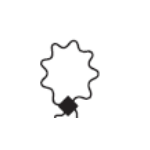}
 \ee
calculated with the full Green function of the boson ($iD^R$ is shown by the bold wavy line). The bold block shows  the effective boson-boson vertex  that contains the vacuum part $-i\Lambda_{\rm vac}$ and an in-medium contribution.  This term is given by the correlator $N(0)$ times the vertex.

The equation for the effective gap $\tilde{\omega}^2$ that after taking into account of fluctuations replaces the value $\omega_0^2$
is given by \cite{Voskresensky:1984rd}
\be
\tilde{\omega}^2=\omega_0^2 +\Lambda N(0, \omega_0\to \tilde{\omega})\,,
\ee
where now $N(0)\propto 1/\tilde{\omega}$ and does not reach zero. In the critical point there arises jump from the branch with $\tilde{\omega}^2>0$ to $\tilde{\omega}^2<0$ that is interpreted as the first-order phase transition to the state with the liquid-crystal or solid-like structured condensate. Only provided the value of the jump is not as large, with some accuracy one may continue to speak about the second order phase transition.

Note that the fluctuations at the phase transition to the state $k_0\neq 0$ can be considered as one of the reasons for the appearance of the $\propto h$ terms in the effective Lagrangian  (\ref{Omoper}), (\ref{Omoperreal}) and in $\hat{D}^{-1}\psi$ in equation of motion, since for $\omega_0\to 0$ one has
$\Lambda N(0, \omega_0\to \tilde{\omega})\psi \sim \psi/(\psi^*\psi)^{1/2}$.

Note that even the superconducting phase transition to the state $k_0=0$ proves to be of a weakly first order transition because of effects of the intrinsic fluctuating magnetic field, cf. \cite{Halperin1974}. In the Lagrangian there appears the term $\propto e^3\phi^3$.  Similar results hold for the phase transition from the A-smectic to a nematic liquid crystal.

For $k_0\neq 0$ the specific heat gets a contribution $C_V\propto 1/|\tilde{\omega}|^3$, i.e. it diverges stronger than in case of $k_0=0$, cf. \cite{Voskresensky:1981zd,Voskresensky:1982vd}, whereas the magnetic susceptibility proves to be convergent, $\chi_H\sim |\tilde{\omega}|$, cf. \cite{Voskresensky:1984rd}.

Besides the fluctuations of the amplitude of the order parameter, there exist fluctuations of the phase related to presence of the Goldstone modes. For $T\neq 0$ these modes destroy the ordering in one-dimensional condensates at very long distances. For condensates at $k_0=0$ it was shown by Peierls and Landau, cf. \cite{LL5}.  References  \cite{Grinstein81,Baym:1982ca}
demonstrated that  at very large distances for condensates at $k_0\neq 0$ also  $\langle\psi\rangle =0$.  However quasi-one dimensional condensates are not prohibited. Thereby, either $k_0 (x)$ should vary with the distance at large length scale \cite{Baym:1982ca}, or due to the phase fluctuations  the domains are formed \cite{Voskresensky:1984rd,MSTV90}.

\subsection{Transition from the order parameter to hydrodynamical variables}

\subsubsection{Schr\"odinger equation in hydrodynamical variables}
First, consider approximate solutions of the Schr\"odinger equation
\be
i\hbar \partial_t \Psi =-\frac{\hbar^2\Delta\Psi}{2m}+U\Psi\,,\label{SchrodSc}
\ee
 with $U(t,\vec{r})$ to be a very smooth function of $t$ and $\vec{r}$. In this case $U$ can be either a scalar field or a zero-component of  electromagnetic  field $eA_0$. Dependence on $\hbar$ is recovered in order to track quantum corrections, which would vanish in the limit $\hbar \to 0$.

Let us search solution in the form
\be
\Psi =Ae^{iS/\hbar}\,,\label{Psisem}
 \ee
 where the amplitude $A(t,\vec{r})$ and the phase  $S(t,\vec{r})$ are real quantities, being smooth functions of $t,\vec{r}$. Substituting  (\ref{Psisem}) in (\ref{SchrodSc}) and separating the real and imaginary parts we have
 \be
 -\partial_t S=\frac{(\nabla S)^2}{2m} +U -\frac{\hbar^2 \Delta A}{2mA}\,,\label{semiclS}
 \ee
 \be
 \hbar \partial_t A=-\frac{\hbar \Delta  S}{2m}A  -\frac{\hbar \nabla A\cdot\nabla S}{m}\,.\label{semiclA}
 \ee

 Introducing mass-density $\rho_\psi =m A^2$, and  the quantity $\vec{v}=\nabla S/m$  and multiplying (\ref{semiclA}) by $A$ we derive the continuity equation
\be
\partial_t \rho_\psi + \mbox{div} (\rho_\psi \vec{v})=0\,.
\ee
 Applying the operator $\nabla$ to Eq. (\ref{semiclS}) we obtain
\be
 \partial_t \vec{v}+ (\vec{v}\nabla)\vec{v}=-\frac{\nabla P_{\rm ext}}{\rho_\psi} =  -\frac{\nabla U}{m} +\frac{\hbar^2}{2m^2}\nabla\frac{\Delta \sqrt{\rho_\psi}}{\sqrt{\rho_\psi}}\,.\label{semiclSv}
 \ee

 A note is in order: $\Psi$ function is in general a complex function and it is not directly measurable thereby. However according to Eq. (\ref{Psisem}) it can be presented via two real functions. The amplitude $A$ is directly measurable, since it is  expressed via the mass-density ($A=\pm\sqrt{\rho_\psi/m}$, ``$-$" sign can be hidden in a constant phase). Flux, associated with  the velocity, $\rho_\psi\vec{v}=\rho\nabla S/m$, and, thereby, $\nabla S$ are also measurable.

 The term
\be\frac{\nabla P_{\rm quant}}{\rho_\psi}=-\frac{\hbar^2}{2m^2}\nabla\frac{\Delta \sqrt{\rho_\psi}}{\sqrt{\rho_\psi}}\label{LaplP}\ee
 is called the Laplace pressure. This surface term is a purely quantum contribution $\propto \hbar^2$.
 Neglecting it we formally arrive at the Euler equation of ideal hydrodynamics.

\subsubsection{Phase transition to the  state $k_0=0$}
Let us simplifying consideration take $k_0=\alpha_1=\alpha_2=\alpha_{32}=\alpha_{02}=\alpha_4 =0$, whereas $\alpha_{01}\neq 0$, $\alpha_{31}>0$, $\Lambda_{12}\neq 0$, $\Lambda_{11}=0$, $\Lambda_3=0$, $\Lambda_0 \neq 0$. i.e.,
\be
D^{-1}_R=-\omega_0^2+\alpha_{01}(i\partial_t -\omega_c)+\alpha_{31}\Delta\,,\quad
\psi^*\hat{\Lambda} \psi =\Lambda_0 |\psi|^2 +\Lambda_{12}i\partial_t |\psi|^2\,.\label{DLambdaexample}
\ee
A more general consideration can be found in \cite{Voskresensky:1993ux}.

 Put $\psi =\psi_0 e^{iS}$ in equation of motion,
where $\psi_0$ and $S$ are real quantities. Separating real and imaginary parts one obtains:
\be
-\alpha_{01}\psi_0 (\partial_t S +\omega_c) +\Lambda_{12}\psi_0\partial_t \psi_0^2 -
\omega_0^2 \psi_0 -\Lambda_0\psi_0^3+\alpha_{31}\Delta\psi_0-\alpha_{31}\psi_0 (\nabla S)^2=0\,,\label{NavierPrelim}
 \ee
\be \alpha_{01}\psi_0\partial_t\psi_0 +\alpha_{31}\mbox{div}[\nabla S\psi_0^2]=0\,.\label{continPrelim}
\ee

Introducing a mass-density-like variable, $\rho_\psi =\alpha_{01}m_{\rm ef}\psi_0^2$, velocity $\vec{v}=\nabla S /m_{\rm ef}$, and an effective mass, $m_{\rm ef}=\alpha_{01}/(2\alpha_{31})$, after applying the gradient operator to Eq. (\ref{NavierPrelim}) one arrives at  equation
\be
\partial_t \vec{v}+\nabla \frac{\vec{v}^{\,2}}{2}=
-\frac{\nabla P}{\rho_\psi}+\nabla \left[\frac{(4\eta/3 +\zeta)}{\rho_\psi}\frac{\mbox{div}(\rho_\psi \vec{v})}{\rho_\psi}\right]\,,\label{NavierHom}
 \ee
 which coincides formally with the ordinary Navier--Stokes equation  provided $\rho_\psi$ is a  very smooth function of $t$ and $\vec{r}$. From Eq. (\ref{continPrelim}) one arrives at the ordinary continuity equation
\be
\partial_t \rho_\psi +\mbox{div} (\rho_\psi\vec{v})=0\,.\label{continHom}
\ee
Here we introduced the shear, $\eta$, and bulk, $\zeta$, viscosity-like notations
\be
({4}\eta/3 +\zeta)/\rho^2_\psi =-\Lambda_{12}/(\alpha_{01} m_{\rm ef})^2\,,\label{eta-zeta-phi}
\ee
and
\be
\nabla P=-\frac{\rho_\psi}{\alpha_{01} m_{\rm ef}}\nabla\left[-\omega_0^2 -\frac{\Lambda_0\rho}{\alpha_{01} m_{\rm ef}}-\alpha_{01} \omega_c  +\frac{\alpha_{31} \Delta\sqrt{\rho_\psi}}{\sqrt{\rho_\psi}}\right]\,.\label{Phom}
\ee
Since the coefficient $4\eta/3 +\zeta$ plays a role of the effective viscosity, it should be positive and thereby one should take $\Lambda_{12}<0$. To avoid a possible misunderstanding   we should stress that derived equations only formally coincide with the hydrodynamical equations.

\subsubsection{Phase transition to the state $k_0\neq 0$. One-axis periodic system}

Take now $\psi =\psi_0 (t,\vec{r})e^{-i\omega_c t+\vec{k}_0 \vec{r}+iS(t,\vec{r})}$ and assume $\vec{k}_0\parallel z$.
Let
\be\hat{D}^{-1}_R =-\omega_0^2 +\alpha_{01}(i\partial_t -\omega_c)+\alpha_{41}(\Delta +k_0^2)^2\,,\label{DinhomOne}
\ee
\be\hat{\Lambda}|\psi|^2=  (\Lambda_0 +\Lambda_{1} i\partial_t) |\psi|^2\,.
\label{LambdaexpOne}\ee
We arrive at the equation of motion for $\chi =\psi_0 (t,\vec{r}) e^{iS(t,\vec{r})}$, cf. \cite{Voskresensky:1993ux},
\be
\alpha_{01}(i\partial_t -\omega_c)\chi -4\alpha_4(\vec{k}_0\nabla)^2\chi +\alpha_4 \Delta^2\chi -\omega_0^2\chi -
\Lambda_0 |\chi|^2\chi +\Lambda_{12}\chi \partial_t |\chi|^2=0\,.
\ee

Neglecting $\Delta^2 \psi_0$ term compared to $(\vec{k}_0 \nabla)^2\psi_0$ we obtain (for $\vec{k}_0\parallel z$)
\be
\partial_t \rho_\psi +\partial_z ( \rho_\psi v_z)\simeq 0\label{continLatHomOne}
\ee
with $\rho_\psi =\alpha_{01}\tilde{m}_{\rm ef}\psi_0^2$,  $\vec{v}=\nabla S /\tilde{m}_{\rm ef}$, where now the effective mass $\tilde{m}_{\rm ef}=\alpha_{01}/(-8\alpha_{41}k_0^2)$, $\alpha_{01}>0$, $\alpha_{41}<0$ and
\be
\partial_t \vec{v}+\nabla \frac{{v}_z^{\,2}}{2}\simeq
-\frac{\nabla P}{\rho_\psi}+\nabla \left[\frac{(4\eta/3 +\zeta)}{\rho_\psi}\frac{\partial_z(\rho_\psi {v}_z)}{\rho_\psi}\right]\,,\label{NavierLatHomOne}
 \ee
 with the kinetic coefficient $(4\eta/3 +\zeta)$ from (\ref{eta-zeta-phi}) but with ${m}_{\rm ef}\to \tilde{m}_{\rm ef}$ and
 \be
\nabla P=-\frac{\rho_\psi}{\alpha_{01} \tilde{m}_{\rm ef}}\nabla\left[-\omega_0^2 -\frac{\Lambda_0\rho_\psi}{\alpha_{01} \tilde{m}_{\rm ef}}-\alpha_{01} \omega_c -4\frac{\alpha_{41}k_0^2 \partial^2_z\sqrt{\rho_\psi}}{\sqrt{\rho_\psi}}\right]\,.\label{PinhomOne}
\ee

\subsubsection{Phase transition to  state $k_0\neq 0$. Anisotropic three-axis crystal}
Now, employing (\ref{coscoscos}) let us take
$\psi = \psi_0 e^{-i\omega_c +iS} \cos (k_{0x}x)\cos (k_{0y}y)\cos (k_{0z}z)$.
%\label{coscoscos}
We arrive at equation of motion for $\chi =\psi_0 (t,\vec{r}) e^{iS(t,\vec{r})}$,  cf. \cite{Voskresensky:1993ux},
\be
\alpha_{01}(i\partial_t -\omega_c)\chi -4\alpha_4(\vec{k}_{0x}^2\partial_x^2+ \vec{k}_{0y}^2\partial_y^2+\vec{k}_{0z}^2\partial_z^2)\chi +\alpha_4 \Delta^2\chi -\omega_0^2\chi -
(\frac{3}{4})^3\Lambda_0  |\chi|^2\chi +(\frac{3}{4})^3\Lambda_{12}\chi \partial_t |\chi|^2=0\,.
\ee

Neglecting $\Delta^2 \psi_0$ compared to $(\vec{k}_0 \nabla)^2\psi_0$ we obtain the continuity equation
\be
\partial_t \rho_\psi +\tilde{\nabla} (\rho_\psi\vec{v})=0\,,\label{continLatHom}
\ee
where $\tilde{\nabla} =(k_{0x}\partial_x, k_{0y}\partial_y, k_{0z}\partial_z)/k_0$,
$\tilde{\vec{v}}=\tilde{\nabla}S/\tilde{m}_{\rm ef}$,
and the Navier--Stokes equation
\be
\partial_t \tilde{\vec{v}}+\tilde{\nabla} \frac{\tilde{\vec{v}}^{\,2}}{2}=
-\frac{\tilde{\nabla} P}{\rho_\psi}+\tilde{\nabla} \left[\frac{(4\tilde{\eta}/3 +\tilde{\zeta})}{\rho_\psi}\frac{\tilde{\nabla}(\rho_\psi \tilde{\vec{v}})}{\rho_\psi}\right]\,,\label{NavierLatHom}
 \ee
 compare Eqs. (\ref{NavierLatHom}), (\ref{continLatHom}) with Eqs. (\ref{NavierHom}) and (\ref{continHom}),  derived in the homogeneous case.
Here
$$(4\tilde{\eta}/3 +\tilde{\zeta})/\rho^2_\psi =-(3/4)^3\Lambda_{12}/(\alpha_{01} \tilde{m}_{\rm ef})^2$$
and
 \be
\tilde{\nabla} P=-\frac{\rho_\psi}{\alpha_{01} \tilde{m}_{\rm ef}}\tilde{\nabla}\left[-\omega_0^2 -\frac{(3/4)^3\Lambda_0\rho_\psi}{\alpha_{01} \tilde{m}_{\rm ef}}-\alpha_{01} \omega_c -4\frac{\alpha_{41}k_0^2 \tilde{\nabla}^2\sqrt{\rho_\psi}}{\sqrt{\rho_\psi}}\right]\,.\label{Pinhom}
\ee
Note that in difference with the isotropic case,  here enters the operator $\tilde{\nabla}$ rather than $\nabla$.

Summarizing, one can consider dynamics of the order parameter employing the hydrodynamical variables and solving then the equations formally coinciding with the equations of the non-ideal hydrodynamics.

\subsubsection{Normal liquid}
 Standard hydrodynamical equations for the potential motion ($\mbox{curl}\,\vec{v}=0$) of the normal liquid can be derived in a similar fassion, cf. \cite{Voskresensky:1993ux}.  For that we may  introduce a physically small volume characterized by the common
collective coordinates of its center of inertia $\vec{r}$ considered at the time moment $t$. So, the $N$-particle $\Psi$ function characterizing particles of this volume can be presented as
\be\Psi_N =\psi_0 (t,\vec{r})\psi (t,|\vec{r}_i-\vec{r}_j|)e^{iS(t,\vec{r})+ i\xi (t,|\vec{r}_i-\vec{r}_j|)}\,,\label{Psimany}
\ee
where $i,j$ run $1,...,N$, $\psi_0$ and $S$ are smooth real functions of $(t,\vec{r})$ and $\psi$ and
$\xi$ are sharp real functions characterizing particles within the physically small volume.
Then  with the  expansion (\ref{DLambdaexample}) for the $\hat{D}^{-1}_R$ and $\Lambda$,   instead of (\ref{NavierPrelim}), (\ref{continPrelim})
we obtain the equations of motion, where operators  $\hat{D}^{-1}_R$ and $\hat{\Lambda}$ act on the function
(\ref{Psimany}).
Multiplying these equations of motion from the left on $\psi (t,|\vec{r}_i-\vec{r}_j|)e^{i\xi (t,|\vec{r}_i-\vec{r}_j|)}$ and averaging over $\int d^3\vec{r}_1... d^3\vec{r}_N$, in case of the isotropic system we recover
the equations, which have the same form as (\ref{NavierPrelim}), (\ref{continPrelim}), but with the coefficients  $\alpha_i$, $\Lambda_i$ replaced to the short-range averaged quantities $\bar{\alpha}_i $,
$\bar{\Lambda}_i $. Finally we arrive at the same Eqs. (\ref{continHom}), (\ref{eta-zeta-phi}), (\ref{Phom}) but with  $\alpha_i\to \bar{\alpha}_i $,
$\Lambda_i\to \bar{\Lambda}_i $, $\rho_\psi\to \rho=\bar{\alpha}_{01}^2\psi_0^2/(2\bar{\alpha}_{31})$. Similarly, for the one-axis crystal we arrive at Eqs. (\ref{continLatHomOne}), (\ref{NavierLatHomOne}), (\ref{PinhomOne})
and in case of the lattice (\ref{coscoscos}) at Eqs. (\ref{continLatHom}), (\ref{NavierLatHom}), (\ref{Pinhom})
with the  modifications   $\alpha_i\to \bar{\alpha}_i $,
$\Lambda_i\to \bar{\Lambda}_i $, $\rho_\psi\to \rho$. The coefficients should be found from the comparison of the model predictions with the data.

\section{First-order phase transitions in slowly evolving  systems}\label{First-order-transitions}
\subsection{Typical pressure--density behavior at  first-order phase transitions }\label{TypPressureSetion}

\subsubsection{Transition of liquid--vapor type}
The first-order phase transitions may occur in various systems. These are the ordinary  liquid--vapor phase transition in water and other liquid- and gaseous systems, the nuclear liquid--vapor phase transition, whose signatures are manifested in low-energy heavy-ion collisions, cf. \cite{SVB,Skokov:2009yu,Skokov:2010dd,Voskresensky:2010gf}, feasibly the quark--hadron phase transition in baryon-rich matter, cf. \cite{Skokov:2009yu}, possible  transition to the $\Delta$-rich  matter in the heavy-ion collisions \cite{Kolomeitsev:2016ptu}, all phase transitions occurring to the states $k_0\neq 0$, such as
 phase transitions to p wave pion- and kaon-  condensate states \cite{MSTV90,Kolomeitsev:1995xz,Kolomeitsev:2002pg}, may be phase transition to the  charged $\rho^-$ condensate state in rather massive neutron stars \cite{Voskresensky:1997ub,KolVosk2005,Kolomeitsev:2017gli}, etc.

In compact stars, appearance of a strong phase transition may result
in a second neutrino burst, if the transition occurred right after a supernova explosion at a  hot stage of the compact star evolution. A possibility of the delayed second neutrino burst in the  event  supernova SN 1987A has been considered in  \cite{Voskresensky:1987ut,Haubold:1988uu,MSTV90}. It might be associated with a significant delay of the heat transport to the neutron-star surface, if the system is close to the pion--condensate phase transition. Recently, new arguments have been expressed for that namely two neutrino bursts were measured during 1987A explosion, one delayed respectively the other one by 4.7 h, cf. \cite{Galeotti:2016uum}.
%This has been added - Pls confirm
The second neutrino burst could then be related to the phase transition of the neutron star to the pion condensate state.  In addition, a phase transition to the pion- or to the kaon--condensate state could also occur during tens-second-period of the neutron star formation or later resulting in a blowing off a star matter \cite{Migdal:1979je,MSTV90}. In old neutron stars, the first-order phase transitions could also result in a strong star-quakes  \cite{MSTV90,Prakash:1996xs}.

In the cases of such  systems as the liquid water -- vapor and the nuclear fireball formed at low-energy collisions of isospin-symmetric nuclei  one deals with the same species, in the latter case  with nucleons, although at different densities.
So, in both cases one deals with the one conserved  charge: the number of particles in the first case and the baryon number in the second case. Such transitions in statistical physics are named the congruent transitions, cf. \cite{Iosilevskiy:2010qr,Hempel2013}. So we will speak about the liquid--vapor (or differently saying liquid--gas) transition, if the variation of the density  can be considered as the order parameter. Then one may deal with the equations of the non-ideal hydrodynamics to describe the dynamics of the phase transition. We will also conjecture, cf. \cite{Skokov:2009yu}, that  the hadron--quark phase transition. as the ordinary liquid water--vapor transition, can be described in $n(t,\vec{r}),T(t,\vec{r})$ local variables.

\begin{figure}\centering
\includegraphics[width=5.8cm,clip]{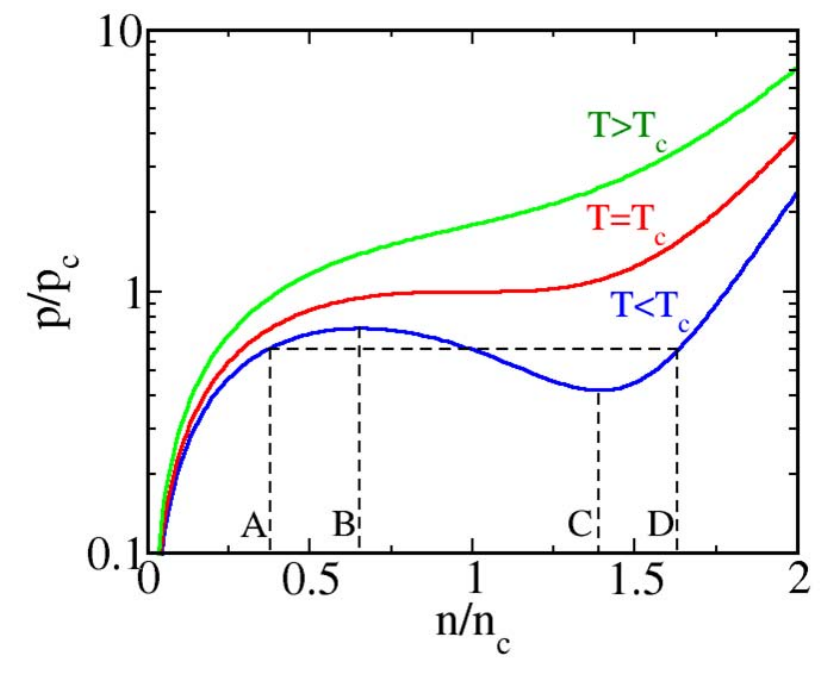}
\caption{Schematic pressure isotherms as  functions  of the number density at a liquid--vapor-like phase transition. $P_c$, $n_c$ and $T_c$ are the pressure, number density and  temperature at the critical point. Horizontal dashed line shows the Maxwell construction.
 }\label{VanderWaals}
\end{figure}

The typical   pressure--density behaviour has a van der Waals form.
The pressure isotherms as  functions of the density   are shown in Fig. \ref{VanderWaals}, cf., e.g., \cite{LL5}. In the thermodynamical  equilibrium, a necessary condition for the stability is that pressure $P$ does not increase with the volume $V_3$.
The interval AB corresponds to a metastable supercooled vapor (SV), when on any horizontal line $\mu_{\rm V}>\mu_{\rm L}$, and the interval CD, relates to a  metastable overheated liquid (OL), when on any horizontal line the chemical potential in liquid state enlarges that one in vapor phase, $\mu_{\rm L}>\mu_{\rm V}$. The~interval BC labels an unstable spinodal region, where  excitations grow exponentially. At the first-order phase transition  in case of one conserved charge, thermally equilibrium configurations  belong to the Maxwell construction (MC) shown by the  horizontal dashed line, on which the chemical potentials $\mu_A$ and $\mu_D$ are equal to each other.  The areas separated by  the curve $P(1/n)$ and the MC horizontal line  prove to be equal \cite{LL5}.

\begin{figure}[h]
\centering
\includegraphics[width=0.35\textwidth]{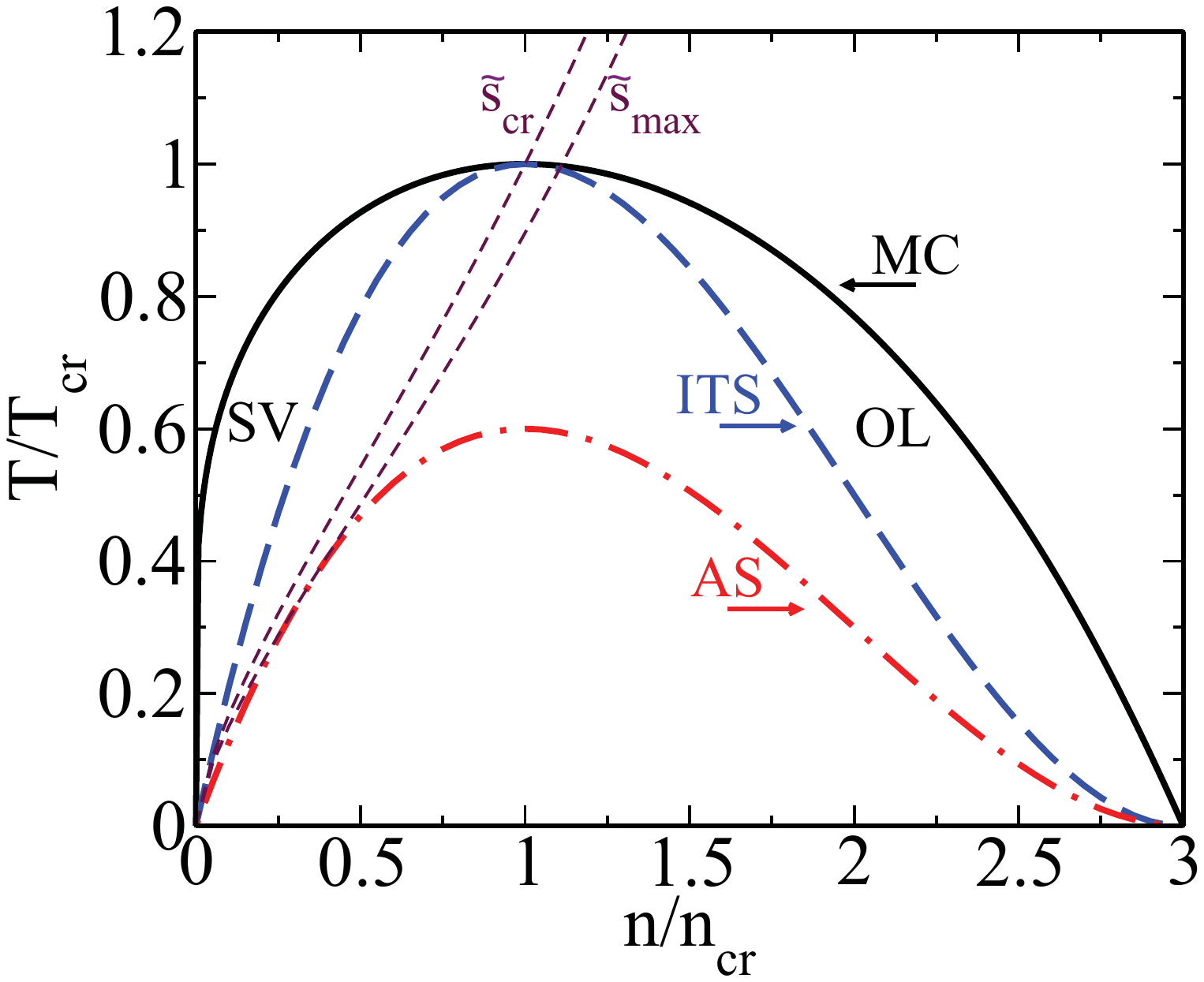}
\caption{ The phase diagram of the van der Waals equation of state on
$T(n)$-plane. The bold solid,  dashed and dash-dotted curves show boundaries of the MC, the spinodal
region at $T=$const and $\tilde{s}=$const, respectively. The short
dashed lines show two adiabatic trajectories of the system evolution: the curve labeled
 $\tilde{s}_{\rm cr}$ passes through the critical point; the curve $\tilde{s}_{\rm max}$, through the maximum pressure point $P(n_{P,\rm max})$ on the
 $P(n)$ plane, cf.
%. Figure is adopted  from
\cite{Skokov:2010dd}.
 }
\label{spinT}
\end{figure}

Fig. \ref{spinT} shows  the plot of $T/T_{\rm cr} = f (n/n_{\rm cr})$ for the van der Waals equation of state.  Trajectories of the system undergoing an approximately adiabatic  cooling are shown by the short dashed lines $s_{\rm cr}$ and $s_{\rm max}$, where $\tilde{s}\equiv s/n \simeq$ const, $s$ is the entropy density.
The upper convex curve (MC, bold solid line) demonstrates the boundary of the MC. The
bold dashed line, ITS, shows the boundary of the isothermal spinodal  region, and the  bold dash-dotted
curve, AS, indicates the boundary of the adiabatic spinodal region.
%At the ITS line $u_T^2=(\partial P/\partial \rho)_T=0$ and at AS line, $u_{\tilde{s}}^2=(\partial P/\partial \rho)_{\tilde{s}}=0$, where $u_T$ and $u_{\tilde{s}}$ are the isothermal and adiabatic sound velocities, respectively,  $n$ is the number density.
The supercooled vapor (SV) and the overheated liquid (OL) regions
are situated between the MC and the ITS curves, on the left and on the right, respectively.
For $\tilde{s}_{\rm cr}>\tilde{s}>\tilde{s}_{\rm MC2}$, where
$\tilde{s}_{\rm cr}$ is the  value  of the specific entropy $\tilde{s}$
at the critical point, and (in the given example) the line
with $\tilde{s}_{\rm MC2}$  passes through the point $n/n_{\rm cr}=3$
at $T=0$, the system traverses the OL state (the region OL in Fig. \ref{spinT}), the ITS region (below the ITS line) and the AS region (below the
AS line). For $\tilde{s}>\tilde{s}_{\rm cr}$ the system trajectory
passes through the SV state (the region SV in Fig. \ref{spinT}) and  the
ITS region. At the ITS line $$u_T^2=(\partial P/\partial \rho)_T=0$$
 and at AS line
 $$u_{\tilde{s}}^2=(\partial P/\partial \rho)_{\tilde{s}}=0\,,$$ where $u_T$ and $u_{\tilde{s}}$ have the meaning of the isothermal and adiabatic sound velocities, respectively,   $n$ is the number density. Below the ITS line (and above the AS line) one has  $u_T^2
<0$, $u_{\tilde{s}}^2
>0$.
After the system enters the region of the fist-order phase transition of the liquid--vapor type the approximation of  constant entropy fails and  the description of the dynamics of the  system  needs solution of non-ideal hydrodynamical equations \cite{Skokov:2009yu,Skokov:2010dd}.
Similarly, the description of the dynamics of the second-order phase transition needs solution of non-ideal hydrodynamical equations in the case, when the density and the temperature (or entropy) can be considered as appropriate order parameters. Moreover, let us note that at the expansion of the nuclear fireball prepared in a heavy-ion collision first the trajectory of the system  crosses the ITS and only at lower $T$ it crosses the AS line. Thus, the  vapor -- liquid phase transition calculated within the non-ideal hydrodynamics starts at a higher $T$ than the transition calculated within  the ideal hydrodynamics. In the latter case, e.g., in the so called three-fluid ideal hydrodynamics  in order to somehow simulate the viscosity effects one introduces the effective friction forces \cite{Ivanov:2005yw}.

 \subsubsection{Mixed phases and pasta}

The liquid -- vapor congruent phase transition, which we discussed, is well known issue in condensed matter physics as well as the non-congruent first-order phase transitions. In the latter case two or more charges are conserved.
In case of the asymmetric nuclear matter occurring in collisions of heavy nuclei the proton and neutron fractions are different in both phases due to significant density dependence of the symmetry energy. Thereby two conserving charges, the baryon charge and isospin, are relevant quantities. In case of the  compact stars the  conserving charges are the baryon and electric charges.    The Coulomb force is long-distant one. As the consequence, the Coulomb  contribution to the energy of the nucleus grows with the mass number $A$ and the charge $Z$   stronger (as $\propto Z^2 /A^{1/3}$) than  the surface energy term $\propto A^{2/3}$ that leads to the preference of the non-spherical form and fission of heavy nuclei. Interplay between the surface tension and the Coulomb interaction, which is screened in the matter on the Debye length,   results in a possibility of appearance of structures of different geometry, as droplets, rods and slabs and configurations of a more whimsical form, cf. \cite{Ravenhall1983,Lorenz1993,Watanabe2000,Voskresensky:2001jq,VYT2003}.

It is commonly accepted that the outer crust of the neutron star, at $\rho< 4\cdot 10^{11}$ g$/$cm$^3$, contains the Coulomb lattice of Fe nuclei and electrons  compensating the charge. At the neutron density $\rho > 4\cdot 10^{11}$ g$/$cm$^3$ neutrons  drip out of nuclei. Nuclei organized in the lattice are surrounded by the neutron gas, cf. \cite{Baym1971,Negele1973}. With increasing baryon density in an interval, $0.1n_0\lsim n<n_m\simeq 0.7n_0$,  the  pasta phase may arise \cite{Ravenhall1983,Lorenz1993,Watanabe2000,Maruyama:2005vb}. In some interval of  densities above $n_m$ the state consists of  the neutron Fermi liquid, and a shallow proton Fermi sea with the charge compensated by  electrons   and $\mu^-$. Then for $n\gsim (2-3)n_0$ there may appear
the pion and
anti-kaon condensates in states  with $k_0\neq 0$, cf. \cite{Migdal78,MSTV90,Kolomeitsev:1995xz,Kolomeitsev:2002pg}, as well as $K^-$ condensate in the state  with $k_0=0$, cf. \cite{TPL,T95,GS99,CGS00},
and probably for a somewhat larger density the phase transition to the quark matter state may occur, cf.  \cite{Heiselberg,VYT2003,Klahn:2006iw,Maruyama2008,Ayriyan:2017nby,Maslov:2018ghi}. It has been argued that all mentioned transitions are, most likely,  the first-order transitions.

Glendenning rised the issue, whether systems composed of charged particles consist mixed phases instead of the configuration described by the Maxwell
construction \cite{G92}. In particular, the possibilities of the hadron ($npe$) -
kaon condensate  ($npeK_{cond}$) and hadron - quark
mixed phases were discussed. The existence of such kind of mixed phases
in dense neutron star interiors would have important consequences
for the equation of state,  glitch phenomena and  neutrino transport and emissivity, cf.
\cite{RBP00,Glendenning2001,Sonoda,Yasutake:2012hy,Alcain:2014fma}. Reference \cite{Yasutake:2012hy} suggested that at the neutrino trapping stage in compact stars the pasta phase may behave similarly to amorphous  matter. Already very low temperatures (typically $T\gsim 0.1 $ MeV) may simulate transitions between different geometrical structures in the pasta phase of the neutron star crusts \cite{Xia2022}. Recall  here that the isospin-symmetric nuclear matter   at $T\neq 0$ already for $n\gsim n_{c}^{(1)}\sim (0.5-0.8)n_0$ begins to behave as an  amorphous  matter due to strong pion fluctuations with $k_0\neq 0$, cf. \cite{Dyg1,MSTV90,Voskresensky:1989sn,Voskresensky:1993ud}.

First works \cite{G92,GS99,CGS00,Glendenning2001} assumed that the MC
is always unstable due to the inequality of the local electron
chemical potentials of two phases $\mu_e^{\rm I}(n^{\rm I})$ and $\mu_e^{\rm II}(n^{\rm II})$ and, thus,  a possibility
for particles to fall down from the  level characterized by the higher electron chemical potential of the one
phase to the lower level of the other phase. It was thought that  existence of a wide region of the mixed phase  determined by fulfilment of the Gibbs conditions is inevitable.  However it has been observed that in some models the Gibbs condition of equality of electron chemical potentials of two phases can't be fulfilled at all \cite{PREPL00,MYTT,ARRW}, whereas
conditions for the MC are fulfilled.
Moreover  \cite{Ravenhall1983}  demonstrated the possibility of
existence of the structures of different geometry  determined by
competition between the Coulomb energy and the surface energy of
droplets. Then  \cite{HPS93}  and \cite{NR00}  applied these ideas to the description of the mixed phase for the hadron--quark and kaon phase transitions, respectively. It was demonstrated that there should exist a critical value of the surface tension, above which structures are not permitted.

References  \cite{Voskresensky:2001jq,VYT2003} clarified that  the Gibbs condition of equality of the charged chemical potentials in two phases $\mu_e^{\rm I}=\mu_e^{\rm II}$, as
it was formulated for spatially homogeneous configurations, has no meaning
in the application to charged systems of a small size, if one does not incorporate properly the electric field effects. It only fixes the level, from which one counts the electric potential. Only the gauge invariant quantity $\mu_e -V$ should enter the equations of motion including the Poisson equation for the electric potential $V$. References \cite{Voskresensky:2001jq,VYT2003,Maruyama:2005vb,MTVTEC,Maruyama2008,Xia2021} demonstrated that the careful consideration of the electric field  including the Debye screening effects allows to resolve the mentioned contradiction.  For the quark--hadron first-order phase transition, it was found that $\sigma_c \simeq 60$ MeV$/$fm$^2$, so for $\sigma >\sigma_c$ the mixed quark--hadron phase is not realized. It was shown that, in  cases of the quark-hadron and kaon condensate pastas, the $P(n)$ behaviour is  close to that given by the MC, cf. \cite{MTVTEC,Maruyama2008,Xia2021,Lugones:2021tee}.

It is interesting to notice that elastic properties of the pasta phases might be similar to those of liquid crystals, cf.
\cite{PethickPotekhin1998,Pethick:2020aey}. Proton superconductivity still complicates
structure of the pasta phase in neutron star crusts \cite{Pethick2021super}.
A possibility of a manifestation of some features of the pasta phase in heavy-ion collisions was recently discussed in \cite{Maslov:2019dep}.

The dynamics of the pasta phase
transition was studied  with the help of the  formalism of molecular dynamics, cf. \cite{Horowitz2005,Horowitz2008,Watanabe2009,Watanabe2012,Horowitz2013,Alcain:2017xbr,Lopez:2020zne}. References \cite{Watanabe2009,Watanabe2012}  demonstrated that in supernova matter a lattice of rod-like nuclei is formed from a bcc lattice by compression. It was demonstrated that in the transition process the system undergoes a zigzag configuration of elongated nuclei, which are formed by a fusion of  original spherical nuclei.
Artificial stretching rates were employed, cf. \cite{Horowitz2013},
whereas  more realistic calculations are still required.

In a more straight way the dynamics of the formation of the pasta structures can be studied using Eq. (\ref{Leqorderparam}). Within relativistic mean-field models the mean meson fields and the electric potential can be considered as the order parameters. Such a way (at ignorance of the white noise terms) was sketched in \cite{Maruyama:2005vb}. However the kinetic rates $\Gamma_i$  cannot be found within the mean-field approach. They are expressed via the transport coefficients.  The latter should follow from the analysis of experimental information and the study of relevant microscopic processes. In application to heavy-ion collisions, different estimates have been done, cf. \cite{Khvorostukhin:2010aj,Khvorostukhin:2009pe,Chakraborty:2010fr,Khvorostukhin:2010cw,
Khvorostukhin:2011mt,Khvorostukhin:2012kw,Albright:2015fpa,Chakraborty:2016ttq,Den:2022abp} and refs therein.  Various evaluations of the shear and bulk viscosities have been also performed in case of the neutron star matter,  cf. \cite{Kolomeitsev:2014epa,Kolomeitsev:2014gfa} and refs. therein. Their knowledge is important, e.g., for the  description of the neutrino transport and the $r$- mode relaxation. In case of the pion condensate phase transition the $\Gamma$- rate is determined by the imaginary part of the pion polarization operator, cf. \cite{Ivanov:2000ma}. However this ambitious program still needs much more effort.

\subsection{Transition  between homogeneous   configurations}
\subsubsection{Phenomenological model}
Let us assume $k_0=0$, $\alpha_{02}=\alpha_1=\alpha_2=\alpha_{32}=\alpha_4 =0$, $\alpha_{31}>0$. Then the part of the  free energy dependent on the order parameter is as follows \cite{Voskresensky:1993ux}
\be
\delta F[\psi]=\int d^3 x [\alpha_{31}|\nabla \psi|^2+\omega^2_0 |\psi|^2 +\Lambda_0 |\psi|^4/2 -he^{im\pi}\psi(\psi^*/\psi)^{1/2}-he^{im\pi}\psi^*(\psi/\psi^*)^{1/2}\,]\,.\ee

Equation (\ref{Leqorderparam}) yields
\be
-\alpha_{31}\Delta \psi +\omega_0^2 \psi +\Lambda_0 |\psi|^2\psi -he^{im\pi}(\psi/\psi^*)^{1/2}=-\Gamma^{-1}\partial_t\psi\,.
 \ee

 Introducing dimensionless variables
 $$\psi =e^{i\phi}\psi_0\chi\,,$$  $\psi_0=\sqrt{-\omega_0^2/\Lambda_0}$, $\phi$ is arbitrary real constant phase, $\tilde{\Delta} =\Delta/l_0^2$,
 $\tilde{h}=-2h/(\omega_0^2 \psi_0)$, $|\tilde{h}|\ll 1$, $l_0=\sqrt{2\alpha_{31}/(-\omega_0^2)}$, $t_0=-2/(\Gamma \omega_0^2)$, for $\omega_0^2<0$      we get equation
 \be
 \chi^{'}_{\tau}=\tilde{\Delta}\chi +2\chi (1-|\chi|^2)+\tilde{h}\,,\label{eqMchi1}
  \ee
 where $\tau=t/t_0$, $\tilde{\vec{r}}=\vec{r}/l_0$.

 \subsubsection{ Limit $h=0$ (second-order phase transition)}
 In case of the second-order phase transition (for $h=0$)   Eq. (\ref{eqMchi1}) simplifies as
 \be
 \chi^{'}_{\tau}=\tilde{\Delta}\chi +2\chi (1-|\chi|^2)\,.\label{eqMchih0}
  \ee
 For $|\chi|\ll 1$ it can be linearized and gets the solution of the form (in dimensionless variables)
\be\chi =\chi_0 e^{i\tilde{k}\tilde{x}+(2-\tilde{k}^2)\tau}\,,\label{link}\ee for arbitrary  $\chi_0\neq 0$.

Equation (\ref{eqMchih0})
 has also the spatially uniform solution
 \be
 \chi^2 (\tau >0)=\frac{\chi_0^2}{\chi_0^2+ (1-\chi_0^2)e^{-4\tau}}\,,\label{secorder}
  \ee
 where $\chi_0$ is the real constant quantity describing an initial uniform  distribution $\chi (0)=\chi_0$. If  $\chi (0)>0$, then  the solution $\chi(\tau >0)$ reaches the equilibrium value  $\chi (\tau\to \infty)= 1$, and $\chi (\tau\to \infty)= -1$ for $\chi (0)<0$.
 Both solutions correspond to the same energy. We see that spatially inhomogeneous small perturbations  satisfying Eq. (\ref{link}) with $\tilde{k}^2<2$  grow in time with a smaller rate compared to the spatially uniform fluctuations. For $\tilde{k}^2>2$ excitations are damped.
Note that solutions (\ref{link}) and (\ref{secorder}) are limiting cases  of a more general solution
\be
 \chi (\tau >0)=e^{i\tilde{k}\tilde{x}}\sqrt{\frac{(2-\tilde{k}^2)\chi_0^2 }{2\chi_0^2+ (2-\tilde{k}^2-2\chi_0^2)e^{-2\tau (2-\tilde{k}^2)}}}\,.\label{secorder1}
  \ee

 Equation (\ref{secorder}) continues to hold for $h\neq 0$ provided $|\chi_0|\gg |h|$. In this case it describes growth of the initially spatially-uniform configurations within the spinodal region
 at the first-order phase transition to the state $k_0=0$.

  In case of the real order parameter  we could start with Eq. (\ref{Omoperreal}) yielding the equation of motion in the form
\be
-\alpha_{31}\Delta \psi +\omega_0^2 \psi +\Lambda_0 \psi^3 -h=-\Gamma^{-1}\partial_t\psi\,.\label{GLcos}
 \ee
 For $h=0$  this equation, being expressed in appropriate dimensionless variables, has the same uniform solution  (\ref{secorder}).

Note here that, as it will be shown in Section \ref{hydro},  equations describing the phase transition of the gas--liquid type within  the non-ideal hydrodynamics, cf. Eqs. (\ref{dimens}), (\ref{v-tS}) below,  are in general case not reduced to the Ginzburg--landau  Eqs. (\ref{eqMchi1}), (\ref{GLcos}).

\subsubsection{Transition from metastable to stable state}
 {\bf{Angle-independent solutions.}}
 For spherically symmetric configurations assuming $\omega_0^2<0$ and using dimensionless variables $\psi =e^{i\phi}\psi_0\chi$, $\psi_0=\sqrt{-\omega_0^2/\Lambda_0}$, $\phi$ is arbitrary real constant phase,
 $\tilde{h}=-2h/(\omega_0^2 \psi_0)$, for $|\tilde{h}|\ll 1$, $l_0=\sqrt{2\alpha_{31}/(-\omega_0^2)}$, $t_0=-2/(\Gamma \omega_0^2)$,    we derive equation
 \be
 \chi^{'}_{\tau}=\chi^{''}_{\tilde{r}}+\frac{n-1}{\tilde{r}}\chi^{'}_{\tilde{r}}+2\chi (1-\chi^2)+\tilde{h}\,,\label{eqMchi}
  \ee
 where $\tau=t/t_0$, $\tilde{r}=r/l_0$, $r=x$ for $n=1$; $r=\sqrt{x^2+y^2}$ for $n=2$ and
 $r=\sqrt{x^2+y^2+z^2}$ for $n=3$. As we have discussed, to get this equation we took $m=0$ for $\chi (r)>0$ and $m=1$ for $\chi (r) <0$. In this case the same equation follows from variation of the functional (\ref{Omoperreal}) introduced from the very initial for the purely real field.

 The free energy  of the seed of the stable phase inside the metastable medium counted from
 the volume free energy, $F_{\rm V}=-\frac{\omega_0^4}{2\Lambda_0}V_3$,
 is as follows
 \be
 \delta F_{\rm seed} =E_0 \int d^3 x\left[\frac{(\nabla_{\tilde{r}}\chi)^2 +(\chi^2-1)^2}{2}-\tilde{h}\chi \right]\,,\quad E_0 =\Lambda_0\psi_0^4 \,.\label{FreeEnSeed}
 \ee
 Solution of Eq. (\ref{eqMchi}), $\chi =1+\tilde{h}/4$, describes stable phase, $\delta F_{\rm seed}= [-E_0 \tilde{h}+O(\tilde{h}^2)]V_{\rm seed}$,  and $\chi =-1+\tilde{h}/4$ describes metastable phase, $\delta F = [E_0 \tilde{h}+O(\tilde{h}^2)](V_3-V_{\rm seed})$. Here $V_3$ is the total volume, $V_{\rm seed}$ is the volume of the seed, $V_{\rm seed}=4\pi r_0^3(t)/3$ for spherical droplets.
 Let us seek solution of Eq. (\ref{eqMchi}) in the form
 \be
 \chi =-\mbox{tanh}(\tilde{r}-\tilde{r}_0(\tau))+\tilde{h}/4\,,\label{apsolchi}
  \ee
  such that solution  for $\tilde{r}_0(\tau))-\tilde{r}\gg 1$ describes the stable phase ($\chi =1+\tilde{h}/4$ ) and for $\tilde{r}-\tilde{r}_0(\tau))\gg 1$, the metastable phase.
Substituting Eq. (\ref{apsolchi}) in Eq. (\ref{eqMchi}), where  we approximate the curvature term  as
$\frac{n-1}{\tilde{r}}\chi^{'}_{\tilde{r}}\simeq \frac{n-1}{\tilde{r}_0}\chi^{'}_{\tilde{r}}$, we arrive at the equation
\be
\frac{d\tilde{r}_0(\tau)}{d\tau}=\frac{3\tilde{h}}{2}-\frac{n-1}{\tilde{r}_0(\tau)}\,.
\label{critsizeseed}
 \ee
For $n=1$ solution
\be
\chi =-\mbox{tanh}(\tilde{r}-\tilde{r}_0(0)-3\tilde{h}\tau/2)+\tilde{h}/4
 \ee
describes growth of the seed heaving the form of a slab, where now $\tilde{r}=|x|/l_0$,
$\tilde{r}_0(0)=|x_0(0)|/l_0$, $2|x_0(0)|$ is the initial size of the slab-seed. A slab of the stable phase (region $\chi >0$, $\tilde{r}-\tilde{r}_0(0)-3\tilde{h}\tau/2<0$) of arbitrary initial  size, being formed  in a fluctuation inside a metastable phase ($\tilde{r}-\tilde{r}_0(0)-3\tilde{h}\tau/2>0$), begins to grow.  Following (\ref{FreeEnSeed}) the gain in the free energy is
\be
\delta F_{\rm seed}^{\rm slab} =\frac{4}{3}E_0 l_0 -2E_0 l_0\tilde{h}\tilde{r}_0(\tau)\,,
 \ee
where $\tilde{r}_0(\tau)=\tilde{r}_0 (0)+3\tilde{h}\tau/2$.

For $n\neq 1$, seeds of stable phase  of the under-critical size, being initially formed inside the metastable phase, are then dissolved, whereas seeds of the overcritical size grow. According to Eq. (\ref{critsizeseed}), the critical size of the seed is given by
\be
r^{\rm cr}_0=\frac{2(n-1)l_0}{3\tilde{h}}=\frac{(2\alpha_{31})^{1/2}2(n-1)}{3|\omega_0|\tilde{h}}\gg 1/|\omega_0|\,.
 \ee
 The critical size of the initial seed also can be found by minimization of the free energy.
For $n=3$ (when seeds are spherical droplets), as it  follows from (\ref{FreeEnSeed}) the gain in the free energy is given by
\be
\delta F_{\rm seed}^{\rm drop} =\frac{16\pi}{3}E_0 l_0^3 \tilde{r}_0^2 (\tau) -\frac{8\pi}{3} E_0 \tilde{h}l_0^3 \tilde{r}_0^3(\tau)\,.\label{freeendrop}
 \ee

 Probability of the formation of a  seed of the stable phase of the critical size  inside the metastable phase is as follows $W\sim  e^{-\delta F(r^{cr}_0)/T}$ and the typical time needed to prepare a droplet  seed is
 \be
 t_W \sim (\Gamma \Lambda_0 \psi_0^2)^{-1}\mbox{exp}\left[\frac{64\pi({2\alpha_{31}})^{3/2}|\omega_0|^7}{3^4\Lambda_0^2 h^2T}
 \right]\,.
  \ee

{\bf{Weakly non-spherical  solutions.}}
Let initial seed of the stable phase is slightly non-spherical. Then
$\tilde{r}_0$ depends not only on $\tau$ but also on angles $\theta$ and $\phi$ in spherical coordinates,
\be
\tilde{r}_0(\tau, \theta, \phi)=\sum_{lm}\tilde{r}_0^{lm}(\tau)Y_{lm}(\theta, \phi)\,,
 \ee
 where $Y_{lm}(\theta, \phi)$ are spherical functions. Assume $\tilde{r}_0^{lm}$ for $l\geq 1$ be small. Then equation of motion for the seed renders
 \be
 \chi^{'}_{\tau}=\chi^{''}_{\tilde{r}}+\frac{2}{\tilde{r}_0 (\tau, \theta, \phi)}\chi^{'}_{\tilde{r}}+2\chi (1-\chi^2)+\tilde{h}-\frac{\hat{l}^2}{\tilde{r}_0^2 (\tau, \theta, \phi)}\,,\label{eqMchiNsp}
  \ee
 where $\hat{l}$ is the operator of the angular momentum.
 The solution is as follows \cite{Patashinsky},
 \be
 \chi =-\mbox{tanh}(\tilde{r}-\tilde{r}_0(\tau,\theta,\phi))+\tilde{h}/4\,
  \ee
and
\be
\frac{d\tilde{r}_0^0(\tau)}{d\tau}=\frac{3\tilde{h}}{2}-\frac{2}{\tilde{r}_0^0(\tau)}\,,\quad
\frac{d\tilde{r}_0^l(\tau)}{d\tau}=\frac{(2-l(l+1))\tilde{r}_0^l(\tau)}{(\tilde{r}_0^0(\tau))^2}\,
\,\label{critsizeseed1}
 \ee
 with $l\neq 0$.
From here
\be
\tilde{r}_0^l(\tau)=\tilde{r}_0^l(0)\mbox{exp}\left(\int\frac{(2-l(l+1)}
{[\tilde{r}_0^0(t')]^2}dt'\right)
 \ee
and modes with $l>1$ are damped with growing time. Thereby initially a weakly non-spherical seed becomes spherical with passage of time. Undamped first harmonic, $l=1$, describes displacement of the seed as a whole. For $\tilde{h}\sim 1$ initially spherical seeds may acquire a non-spherical form, cf. \cite{Devyatko1987,Voskresensky:1993ux}.

{\bf{Strongly non-spherical configurations.}}
Some configurations, which conserve their form during the first-order phase transition from metastable phase to the stable one, were found in \cite{Voskresensky:1993ux}.
For example, the seeds of the pyramidal form are growing following the law
\be
\chi =-\mbox{tanh}\left[
\frac{|\tilde{x}|+|\tilde{y}|+|\tilde{z}|}{\sqrt{3}}-\frac{3\tilde{h}\tau}{2}\right]
+\frac{\tilde{h}}{4}\,.
 \ee
Such seeds grow with $\sqrt{3}$ times higher velocity than
spherical droplets and slabs and, as slabs, pyramidal seeds have no critical size.

The cone-like seeds grow as
\be
\chi =-\mbox{tanh}\left[
\frac{|\tilde{z}|+\tilde{\rho}}{\sqrt{2}}-\frac{3\tilde{h}\tau}{2}\right]
+\frac{\tilde{h}}{4}\,,
 \ee
 where $\tilde{\rho}=\sqrt{\tilde{x}^2+\tilde{y}^2}$. These seeds grow
with $\sqrt{2}$ times higher velocity than
spherical droplets and slabs and, as slabs and pyramidal seeds, they have no critical size.

Rod-like seeds grow as follows
\be
\chi =\mbox{tanh}\left[|\tilde{z}|-\tilde{z}_0-\frac{3\tilde{h}\tau}{2}\right]
\mbox{tanh}\left[\tilde{\rho}-\tilde{\rho}_0(\tau)\right]
+\frac{\tilde{h}}{4}\,,
 \ee
where
\be
\tilde{\rho}_0(\tau)+\frac{2}{3\tilde{h}}\ln \frac{\tilde{\rho}_0(\tau)-\tilde{\rho}_c}{\tilde{\rho}_0(0)-\tilde{\rho}_c}=\tilde{\rho}_0(0)+
\frac{3\tilde{h}\tau}{2}\,,
 \ee
$\tilde{\rho}_c =2/(3\tilde{h})$, whereas it follows that ${\rho}_0(\tau) \simeq \tilde{\rho}_0(0) +\frac{3\tilde{h}\tau}{2}$
for $\rho_0(0)\gg \tilde{\rho}_c$. This solution describes the regions of  stable phase
$|\tilde{z}|-\tilde{z}_0-\frac{3\tilde{h}\tau}{2} <0, \tilde{\rho}-\tilde{\rho}_0(\tau)<0$, and metastable phase
$|\tilde{z}|-\tilde{z}_0-\frac{3\tilde{h}\tau}{2} <0, \tilde{\rho}-\tilde{\rho}_0(\tau)>0$,
and $|\tilde{z}|-\tilde{z}_0-\frac{3\tilde{h}\tau}{2} >0, \tilde{\rho}-\tilde{\rho}_0(\tau)<0$.

The solution describing evolution of a parallelepiped-like seed is given by
\be
\chi =-\mbox{tanh}\left[|\tilde{x}|-\tilde{x}_0-\frac{3\tilde{h}\tau}{2}\right]
\mbox{tanh}\left[|\tilde{y}|-\tilde{y}_0-\frac{3\tilde{h}\tau}{2}\right]
\mbox{tanh}\left[|\tilde{z}|-\tilde{z}_0-\frac{3\tilde{h}\tau}{2}\right]+\frac{\tilde{h}}{4}\,,
\ee
for $\tilde{x}_0, \tilde{y}_0, \tilde{z}_0\gg 1$. This solution is applicable everywhere except the region where simultaneously $|\tilde{x}|-\tilde{x}_0-\frac{3\tilde{h}\tau}{2}>0$,
$|\tilde{y}|-\tilde{y}_0-\frac{3\tilde{h}\tau}{2}>0$ and $|\tilde{z}|-\tilde{z}_0-\frac{3\tilde{h}\tau}{2}>0$, and, as  other quasi-one dimensional configurations, this solution  has no critical size.

There are also other approximate solutions describing evolution of seeds of a more whimsical form, which keep their initial form during the time evolution. Certainly, formation of such configurations in fluctuations is tiny, cf. \cite{Hohenberg1995}, but in presence of various defects forming the centers of  condensation  these seeds may grow to the new phase.

\subsection{Transitions between inhomogeneous  configurations}
\subsubsection{Simple phenomenological model of  phase transition to the state $k_0\neq 0$}
Let us consider the first-order phase transition described by the complex order parameter and simplifying consideration we take $\alpha_{01}=\alpha_{1}=\alpha_{2}=0$ but $\alpha_{02}>0, \alpha_4 <0$. Then the equation describing dynamics of the order parameter is of the form given by Eq. (\ref{Leqorderparam}). Let the free-energy density be
\be
\delta{\cal F}=\psi^*\hat{L}\psi +\Lambda_0|\psi|^4/2 -he^{im\phi}\psi(\psi^*/\psi)^{1/2}- he^{im\phi}\psi^*(\psi/\psi^*)^{1/2}\,.
 \ee
For inhomogeneous configurations
\be
\hat{L}=\omega_0^2 -\alpha_4 (\Delta +k_0^2)^2\,. \label{Linhom}
\ee
If transition arises between two inhomogeneous configurations then  in the whole space $\hat{L}$ has the form
(\ref{Linhom}).

\subsubsection{Transition between two one-axis configurations}
In order to describe first-order phase transition between two one-axis configurations we take $\psi$ in the form
\be
\psi =\psi_0\chi (t,\vec{r})e^{i\vec{k}_0\vec{r}+i\phi}\,,\quad \psi_0^2 =-\omega_0^2/\Lambda_0\,,\label{inhomchitr}
  \ee
  where $\phi$ is as before the arbitrary constant phase, $\omega_0^2<0$, $\Lambda_0>0$.
Then employing (\ref{Linhom}), (\ref{inhomchitr}) and (\ref{eqMotPhi}) we obtain equation of motion
\be
\partial_{\tilde{t}}\chi=2\chi (1-\chi^2)+\tilde{h}+\tilde{\hat{L}}\chi\,,\quad \tilde{\hat{L}}\simeq(\vec{n}\tilde{\nabla})^2-\tilde{\alpha}\tilde{\Delta}^2\,,\label{dimenstime}
 \ee
where $\tilde{\alpha}=\omega_0^2/(32k_0^4\alpha_4)$,  $\tilde{h}=-2h/(\omega^2_0\psi_0)$, $\tilde{\nabla}=(\partial_{\tilde{x}},
\partial_{\tilde{y}},\partial_{\tilde{z}})$, $\tilde{x}_i = x_i /\sqrt{8k_0^2\alpha_4/\omega_0^2}$, $\tilde{t}=-t\omega_0^2/(2\alpha_{02})$.
If, as we suppose, $|\omega_0^2|\ll k_0^2$ and thereby $\tilde{\alpha}\ll 1$, the term
 $\tilde{\alpha}\tilde{\Delta}^2\chi$ can be neglected.

 Let us consider now the case when the seed of the stable phase inside the metastable one has a form of the slab of the transverse size (in dimensionless units)  $2\tilde{x}_0\gg 1$. Then for $\tilde{h}\ll 1$ the required solution of Eq. (\ref{dimenstime}) is
 \be
 \chi =-\mbox{tanh}[|\tilde{x}|-\tilde{x}_0-3\tilde{h} \tilde{t}/2]+\tilde{h}/4\,.\label{inhomslab}
  \ee
As we have discussed, in this approximation  there is no critical size for slab-seeds to grow from metastable to stable phase.   Generally speaking, a tiny critical size appears provided one takes into account a small $\tilde{\Delta}^2$ contribution, however this critical size is much smaller than that one  ($\propto 1/h$) we considered above. So, we will ignore this effect. The solution (\ref{inhomslab}) describes growing process of a  slab of the stable phase ($\psi =a(1+\tilde{h}/4)e^{i\vec{k_0}\vec{r}}$) formed within the metastable phase with $\psi =a(1-\tilde{h}/4)e^{i\vec{k_0}\vec{r}}$. In the dimensionless variables the slab boundary grows with the speed $\tilde{{v}}_{\tilde{x}} = 3\tilde{h}/2$ in the $\vec{k}_0\parallel x$ direction.

Let us assume now that an initially  spherical  seed was formed in a fluctuation inside the metastable one-dimensional system characterized by the order parameter $\psi =a(1-\tilde{h}/4)e^{i\vec{k_0}\vec{r}}$, $\vec{k_0}\parallel z$.
Then the solution describing growth of the seed for $t\geq 0$ is as follows
\be
\chi \simeq - \mbox{tanh}[|\tilde{z}|-\sqrt{|\tilde{R}_0^2-\tilde{\rho}^2|}\mbox{sgn} (\tilde{R}_0-\tilde{\rho})-3\tilde{h} \tilde{t}/2]+\tilde{h}/4\,,
 \ee
$\mbox{sgn}(x)=1$ for $x>0$ and $-1$ for $x<0$. In this case there is no critical size of the seed $\propto 1/h$. The speed of the growth of the boundary (labeled below by subscript  $\rm b$) of the seed ($|\tilde{z}_{\rm b}|-\sqrt{|\tilde{R}_0^2-\tilde{\rho}^2_{\rm b}|}\mbox{sgn} (\tilde{R}_0-\tilde{\rho}_{\rm b})-3\tilde{h} \tilde{t}/2\simeq 0$) is different in $z$ and $x,y$ directions:
\be
(\tilde{v}_{\tilde{z}})_{\tilde{\rho}_{\rm b}}=3\tilde{h}/2\,, \quad (\tilde{v}_{\tilde{\rho}})_{\tilde{z}_{\rm b}} \simeq \frac{(3\tilde{h}/2)(3\tilde{h}\tilde{t}/2-|\tilde{z}_{\rm b}|)}{
\sqrt{(3\tilde{h}\tilde{t}/2 -|\tilde{z}_{\rm b}|)^2+\tilde{R}_0^2}}\,.
 \ee
%for $|\tilde{z}|<3\tilde{h}\tilde{t}/2$.
Thus $\tilde{v}_{\tilde{z}}>\tilde{v}_{\tilde{\rho}}$,
for all $t$.
The elongation of the seed occurs in the $\vec{k}_0$ direction. Note that similar  ``battonets''
have been observed in the A-smectic liquid crystals, cf. \cite{Chandrasekhar}.

The rod with the order parameter of the one-dimensional symmetry (\ref{inhomchitr}) evolves
as
\be
\psi = ae^{ik_0 z}[-\mbox{tanh}[|\tilde{z}|-\tilde{z}_0-3\tilde{h} \tilde{t}/2]+\tilde{h}/4
-C_1 \nu_3 \mbox{exp}[(\tilde{\rho}-\tilde{\rho}_0)/\tilde{l}_{\perp}]
\cos[(\tilde{\rho}-\tilde{\rho}_0)/\tilde{l}_{\perp}+C_2]]\,,
 \ee
where $\nu_3 = \mbox{sgn}[|\tilde{z}_0+3\tilde{h} \tilde{t}/2-\tilde{z}|$, $\tilde{l}_\perp = \tilde{\alpha}^{1/4}$, $\tilde{\rho}<\tilde{\rho}_0$, i.e. it elongates only in the $z$ direction.

 Initially spherical seed of inhomogeneous stable phase of the symmetry (\ref{infspher})
 placed in the metastable medium of the same symmetry grows to the stable phase  as
\be\psi =ae^{i{k}_0{r}}[-\mbox{tanh}(\tilde{r}-\tilde{r}_0-3\tilde{h}\tilde{t}/2)+\tilde{h}/4]\,.
\ee
Initially spherical seed of inhomogeneous stable phase of the symmetry $e^{ik_0\rho}$
evolves to the expanding disk
\be\psi =ae^{i{k}_0{\rho}}[-\mbox{tanh}(\tilde{\rho}-\sqrt{|\tilde{R}_0^2-\tilde{z}^2|}\mbox{sgn} (\tilde{R}_0-|\tilde{z}|)
-3\tilde{h}\tilde{t}/2)+\tilde{h}/4]\,.
\ee

 A rod of inhomogeneous stable phase of the symmetry (\ref{infrod})
 placed in the metastable medium of the same symmetry grows to the stable phase  as
\be\psi =ae^{i{k}_0{\rho}}[-\mbox{tanh}(\tilde{\rho}-\tilde{\rho}_0-3\tilde{h}\tilde{t}/2)+\tilde{h}/4]\,.
\ee
In all considered cases there is no critical size  $\propto 1/h$ for the initial seed.

Above we considered the dynamics of the first-order phase transition from metastable to stable phase in case of the complex field with the effectively one-dimensional ordering (purely one-dimensional $e^{ik_0z}$, spherical $e^{ik_0 r}$, or  cylindric $e^{ik_0 \rho}$) in both phases. The same consideration holds in case of the real order-parameter heaving the same ordering in both phases,
however provided the $h$-term in (\ref{Omoperreal}) is also periodic with the same ordering, e.g., for $\phi=a_{\rm stan}\chi (t,\vec{r})\cos (\vec{k}_0\vec{r})$ we choose
$$h=h_0\cos (\vec{k}_0\vec{r})$$
 for $h_0=const$. In these cases the equation for $\chi(t,\vec{r})$ has the same form (\ref{dimenstime}) and thus solutions for $\chi$ we have considered for the case of the complex field continue to hold after the replacement $h\to h_0$.

%\subsubsection{Spinodal instabilities and glassing}
 Above we focused on dynamics of the order parameter in case of the transition of the system from the metastable state to the stable one.
Within the unstable, so called spinodal region, the  perturbations grow exponentially. Spinodal instabilities  may manifest themselves
in experiments with heavy ions in some collision energy interval that corresponds to the first-order phase transition region of the QCD phase diagram. One of the possible signatures is
a manifestation of  fluctuations with a typical size $r\sim 1/p_m$, $p_m \lsim m_\pi$, in the rapidity spectra, see  Ref.~\cite{Skokov:2010dd} and discussion in Section \ref{hydro}. Also note that in condensed matter physics  a glassing transition from a liquid to a glass state can be interpreted as the first-order phase transition occurring  at a very high viscosity, when the system passes a  spinodal region,  or it is very rapidly overcooled \cite{Voskresensky:1993ux}. Then, there may appear an order at several angstr\"om- scale, which transforms in a disorder at larger distances. Spinodal instabilities will be discussed in a more detail in next  Section.

\subsection{Transitions between homogeneous and inhomogeneous phases}

If a phase transition arises between two homogeneous configurations then  in the whole space $\hat{L}$ has the form
\be\hat{L}=\omega_0^2 -\alpha_3\Delta\,,\label{Lhom}
\ee
with $\alpha_3>0$. Situation is a more involved, if the transition occurs between homogeneous and inhomogeneous states. Then in a part of space $\hat{L}$ should be taken in the form
(\ref{Lhom}) and in the other part of space it is given by
(\ref{Linhom}). The transition region from these regions
is rather narrow $\delta r\sim 1/k_0$ being characterized by the minimal length scale in the given problem. Using this, instead of explicit description of the transition region one can employ the appropriate boundary conditions, cf. \cite{Voskresensky:1980nk,Voskresensky:1984rd}.

For the description of the first-order phase transition to homogeneous state one arrives at the same Eq. (\ref{dimenstime}) where now
\be\tilde{\hat{L}}=\tilde{\Delta}\,, \quad \tilde{x}_1 =x_i \sqrt{(-\omega_0^2)/(2\alpha_3)}\,,
\label{tildeLhom}
\ee
and $\psi_0=\sqrt{-\omega_0^2/\Lambda_0}$, $\tilde{h}=2h/(-\omega_0^2\psi_0)$ and
$\tilde{t}=-\omega_0^2/2\alpha_{02}$ have the same values as for the inhomogeneous phase transition, $\omega_0^2<0$, $\Lambda_0>0$.

For the three-axis system with the real order parameter of the form (\ref{coscoscos}) we may also obtain Eq. (\ref{dimenstime}), but
 where now
 \be\tilde{\hat{L}}=-\tilde{\alpha}\tilde{\Delta}^2+\tilde{\Delta}\,,
 \quad \tilde{x}_i = x_i /\sqrt{8k_{0i}^2\alpha_4/\omega_0^2}\,,\quad x_i =(x,y,z)\,,\label{tildeLlattice}
 \ee
$\alpha_4<0$, $\omega_0^2<0$.

For $\tilde{\alpha}\ll 1$, i.e. in the vicinity of the critical point, the operators
$\tilde{\hat{L}}$ given by Eq. (\ref{tildeLlattice}) and (\ref{tildeLhom}) are of approximately the same form although presented in different variables. Thus using results obtained with (\ref{tildeLhom}) one may conclude that the seed of the metastable phase (\ref{coscoscos}) of ellipsoidal form with $\tilde{\hat{L}}$ determined by Eq. (\ref{tildeLlattice}) does not change the form during its expansion into the stable phase of the same type of the symmetry. It is described by Eq. (\ref{eqMchi}) although in other variables. Initially spherical seed of the phase (\ref{tildeLlattice})  gets ellipsoidal form during expansion.
%The process is described by Eq.   (\ref{sumcos}), now with $\tilde{\hat{L}}$ from Eq. (\ref{tildeLlattice}).
Similar description of the  evolution of the system takes place, if the field is of the type (\ref{sumcos}). In this case $\tilde{\hat{L}}=\tilde{\Delta}-\tilde{\alpha}(\tilde{\partial}_x^4+ \tilde{\partial}_y^4+\tilde{\partial}_z^4)\simeq \tilde{\Delta}$. Transitions from the metastable phase of the symmetry (\ref{sumcos}) to the stable one of the symmetry (\ref{coscoscos}),
as well as from the metastable configuration of the type (\ref{coscoscos}) to that of the symmetry (\ref{sumcos}), are described in the same manner, if parameters $\Lambda_0$ are the same in both cases, since $\tilde{\hat{L}}$ operators in both cases approximately coincide.

Now let us consider more specifically the case of the phase transition from the metastable homogeneous phase to the stable inhomogeneous phase.  The case of equal surface energies of homogeneous and inhomogeneous phases can be studied analytically.
The boundary conditions
\be\Re D^{-1}(k_0^2)=\Re D^{-1}(k^2=0)\,, \label{firstcond}\ee
\be (d\Re D^{-1}(k^2)/d k^2)_{k=k_0}=0\,, \label{secondcond}\ee
and\be (d^2\Re D^{-1}(k^2)/d(k^2)^2)_{k=k_0}=(d^2\Re D^{-1}(k^2)/d (k^2)^2)_{k=0}\, \label{thirdcond}
\ee should be fulfilled.
The conditions (\ref{firstcond}), (\ref{secondcond}) mean that $\Re D^{-1}(k^2)$ has the same minimal value at $k=k_0\neq 0$ and $k_0=0$ and condition (\ref{thirdcond}) means that surface energies of seeds of both phases are equal. Employing (\ref{dimenstime}), (\ref{tildeLhom}) and (\ref{tildeLlattice})
we  need to require $\alpha_3 =-4k_0^2 \alpha_4>0$.
Further assume that  conditions (\ref{firstcond})--(\ref{thirdcond})
are fulfilled.

Let  initially the metastable homogeneous phase occupies the half  space $x>0$ and the  one-axis inhomogeneous stable phase  $\psi=ae^{ik_0x}$ occupies the half  space $x<0$ and let
\be
\psi_{<}=a\chi_{<}(0,x) e^{ik_0x}\,\quad x<0\,,\ee
\be
\psi_{>}=a\chi_{>}(0,x) \,\quad x>0\,,\ee
Both solutions $\chi_{<}(t,x)$ and $\chi_{>}(t,x)$ in the considered case have the same form
\be
\chi =-\mbox{tanh}(\tilde{x}-3\tilde{h}\tilde{t}/2 -\tilde{x}_1)+\tilde{h}/4\,.
 \ee
The order parameters are matched at $x=0$ provided $\tilde{x}_1\simeq -\tilde{h}/4$. Then
$\psi_{<}(0,0)=\psi_{>}(0,0)=0$, $\psi^{\prime}_{<}(0,0)=\psi^{\prime}_{>}(0,0)=0$, since a smooth continuation of $\chi$ and since $ik_0\psi_{<}(0,0)=0$. The time dependent solution becomes
\be
\psi_{<}=a e^{ik_0x}[-\mbox{tanh}(\tilde{x}-3\tilde{h}\tilde{t}/2 +\tilde{h}/4)+\tilde{h}/4]\,,
\label{latticeOnetimeneg}
 \ee
\be
\psi_{>}=a [-\mbox{tanh}(\tilde{x}-3\tilde{h}\tilde{t}/2 +\tilde{h}/4)+\tilde{h}/4]\,.
 \ee
The one-axis crystal, $\psi_{<}\propto e^{ik_0x}$, grows  from the surface with the passage of time  following the law $x=3{h}{t}/2$. The solution (\ref{latticeOnetimeneg}) continues to hold for the case, when    the metastable  phase occupying the half  space $x>0$  has the same symmetry as the stable phase  $\psi=ae^{ik_0x}$ at $x<0$ at $t=0$.

Initially spherical seed with the order parameter $\psi_{<}\propto e^{ik_0r}$ appeared inside a  metastable uniform matter evolves as
\be
\psi_{<}=a e^{ik_0r}[-\mbox{tanh}(\tilde{r}-3\tilde{h}\tilde{t}/2 -\tilde{r}_0 +\tilde{h}/4)+\tilde{h}/4]\,,
\label{latticeOnetimeneg1}
 \ee
 and
 \be
\psi_{<}=a [-\mbox{tanh}(\tilde{r}-3\tilde{h}\tilde{t}/2 -\tilde{r}_0 +\tilde{h}/4)+\tilde{h}/4]\,.
\label{latticeOnetimenegout}
\ee
At the phase boundary $r=r_0+3ht/2+h/4$ the order parameter $\psi$ and its $r$-derivative are continues quantities.

 Therefore,  provided conditions (\ref{firstcond})-(\ref{thirdcond}) are fulfilled, in dependence on the structure of the order parameter in the seed the metastable uniform phase either undergoes the first-order phase transition to the phase of the same symmetry or to the lattice.

\section{Hydrodynamics   of liquid--gas-type  transition at small overcriticality. Nuclear liquid--gas and hadron--quark transitions}\label{hydro}

 Further assume that the dynamics of a  phase transition  can be described  using the variables $n$ and $s$ (or $T$), where  $n$ is the local baryon density, $s$ is the local
entropy density, $T$ is the local temperature, cf.  \cite{Skokov:2008zp,Skokov:2009yu,Skokov:2010dd,Voskresensky:2010gf}.
 Also, to proceed analytically assume that the system is rather close to the critical point of the phase transition.  Refs.
\cite{Berdnikov,NA,NA1} demonstrated effect of the critical slowing down
that limits the growth  of the $\sigma$-field correlation length in
the vicinity of the critical point. Some models speculate about
explosive  freeze-out assuming an increase of the bulk viscosity close to
the critical point, see \cite{Scavenius,Torrieri}. Since all the processes   in the vicinity of the critical point  are slowed down, the velocity of a seed of a new phase prepared in the old phase, $\vec{u}$, is much smaller than the mean thermal velocity and one may use
equations of non-relativistic non-ideal hydrodynamics:
the Navier--Stokes equation, the continuity equation,
and equation for the heat
transport, even if one deals with violent heavy-ion collisions,
\begin{eqnarray}
\label{Navier} m^{*} n\left[ \partial_{t} {u}_i + (\vec{u}\nabla)
{u}_i \right] &=& -\nabla_i P   + \nabla_k \left[ \eta \left(
\nabla_k u_i + \nabla_i u_k -\frac{2}{\nu} \delta_{ik} \mbox{div}
\vec{u}   \right)   + \zeta \delta_{ik} \mbox{div} \vec{u}
\right], \\
\label{contin}
\partial_{t}n +\mbox{div} (n \vec{u})&=&0, \\
 \label{therm}   T\left[\frac{\partial
s}{\partial t} +\mbox{div}(s\vec{u} )\right]&=&\mbox{div}(\kappa
\nabla T) +\eta \left(\nabla_k u_i + \nabla_i u_k -\frac{2}{\nu}
\delta_{ik} \mbox{div} \vec{u} \right)^2 +\zeta (\mbox{div}
\vec{u})^2\,.
\end{eqnarray}
Here  $m^{*}$ is the baryon quasiparticle mass, $P$ is the pressure. The quantities   $\eta$
and $\zeta$ are the shear and bulk viscosities,    $\kappa$ is the thermal conductivity, $\nu$ shows the
geometry of the seed under consideration  (droplets, rods, slabs).

All thermodynamical quantities can be expanded  near  a  reference point
$(n_{\rm r}, T_{\rm r})$,  which is assumed to be rather close to the critical point but  outside the  fluctuation region (determined by the value of the $Gi$ number).   The latter circumstance is important for the determination of the specific heat  $c_{V ,\rm r}$ and  transport coefficients, which may diverge in the critical point, whereas other quantities are  smooth functions of $n,T$ and calculating  them one can put $n_{\rm r}=n_{cr}, T_{\rm r}=T_{cr}$.

The Landau free energy counted from the value at $n_{\rm r}\simeq n_{cr}, T_{\rm r}\simeq T_{cr}$
in the variables $\delta n
=n-n_{cr}$, $\delta T=T-T_{cr}$,
$\delta (\delta F_L)/\delta(\delta n) = P - P_{f}+P_{\rm MC} $ can be presented as \cite{Skokov:2008zp,Skokov:2009yu,Skokov:2010dd,Voskresensky:2020yrd}
\begin{eqnarray}
&&\delta F_{\rm L} = \int \frac{d^3 x}{n_{cr}}\left\{ \frac{cm^{*}[\nabla
(\delta n)] ^2}{2}+\frac{\lambda m^{*\,3} (\delta
n)^4}{4}-\frac{\lambda v^2 m^{*}(\delta n) ^2}{2}-\epsilon \delta n
\right\}+ \delta F_{\rm L} (k_0) ,\label{fren}
\end{eqnarray}
where $\epsilon = P_f-P_{\rm MC}\simeq n_{\rm cr}(\mu_{i}-\mu_f)$ is expressed through the (final)  value of the
pressure after the first-order phase transition has occurred, $P_f$, and the value of the pressure at the Maxwell construction, $P_{\rm MC}$; $\mu_i$ and $\mu_f$ are the chemical potentials of the initial and final configurations (at fixed $P$ and $T$).
 The quantity  $\epsilon\neq 0$, if one deals with a first-order phase transition, and $\epsilon =0$, if a transition is of the second order. The maximum of the quantity
$\epsilon$ is ${\epsilon}_{m}= 4\lambda v^3 /(3\sqrt{3})$.
For the description of phase transitions to the uniform state, $k_0=0$, one may retain only the term $\propto c[\nabla (\delta n)] ^2$ in the expansion of the free energy in the density gradients, using $c>0$.
For the description of phase transitions to the inhomogeneous state, $k_0\neq 0$, one should perform expansion  retaining at least terms up to   $\propto d[\Delta (\delta n)] ^2$, assuming $c<0$ and $d>0$. Therefore the last term in (\ref{fren}) appears only, if  $k=k_0 \neq 0$ \cite{Voskresensky:1993ux,Voskresensky:2020yrd}, like for the phase transition to the solid state, liquid crystal state, or a pion condensate state in a dense nuclear matter.
Then for $k_0\neq 0$ and $c<0$, $d>0$ we have
\begin{eqnarray}
\delta F_{\rm L} (k_0)=\int \frac{d^3 x}{n_{cr}}\left\{ \frac{dm^*}{2} (\Delta \delta n)^2-\left(\frac{cm^* k_0^2}{2}+ \frac{dm^* k_0^4}{2}\right) (\delta n)^2 \right\}\,,
\end{eqnarray}
where $k_0^2=-\frac{c}{2d}>0$ follows from minimization of $\delta F_{\rm L} (k_0)$. In case of the phase transition to the homogeneous state  one should put $k_0 =0$, $d=0$ (then $\delta F_{\rm L} (k_0)=0$)  and $c>0$. Then the first term $\propto c$ in Eq. (\ref{fren}) is associated with the positive surface tension, $\delta F_{\rm L}^{\rm surf}=\sigma S$, where $S$ is
the surface area of the seed.

\begin{figure}
\centerline{%
\rotatebox{0}{\includegraphics[height=5.0truecm] {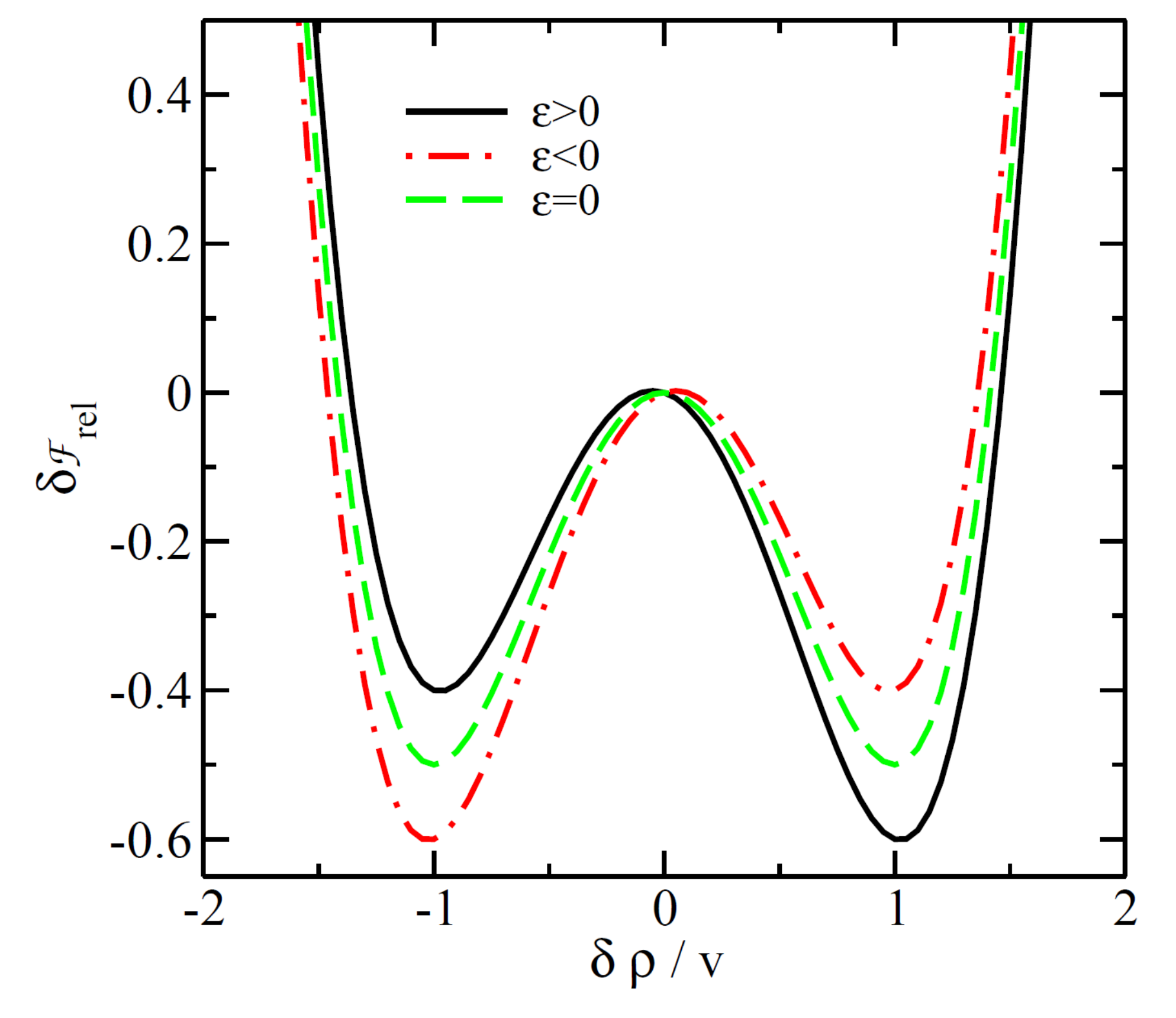}
} \rotatebox{0}{\includegraphics[height=5.1truecm]{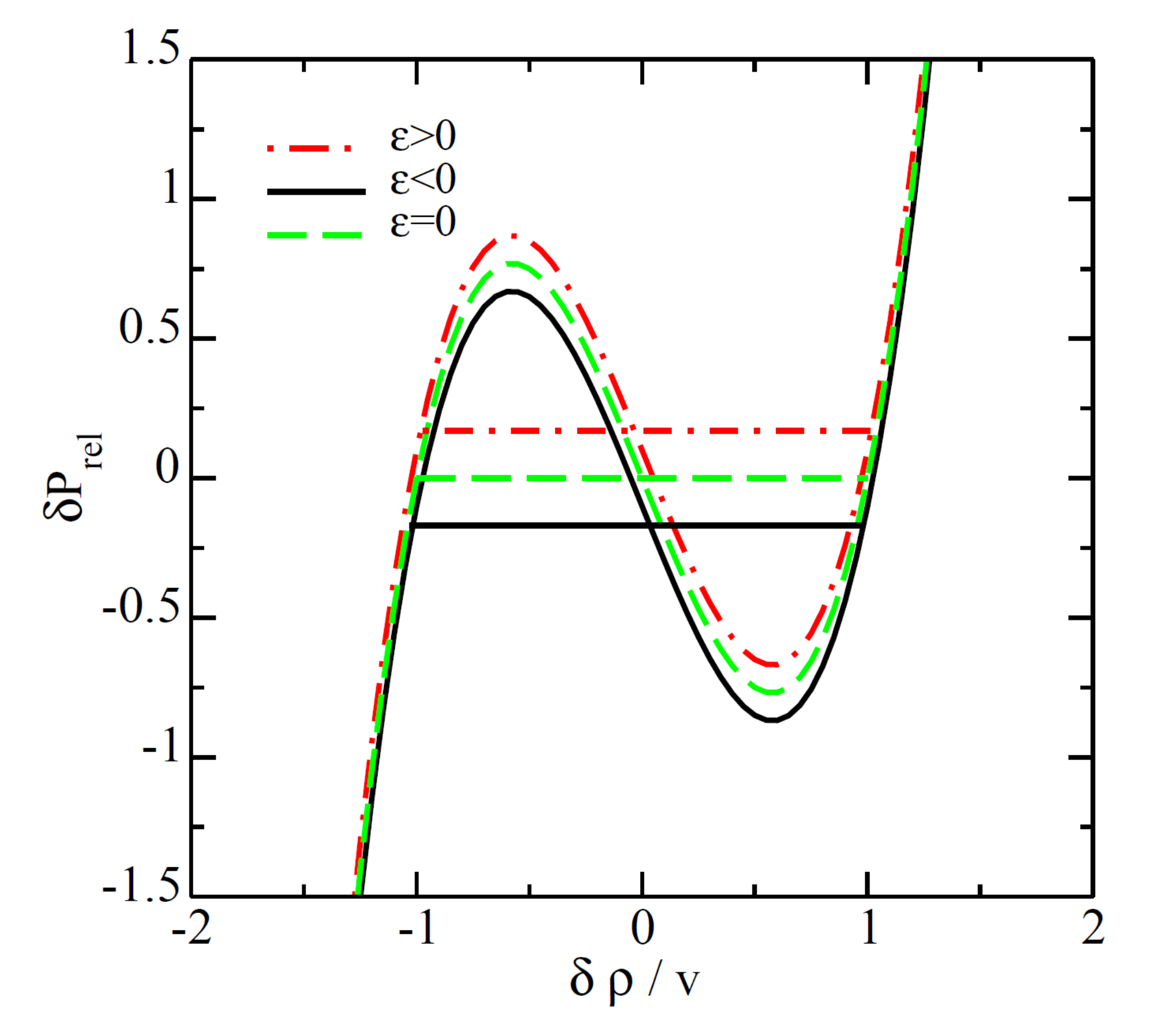}   }  } \caption{ The Landau free-energy density
$\delta {\cal F}_{\rm rel} =  \delta {\cal{F}}_{\rm L} /{\cal{F}}_{\rm L} (T_{cr},\rho_{cr})$ and the value
$ \delta { P}_{\rm rel} = \rho_{cr}\frac{\delta [F_{\rm L} (T,\delta\rho )]}{\delta
(\delta\rho)}|_{T}/P (T_{cr},\rho_{cr})$, as functions of the order parameter $\delta
\rho =m^*\delta n$ for the equation of state determined  by Eq. (\ref{fren}), at $T<T_{cr}$.
Dashed horizontal line ($\epsilon =0$) in the right panel shows
MC. Figure is adopted from \cite{Skokov:2009yu}.}\label{pres}
\end{figure}

The Landau free-energy density and pressure as functions of the order parameter $\delta
\rho$ for the equation of state determined  by Eq. (\ref{fren}) at $T<T_{cr}$ are shown in Fig. \ref{pres}, cf. schematic Fig. \ref{VanderWaals} demonstrating isotherms $P(n)$ and Fig. \ref{spinT} showing $T(n)$ dependence for the van der Waaals equation of state discussed above in Section \ref{TypPressureSetion}. For $\epsilon =0$ two
minima  of the Landau free energy coincide
and lie on the  MC line  on the plot $\delta P
(1/\rho)$ (dashed horizontal line  in the plot $\delta P
(\delta\rho)$ in the right panel) describing thermal equilibrium of phases. At equilibrium the baryon chemical potentials of the vapor and liquid phases are equal, $\mu_{\rm eq,V}=\mu_{\rm eq, L}$. For $\epsilon \neq 0$ the interval of values $P(\delta \rho)$ from the  point   corresponding to the left local minimum of the Landau free energy to the point of the local maximum of the pressure  describes a metastable supercooled vapor (SV) with $\mu_{\rm SV} >\mu_{\rm eq, L}$. The interval of values $P(\delta \rho)$ from the point  corresponding to the right  local minimum of the Landau free energy in the left panel to the point of the local  minimum of the pressure in the right panel    relates to a  metastable overheated liquid (OL). In this interval  $\mu_{\rm OL} >\mu_{\rm eq, V}$. The interval from maximum to minimum of $P(\delta\rho)$  shows  unstable isothermal spinodal (ITS) region. If  in the initial state  $(\delta\rho)_i = \rho_i -\rho_{cr}=0$, one deals with the spontaneous symmetry breaking and the second-order phase transition to the new state. For $(\delta \rho)_i =\rho_i -\rho_{cr}\neq 0$, $\epsilon
>0$ or $\epsilon <0$, one deals  with the first-order phase transition either from the metastable to the stable state or with the transition from spinodal region.
For $\epsilon
>0$ (solid lines) the liquid state is stable and the vapor state is metastable (SV),
and for $\epsilon <0$ (dash-dotted lines) the liquid state is
metastable (OL), whereas the vapor state is stable. The dynamics of the transition starting from a point within spinodal region for $\epsilon \neq 0$ (but small) is described similarly to that for the second-order phase transition, for $\epsilon =0$.
For the purely van der Waals equation of state (in this case $k_0=0$) one gets \cite{Skokov:2009yu}:
\begin{eqnarray}\label{parame}
v^2 (T) =- 4 \frac{\delta{{T}} n_{cr}^2 m^{*\,2}}{T_{cr}} ,
\quad \sigma =\sigma_0 \frac{|\delta
T|^{3/2}}{T_{cr}^{3/2}}\,,\quad  \sigma_0^2
 =32 m^{*}n_{cr}^2T_{cr} c.
\end{eqnarray}

Applying operator $\mbox{div}$ to  Eq. (\ref{Navier}) and expressing $\mbox{div}\,\vec{u}$ in Eq.~(\ref{contin}) via $\partial_t n$, for small $\delta n$ and $u$, keeping only linear terms in ${u}$, that is legitimate, since near the critical point processes develop slowly ($v^2 \propto -\delta T$),  we rewrite Eq.~(\ref{Navier}) as
 \begin{eqnarray}\label{v-t1}
-\frac{\partial^2 \delta n}{\partial t^2}&=\Delta\left[c\Delta
\delta n +\lambda v^2 \delta n -\lambda m^{*2} (\delta n)^3
+\epsilon/m^*  -(m^*n_{cr})^{-1}\left(\widetilde{\nu}{\eta_{cr}}
+{\zeta_{cr }} \right)\frac{\partial \delta n}{\partial
t}\right]\\
&+\Delta \left[ -d\Delta^2 \delta n +(ck^2_0 +dk_0^4)\delta n \right] \,,\nonumber
 \end{eqnarray}
 $\widetilde{\nu}={2(\nu-1)}/{\nu}$, cf. \cite{MSTV90,Voskresensky:1993ux,Skokov:2009yu,Voskresensky:2020yrd}.
Second line  in Eq. (\ref{v-t1}) yields  non-zero term only for the description of the condensation to the  state $k_0\neq 0$. With the help of the notation
\begin{eqnarray}
\widetilde{\omega}^2 (\hat{k}^2)=-\lambda v^2 +c\hat{k}^2 +d\hat{k}^4-ck_0^2 -dk_0^4\,\label{omtilde}
\end{eqnarray}
Eq. (\ref{v-t1}) can be rewritten as
 \begin{eqnarray}\label{v-t2}
-\frac{\partial^2 \delta n}{\partial t^2}&=\Delta\left[-\tilde{\omega}^2(\Delta) \delta n -
\lambda m^{*2} (\delta n)^3
+\epsilon/m^*  -(m^*n_{cr})^{-1}\left(\widetilde{\nu}{\eta_{cr}}
+{\zeta_{cr }} \right)\frac{\partial \delta n}{\partial
t}\right]\,.
 \end{eqnarray}

Let us consider $T<T_{cr}$. In  the dimensionless
variables $m^* \delta n =v \psi$, ${\tau}=t/t_0$, $\xi_i =x_i /l$, $i =1 ,\cdots ,
\nu$, $\nu =3$ for seeds of spherical geometry, Eq. (\ref{v-t1}) is presented as
 \begin{eqnarray}\label{dimens}
 &&- \beta \frac{\partial^2 \psi }{\partial
{\tau}^2} =\Delta_{\xi}\left(\Delta_{\xi}\psi +2\psi
 (1-\psi^2)+\widetilde{\epsilon}- \frac{\partial \psi}{\partial
 {\tau}}-  \frac{\lambda v^2 d}{2c^2}\Delta_{\xi}^2\psi +\frac{2(ck_0^2+dk_0^4)}{\lambda v^2}\psi   \right),\\
 &&l=\left(\frac{2c}{\lambda v^2}\right)^{1/2}\,,\,\,\,
  t_0
 =\frac{2\tilde{\eta}_{\rm r}}{\lambda v^2
 }\,,\,\,\,
 \widetilde{\epsilon}=\frac{2\epsilon}{\lambda v^3}\,, \,\,\, \beta
 =\frac{c }{\tilde{\eta}_{\rm r}^2 }\,,\,\,\,\tilde{\eta}_{\rm r}=\frac{(\tilde{\nu}\eta_{\rm r} +\zeta_{\rm r}) }{m^* n_{ cr}}.\nonumber
  \end{eqnarray}

It is important to notice that even for $k_0=0$
Eq.~(\ref{dimens}) differs in the form from the standard Ginzburg--Landau  equation broadly exploited   in  the condensed matter physics, since Eq.~(\ref{dimens}) is of the second-order in time derivatives, whereas the standard Ginzburg--Landau  equation is of the first order in time derivatives.
The difference  disappears, if one puts   the  bracketed-term in
the r.h.s. of Eq.~(\ref{dimens}) to zero.  This procedure is however not legitimate at least for description  of the order parameter on an initial stage, since two initial conditions, such as $\delta n (t=0,\vec{r})=0$ and
$\partial_t \delta n (t,\vec{r})|_{t=0}\simeq 0$, should be fulfilled to describe a seed  formed  in a fluctuation. Thereby, at least there exists an initial stage of the dynamics of seeds ($t\lsim t_{\rm init}$), which  is not described by the standard Ginzburg--Landau equation \cite{Skokov:2008zp,Skokov:2009yu}. The  bracketed-term in
the r.h.s. of Eq.~(\ref{dimens}) can indeed be put  zero  at  large time, $\tau \gg 1$, if one considers   an effectively very viscous medium, see below. In this case we arrive at Eqs. (\ref{eqMchih0}), (\ref{dimenstime}) studied in Section \ref{First-order-transitions}.
% Also note that Eq.~(\ref{dimens}) with the   bracketed-term in the r.h.s. equal zero can be derived from the first-gradient order kinetic equation of Kadanoff--Baym \cite{Voskresensky:2010qu}.

 The  parameter
$\beta$ characterizes an inertia on an initial stage of evolution of the seeds, for $t\lsim t_0\sqrt{\beta}$, cf. \cite{Skokov:2009yu}. It is expressed in terms
of the surface tension and the viscosity as
 \begin{eqnarray}
\beta
 = (32T_{cr})^{-1}[\widetilde{\nu}\eta_{\rm r} +\zeta_{\rm r} ]^{-2}\sigma_0^2 m^{*}.\label{betasigma}
   \end{eqnarray}
The larger viscosity and the smaller surface tension, the
effectively more viscous (inertial)  is the fluidity of seeds.
For $\beta \ll 1$ one deals with the regime of effectively viscous (inertial)
fluidity and at $\beta \gg 1$ one deals  with the regime of almost perfect fluidity. Estimates \cite{Skokov:2009yu} show that for the nuclear liquid--vapor phase transition typically $\beta\sim 0.01$.  For the
 quark--hadron   transition $\beta\sim 0.02-0.2$, even  for  very low
value of the ratio $\eta/s\simeq 1/(4\pi)$. The latter quantity characterizes fluidity of the matter at ultra-relativistic heavy-ion collisions \cite{Romatschke:2007mq}.  Thus, as we argued, in case of the baryon-rich matter one deals with
effectively very viscous (inertial) evolution of density fluctuations  both
in cases of the nuclear liquid--vapor and quark--hadron phase transitions.

In a process of a neutron star formation and cooling at $T>T_{\rm opac}\sim 1$ MeV, when neutrino mean free path $\lambda_\nu <R_{\rm star}\sim 10$ km,   the viscosities $\eta$, $\zeta$ and the heat conductivity $\kappa$ are determined by the most long-range  neutrino processes  \cite{Haubold:1988uu,MSTV90}.
An overcritical pion-condensate drop reaches a size $R\sim 0.1$ km for $t\sim 10^{-3}$ sec. by the growth of the density mode. Then it may reach $R\sim (1-10)$ km for typical thermal transport time $t_T$ varying from $\sim 10$ sec.  up to several hours (rather than for typical collapse time $\sim 10^{-3}$ sec). A delay may appear owing to the neutrino heat transport to the surface
(effect of thermal conductivity) that strongly depends on the value of the  softening of the pion mode responsible for the efficiency of the nucleon-nucleon interaction at $n>n_0$, being stronger for heaviest neutron stars \cite{MSTV90}.  One should also take into
account that  the bulk viscosity is significantly increased in presence of the soft modes \cite{ML37,LL06}, e.g., near the pion condensation critical point \cite{Kolomeitsev:2014gfa}.
Also notice that description of the dynamics of the pion-condensate phase transition  is specific, since the transition occurs to the state $k_0\neq 0$, see further discussion in Section \ref{Pion-section}.

Still, Eq. (\ref{dimens}) should be supplemented by  Eq. (\ref{therm}) for the
heat transport, which owing to
Eq.~(\ref{contin})  after its linearization reads as
\begin{equation}\label{v-tS}
T_{cr}\left[
\partial_t \delta s
-s_{cr}(n_{cr})^{-1}
\partial_t \delta n
\right]=\kappa_{\rm r}\Delta \delta T\,.
\end{equation}
The variation of the temperature is related to the variation of
the entropy density $s[n,T]$ by %\begin{equation}\label{T-s}
\begin{equation}\label{v-tS1}
\delta T \simeq T_{cr} (c_{V ,\rm r})^{-1}\left(\delta s
-({\partial s}/{\partial n})_{T, cr}\delta n\right)\,,
\end{equation}
$c_V$ is the density of the heat copacity.

The time scale  for the
relaxation of the entropy/temperature mode, following (\ref{v-tS}), is
 \begin{eqnarray}\label{kT}
t_T = R^2_{\rm seed} c_{V,\rm r} /\kappa_{\rm r} \propto R_{\rm seed}^2,
 \end{eqnarray}
 i.e., relaxation time of the temperature/entropy is proportional to the surface area of the seed.
Thus,
interplay between viscosity, surface tension, and thermal
conductivity effects is responsible for the typical time- and size- scales of
fluctuations.

\subsection{Stationary solutions}

Now let us find stationary solutions of Eq. (\ref{v-t1}). For the condensation in the  state $k_0\neq 0$, $c<0$, $d>0$, and the gap $\widetilde{\omega}^2 (k^2)$ has a minimum for $k=k_0$.
The phase transition arises for $\widetilde{\omega}^2 (k_0^2)<0$. We find solution in the form \cite{Voskresensky:2020yrd}
\begin{eqnarray}
m^* \delta n=a_{\rm stan}[\cos (k x+\chi) +\frac{c_1 \widetilde{\omega}^2 (k^2)}{\widetilde{\omega}^2 (9k^2)}\cos (3k x +\chi) +...]+O(\epsilon)\,,\label{sinExp}
\end{eqnarray}
where $\chi$ is arbitrary constant phase.

Setting (\ref{sinExp}) in Eq. (\ref{v-t1}) we find
\begin{eqnarray}
a_{\rm stan}^2 =-\frac{4}{3} \widetilde{\omega}^2 (k_0^2)/\lambda>0, \quad c_1 =-1/3\,.
\end{eqnarray}
Minimization of the free energy in $k$ yields $k=k_0$ and $\widetilde{\omega}^2 (k_0^2)=-\lambda v^2$. We have
$\widetilde{\omega}^2 (k_0^2)>0$ for $T>T_{cr}$ and $\widetilde{\omega}^2 (k_0^2)<0$ for $T<T_{cr}$, $\widetilde{\omega}^2 (9k_0^2)=-\lambda v^2 +16 c^2/d\gg |\widetilde{\omega}^2 (k_0^2)|$. Thereby with appropriate accuracy we may use $m^*\delta n\simeq a_{\rm stan}[\cos (k_0 x+\chi)$ that yields
$$\delta F_L (k_0)\simeq -\lambda v^4 V_3/(6m^*n)+O(\epsilon^2)\,,$$
 where $V_3$ is the volume of the system.

For the condensation in the uniform state $k_0=0$ we have \cite{Skokov:2009yu}
\begin{eqnarray}
\widetilde{\omega}^2 (k^2)=-\lambda v^2 +ck^2\,,\quad  k^2<\lambda v^2/c\,,\quad c>0\,.
\end{eqnarray}
 Two spatially constant stationary solutions minimizing the free energy for $T<T_{cr}$ describe metastable and stable states:
\begin{eqnarray}
\delta n_{\rm st}\simeq \pm v/m^* +\epsilon/(2\lambda v^2 m^*)\,.\label{solConst}
\end{eqnarray}
The free energy corresponding to these solutions is given by
\begin{eqnarray}
\delta F_{\rm L} (k=0, k_0=0)\simeq -\frac{\lambda v^4 V_3}{4m^*n_{cr}} \left(1\pm \frac{4\epsilon}{\lambda v^3}\right)\,.\label{FLknul}
\end{eqnarray}
For $k\neq 0$ solutions in the form (\ref{sinExp}) are valid  for $|\widetilde{\omega}^2 (k^2)|\ll \widetilde{\omega}^2 (9k^2)$, and $\widetilde{\omega}^2 (9k^2)>0$, and they yield for $k_0=0$:
\begin{eqnarray}
  \delta F_{\rm L} (k\neq 0, k_0=0)\simeq -\frac{\lambda v^4 (1-ck^2/(\lambda v^2))V_3}{6m^*n_{cr}}\,.\label{FLknulk}
  \end{eqnarray}
 Although the minimum of the free energy for $k_0=0$  is given by (\ref{FLknul}) corresponding to  solutions (\ref{solConst}) obtained for  $k=0$,  rather than to solutions of (\ref{sinExp}) corresponding to the free energy  (\ref{FLknulk}) at $k\neq 0$, nevertheless, as we will demonstrate below, solutions (\ref{sinExp}) characterized by $k\neq 0$ have a physical meaning in description of fluctuations in the spinodal region.

\subsection{Dynamics of seeds at first-order phase transition from metastable state to stable state}\label{metastable}
Consider the limit of a high thermal conductivity, when in Eq. (\ref{v-t1}) the temperature can be put constant. Solution of Eq. (\ref{v-t1}) describing
dynamics of the density  fluctuation  developing from the metastable state to the stable state is then presented in the form \cite{Skokov:2010dd}
 \begin{equation}\label{delr} \delta n (t,r)\simeq
\frac{v(T)}{m}\left[\mp\mbox{tanh} \frac{r-R_{\rm seed}
(t)}{l}+\frac{{\epsilon}}{2\lambda v^3(T)}\right]+(\delta
n)_{\rm cor}.
\end{equation}
The
solution is valid for $|\epsilon/(\lambda v^3(T))|\ll 1$.
 Compensating correction  $(\delta
n)_{\rm cor}$ is introduced to  fulfill the   baryon
number conservation. Considering spatial coordinate $r$ in the vicinity of a
bubble/droplet boundary we get  equation describing evolution of the
seed size \cite{Skokov:2009yu,Skokov:2010dd}:
\begin{equation}\label{dim}
\frac{m^{*2}\beta t_0^2}{2l}
\frac{d^2R_{\rm seed}}{dt^2}=m^{*2}\left[\pm\frac{ 3\epsilon}{2\lambda v^3 (T)
}-\frac{2l}{R_{\rm seed}}\right]-\frac{m^{*2} t_0}{l}\frac{d R_{\rm seed}}{dt}.
\end{equation}
This equation reminds the Newton second law for a one-dimensional system, where the quantity $M=\frac{m^{*2}\beta t_0^2}{2l}\propto {(T_{cr}-T)^{-3/2}}$ has a meaning of a mass, $m^{*2}[\frac{3|\epsilon|}{2\lambda v^3 (T)
}-\frac{2l}{R_{\rm seed}}]$ is an external force and $-\frac{m^{*2}t_0}{l}\frac{d R_{\rm seed}}{dt}$ is the friction force, with a viscous-friction coefficient  proportional to an effective viscosity and inversely proportional to $\sqrt{T_{cr}-T}$. Following Eq. (\ref{dim})
 a bubble of an overcritical  size $R_{\rm seed}>R_{cr}=4l\lambda v^3(T)/(3|\epsilon|)$
 of the stable ``vapor''
phase, or respectively a droplet of the stable ``liquid''  phase, been initially
prepared in a fluctuation inside a metastable phase,  then grow with increasing time. On an early stage of the evolution  the size of
the overcritical bubble/droplet  $R_{\rm seed}(t)$ (for $R_{\rm seed}>R_{cr}$) grows with an acceleration.
Then it reaches a steady growth regime with a constant velocity
$u_{\rm as}=\frac{3\epsilon l}{\lambda v^3 (T) t_0}\propto |(T_{cr} -T)/T_{cr}|^{1/2}$.
 In the interior/exterior of the
seed $\delta n \simeq \mp v(T)/m^*$. The
correction $(\delta n)_{\rm cor} \simeq v(T)R_{\rm seed}^3(t)/ (m^* R^3)$
is very small for $R_{\rm seed}(t)\ll R_{\rm syst}$, where $R_{\rm syst}$ is the radius of the whole system. In cases of the quark--hadron and nuclear liquid--vapor phase transitions in heavy-ion collisions  $R_{\rm syst}(t)$ is the radius of the expanding fireball. Usage of the isothermal approximation in Eq.~(\ref{delr}) needs fulfillment of the inequality $t_{\rho}\sim \frac{R_{\rm seed} }{u_{\rm as}}\gg t_T$. For  $\epsilon \sim \epsilon_m$ we get $t_{\rho}\sim (R/l_0) t_0$, and isothermal approximation is valid for $R_{\rm seed}<R_{\rm fog}$ ($R_{\rm fog}$ is typical radius of the seed at which  $t_T\sim t_\rho$). For seeds with sizes $R>R_{\rm fog}$, $t_T\propto R^2$ exceeds
$t_{\rho}\propto R$ and growth of seeds is slowed down. Thereby,
the number of seeds with the size $R\sim R_{\rm fog}$ may increase with
time (stage of a nuclear fog). Estimates \cite{Skokov:2009yu,Skokov:2010dd}
show that for the hadron--quark phase transition  $R_{\rm
fog}\sim 0.1-1$ fm and for the nuclear liquid--gas transition $R_{\rm fog}\sim
1-10$ fm $\lsim R_{\rm syst}(t_{\rm f.o.})$, where $R_{\rm syst}(t_{\rm f.o.})$ is the fireball size at the freeze-out  time.
 Thus, thermal conductivity effects may manifest themselves  in heavy-ion
collision dynamics. In the clouds in terrestrial conditions the value $R_{\rm fog}$, continuing to grow,  may become so large that the gravity comes into play and there may arise the rain.

Substituting Eq.~(\ref{delr}) to Eq.~(\ref{v-tS}) for $T\simeq$ const (that is correct in linear approximation) we obtain
 \begin{eqnarray} \label{deltasinf}
&&\delta s = \left(\frac{\partial s}{\partial
n}\right)_{T}\left\{\frac{v(T)}{m}\left[\pm\mbox{tanh} \frac{r-R_{\rm seed}
(t)}{l}+\frac{{\epsilon}}{2\lambda_{cr} v^3 (T)}\right]+(\delta
n)_{\rm cor}\right\}.
 \end{eqnarray}

Note that for the description of the expanding fireball formed in heavy-ion collisions  the approximation of a quasi-adiabatic expansion  can be used  even in presence of the weak first-order phase transition (for $\delta s \ll s$ and $\delta n \ll n$). The evolution of droplets/bubbles in metastable region can be considered at fixed size of the fireball  provided  expansion time $t_{\rm f.o.}\gg (t_\rho, t_T)$.

The  dynamics is characterized by the relations between various time-scales. The total time of the phase transition is given by
\be
t_{\rm tr}=t_W +t_\rho +t_T\,,
 \ee
 where $t_W\sim t_0 e^{\delta F(r_{cr})/T}$ is the typical time of the cooking of the initial seed of the overcritical size.

\begin{figure}
\centerline{%
{\includegraphics[height=5.0truecm] {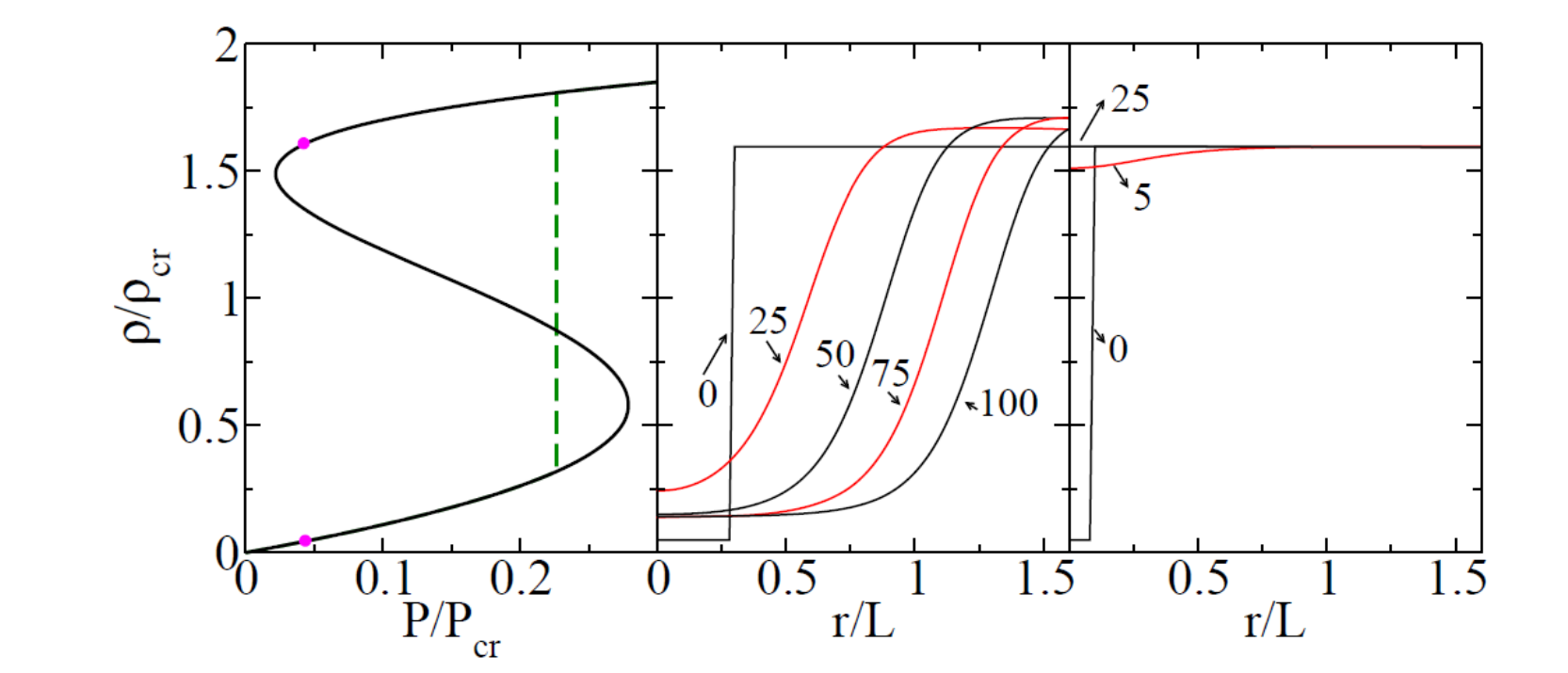}   } }
%\vspace{9.0mm}
 \caption{Isotherm for the pressure as a function of the density, with initial and final configurations
shown by dots  (left column).
 Dash vertical  line corresponds to the
MC on the curve $P(1/\rho)$.
Initial state relates to stable  vapor-phase disk in metastable overheated liquid.
Middle column shows time evolution of the density profiles
for the overcritical  vapor disk.
 Right column, the
same for initially undercritical vapor disks.
Numbers near curves (in $L$) are time snapshots; $r=\sqrt{x^2
+y^2}$,  $L= 30$ fm,   $T_{cr}=162$~MeV, $T/T_{cr}=0.85$, $n/n_{0}=1.3$,
$\eta \simeq 45$MeV$/\mbox{fm}^2$, for $\beta =0.2$. Figure is adopted from
\cite{Skokov:2009yu}.}
%\vspace{0.0mm}
\label{EoS_dynamicsQ}
\end{figure}

 In Fig. \ref{EoS_dynamicsQ}  we show the  dynamics of the stable vapor-phase disk in the metastable liquid surrounding for the parameter
choice $T_{cr}=162$~MeV, $n/n_{sat}=1.3$, relevant for the
hadron--sQGP phase transition. We take $T/T_{cr}=0.85$
and compute the configuration for $\eta \simeq
45$MeV$/\mbox{fm}^2$ and for $\beta =0.2$ (effectively large
viscosity). The seed-vapor disk of an overcritical size is increasing in size (see the middle column), whereas the disk of an undercritical size (see the right column) is decreasing in size.
Further details of calculations see in \cite{Skokov:2009yu}.
\begin{figure}
\centerline{%
\rotatebox{0}{\includegraphics[height=5.0truecm] {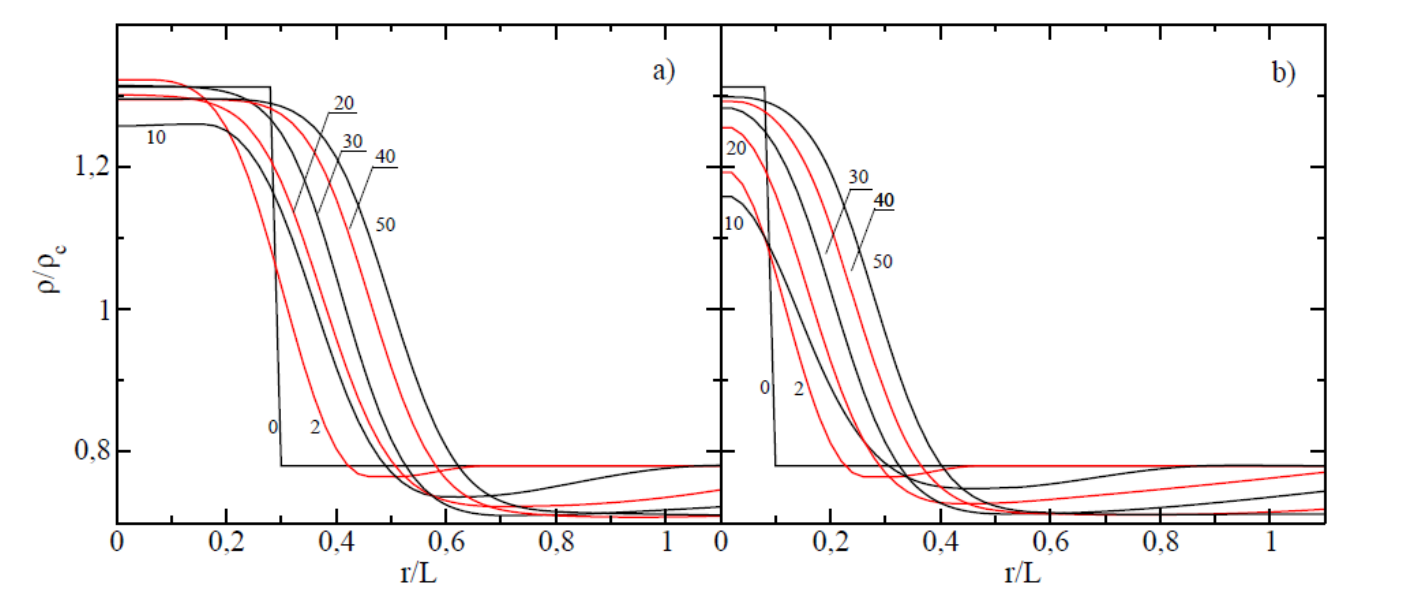}   }  }
\caption{Evolution of  bands ($d_{\rm sol}=1$) of initially large
(left) and  small (right) sizes, $r =|x|$. Parameters: $T/T_{cr}=0.98$, $L=30$~fm,
$T_{cr}=18.6$~MeV, $n_{cr}/n_{0}=0.42$, $\eta \simeq
3.2$~MeV$/$fm$^{2}$, for $\beta \simeq 12.6$ (effectively small
viscosity). Figure is adopted from
\cite{Skokov:2009yu}.
%\vspace{20mm}
}
\label{cracks}
\end{figure}

\subsection{Evolution of bands ($d_{\rm sol}=1$) and two-dimensional  seeds ($d_{\rm sol}=2$)}
 In Fig.
\ref{cracks} we show dynamics of liquid bands ($d_{\rm sol}=1$) in metastable vapor
phase. Values $T/T_{cr}=0.98$, $L=30$~fm,
$T_{cr}=18.6$~MeV, $n_{cr}/n_{0}=0.42$, $\eta \simeq
3.2$~MeV$/$fm$^{2}$ and $\beta \simeq 12.6$ (case of effectively small
viscosity). These solutions are similar to  slabs in $d=3$.
 Left panel
shows time evolution of a  band of a large initial size ($R_0
=0.3L$), whereas right panel demonstrates evolution of a band
having initially rather small size ($R_0 =0.1L$). In difference
with disks (solutions with $d_{\rm sol}= 2$) in both cases
(for large and small initial sizes of bands) dynamics looks
similar: bands of the stable phase, being prepared in the
metastable phase, undergo growth to the new phase.
Nevertheless, we also see that during the shape reconstruction the
slab first begins to dissolve and then  grows. This peculiarity
appeared since initial form of the density distribution  that was
exploited in numerical calculations deviates from the form given by
analytical solution. Thus actually even for slabs there
might exist a small critical size, that depends on peculiarities
of the initial density profile.  Slabs having sizes smaller than
this critical size could then completely dissolve.

 The
probability to prepare a band in a fluctuation is tiny. However,
 bands of the stable phase could be
formed near the system boundary, provided the latter is flat.

\begin{figure}
\centerline{%
\includegraphics[height=5.0truecm] {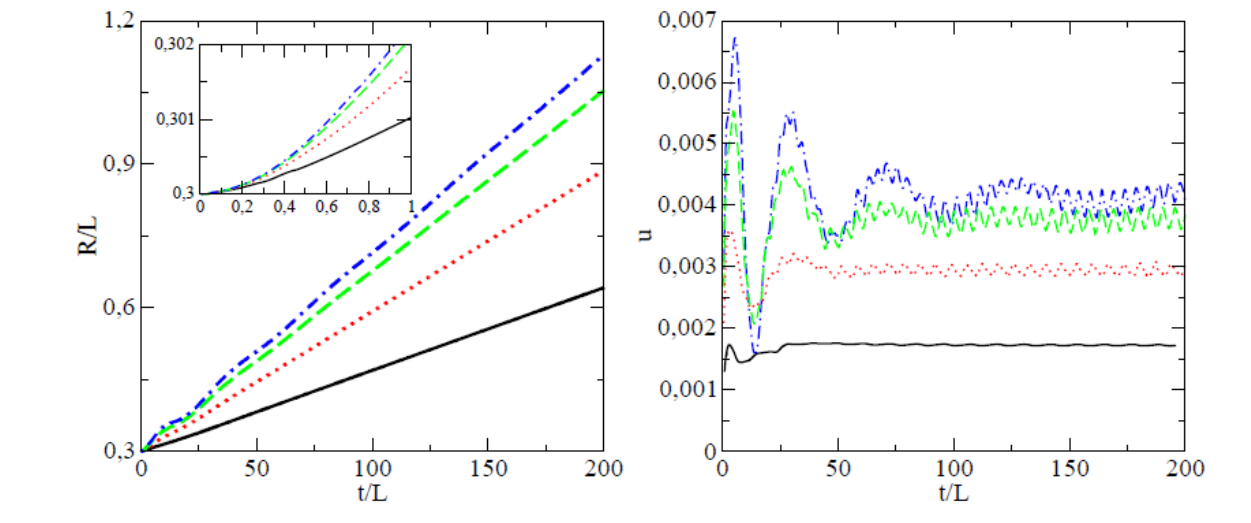}
 } \caption{Solution $R(t)$ (left panel) and $u=dR/dt$ (right
panel) for the band boundary for several values of the viscosity.
Solid, dotted, dashed and dash-dotted lines are calculated for $\beta=10^{-1}; 1; 10; 100$, $\delta{\cal T}=0.02$, $L=30$ fm. Figure is adopted from
\cite{Skokov:2009yu}.
%and other parameters are taken the same as in  left panel of Fig. \ref{interface}
}\label{band}
\end{figure}

In Fig. \ref{band}    we demonstrate the law for the growing with
time of the band boundary $R(t)$ (in the left panel) and the
velocity of the boundary $u=dR/dt$ (in the right panel) for
different values of the viscosity. As for discs, the band boundary
is specified as the point, where the density reaches the critical
value ($\rho =\rho_{cr}$). Results are presented for
$T/T_{cr}=0.98$, $L=30$~fm. For small values of time (see Figure
insertion)  $R(t)$ obeys the quadratic law.
Solid curve (case of effectively large viscosity, $\beta
=0.1$) shows the evolution of the slab-seed
for $t>t_{\rm rec}\sim
100$~fm,
the reconstruction time increases
as $t_{\rm
rec}\propto \sqrt{\beta}$. It is clearly demonstrated in the right
panel, where  the time dependence of  the velocity of
the seed growth is presented. As follows from the Figure, even for large times
the velocity $u$ does not obey the scaling law.
The velocity of the seed surface still slowly increases
with time. The asymptotic regime is reached at  larger values of
time (or for smaller  $\beta$ at values of time shown in Fig.  \ref{band}).
Another important issue is presence of the damped  long-wave
oscillations, which are clearly seen for all values of the
effective viscosity. They occur at $t\lsim t_{\rm rec}$. Moreover,
in case of an effectively small viscosity the short-wave quasi-periodic oscillations are
clearly seen.

Finally, we would like to notice that obtained solution for $\beta\gg 1$ describing accelerated expansion of the  seed and oscillations might be useful in application to the modeling of the accelerated expansion of the Universe with the formation of inhomogeneities.

\subsection{Dynamics of fluctuations in unstable region}\label{unstable}
\subsubsection{Growth of fluctuations of small amplitude}
The  phenomenon illustrating exponential
growth of fluctuations in the spinodal region is the opening  of the bottle of  champagne.
The dynamical trajectories of the expanding baryon-rich matter in the  heavy-ion collisions and of the falling matter in supernova explosions, until a phase transition did not occur, can be  characterized by approximately conserved entropy, whereas the volume $V_3$ and the temperature $T$ are changed with the time.
 Adiabatic trajectories of the matter and different possibilities of occurring of the liquid--vapor type phase transition
were
shown in Fig. \ref{spinT} on the plot of $T/T_{cr} = f (n/n_{cr})$.
Note that in reality for the quark--hadron first-order phase transition the plot looks a bit different, since then $T_{cr}$ increases with a decrease of the baryon density \cite{RandrupGQP,Steinheimer:2013gla,Steinheimer:2016bet}. However  this peculiarity   does not change a general analysis.

Now let us consider evolution of fluctuations in the spinodal region.
First, we find solutions of the linearized hydrodynamical equations, cf. \cite{Skokov:2010dd}. Now   ``r''-reference point can be placed at arbitrary  distance from the critical point, and assumption of a small over-criticality is not as essential. So,  let us suppress the subscript   ``r''. Let us introduce
\begin{eqnarray}\label{delns}
\delta n =\delta n_0 \mbox{exp}[\gamma t +i \vec{k}\vec{r}] -\frac{\epsilon}{m^*\lambda v^2}\,,
\quad \delta s= \delta s_0 \mbox{exp}[\gamma t +i
\vec{k}\vec{r}],
\quad T =T_{>}+\delta T_0
 \mbox{exp}[\gamma t +i \vec{k}\vec{r}],
\end{eqnarray}
where $T_{>}$ is the temperature of the uniform matter. For $|\delta n|\gg |\frac{\epsilon}{m^*\lambda v^2}|$, i.e. for $|\epsilon| \ll |\epsilon_m|$,  description of a fluctuation  in spinodal region at the first-order phase transition and at the second-order phase transition are  similar  and we may put $\epsilon \to 0$.

 For the case of a finite
thermal conductivity, $\kappa \neq 0$, from  Eqs. (\ref{v-tS}),
(\ref{v-tS1}) we obtain
\begin{eqnarray}\label{123}
&&\delta s_0 =\delta n_0\frac{ s  }{ n \left[1+\kappa p^2/(c_{V
}\gamma)\right]} \left[1+\frac{ n \kappa
p^2(\partial s/\partial n)_{T} }{\gamma s c_{V }}\right].
\end{eqnarray}
 The increment,
$\gamma (p)$, is determined by the linearized Eq. (\ref{v-t1}), where we put
 \begin{equation}\label{varP}
 \delta P= \left( \frac{\partial P}{\partial n} \right)_{T}\delta n +\left( \frac{\partial P}{\partial T} \right)_{n}\delta
 T .
 \end{equation}
Thus from
% Then from
linearized equations of the non-ideal hydrodynamics
%(\ref{v-t1}),  (\ref{v-tS})
we find
 the increment, $\gamma (k)$,
 \begin{eqnarray}\label{incr} &&\gamma^2 =
-k^2 \left[\widetilde{\omega}^2 (k^2) +\tilde{\eta}\gamma
%+ck^2
+\frac{u_{\tilde{s}}^2 -u_T^2 }{1+\kappa k^2
/(c_{V}\gamma)}\right], \end{eqnarray}
where $\tilde{\eta}=\frac{(\tilde{\nu}\eta +\zeta) }{m^* n}$ and quantity $\omega_0(k^2)$ is determined in Eq. (\ref{omtilde}).
Eq. (\ref{incr}) has three solutions corresponding to the growth, damping and oscillation  of the density- and thermal modes. It differs from the equation derived in \cite{Pethick-Ravenhall88} by presence of the surface tension term. General solutions can be found in \cite{Skokov:2009yu,Skokov:2010dd,Voskresensky:2020yrd}.

\subsubsection{Limit of a rather high thermal conductivity}
Let us focus on the limit of a rather high thermal conductivity, cf. \cite{Skokov:2009yu,Skokov:2010dd},  when  the temperature of the seed can be put constant and we may deal with only one equation for the density mode (\ref{v-t1}). Thereby, let us present solutions for  $\kappa k^2
/(c_{V}|\gamma|)\gg 1$.  Then
\begin{eqnarray}\label{incrn} &&\gamma^2 =
-k^2 \left[\widetilde{\omega}^2 (k^2) +\tilde{\eta}\gamma
\right]\,, \end{eqnarray}
from where we find two  solutions for the density-modes,
\begin{eqnarray}\label{Solincr}
\gamma_{1,2} =-\frac{k^2\tilde{\eta}}{2}\pm \sqrt{\frac{k^4\tilde{\eta}^2}{4}-k^2 \tilde{\omega}^2 (k^2)}\,.
\end{eqnarray}
For $\tilde{\omega}^2 (k^2)<0$, that corresponds to  the region of the phase transition,  the upper-sign solution, $\gamma_1 >0$,  describes the growing mode and the lower sign solution, $\gamma_2 <0$, describes the damping mode.
For $k^2 \tilde{\eta}^2/|\tilde{\omega}^2 (k^2)|\ll 1$, we have
\begin{eqnarray}\label{gam1lowvisk}
\gamma_1 \simeq \sqrt{-k^2\tilde{\omega}^2 (k^2)}-\frac{k^2\tilde{\eta}}{2}+O(k^3)
 \end{eqnarray}
 for the growing mode.  In the  limit $k^2 \tilde{\eta}^2/|\tilde{\omega}^2 (k^2)|\gg 1$, we obtain
\begin{eqnarray}\label{gam1highvisk}
\gamma_1 \simeq -\tilde{\omega}^2 (k^2)/\tilde{\eta}+O(\tilde{\omega}^4 (k^2)/(k^2\tilde{\eta}^3))\,.
\end{eqnarray}

{\bf In case $k_0=0, c>0$,}  for the most rapidly growing mode of a small amplitude (for $\gamma_m =\mbox{max}\{\gamma_1\}$ corresponding to $k=k_m$)  from  (\ref{Solincr}) we find
  \begin{eqnarray}\label{gam12m}
 \gamma_{m} \simeq \frac{\lambda v^2  }{(2\sqrt{\beta}+1)\tilde{\eta}}\,, \quad k_m^2 \simeq \frac{\lambda v^2\sqrt{\beta}}{(2\sqrt{\beta}+1)c}\,.
 \end{eqnarray}
Note that the value $k_m\neq 0$ exists only because the dynamics is determined by the nonideal hydrodynamical equations, which resulted in the second-time-derivative Eq. (\ref{v-t1}), instead of the first-time-derivative   Ginzburg--Landau equation, which yields $k_m=0$, see also Eqs. (\ref{eqMchih0})-(\ref{secorder1}) above.

{\bf{ For $k_0\neq 0, c<0, d>0$}} the most rapidly growing mode corresponds to $k \simeq k_0$, then $\tilde{\omega}^2 (k_0^2)<0$ and $|\tilde{\omega}^2 (k_0^2)|$ as a function of $k^2$ is the largest.

Let us continue to study limit of a high thermal conductivity
and consider now fluctuations in the system close to
equilibrium.  Assuming that $\psi$ is
close to
 its new equilibrium value, $\psi_{eq}\simeq \pm 1 +\widetilde{\epsilon}/4$, we put $\psi
 =\psi_{eq}
 +\delta\psi$ in Eq. (\ref{dimens}) and linearize the latter equation in
 $\delta\psi$:
 \be\label{newst}
  -\beta\frac{\partial^2 \delta\psi}{\partial{\tau}^2} =\Delta_{\xi} \left(
\Delta_{\xi} \delta\psi -4 \delta \psi -\frac{\partial
\delta\psi}{\partial{\tau}} -\frac{\lambda v^2 d}{2c^2}\Delta^2_\xi \delta\psi+\frac{2c k_0^2 +2dk_0^4}{\lambda v^2}\delta\psi\right).
 \ee
Setting $\delta\psi= \mbox{Re}\{\psi_0  e^{\gamma_{\psi} \tau
+i\vec{k}_\psi\vec{\xi}}\},$
where $\psi_0$ is an arbitrary but small real constant, we find
 \be\label{gamst} \gamma_{\psi}(k_\psi)=({2\beta})^{-1}\left(-{k^2_\psi}\pm \sqrt{k^4_\psi (1-4\beta)
 -{16\beta (k^2_\psi -k^2_{0\psi})+2\beta^2 d(k^4_\psi - k^4_{0\psi})/c^2}}\right)\,,
  \ee
  where $k^2_{0\psi}=2ck_0^2/(\lambda v^2)$.

In case $k_0=0$, $d=0$ and for effectively  large viscosity ($\beta \ll 1$) and for
$k^2_\psi >16\beta$ there are only  damped solutions. For $k^2_\psi \gg 16\beta$:
 \be\label{large-gst}
\gamma_{\psi}^{(1)}(k_\psi)\simeq -(k^2_\psi +4),\quad
\gamma_{\psi}^{(2)}(k_\psi)\simeq -{k^2_\psi}/{\beta} .
 \ee
Fluctuations with large $k$ rapidly dissolve with  time. Existence of long living short-wave excitations  is unlikely
in the viscous medium. However there remain long-wave damped
oscillations, for  $k^2_\psi < 16\beta$.

 In  case of effectively small viscosity ($\beta \gg 1$) we get
 \be\label{smeq}
\gamma_{\psi}(k_\psi)= -({2\beta})^{-1}\left({k^2_\psi}\pm {4i \sqrt{
\beta}\,k_\psi} \sqrt{1+{k^2_\psi}/{4}}\right),
 \ee
 that corresponds to oscillating  slowly damped modes
near the  equilibrium state.
Since $t_0 \sqrt{\beta}$ does not depend on the viscosity and $t_0
\beta \rightarrow \infty$ for $\eta , \zeta\rightarrow 0$, in case
of the ideal fluid rapid oscillations  continue till the energy is
transported to the surface of the system (the process is governed
by the heat transport) or till the energy is radiated away in the
course of direct reactions. Thus {{in case of effectively small
viscosity  the stable phase is covered by fine ripples  during
some rather long period of time.}}

For $k_0\neq 0$, $d>0$ putting $k\simeq k_0$ in (\ref{gamst}) we obtain
\be\label{gamst1} \gamma_{\psi}(k_\psi)=({2\beta})^{-1}\left(-{k^2_\psi}\pm \sqrt{k^4_\psi (1-4\beta)}\right)\,,
  \ee
so both solutions are damped. Also, for $\beta >1/4$ (small viscosity) becides the damping there are oscillations.

Now let us consider a general case ($k_0=0$ either $\neq 0$), and seek the solution of nonlinear equations  (\ref{v-t1}), (\ref{dimens}))
in the form \cite{Voskresensky:2020yrd}
\begin{eqnarray}
m^* \delta n=a_{\rm stan} f(t) \left[\cos (k x+\chi) +\frac{c_1 \widetilde{\omega}^2 (k^2)}{\widetilde{\omega}^2 (9k^2)}\cos (3k x +\chi) +...\right]+O(\epsilon)\,,\label{sinExpnonlin}
\end{eqnarray}
 with $a_{\rm stan}^2 =-\frac{4\widetilde{\omega}^2}{3\lambda}>0$ as in (\ref{sinExp}) but now with  $f(t)$ satisfying equation
\begin{eqnarray}\label{ft}
\partial_t^2 f =-k^2 \widetilde{\omega}^2 (k^2) f(1-f^2) -k^2 \tilde{\eta} \partial_t f\,.
\end{eqnarray}
For $k^2 \tilde{\eta}^2/|\tilde{\omega}^2 (k^2)|\gg 1$,  for $\beta \ll 1$ or $\tilde{\eta}\gg \sqrt{c}$,   the term  $\partial_t^2 f$ in the l.h.s. of Eq. (\ref{ft}) can be dropped and the amplitude
\begin{eqnarray}\label{finterp}
 f(t)=\frac{f_0 e^{\gamma t}}{\sqrt{1+f_0^2 e^{2\gamma t}}}\,
\end{eqnarray}
 fulfils  the resulting  Eq. (\ref{ft}) with $\gamma =-\tilde{\omega}^2 (k^2)/\tilde{\eta}$,  $f_0/\sqrt{1+f_0^2}$ shows the amplitude of the  fluctuation at $t=0$,  $f_0$ is an arbitrary constant. For  $k\sim k_m$ at $k_0=0$ this solution  holds for $k_m^2 \tilde{\eta}^2/|\tilde{\omega}^2 (k_m^2)|\gg 1$. For $k=k_0\neq 0$ the criterion of applicability is $k_0^2 \tilde{\eta}^2/|\tilde{\omega}^2 (k_0^2)|\gg 1$. In both cases $k_0=0$ and $k_0\neq 0$ with the density distribution given by  (\ref{sinExpnonlin}), (\ref{finterp})
 the free energy renders
\begin{eqnarray}\label{freeent}
\delta F_{\rm L} (t) =-\frac{V_3\widetilde{\omega}^4(k^2)}{6\lambda m^* n}f^2(t)\left(2-f^2(t)\right)\,.
\end{eqnarray}
For $t\to \infty$ we have $f(t\to\infty)\to 1$ and $\delta F_{\rm L}$ reaches the minimum. For $k=k_0$ this value coincides with (\ref{FLknulk}) given by the stationary solution.

In general case, Eq. (\ref{finterp}) yields an interpolation between  two approximate  solutions of  Eq. (\ref{ft}) valid for the limit cases  $\gamma t\ll 1$ and  $\gamma t \gg 1$. Replacing (\ref{finterp}) in Eq. (\ref{ft}) we obtain then the same solutions (\ref{Solincr}) as in linear case.

Let first $k_0 =0$. For $t\to \infty$ employing solution (\ref{finterp}) at $\gamma =\gamma_m =\gamma(k_m)$ we find
\begin{eqnarray}\label{FLgamm}
\delta F_{\rm L} (t\to \infty)=-\frac{\widetilde{\omega}^4 (k_m^2)V_3}{6\lambda m^* n}.
\end{eqnarray}
 For the case  of a large effective viscosity/inertia, $\beta \ll 1$, we obtain $\delta F_{\rm L} (t\to \infty)\simeq -\frac{\lambda v^4 V_3}{6 m^* n}$ that coincides with (\ref{FLknulk}) but is still larger than the value given by (\ref{FLknul}). For the case   of a  small  effective viscosity/inertia, $\beta \gg 1$, we find $\delta F_L (t\to \infty)\simeq -\frac{\lambda v^4 V}{24 m^* n}$ that is much higher than the free energy given by both stationary solutions (\ref{FLknul}), (\ref{FLknulk}).  Thus one may expect that expression (\ref{FLgamm}) either describes  a metastable state or a state, which slowly varies on a time  scale $t_k \gg t_\gamma \sim 1/\gamma_m$, reaching for $t\gg t_k$ the stationary state with the free energy given by (\ref{FLknul}).

  To show the latter possibility consider the case $\beta \gg 1$ and assume $k$ in solution (\ref{sinExpnonlin}) to be a slow function of time, i.e. $k=k(t)$, for typical time scale $t_k\gg t_\gamma$.
One can see that for   $R_{\rm seed}\ll 1/k_m$  the quantity  $k(t)$ satisfies equation
$(d^2k/dt^2) =-k^2 \tilde{\eta}(dk/dt)$ with the solution
\begin{eqnarray}\label{kt}
k(t)==k_{00}[1+{2}\tilde{\eta}k^2_{00}t/3]^{-1/2} =k_{m}[1+\tilde{\eta}\lambda v^2 t/(3c)]^{-1/2}
\end{eqnarray}
 for $k_{00}=k_m$ from (\ref{gam12m}), such as $k(t\to \infty)\to 0$ and the free energy for $t\to \infty$ indeed reaches the limit (\ref{FLknul}) provided we set $\cos \chi \simeq \frac{\sqrt{3}}{2} -\frac{\sqrt{3}\,m^*\tilde{\epsilon}}{8}$.
From Eq. (\ref{kt}) we easily find that the typical time scale is
 $t_k \sim \beta t_0$ and we check that indeed $t_k\gg t_\gamma$. For $R_{\rm seed}\gsim 1/k_m$ the solution (\ref{sinExpnonlin}) with (\ref{kt}) does not hold and should be modified.

 For $\beta \ll 1$, $k_0=0$,  Eq. (\ref{kt}) with slowly varying $k(t)$ does not hold.  At  realistic conditions,  convection and sticking processes (at sizes $\sim l$) may be allowed, which destroy periodicity, and owing to these processes the system may finally reach the ground  state with the free energy given by  (\ref{FLknul}).
 Thus one possibility is that for the typical  time $t\sim t_\gamma\sim t_0$ the  quasiperiodic solution
 is formed with typical $k\simeq k_m$, corresponding to a metastable state with the free energy given by (\ref{FLknulk}). Such a distribution is formed most rapidly.    Another possibility is that for the  typical time scale $t_{hom}> t_\gamma$ in the system of a large size an approximately  homogeneous solution
 is developed.
 In the latter case to proceed consider the case $k\sim 1/R_{\rm syst}\ll k_m$, where $R_{\rm syst}$ is the typical size of the system ($R_{\rm syst}=R_{\rm f.o.}$ for
 the fireball formed in heavy-ion collisions). The spatially homogeneous solution of equation
 $$\Delta_\xi \psi+2\psi (1-\psi^2)+\tilde{\epsilon}=\partial_\tau \psi\,,$$
  that follows from (\ref{dimens}) in this case (as well as for seeds of a  size $R_{\rm seed}\ll R_{\rm syst}$ at  $\beta \ll 1$, as we have argued above), is given by
 \begin{eqnarray}\label{unif}
 \psi (t) =\pm 1/\sqrt{1+e^{-\tau}{(1-\psi_0^2)}/{\psi_0^2}}\,,\end{eqnarray}
 where we for simplicity have put $\tilde{\epsilon} \to 0$. Typical time needed for the initial amplitude $\psi_0\ll 1$ to grow to $\psi(t\to \infty)\simeq \pm 1$ is  $t_{hom}\sim t_0 \ln (1/\psi_0^2)\gg t_\gamma$.

 Thus,
 additionally  to the homogeneous solution (\ref{unif}) we found some novel solutions describing evolution of fluctuations in the region of the instability. For $k=$const$\neq 0$ we found  periodic solutions  given by (\ref{sinExpnonlin}), (\ref{finterp}). For $k=k_0\neq 0$ the solution yields minimum of the free energy for $t\to \infty$. For $k_0=0$, $\beta \gg 1$,  we found quasiperiodic  solutions (\ref{sinExpnonlin}), (\ref{finterp}) with    $k=k(t)$ from (\ref{kt}), yielding minimum of the free energy for $t\to \infty$.

Above we considered the case of a large thermal conductivity. Discussion of  cases of a small and zero thermal conductivity can be found in \cite{Skokov:2009yu,Skokov:2010dd}.

\subsubsection{First-order phase transition in  the process of cooling}

Although different aspects of the nucleation processes have been enlightened  in many  textbooks and reviews, cf. \cite{LifshizPitaevskii1981,Onuki,Schmelzer2005,Kapusta2006,Slezov2009} and refs. therein, active investigations in this field are continued.
If  during the process of the cooling of the system characterized by the time $t_{\rm cool}$ the first-order phase transition to the inhomogeneous state starts from the metastable phase and chemical potentials in both phases are close to each other, the time for the preparation of the overcritical seed of stable phase, $t_W$, is very large. In  case of a slow cooling, when  $t_{\rm cool}\gg t_W$,  a monocrystal is formed. At a smaller value of $t_{\rm cool}$, the resulting state contains crystal domains with different orientations of the wave vector $\vec{k}_0$. After this state is formed, it undergoes an aging process, in which the domains coalesce to larger domains with a more energetically profitable structure.
In the limit case,  $t_{\rm cool}\ll t_W$, the glassing transition either starts from a metastable region near the top of the energy barrier, cf. \cite{Schmelzer2022}, or the system trajectory
passes to the spinodal region \cite{Voskresensky:1993ux}  and then the polycrystal or glass-like state are formed, in which
the directions of $\vec{k}_0$ change already at a short space-scale $l\sim 1/k_0$.
The viscosity of the amorphous  state is very large and after a long time in the aging process the matter transforms to the monocrystal.

In case of the nuclear matter, being  of the main interest of the present review, a similar increase of the bulk viscosity should occur at  densities $n>n_c^{(1)}\sim (0.5-0.8)n_0$ (in the liquid phase of the pion condensation) due to occurrence of the pion fluctuations with finite momenta $k\sim k_0\sim p_{\rm F,N}$, cf. \cite{Voskresensky:1993ud,Voskresensky:2020yrd}. With increasing density, the bulk viscosity still increases reaching the maximum in the critical point of the pion condensation to the liquid-crystal or solid-like state (at $n=n_c^\pi$).  In neutron star matter these effects may help to explain $r$-mode damping in rapidly rotating pulsars, cf. \cite{Kolomeitsev:2014epa,Kolomeitsev:2014gfa}.
Also note that the bulk viscosity should increase near the critical point of any first-order phase transition and it diverges in the critical point of the second-order phase transition. Such a critical opalescence phenomena should occur at the hadron--quark phase transition, cf. \cite{Kerbikov2018}.
In addition, Ref. \cite{Torrieri2008} suggested  that in heavy-ion collisions the spike of the bulk viscosity near the critical point  could trigger instabilities that rapidly break the system into evaporating clusters.

\subsubsection{Sticking of domains}
A supply of the material across the border of the seed occurring in the diffusion process during the seed growth to the stable phase from the metastable one requires a long time. So, near the seed there may appear a sparse   space. In this case two domains of the stable phase formed nearby each other may undergo a sticking. Indeed the surface free energy  due to the presence of the domain-domain boundary is estimated as
$\delta F_{\rm d-d}\sim R^2 l_\psi \omega_0^4 \tilde{h}/\Lambda_0$, at $\tilde{h}\ll 1$, whereas the surface energy at the domain-vacuum boundary is
$\delta F_{\rm d-v}\sim R^2 l_\psi \omega_0^4 /\Lambda_0\gg \delta F_{\rm d-d}$, cf. \cite{Voskresensky:1993ux}. Thus the domain-domain surface energy is less than the domain-vacuum plus domain-vacuum surface energy.
Thereby, there appears attractive sticking force $\sim (\delta F_{\rm d-d}-\delta F_{\rm d-v})/l_\psi$, which permits sticking of the matter. Domains of a whimsical form can be formed during the sticking process. In case when quasi-one dimensional configurations are energetically favorable the domains are elongated.
The processes of the sticking of seeds, formation of a larger-size domains and formation of monocrystal take a very long-time.
%At the end, let us pay attention to

Finally, note that domain structures naturally appear in systems described by the vector-boson condensates such as ferromagnets and ferromagnetic superfluids and superconductors, cf. \cite{Voskresensky:2019zcp}.

\section{Pion condensation in dense and not too hot nuclear matter}\label{Pion-section}
\subsection{Fermi liquid description of $NN$ and $\pi N$ interactions}
Usually one supposes that nucleons and pions interact via
pseudovector $\pi NN$ coupling  described by the
Hamiltonian
 \cite{Migdal78,Ericson:1988gk,MSTV90}:
\begin{equation}\label{f-vertex}
H_{\pi NN}^{\rm p.v.}  =
f_{\pi NN}\bar{\psi}_N \gamma^{\mu}\gamma_5
\partial_{\mu}(\vec \tau \vec \pi )\psi_N ,\,\,\,\, \mu =0, 1, 2, 3.
\end{equation}
Within the $\sigma$ model one  also uses  pseudoscalar coupling \cite{Migdal78,Ericson:1988gk,MSTV90}
\begin{equation}\label{f-sc-vertex}
H_{\pi NN}^{\rm p.s.} =
-i g \bar{\psi}_N \gamma_5\vec \tau \vec \pi \psi_N  ,\,\,
g/2m_N \simeq f_{\pi NN}\simeq 1/ m_{\pi}.
\end{equation}
 Both variants
correspond to the same non-relativistic limit expression for the vertex
\begin{equation}\label{f-non-vertex}
V^{n.rel}_{\pi NN} =-i f_{\pi NN} \sigma_{\alpha}
\tau_{\beta} \partial_{\alpha} \pi_{\beta}, \,\,\,\,\alpha , \beta
=1, 2, 3,
\end{equation}
$\gamma^{\mu}$ are Dirac matrices and $\vec \sigma$ and $\vec \tau$
are spin and isospin Pauli matrices.

Hamiltonian of $\pi N\Delta$ interaction is usually taken as
\begin{equation}\label{fdelta-vertex}
H_{\pi N\Delta}=
f_{\pi
N\Delta}\bar{\psi}_{\Delta}^{\mu}\vec{T}\partial_{\mu}\vec{\pi}\psi_N ,
\,\,\,\, f_{\pi N\Delta}\simeq
2.15~
f_{\pi NN},
\end{equation}
with appropriate non-relativistic limit expression for the vertex
\begin{equation}\label{f-non-vertex1}
V^{n.rel}_{\pi N\Delta}
=-i f_{\pi N\Delta} S_{\alpha}\partial_{\alpha}
T_{\beta} \pi_{\beta}   , \,\,\,\,\alpha , \beta
=1, 2, 3,
\end{equation}
where $\vec S$ and $\vec T$ denote transition spin and isospin operators.

A more local part of the nucleon--nucleon
interaction is due to exchanges  of heavy mesons, like
$\sigma$, $\rho$, $\Omega$, being characterized by large ($\gg 1$) values of the coupling
constants, cf. \cite{Glendenning2000}. Methods of the perturbation  theory do not work in this case.
The Fermi-liquid
approach with the explicit separation of the in-medium pion exchange was formulated  by A. B. Migdal for description of the cold nuclear matter, cf.  \cite{Migdal78}, and then it was  generalized for description of  equilibrium systems at finite temperatures ~\cite{Voskresensky:1982vd,MSTV90} and  for nonequilibrium systems, cf.  ~\cite{Voskresensky:1987hm,Voskresensky:1988pxw,Voskresensky:1989sn,
MSTV90,Voskresensky:1993ud,Kolomeitsev2011}.
In this approach the long-range (small momenta) processes
are treated explicitly, whereas short-range (large momenta) ones are described
by the local quantities approximated with the help of  the phenomenological Landau--Migdal  parameters.

At low excitation energies the retarded $NN$ interaction amplitude is
presented
as follows
%--------
\begin{eqnarray} \label{TKFS}
\parbox{7.3cm}{\includegraphics[width=10.3cm]{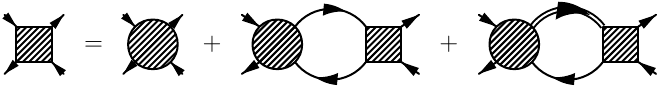}}
\end{eqnarray}
\begin{eqnarray}
\parbox{4.5cm}{\includegraphics[width=6.5cm]{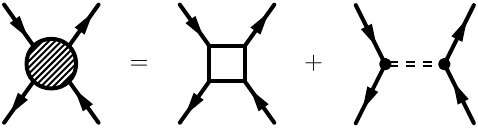}}\,
\label{irred}
\end{eqnarray}
%--------
The solid line, $iG_N$, stands for the nucleon quasiparticle, the
double line, $iG_\Delta$, for the dressed $\Delta$ isobar. For a higher excitation energy,  the $\Delta\Delta^{-1}$ diagrams should be included ~\cite{Voskresensky:1982vd,MSTV90}.  The doubly-dashed line shows the exchange by the free pion
with inclusion of the  residual s-wave $\pi NN$ interaction and $\pi\pi$
scattering. The empty block in Eq.~(\ref{irred})
is irreducible with respect to the particle--hole, $\Delta$--nucleon hole and pion states.   It is expressed via Landau--Migdal parameters as follows,
\begin{equation}
\label{localint}
\Gamma^{\omega}_{{\vec{n}}\alpha\beta,{\vec{n}}'\gamma\delta}=F_{{\vec{n}},{\vec{
n}}'}\delta_{\alpha\beta}\delta_{\gamma\delta}+G_{{\vec{ n}},{\vec{
n}}'}\vec{\sigma}_{\alpha\beta}\vec{\sigma}_{\gamma\delta}\,.
\end{equation}
 Due to the locality of the exchanges by massive mesons the momentum dependence of the values $F$ and $G$ is rather smooth. So, one usually assumes that  $F$ and $G$ are functions dependent only on  the direction of
the momenta of incoming and outgoing nucleon and hole at the Fermi surface. The dimensionless amplitudes are $f_{{\vec{ n}},{\vec{ n}}'}=C_0^{-1} F_{{\vec{
n}},{\vec{n}}'}$ and $g_{{\vec{ n}},{\vec{ n}}'}=C_0^{-1} G_{{\vec{
n}},{\vec{ n}}'}$,  $C_0^{-1}= m_N^* (n_0)p_{{\rm
F},N}(n_0)/\pi^2$ is the normalization factor -- the density of states at the Fermi surface
for $n=n_0$, $m_N^*$ is the nucleon effective mass, $p_{{\rm
F},N}$ is the nucleon Fermi momentum, $\vec{n}$ and $\vec{n}\,'$ are the directions of the fermion momenta
before and after scattering.
The amplitudes are  expanded in the Legendre polynomials. In most cases it is  sufficient
to deal  with zero and first harmonics, which can be
extracted from the comparison with the data or they should be computed within some models.
Most important role in nuclear physics problems is played by  the zero harmonics.

The $N\Delta$ interaction is constructed analogously to that for $NN$.
Information on the local part of the $N\Delta$ interaction is rather scarce. One usually supposes that the only important channel is the
spin-isospin channel. Thereby,
\begin{equation} \label{gam-d-l}
(-i)\parbox{1cm}{\includegraphics[width=1cm]{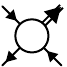}} \simeq C_0 g^{\prime}_{N\Delta}
(\vec{\tau}_1 \vec{\tau}_2 )
(\vec{\sigma}_1 \vec{\sigma}_2 )
\end{equation}

Resummation of the
diagrams shown in (\ref{TKFS}) in the pion channel  yields the Dyson
equation for the retarded pion full Green function $D^R =\mbox{Re}D^R -iA_\pi /2$,
%--------
\begin{eqnarray}
\includegraphics[width=9cm]{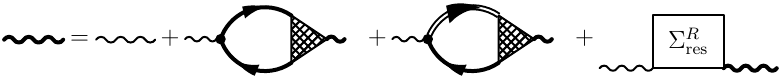}\,
\label{pion-l}
\end{eqnarray}
%--------
Here  the bold wavy line is the full pion Green function $iD^R$, thin wavy line is bare pion Green function $iD_0^R$, $-i\Sigma_{\rm res}^{R}$ is a residual retarded pion self-energy   including the contribution
of all diagrams, which are not presented explicitly in (\ref{pion-l}), like  the s-wave $\pi N$ and
$\pi\pi$ scattering. The full (retarded) vertex takes into
account $NN$ correlations
\begin{eqnarray}
\includegraphics[width=6cm]{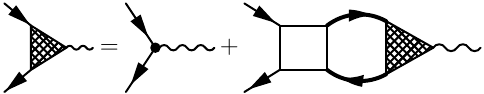}\,
\label{dressed}
\end{eqnarray}

Note that  Eqs. (\ref{TKFS})--(\ref{dressed}) presented for the retarded quantities are valid not only for the ground state but also for the equilibrium matter at $T\neq 0$, as well as for the  nonequilibrium matter, cf. \cite{Voskresensky:1993ud}.

In the matter of an arbitrary isospin composition the resummed $NN$ interaction amplitude reads
\begin{equation}\label{g-ampl}
\Gamma^{R}_{NN}=F+F^{\prime} \vec{\tau}_1 \vec{\tau}_2 +(G+G^{\prime}
\vec{\tau}_1 \vec{\tau}_2 )
\vec{\sigma}_1 \vec{\sigma}_2  +f_{\pi NN}^2 T_{\pi}\vec{\sigma}_1 \vec{k}
\cdot \vec{\sigma}_2 \vec{k}\,,
\end{equation}
$\vec{k}$ is the momentum transferred in the particle-hole channel. First
three  terms in Eq. (\ref{g-ampl}) arise due to the loop resummation of the empty block graphs in Eq.
(\ref{irred}), the last term in Eq. (\ref{g-ampl})
 is due to the contribution of the second diagram
of Eq. (\ref{irred}).
Amplitudes $F$, $F^{\prime}$, $G$, $G^{\prime}$,
and $T_{\pi}$ are expressed with the help of the dimensionless
Landau--Migdal parameters $f$, $f^{\prime}$, $g$, $g^{\prime}$,
related to the local $NN$ interaction in the particle--hole channel \cite{MSTV90}
\begin{eqnarray} \label{loc-const}
F&=&C_0 f\Gamma (f), \,\,\,F^{\prime} =C_0 f^{\prime}\Gamma (f^{\prime}),
\nonumber \\
G&=&C_0 g\Gamma (g), \,\,\,G^{\prime} =C_0 g^{\prime}\Gamma (g^{\prime}),\\
T_{\pi}&=& \Gamma^{2} (g^{\prime}) D_{\pi}^{R} (\omega , k)=
\Gamma^{2} (g^{\prime})/[\omega^{2} -m_{\pi}^{2} -k^{2} -
\Sigma^{R}(\omega , k)]. \nonumber
\end{eqnarray}
The Landau--Migdal parameters $f=(f_{nn}+f_{np})/2$, $f^{\prime}=(f_{nn}-f_{np})/2$ in the scalar channel are expressed via the nucleon incompressibility and can be calculated already on the mean-field level, e.g., cf. the calculation in \cite{Maslov:2015wba}. Calculation of the spin parameters $g=(g_{nn}+g_{np})/2$, $g^{\prime}=(g_{nn}-g_{np})/2$ needs to go beyond the mean-field level, cf. \cite{Backman:1984sx,Holt,Gambacurta,Speth}.

Resummation of Eq. (\ref{TKFS})
yields in the spin--isospin channel
%--------
\begin{eqnarray}
\includegraphics[width=6cm]{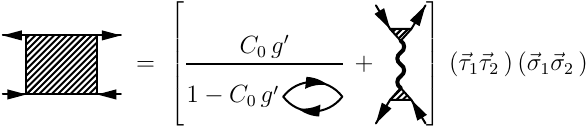}\,.
\label{gam-res}
\end{eqnarray}

Empirical value of $g_{N\Delta}^{\prime}\simeq 0.1-0.2$ is rather small,  cf.
\cite{Troitsky,MSTV90,Suzuki1999}, whereas   $g_{\Delta\Delta}^{\prime}$ was estimated as $\simeq
0.8$, cf. \cite{Troitsky,MSTV90}.
Thus with some accuracy one  may for simplicity drop the contribution of the local $N\Delta$ interaction.

 For $\omega\ll m_\pi$ and $k\lsim p_{{\rm F},N}$ the main contribution to the pion self-energy is determined by the second diagram in r.h.s. of (\ref{pion-l}). Therefore one has
\begin{equation} \label{gam-x}
\Gamma (x) =\left[ 1+2xp_{{\rm F},N}\Phi^R (\omega,k,T) /p_{{\rm F},N} (n_0 )
\right]^{-1}\,,
\end{equation}
$x=f,\, f^{\prime},\,
g,\, g^{\prime}$, $\Phi^R$ is the retarded Lindhard function
\begin{equation}
\Phi^R (\omega, k, T)=\Phi_1^R (\omega, \vec{k}, T)+
\Phi_1^R (-\omega, -\vec{k}, T)\,,\label{PhiT}
\end{equation}
and at zero temperature \cite{Migdal78}
\begin{eqnarray}\label{Pht}
\Phi_1^R (\omega, k, T=0)&=&-\frac{m^{\ast 2}_N}{2p_{{\rm F}.N}k^3}
\left[\frac{a^2-b^2}{2}\mbox{ln}\left( \frac{a+b}{a-b}\right) -ab
\right], \,\, \\
a&=&\omega -\frac{q^2}{2m^{\ast}_N},\,\,
b=kv_{{\rm F},N}\,,\quad q^2=\omega^2 -k^2\,, \nonumber
\end{eqnarray}
$v_{{\rm F},N}=p_{{\rm F},N}/m_N^*$.  For low $T$ the real part of the full Lindhard function in the limiting case of low frequences yields \cite{Voskresensky:1982vd,MSTV90}
\begin{equation}\label{rephi-lim-full}
\mbox{Re}\Phi (\omega \ll kv_{{\rm F},N} , k\ll 2p_{{\rm F},N} , T\ll \epsilon_{{\rm F},N}
)\simeq 1-\frac{\omega^2 m_N^{\ast 2}}{k^2 p_{{\rm F},N}^{2}}-\frac{\pi^2}{12}
\frac{T^2}{\epsilon_{{\rm F},N}^{2}}-\frac{k^2}{12p_{{\rm F},N}^{2}}\,,
\end{equation}
$\epsilon_{{\rm F},N}=p^2_{{\rm F},N}/m^*_N$.
%$\Re \Phi^R (\omega\ll m_\pi ,k\lsim p_{\rm F},T\ll \epsilon_{\rm F})\simeq 1$, cf. \cite{Voskresensky:1982vd,MSTV90}.

There were found two sets of the parameters fitted to atomic-nucleus experiments, cf.~\cite{M67a}, being~$f\simeq 0.25$, $f^{\prime}
\simeq 1$, $g\simeq 0.5$, $g^{\prime} \simeq 1$ and $f\simeq 0$, $f^{\prime}
\simeq$ 0.5--0.6, $g\simeq 0.05\pm 0.1$, $g^{\prime} \simeq 1.1 \pm 0.1$.
Uncertainties in numerical values appear due to attempts to
get the best fit to
experimental data in each specific case  modifying
the parametrization used
for the
residual part of the $NN$ interaction.
All of these numerical values of the parameters relate to the  isospin-symmetric matter and the density $n \simeq
n_0$, whereas there is no direct experimental information on their
values for $n >n_0$ and for isospin-asymmetric matter. There exist various calculations of the Landau--Migdal  parameters as functions of the density for the isospin-symmetric nuclear matter and for the purely neutron matter, cf.  \cite{Backman:1984sx,Holt,Gambacurta,Maslov:2015wba} and references  therein.

The spectrum of excitations with pion quantum numbers is determined
by the spectral function $A_\pi =-2\Im D^R_\pi$. In the quasiparticle approximation the spectrum is given by
$$\Re D^{R-1}_{\pi}(\omega  ,k)\simeq \omega^2 -k^2 -\Re\Sigma^R (\omega,k)=0\,.$$

 In vacuum  the pion dispersion law is $\omega (k)\equiv
\omega_k
=\sqrt{m_{\pi}^2 +\vec{k}^2 }$.
But the pion atom data can't be described with the free spectrum
indicating that the pion branch of the spectrum behaves like
$\omega (k) \simeq \sqrt{m_{\pi}^2 +\alpha_0 \vec{k}^2 }$,
$\alpha_0 \simeq 0.4$ for $n \simeq n_0$ and for small
momenta $k\lsim m_{\pi}$, cf. \cite{Troitsky,Delorme92,Ericson94,Kolomeitsev:2002gc,Kolomeitsev:2002mm,FriedmanGal20}. The main contribution to the pion polarization at $\omega \sim m_\pi$ is determined by the $\Delta$ term given by the third diagram in r.h.s. of (\ref{pion-l}). Besides the pion branch there is the  $\Delta$ branch of the excitation with the pion quantum numbers. In  vacuum it degenerates into the relation $\omega \simeq m_\Delta -m_N +k^2/(2m_\Delta)$. In the medium the $\Delta$ branch is modified.
Analysis of the $^3\mbox{He}$--$t$ reaction on nuclei, cf. \cite{Dmitriev:1984ud,Dmitriev:1986ud,Hennino:1993sx,Sneppen:1994zx}, demonstrated
agreement with  the pion spectrum consisting of the medium-modified  pion and $\Delta$ branches and with the assumption of the softening of these branches
for $k\gsim 2 m_{\pi}$.
%--------
\begin{figure}
\centerline{
\includegraphics[width=5cm]{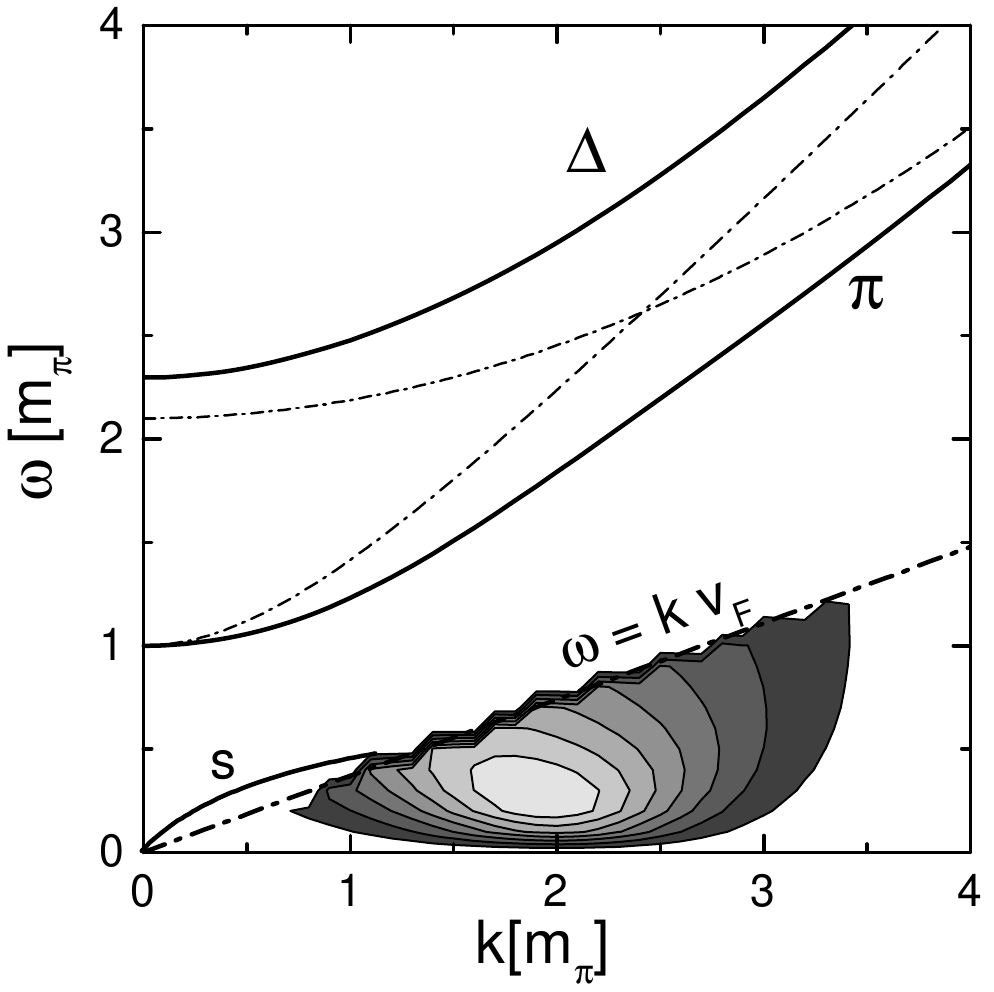}
} \caption{Pion spectrum in isospin-symmetric matter for $n=n_0$.
%, cf.  \cite{Kolomeitsev:2000ie}.
 Solid lines show
quasiparticle branches of pion excitations in medium, computed neglecting  effects of the particle widths. The contour plot depicts the spectral density of virtual pions. Thin dash-dotted lines are
the dispersion laws for pions and $\Delta$ in the free space, thick dash-dotted line shows $\omega =k v_{{\rm F}}$.}
\label{Spectrum}
\end{figure}

The spectrum of quasiparticles   $\pi^{\pm,0}$ in the isospin-symmetric matter and of $\pi^{0}$ also in asymmetric matter ($N\neq Z$) possesses three branches. For $n=n_0$ in the cold  isospin-symmetric matter the spectrum is shown in
Fig.~\ref{Spectrum}. For $\omega\gsim m_\pi$ there exist mentioned above two quasi-branches:  the pion branch and the $\Delta$ branch. For $\omega < m_{\pi}$ there is still the spin-sound branch (with $\omega \rightarrow
0$ when $k\rightarrow 0$). This branch appears owing to the nucleon-nucleon repulsive interaction determined in the Fermi liquid theory by the non-zero Landau parameters (see presence of the pole in (\ref{gam-res}) for $g'\neq 0$).  In the lower hatched region,  at $\omega < kv_{{\rm F},N}$ and $k\sim p_{{\rm
F},N}=m_N^* v_{{\rm
F},N}$,  the pion width cannot be neglected. This is the region of the Landau damping. The pion spectral function is enhanced in this region of $\omega$ and $k$ for $n>n_{c}^{(1)}\simeq (0.5$--$0.8)n_0$.  Calculations \cite{Korpa,Korpa:2008ut} taking into account pion and $\Delta$ width effects in a self-consistent approach resulted in some differences with the simplified quasiparticle  spectrum shown in Fig. \ref{Spectrum}, compare this figure and Fig. 5 in \cite{Korpa:2008ut}. These complications are however   not of our  interest in the present work.

The spectrum of $\pi^{+}$ and $\pi^{-}$ in isospin-asymmetric matter is somewhat different.
For purely neutron matter it can be found in \cite{Migdal78,MSTV90}.

The pion excitations may decay to the nucleon particles and holes
and $\Delta$ isobars and nucleon holes. For $\omega\lsim m_\pi$ the main contribution to the pion self-energy is given by  the nucleon particle--hole term. One obtains \cite{Voskresensky:1984zzn,MSTV90,Vexp95}
\be
(\Im D^R_\pi)^{-1}= \beta T
\ln \frac{e^\kappa +1}{e^\kappa +e^{-\omega/T}}\,,\quad \beta =-\frac{f_{\pi NN}^2 q^2 m_N^{*\,2}\Gamma^2 (g^{\prime})}{k\pi }\,,\label{ImpiT}
\ee
$\kappa =(\omega +q^2/(2m^*_N)^2 m^*_N/(2k^2 T)-\mu_N/T$. For $|\kappa|\gg T$, $\kappa <0$, this expression simplifies as $-\Im \Sigma^R=\beta \omega$.
Here $\mu_N$ is the nucleon chemical potential.
Simplifying, one presents the spectral
function for pions in the isospin-symmetric matter and for $\pi^0$ at arbitrary isospin composition as
%---------
\begin{eqnarray}\label{Api}
A_{\pi}(\omega,k)\simeq \sum_{i}
{2\pi\,Z_i \delta\big(\omega -\omega_i(k)\big)\phantom{_{\omega_i (k)}}}
+\frac{2\beta_0 k\omega}{\tilde{\omega}^4 (k)+\beta_0^2 k^2\omega^2}\theta (\omega <p_{\rm F} k/m_N)\,,
\end{eqnarray}
%---------
$Z_i ={\Big[2\omega -\frac{\partial \Re\Sigma^R_{\pi} (\omega ,k)}{\partial \omega }\Big]^{-1}_{\omega_i (k)}}$, and the sum is over the quasiparticle-like branches. The second term is due to virtual particle-hole modes. For~$n=n_0$ in isospin-symmetric matter, with the help of Eq. (\ref{ImpiT})
one estimates $\beta(n_0)=\beta_0 \simeq 0.7$.

\subsection{{Pion softening and pion condensation}}

The quantity \cite{MSTV90,Voskresensky:2001fd}
\begin{eqnarray}
-\Re D^{R\,-1 }_\pi (\mu_\pi, k^2,n,T)=\tilde{\omega}^2 (k)=k^2 +m_\pi^2  +\Re\Sigma_\pi (\mu_\pi ,k)-\mu_\pi^2\,,\label{omtild}
\end{eqnarray}
has the meaning of the  squared effective pion gap.
For
systems of a small size like atomic nucleus one should put $\mu_{\pi^{+}}=\mu_{\pi^{-}}=\mu_{\pi^{0}}= 0$. In case of the isospin-symmetric matter for $n=n_0$  one estimates $\tilde{\omega}^2(k_0(n_0))\simeq 0.8-0.9$, cf. \cite{MSTV90}.
In the beta-equilibrium matter the pion chemical potentials  ($\mu_{\pi^{+}}\neq \mu_{\pi^{-}}\neq 0$, $\mu_{\pi^{0}}=0$) are
determined from equilibrium conditions for the reactions involving  pions \cite{Voskresensky:2001fd}. The value  $\mu_{\pi^-}$ follows from the condition of the chemical equilibrium
with respect to the reactions $n\leftrightarrow p\pi^-$ and $n \leftrightarrow pe\bar{\nu}$:
$\mu_{\pi^{-}}=\mu_e =\varepsilon_{{\rm F}n}-\varepsilon_{{\rm F}p}$, where $\epsilon_{{\rm
F}n}$, $\epsilon_{{\rm F}p}$ are the Fermi energies of the neutron and proton.

\begin{figure}
\centerline{\includegraphics[height=6.5cm,clip=true]{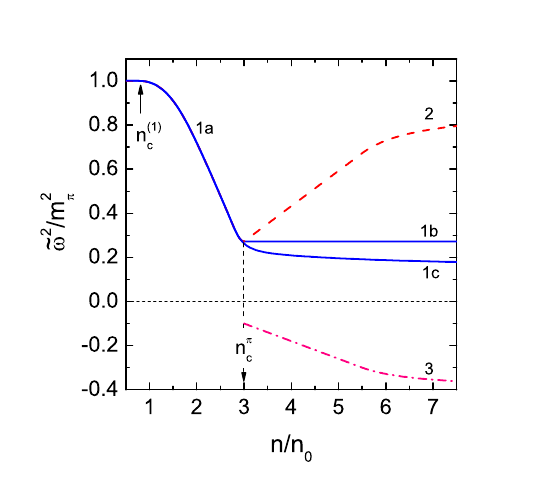}}
\caption{Typical density dependence of the effective pion gap for $\pi^{\pm,0}$ in isospin-symmetric matter and for $\pi^0$ in neutron star matter. The line { 1a} shows
$\widetilde{\omega}^2 (k_0(n))$ for $n_{c}^{(1)}<n<n_c^{\pi}$. For $n>n_c^{\pi}$  the line { 3} demonstrates the stable phase. Line { 2}  characterizes  a  softening of the  spectrum of the pion excitations in presence of the pion condensate. Line { 1c} shows a metastable phase in the system, where the ground state contains the pion condensate and line { 1b} demonstrates artificially introduced saturation of the pion softening effect that could be, if the Landau--Migdal parameter $g^{'}$ increased with an increase of the density.}
\label{piongap1}
\end{figure}

In case of the isospin-symmetric matter the pion gap  at $n>n_{c}^{(1)}$,  $n_{c}^{(1)}\simeq (0.5-0.7)n_0$, gets a roton-like minimum at $k=k_0\neq 0$ and near the minimum it behaves as
 \be\tilde{\omega}^2 (k)\simeq \tilde{\omega}^2 (k_0) +\gamma_0 (k^2-k_0^2)^2/(4k_0^2)\label{tildeom}
 \ee
with  $\gamma_0 \sim 1$, $k_0\simeq p_{{\rm F},N}$, cf. Eqs. (\ref{Dexp}), (\ref{eqMphistat}) used above. A typical density behavior of $\widetilde{\omega}^2 (k_0(n))$ for $\pi^{\pm},\pi^{0}$ in isospin-symmetrical matter and
for $\pi^0$ at $N\gg Z$  is demonstrated  in Fig. \ref{piongap1}.
For $n>n_{c}^{(1)}$, when typical correlations in the spin channel are characterized by $k\sim k_0$,  the  nuclear matter   can be treated as a liquid (or amorphous)  phase of a quantum pion condensate \cite{Voskresensky:1982vd,Dyugaev:1982gf,Voskresensky:1989sn,MSTV90,Voskresensky:1993ud}.
 With increasing density,  $\tilde{\omega}^2 (k_0(n))$ decreases, see line 1a. Various calculations done within the mean-field approximation (dropping fluctuation contribution to the pion self-energy) demonstrate that   at $n>n_c^\pi$ the pion condensate with a liquid-crystal-like or a solid-like structure  may occur by the second-order phase transition. In the isospin-symmetric matter the value $n_c^\pi$ is the same for $\pi^{\pm}$ and $\pi^0$. In the isospin-asymmetric matter the value $n_c^\pi$ depends on the sort of the pion.
It is curious that numerical analysis~\cite{Ph} performed within the so called ``variational theory of nuclear matter'', which  employs  the realistic  two-nucleon potential in vacuum and then introduces  the three-nucleon interactions, gave the value $n_c \simeq 2n_0$ for the critical density
of the $\pi^+ ,\pi^- ,\pi^0$ condensation in isospin-symmetric nuclear matter and the value $n_c^\pi
\simeq 1.3n_0$ for the $\pi^0$ condensation in the purely neutron matter. Cooling of neutron stars is appropriately described in the ``Nuclear medium cooling scenario" with taking into account of the pion softening effects \cite{Voskresensky:1986af,Voskresensky:1987hm,MSTV90} provided $n_c^\pi \gsim (2-2.5)n_0$, cf. \cite{Voskresensky:2001fd,Blaschke:2004vq,Grigorian:2005fn,
Grigorian:2016leu,Grigorian:2018bvg} and references therein.  At $n=n_c^\pi$, in the first-order phase transition due to the effect of fluctuations with $k\sim k_0\sim p_{{\rm F},N}$ the classical pion field is developed. Line { 1c} shows a metastable phase in the system, where the ground state contains the pion condensate and line { 1b} demonstrates a possible saturation of the pion softening effect that could be, if the Landau--Migdal parameter $g^{\prime}$ increased with increase of the density.
Line 3 correspods to the presence of the pion condensate with a liquid-crystal-like or a solid-like structure and line 2 demonstrates behavior of the effective pion gap in presence of the condensate.
  %--------

In the mean-field approximation the pion condensation for $n=n_c^\pi$ and $T=0$ results in a negative contribution to the energy density  $$\delta E_\pi\simeq \widetilde{\omega}^2 (k_0 (n))|\phi|^2 +\Lambda  |\phi|^4/2 = -\frac{\tilde{\omega}^4 (k_0 (n))}{2\Lambda}\theta (-\widetilde{\omega}^2)\,,  $$ where $\theta (x)$ is the step function, $\Lambda$ is the constant of the effective $\pi\pi$ interaction in the medium \cite{Migdal78}. Near the critical point $\widetilde{\omega}^2 (k_0(n))\sim (n_c^\pi -n)/m_\pi\,,$ if one neglects a   jump of the effective pion gap due to quantum fluctuations, cf. \cite{Dyugaev:1975dk,MSTV90}, and   using $\Lambda \sim 1$ one estimates
    $\delta E_\pi\simeq  -\alpha (n-n_c^\pi )^2/(n_c^\pi)^2$
    with $\alpha \sim m^2_\pi$.

Finally, let us mention here that there exist two ways in the literature how one  treats the pion condensation. One choice   is to perform averaging of the initial Lagrangian over baryon degrees of freedom   separating in such a way dressed pions, as we have done it above. Another in principle  equivalent way is to perform averaging over the pion degrees of freedom to deal with the dressed baryons. In the latter case the pion condensation manifests in the appearance of the solid-like structure of baryons. Then one should find the structure yielding minimum of the free energy. Namely in such a way in \cite{Takatsuka:1978ku,Takatsuka:1993pv}
it was demonstrated that the most energetically profitable among several structures, which were considered, is the so called (ALS) alternating-layer structure.

\subsection{{  Fluctuations with $k_0\neq 0$ and where soft pions are in atomic nuclei}}

  \subsubsection{Pion fluctuations}\label{PionFlSect}
%\subsubsection{Fluctuation part of the pion self-energy}

To be specific consider case of the isospin-symmetric matter.
For $T\ll \epsilon_{{\rm F},N}$ and for $\omega<kv_{{\rm F},N}$, $k<2p_{{\rm F},N}$, we have,
see  the tadpole diagram (\ref{ladprtad1}),
%making use of eqs. (\ref{varphitherm}), (\ref{Imr}), we have
\begin{equation}  \label{phi-beta}
\frac{1}{2}<\widehat{\phi}^{\dagger}_{\pi^-}(\vec{k})\widehat{\phi}_
{\pi^-}(\vec{k})
+h.c.> =V_3
\int_{0}^{\infty}
\frac{\beta \omega d\omega}{\pi [\widetilde{\omega}^4(k)+\beta^2\omega^2 ]}
\left(\frac{1}{\mbox{exp}(\omega/T)-1 }+\frac{1}{2}\right).
\end{equation}
Here  the contribution of  fluctuations is included, and
expansion of the Green function for small $\omega$ is as follows
\begin{equation}  \label{gap-}
-D^{R, -1}(\omega\ll m_\pi, k)\simeq \widetilde{\omega}^2(k)-i\beta \omega\,,
\end{equation}
  $\beta = i(\partial \Sigma^R/\partial\omega)_{0,k_0} =\alpha_{02}(\omega_c =0)$, see (\ref{Dexp}). The value $\beta$ can be evaluated with the help of Eq. (\ref{ImpiT}).
Eq. (\ref{gap-}) shows that the spectrum has the diffusion branch, like for paramagnon excitations in $^{3}$He, cf. \cite{Dyg3},
(on complex plane)  $\omega =-i \widetilde{\omega}^2(k)/\beta$. For
positive values of $\widetilde{\omega}^2(k)$ this excitation branch
corresponds to the Landau damping, since $\psi \sim \exp (-i\omega
t)\rightarrow 0$ for $t\rightarrow \infty$ for such excitations.
Due to this "complex plane branch", there arises the
enhancement of the virtual pion sea resulting in  enhancement
of the  pion distributions. Indeed, as one can see from Eqs. (\ref{Api}), (\ref{tildeom}),
the pion spectral function has a pronounced maximum at pion energies
$\omega \lsim\widetilde{\omega}^2(k)/\beta$.

In $^4$He at the critical point one has $\omega_0^2=\Lambda_0=0$, cf. \cite{Ginzburg1976}. This condition is required for presence of the so called Lifshitz point.
Fluctuations in the phenomenological model  of the phase transition to the state $k_0\neq 0$ in the condensed matter systems described by the spectrum $\omega (p)=\Delta +(k-k_0)^2/m$
were first considered in \cite{Brazovskii1975}. The  quantum fluctuations were studied in   \cite{Dyugaev:1975dk}. For nonrelativistic media usually the limit $T\gg |\omega_0^2|/\beta$ is realized.   In relativistic systems, e.g., for consideration of the  pion excitations in heavy-ion collisions, both the temperature regimes  $T\lsim |\omega_0^2|/\beta$ and $T>|\omega_0^2|/\beta $ are relevant \cite{Voskresensky:1981zd,Voskresensky:1982vd,Dyugaev:1982ZHETF,Dyg1,MSTV90}.   With account of pion fluctuations with $k\sim k_0\sim p_{{\rm F},N}$ the phase transition, being of the second order within the mean-field consideration,  becomes of the first order.

As example, asuming
\begin{equation}  \label{omlessT}
 0<\widetilde{\omega}^2(k_0)/\beta (k_0)\ll T,
\end{equation}
 from Eq. (\ref{phi-beta}) we
obtain
\begin{equation}\label{phi-f}
<\widehat{\phi}^{\dagger}_{\pi -}
(\vec{k})\widehat{\phi}_{\pi -}(\vec{k})>_T
\, \simeq V T/[2\tilde{\omega}^2(k)].
\end{equation}
Thus, the pion fluctuation contribution to the pion self-energy,
being proportional to (\ref{phi-f}), has a sharp maximum at
$k\simeq k_0\sim p_{{\rm F},N}$. Integrating Eq. (\ref{phi-f}) over the phase
space, in the approximation (\ref{omlessT}), one obtains
\begin{equation}  \label{piflT}
\Sigma^{R}_{\rm Fl} \simeq 5\Lambda (\omega \simeq 0, k_0 )
\int \frac{d^3 k}{(2\pi)^3 V}<\widehat{\phi}^{\dagger}_{\pi -}(\vec{k})
\widehat{\phi}_{\pi -}(\vec{k})>\simeq
\frac{5\Lambda (\omega \simeq 0, k_0 )k_0 T}{4\pi
\sqrt{2\gamma}\widetilde{\omega}(k_0 )},
\end{equation}
where $\Lambda$ is the effective $\pi \pi$ in-medium interaction
coupling constant and factor $5$ includes possibility of $\pi^{+}$,
$\pi^{-}$ and $\pi^{0}$ intermediate states.

As we have mentioned, a strong enhancement of the virtual pion sea at  $k\sim k_0$ for $n >n_{c}^{(1)}$ can be treated as a liquid phase or as an amorphous
phase of the pion condensation, cf. \cite{Dyg1,Voskresensky:1989sn,Voskresensky:1993ud}. Due to
this enhancement at $T\neq
0$ there arise essential contributions to
thermodynamical characteristics such as the entropy, energy, pressure,
etc., with important consequences for the description of heavy-ion collisions, cf.
\cite{Voskresensky:1988pxw,Voskresensky:1989sn,Voskresensky:1993ud}.
Contribution of a non-quasiparticle nature from thermal virtual
pions of small energies $\omega \ll m_{\pi}$ and of momenta $k\sim
p_{{\rm F},N}$ arising for $n >n_{c}^{(1)}$  can be estimated as follows \cite{Voskresensky:1989sn,Voskresensky:1993ud}
\begin{eqnarray} \label{E-S1}
E_{\pi}^F \simeq \xi T^{3/2} (n -n_{c1})^2 \theta (n -n_{c1})
(n m_{\pi}^{7/2})^{-1},\nonumber \\
S_{\pi}^F /A \simeq 3\xi T^{1/2}
(n -n_{c1})^2 \theta (n -n_{c1})
(n m_{\pi}^{7/2})^{-1}.
\end{eqnarray}
These simplified expressions were introduced in Ref. \cite{Voskresensky:1988pxw} basing on the model \cite{Dyugaev:1982gf,Dyugaev:1982ZHETF,Dyg1}.
 The value $\Lambda (\omega \simeq 0,k_0)$ was estimated in \cite{Migdal78}
and \cite{Dyg1}.
It was shown that $\Lambda (\omega \simeq 0,k_0)$ crucially depends on the
baryon--baryon correlation factor $\Gamma (g^{\prime})$, which
precise value is not too well known, especially for $n >n_0$.
Therefore in \cite{Voskresensky:1989sn,Voskresensky:1993ud} this factor $\Lambda$ was considered
as extra phenomenological parameter, which enters the pre-factor  $\xi$ in Eq. (\ref{E-S1}) chosen to be $\simeq 0.4$ in  \cite{Voskresensky:1988pxw,Voskresensky:1989sn,Voskresensky:1993ud}, that allows to get the best fit of the heavy-ion collision data (differential pion and nucleon cross sections and some other characteristics)
at  energies $\lsim 2$ GeV $/ A$.

For $n >n_c^\pi $  the effective pion gap  $\widetilde{\omega}^2(k_0)$, being negative in a pion condensate  state, depends on the assumed structure of the condensate field. Therefore for various  pion
condensate structures one should calculate $\widetilde{\omega}^2(k_0)$, then find the
structure, which corresponds to the minimum of the free energy,
$F(n )$, for $\tilde{\omega}^2 (k_0)<0$, and to compare the resulting free energy
with the free energy $F(n )$ of the state,
corresponding to the branch $\widetilde{\omega}^2 (k_0)
>0$ at the given density. The liquid crystal-like or solid-like pion condensation starts at the density,
when both the free energies coincide. It occurs by the first order phase
transition for $n>n_c^\pi$, cf. \cite{Voskresensky:1981zd,Voskresensky:1982vd,Dyugaev:1982ZHETF,Dyg1,MSTV90}. Although for $n >n_c^\pi$
the stable system  exists in the state $\tilde{\omega}^2 (k_0)<0$, it can remain for a while in the metastable state $\tilde{\omega}^2 (k_0)>0$. Which state is realized at the  given time moment depends on the peculiarities of the dynamics of the system.

Above we considered the case of  the p-wave pion condensation which appears owing to the strong   p-wave pion-nucleon and pion-$\Delta$ isobar attractions increasing with increase of the baryon density, cf. \cite{Migdal78,MSTV90}. Also in some models, for instance in the model described by the Manohar--Georgi  Lagrangian,  there  is  a possibility that the pion condensate in dense isospin-symmetric baryon matter appears in the s-wave, cf. \cite{Voskresensky2022}. The kaon condensate  may appear both in the s-wave, cf. \cite{Glendenning2001,Glendenning2000}  and the p-wave, cf.
\cite{Kolomeitsev:1995xz,Kolomeitsev:2002pg,Kolomeitsev:1996bh}. In this case the s-wave interaction is stronger than in pion case. Possibilities of the s- and p-wave pion and kaon condensate phases make the $T-n$ diagram of the strongly interacting matter still more complicated.

Within the sigma model the $k_0\neq 0$ condensates began to be studied in \cite{Campbell:1974qt,Campbell:1974qu,Baym:1975tm,Voskresensky:1978cb,Dautry1979,
Voskresensky:1982vd,Muto:1992gb,Nakano:2004cd}. References \cite{Campbell:1974qt,Campbell:1974qu,Baym:1975tm,Voskresensky:1978cb,Voskresensky:1982vd} considered  the  $\sigma$ field together with the running  $\pi^{\pm}$ condensate wave, whereas Ref. \cite{Dautry1979} proposed the $\sigma,\pi^{0}$ condensate standing wave. Reference \cite{Muto:1992gb} studied alternating layer structure of the pion condensate. Reference \cite{Hashimoto2015} suggested to consider possibility of the  ferromagnetic neutron matter with a magnetization and a neutral pion condensation like the $\sigma,\pi^{0}$ condensate  of Dautry and Neyman \cite{Dautry1979}. However, compared to the inhomogeneous phase with the alternating layer structure, the ferromagnetic phase turned out to be unfavored.

As it has been   noted, cf. \cite{Nickel2009a,Carignano2014},    the QCD critical endpoint should be a Lifshitz point, where the normal, homogeneous and inhomogeneous chiral condensed phases meet. This configuration is similar to $\sigma\pi^0$ condensation, obtained in neutron matter within the sigma model of Dautry and Neyman  \cite{Dautry1979}, and  with these phases  one could  expect a smooth transition from description of  the nuclear $\sigma\pi N$ to  the quark matter. However fluctuations with $k_0\neq 0$  lead to the change of the phase transition from the second order to the first order, as we have discussed, cf. \cite{Tatsumi2016}.

\subsubsection{Where  soft pions are in atomic nuclei}\label{wherepions-sect}
References \cite{Bertsch:1993vx,Brown:1993sua} analyzing  experiments on  $(\vec{p},\vec{n})$  quasielastic
polarization transfer and EMC and Drell--Yan ratios did not find any manifestation of  pion softening effects and raised a principal question ``where are the nuclear pions?'' The answer is probably as follows, see \cite{Voskresensky:2001fd} and also \cite{Voskresensky:2018ozf,Voskresensky:2020shb}.

The quantity $\widetilde{\omega}^2 (k_0)$ demonstrates how much the virtual nucleon particle--hole mode with the pion quantum numbers is softened at the given density. Typical momentum transferred in the two-nucleon reaction is $k\simeq p_{{\rm F},N}$. The ratio of the $NN$ cross sections calculated with the help of the model of the free one-pion exchange
(FOPE) and the model of the medium one-pion exchange (MOPE)  is estimated as
%--------
\begin{equation}\label{MOPER}
R=\frac{\sigma [\mbox{FOPE]}}{\sigma [\mbox{MOPE}]} \simeq
\frac{\Gamma^4 (\omega\simeq 0,k\simeq p_{{\rm F},N} )
(m^2_{\pi}+p_{{\rm F},N}^2)^2}{\widetilde{\omega}^4 (p_{{\rm F},N}
)}\,,
\end{equation}
%--------
where $\Gamma$ is the vertex dressing factor determined in (\ref{gam-x}). We estimate
$|D_\pi (\omega \simeq 0,k\simeq p_{{\rm F},N})/(m_\pi^2 +p_{{\rm F},N}^2)|^2 \simeq\,\, 40$ for isospin-symmetric matter at $n=n_0$, whereas $\Gamma (n_0)\simeq 0.4$. Thereby, for $n \lsim n_0$ one has $R\lsim 1$,
whereas already for $n =2n_0$ this estimate yields $R\sim 10$.  The contribution of a purely local interaction to the $NN$ cross section is less than that of the soft pion.
Thus, following (\ref{MOPER}) one can evaluate the $NN$ interaction
for $n> n_0$ with the help of the MOPE and with increasing density the MOPE contribution is increased.
 For small frequences for $n_{c}^{(1)}<n<n_c^\pi$ one may use the presentation $D_{\pi}^{R,-1} (\omega \ll m_\pi , k\sim k_0) \simeq -\widetilde{\omega}^2 (k_0)-\gamma_0 (k-k_0)^2 +i\beta \omega\,,$ cf. Eq. (\ref{tildeom}). With the help of this expression we find that soft pions (with $\omega \ll m_\pi$) yield only a small contribution,
\begin{equation}
\delta S =\int_0^{k p_{{\rm F},N}/m_N^*}2\omega A_{\pi}\frac{d\omega}{2\pi}=\frac{2}{\pi \beta}\left[\frac{k p_{{\rm F},N}}{m_N^* } -\mbox{arctan}\frac{\beta k p_{{\rm F},N}}{m_N^* \widetilde{\omega}^2 (k)}\right]\,,
\end{equation}
to the full sum-rule, $S=\int_0^{\infty} 2\omega A_{\pi}\frac{d\omega}{2\pi}=1$.
An estimation yields $\delta S\simeq 0.05$ for $n=n_0$ and $\delta S$ remains much less than unity ($\sim\!\!0.1$)  even for $\widetilde{\omega}^2 (k_0)=0$. So, pions are soft for $\widetilde{\omega} (k_0)\ll m_\pi$, but their contribution to the sum-rule remains small.

On the other hand, a successful description of the data on the pion atoms \cite{Troitsky,MSTV90,Delorme92,Ericson94,Kolomeitsev:2002gc,
 Kolomeitsev:2002mm,FriedmanGal20,Voskresensky2022}  requires a strong modification of the pion dispersion law   $\omega \simeq \sqrt{m^2_\pi +\alpha_0 k^2}$ with $\alpha_0\simeq 0.4$ for $\omega \simeq m_\pi$ at  $n=n_0$ involved in this problem.
  Analysis of the $^3\mbox{He}$--$t$ reaction on nuclei, cf. \cite{Dmitriev:1984ud,Dmitriev:1986ud,Hennino:1993sx,Sneppen:1994zx}, demonstrates
agreement with two-branch pion spectrum consisting of the pion branch and the $\Delta$ branch and with
the assumption of the softening of these branches
for $k\gsim 2 m_{\pi}$.

Summarizing, there are soft pions in atomic nuclei  but the pion  contribution to various quantities largely depends on the quantity under consideration, the density, the frequency and the momentum involved in the problem.
For instance, virtual pions significantly contribute for $T\neq 0$. The  momentum distribution function of $\pi^-$ entering the detector at a sudden (prompt) break up of the piece of a hot nuclear matter is given by \cite{Senatorov:1989cg,Voskresensky:1989sn,MSTV90,Voskresensky:1993ud,
Voskresensky:1995tx}:
\begin{equation}
n^{\pi^-}_k ({\rm virt}) =\int_{0}^{\infty} \frac{2\omega_k A_\pi d\omega}{2\pi(e^{\omega/T}-1)}=
2\omega_k\int_{0}^{\infty} \frac{\beta\omega}{\widetilde{\omega}^4 (k)+\beta^2 \omega^2}\frac{d\omega}{\pi(e^{\omega/T}-1)}\simeq \frac{T\sqrt{m^2_\pi +k^2}}{\widetilde{\omega}^2 (k)}\,,\label{npifreevirt}
\end{equation}
where the second equality is valid for $0<\widetilde{\omega}^2 (k_0)\ll \beta T$, cf. Eq. (\ref{Matsum}), $\omega_k=\sqrt{m^2_\pi +k^2}$.
This result shows a strong effect of thermal pion fluctuations with $k\sim k_0$ for $n_c^\pi>n>n_{c}^{(1)}$, i.e. in the liquid, better saying, the amorphous phase of the pion condensation. Also, such a fluctuations yield a large  contribution to the  bulk viscosity $\propto 1/\tilde{\omega}^3(k_0)$, cf. \cite{Kolomeitsev:2014gfa}. Thereby, in the vicinity of the critical point $n_c^\pi$, the nuclear matter for $T\neq 0$ behaves similarly to  a glass, here, a pionic glass.

However, we should note that the result (\ref{npifreevirt}) exceeds  significantly  not only the   free pion distribution
\be n^{\pi^-}_k ({\rm free})=1/(e^{\sqrt{m^2_\pi +k^2}/T}-1)\,,\label{freepiondistr}
\ee
but also    the experimental result that follows from the study of the heavy-ion collisions. The solution of the puzzle \cite{Senatorov:1989cg,MSTV90,Voskresensky:1989sn,Voskresensky:1993ud} is as follows: (i) At realistic conditions the freeze out for virtual pions with typical frequencies $\omega\ll m_\pi$ cannot be considered as sudden. Most of  such virtual  pions, existed for $n>n_{c}^{(1)}$, are ``eaten'' by nucleons during the freeze out  stage, since the typical time of the latter is $\tau_{\rm f.o.}>1/m_\pi$. (ii) The freeze out density is estimated  as $n_{\rm f.o}\lsim n_{c}^{(1)}$, cf. \cite{Voskresensky:1989sn,Voskresensky:1993ud}.
Thus contribution of the  second term in (\ref{Api}) to the observable   pion yield should be  strongly suppressed, whereas the first term  contributes significantly. This term describes the  pions emitted from the  $\pi$- and  $\Delta$- branches of the pion spectrum in Fig. \ref{Spectrum}. For them $\tau_{\rm f.o.}<1/|\omega (k,n_{\rm f.o.}) -\sqrt{m^2_\pi +k^2}|$, and the freeze out can be considered as sudden (at $n=n_{\rm f.o.}$,
$T=T_{\rm f.o.}$) and thereby
\be
n^{\pi^-}_k ({\rm branches})\simeq \sum_i \frac{2\sqrt{m^2_\pi +k^2} \,\,Z_i }{(e^{\omega_i (k)/T_{\rm f.o.}}-1)}\,.\label{piondistrbr}
 \ee
As it was demonstrated in \cite{Voskresensky:1989sn,Voskresensky:1993ud},  the latter result   well describes  the experimental pion momentum distributions at heavy-ion collision energies $\lsim 2 \,$GeV $/ A$. It significantly differs from (\ref{freepiondistr}), especially for low $T$.
It  should be noted that   at the sudden transition the momentum of the pion excitation  is conserved, whereas the energy is not conserved. Nevertheless the  energy and momentum of the system as a whole is certainly  conserved. The excess or deficit  of the  energy of the pion sub-system is compensated by the nucleon sub-system.

Note that the ideal pion gas approximation is up to now  used in different kinetic and hydrodynamic codes. For example, the ideal hydrodynamical models employ  both   the sudden freeze-out and the ideal pion gas approximations, which are, generally speaking, incompatible. The equation of state including pion softening effects for $T\neq 0$ and $n>n_c^{(1)}$ unified with  the model of  the sudden freeze-out for nucleons and pions allowed to appropriately describe the Bevalac Berkeley and SIS GSI heavy-ion collision inclusive data at collision energies $\lsim 2 \,$GeV $/ A$, cf. \cite{Voskresensky:1989sn,Voskresensky:1993ud}.

A good moment to remind phrase of Albert Einstein ``Raffiniert ist der Herr Gott, aber boshaft ist er nicht.''

\subsection{Dynamics of pion-condensate phase transition}

Let us now explain how the instability, which results in the  pion
condensation,
develops dynamically.  In a semiclassical scheme the dynamics of a mode is described by the solution of the quantum kinetic Kadanoff--Baym equation derived from the nonequilibrium Green function technique in the first-gradient order. The Boltzmann equation
follows from the Kadanoff--Baym equation only in the  limit of a weak interaction, cf. \cite{KadanoffBaym1962,LifshizPitaevskii1981}. The hydrodynamical equations with the transport coefficients expressed in terms of the retarded Green functions also follow from the Kadanoff--Baym kinetic equation \cite{Voskresensky:2010qu}.
Actually, the first-gradient order quantum kinetic equation can be written in three forms: the  original Kadanoff--Baym  form, the Bottermans--Malfliet form \cite{BM}, which is derived from  the latter for configurations close to the thermal equilibrium and the non-local form \cite{Ivanov:2009wc,Kolomeitsev:2013du}, which up to the second-gradient order coincides with the  Kadanoff--Baym equation and up to the first-gradient order  with the Bottermans--Malfliet form. In presence of the condensate, one needs to add equation of motion of the condensate field. Both equations are derived from the generating functional formulated on the Schwinger--Keldysh contour, cf. \cite{Ivanov:1998nv,Ivanov:1999tj,Knoll:2001jx,Ivanov:2003wa}.

Two Wigner transformed quantities, $F(t, \vec{r}, \omega, \vec{k})=A_\pi (t, \vec{r}, \omega, \vec{k})f_\pi (t, \vec{r}, \omega, \vec{k})$, and
\begin{equation}\label{spec}
A_\pi (t, \vec{r}, \omega, \vec{k}) =\frac{\Gamma_\pi (t, \vec{r}, \omega, \vec{k})}
{[\omega^2-k^2-m_{\pi}^2-\Re\Sigma^{R} (t, \vec{r}, \omega,
\vec{k})]^2+ \Gamma^2_\pi (t, \vec{r}, \omega, \vec{k}) /4},
\end{equation}
where $\Gamma_\pi=-2\Im\Sigma^R_\pi$,
completely determine the kinetic evolution of the given species. The spectral function given by (\ref{spec}) satisfies a general
sum-rule \cite{BM}
\begin{equation}\label{sumrule}
\int_{-\infty}^{\infty} \omega A_\pi d\omega /2\pi =1.
\end{equation}
In the isospin-symmetric nuclear matter
$\pi^{+}$, $\pi^{-}$ and $\pi^0$ mesons have essentially the same
distributions, $A_\pi (\omega,k)=-A_\pi (-\omega,k)$ that simplifies the scheme of the separation of
particle and antiparticle contributions. Thereby one can work with the
quantities related to one particle species at
positive frequencies.

Above we extensively studied the dynamics of the condensate field. Now let us focus on the dynamics of the soft modes. Importance of taking into account width effects in description of the soft modes was emphasised in \cite{Knoll:1995gs,Knoll:1995nz}. As a specific example, let us
consider the pion sub-system as a light admixture in a heavy nucleon
environment, neglecting the feedback of the pions onto the nucleons ($m_\pi/m_N\simeq 1/7\ll 1$), i.e., neglecting the contribution from
in-medium pion fluctuations. Within the above
approximations the quantum kinetic equation  for the pion
distribution, $f_\pi\ll f_N$, in homogeneous and equilibrated nucleon  environment
becomes \cite{Ivanov:2000ma}:
\begin{equation}\label{bath}
\frac{1}{2} \Gamma_\pi B^\pi_{\mu}\partial^{\mu}_X f_\pi =\Gamma^\pi_{in}-\Gamma_\pi f_\pi.
\end{equation}
Here $\mu  =0,1,2,3$, $X=(t,\vec{r})$, $\Gamma^\pi_{in}=i\Sigma^{-+}_\pi(f_N)$, $B^\pi_{\mu}$ is defined as
\begin{eqnarray}\label{B-j}
B_{\mu}^\pi = A_\pi \left[
\left(
2k_{\mu} - \frac{\partial \Re\Sigma^R_\pi }{\partial k^\mu}
\right)
- \Re D^{R,-1}_\pi \Gamma^{-1}_\pi\frac{\partial \Gamma_\pi}{\partial k^\mu}
\right]\,.
\end{eqnarray}

The instability of the system can be discussed
considering a weak perturbation, $\delta f_\pi$, of the equilibrium pion distribution,
$f^{(0)}_\pi=\left[\exp(\omega/T)-1\right]^{-1}$, in
the rest frame of the system. Linearizing Eq. (\ref{bath}) we find
\begin{equation}\label{rel}
\frac{1}{2} B^\pi_{\mu}\partial^{\mu}_X \delta f_\pi
+ \delta f_\pi =0,
\end{equation}
with the solution
\begin{equation}\label{growth}
\delta f_\pi (t,\omega,k) = \delta f_0 (\omega,k) \exp\left[-2t/B^\pi_0(\omega,k)\right],
\end{equation}
where for simplicity the initial fluctuation $\delta f_0 (p)$ of the pion
distribution is assumed to be space-independent.  Let us put $\omega\to 0$ and ${k}\sim k_0\sim  p_{{\rm F},N}$. This four-momentum region,
being far from
the pion mass shell, is  the region, where the pion instability is
expected to occur in isospin-symmetric nuclear matter. The real part of the pion
self-energy, $\Re\Sigma^R_\pi$, is an even function of the pion energy $\omega$ while
the width is an odd function  proportional to $\omega$ for $\omega\rightarrow
0$, and, cf.  \cite{Migdal78,MSTV90}, one has $2\omega - \partial \Re\Sigma^R_\pi
/ \partial \omega \rightarrow 0$, $\Gamma_\pi =\beta ({k}) \omega$ and $\beta
({k})\sim m_{\pi}$ for $\omega \rightarrow 0$, $k\sim p_{{\rm F},N}$. Thus  from Eq.  (\ref{B-j}) we get $B_0^\pi =\beta ({ k})/\widetilde{\omega}^2 ({k})$ and therefore
\begin{equation}\label{growth1}
\delta f_\pi (t,\omega =0,{\vec k}) = \delta f_0 (\omega =0,\vec{k})
\;\exp\left[-2\widetilde{\omega}^2({ k})t/\beta ({k})\right].
\end{equation}
This solution shows that for $\widetilde{\omega}^2 >0$ initial
fluctuations are damped, whereas for $\widetilde{\omega}^2 <0$ they grow.  Thus,
the change of sign of $\widetilde{\omega}^2 ({k}_{0c})$ at $n=n_c^\pi$, $k_0={k}_{0c}$ leads to an
instability of the virtual pion distribution at low energies and for momenta
$k\simeq {k}_{0c}$.  The solution (\ref{growth1}) illustrates the important
role of the width, $\Im \Sigma^R_\pi \neq 0$, in the quantum kinetic description. At
neglected width in the quantum kinetic equation, one would fail to find the above instability.

Increase of the pion distribution $\delta f_\pi$ is accompanied by a
growth of the mean condensate field $\varphi_\pi$.  Due to the latter, the increase
of the virtual pion distribution slows down and finally stops, when the mean
field reaches its stationary value.  Therefore, a consistent treatment of the
problem requires the solution of the coupled system of the quantum kinetic
equation for the particle distribution function, the equation for the spectral function
and the  equation for the mean field.
In presence of the condensate,
$\Sigma^R_\pi$ acquires an additional contribution $\Sigma^R_\pi
(\varphi_\pi)=\Sigma^R_\pi (\varphi_\pi=0)+\Lambda_{\rm{eff}}
|\varphi_\pi|^2 +O(\varphi^3_\pi)$, where $\Lambda_{\rm{eff}}$ denotes the total
in-medium pion-pion interaction. Within the same order, the mean-field equation
becomes
\begin{equation}\label{m-f0}
\left[\widetilde{\omega}^2({k}_{0c} ) +\lambda_{\rm{eff}}
|\widetilde{\varphi}_\pi|^{2} (t) +\frac{1}{2}\beta ({k_{0c}})\partial_t\right]
\widetilde{\varphi}_\pi (t)  =0.
\end{equation}
Here we have assumed the simplest structure for the condensate field
$\varphi_\pi =\widetilde{\varphi}_\pi (t)\mbox{exp}(i \vec{k}_{0c}\vec{r})$,
where $\widetilde{\varphi}_\pi (t)$ is a space-homogeneous real function, which
varies slowly in time. Also one should do the replacement
$$\widetilde{\omega}^2({k}_{0c})\rightarrow \widetilde{\omega}^2(\varphi_\pi,
{k}_{0c})\equiv \widetilde{\omega}^2({k}_{0c})+ \Lambda_{\rm{eff}}
\mid\varphi_\pi\mid^2$$ in Eqs. (\ref{bath}) - (\ref{growth1})
for the pion distribution.

The time dependence of $\widetilde{\varphi}$ can qualitatively be understood
inspecting the two limiting cases.
At a short time scale the mean field is still small and one can neglect the
$\Lambda_{\rm{eff}} \widetilde{\varphi}^{2}(t)$ term in
Eq. (\ref{m-f0}). Then the mean field
\begin{equation}\label{phi-grow}
\widetilde{\varphi}_\pi (t)=\widetilde{\varphi}_\pi (0)
\exp\left[-2\widetilde{\omega}^2 ({k}_{0c})t/\beta ({k}_{0c})\right]
\end{equation}
grows exponentially with time, as well as the distribution function
(\ref{growth1}). Here $\widetilde{\varphi}_\pi (0)$ is an initial small
fluctuation of the field, which value is generated by the white noise.

At a larger time, the solution of Eq. (\ref{m-f0}) approaches the stationary
limit
$\widetilde{\varphi}_\pi \rightarrow \widetilde{\varphi}_\pi^{\rm{stat}}$
with
\begin{equation}\label{phi-stat}
(\widetilde{\varphi}_\pi^{\rm{stat}})^2 =- \widetilde{\omega}^2({k}_{0c} )
/\Lambda_{\rm{eff}} ({k}_{0c} ).
\end{equation}
Since simultaneously
$\widetilde{\omega}^2(\varphi_\pi, {k}_{0c})=
 \widetilde{\omega}^2({k}_{0c})+
\Lambda_{\rm{eff}} \mid\varphi_\pi\mid^2 \rightarrow 0$, the change in the
pion distribution  $\delta f_\pi$ will saturate.
This stationary solution  $\widetilde{\varphi}_\pi^{\rm{stat}}$
is stable. It can be seen from  Eq. (\ref{m-f0}) linearized near the value
$\widetilde{\varphi}_\pi^{\rm{stat}}$. Then we have
\begin{equation}
\widetilde{\varphi}_\pi (t)=\widetilde{\varphi}_\pi^{\rm{stat}}
-\delta\widetilde{\varphi}_0
\exp\left[4\widetilde{\omega}^2({k}_{0c})t/\beta({k}_{0c})
\right],
\end{equation}
where the factor in the exponent is now negative. As $\widetilde{\varphi}_\pi (0)$, the  value  $\delta\widetilde{\varphi}_0$ is   generated by the white noise term in the equation.

\section{Fermion blurring and  hot Bose condensation}\label{Blurring}

\subsection{System of strongly interacting light bosons -- heavy fermions
at $\mu_{\rm f}=0$}\label{Quasi}
References \cite{Dyugaev:1993mn,Voskresensky:2004ux,Voskresensky:2008ur} studied behavior of
a system of strongly interacting heavy fermions at   small (or zero)  baryon chemical potential, $\mu_{\rm f}$, and light bosons, avoiding quasiparticle approximation. The problem has application to the description of heavy-ion collisions at top RHIC and LHC energies.
Description  of the heated pion-nucleon vacuum (at zero total
baryon charge) is in a sense analogous to the description of the electron-phonon
interaction in doped semiconductors. In the latter case, even at
zero temperature the tail
of the electron wave function  penetrates deeply into the band gap
due to multiple electron-phonon collisions \cite{E,E1}. In the nuclear problem, heavy nucleons and antinucleons may play the same role as electrons and holes, and light
pions play role of massless phonons.

Let us first consider a  hot system of two particle species.  References \cite{Voskresensky:2004ux,Voskresensky:2008ur} treated the system in terms of the
self-consistent two-particle irreducible (2PI) $\Phi$ derivable approximation scheme suggested by Baym \cite{Baym}. The
$\Phi$ functional  is given by diagrams
 \be\label{phi}
\includegraphics[width=7cm,clip=true]{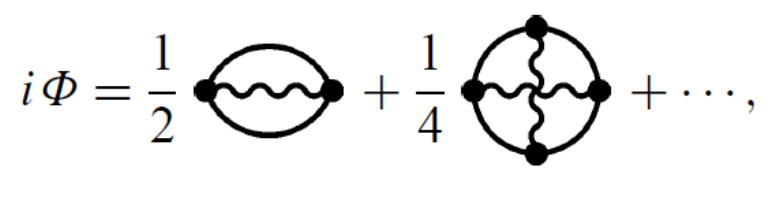}
 \ee
Here both fermion (solid line) and boson (wavy line) Green
functions are full Green functions, whereas vertices are bare. Simplifying consideration let us restrict ourselves by consideration of the simplest $\Phi$ (the
first diagram (\ref{phi})). Then the fermion self-energy
 \be\label{selfzf}
\includegraphics[width=2cm,clip=true]{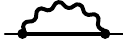}
 \ee
 enters the Dyson equation
  \be\label{selfzfDyson}
\includegraphics[width=10cm,clip=true]{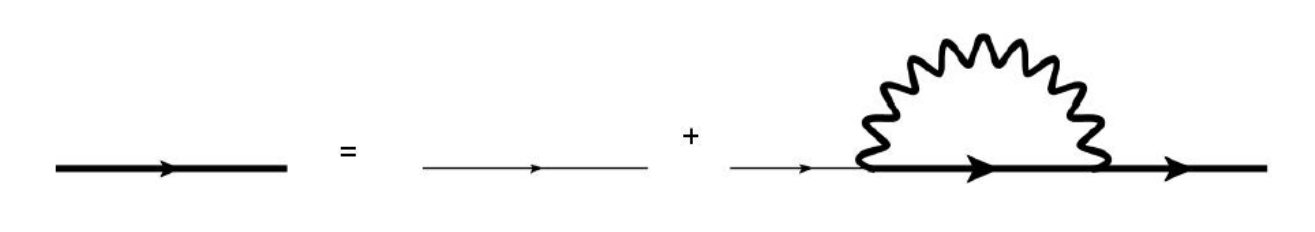}
 \ee
 and the boson self-energy
 \be\label{selfzb}
\includegraphics[width=2cm,clip=true]{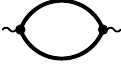}
 \ee
  enters the Dyson equation
    \be\label{selfzfDyson1}
\includegraphics[width=10cm,clip=true]{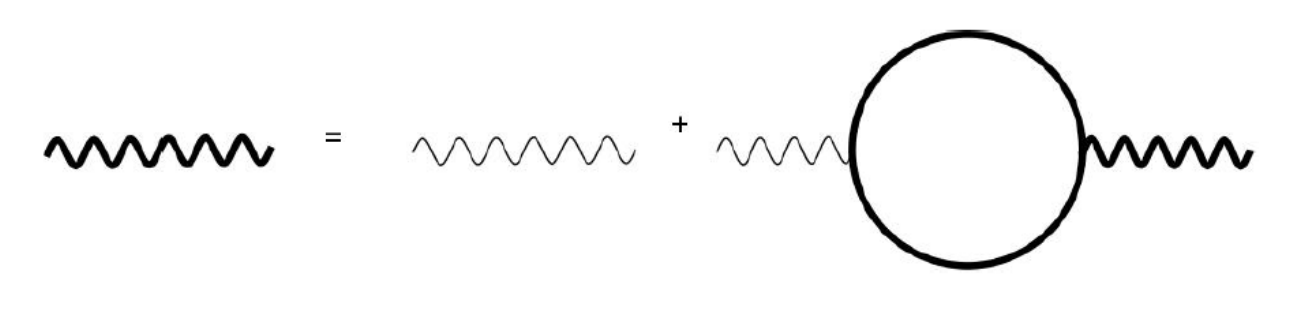}
 \ee
Note that in a sense  similar self-consistent approach was applied   in \cite{Fischer2011} to the quark-gluon matter. For the chiral transition the authors  found a crossover turning into a first order transition at a critical endpoint  at the quark chemical potential $\mu_q /T_c \simeq 3$.

All the multi-particle re-scattering processes
 \be\label{ladpr}
\includegraphics[width=10cm,clip=true]{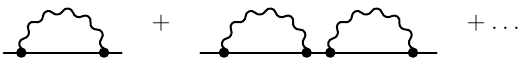}
 \ee
are  included in (\ref{selfzfDyson}), whereas processes with  crossing of boson lines
(correlation effects) like
 \be\label{cr}
\includegraphics[width=5cm,clip=true]{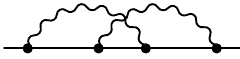}
 \ee
 are disregarded. Note that for $\pi NN$ interaction each diagram with the
one crossed boson line
brings the suppression factor $\nu =\nu_{\tau}\nu_{\sigma}$ compared to the
corresponding
diagram with the same number of boson lines but without their crossing,
cf. \cite{Dyugaev:1993mn,Dyg3,Voskresensky:2004ux,Voskresensky:2008ur}.
In the non-relativistic approximation for nucleons,
$\nu_{\tau}=1/3$ is due to the
non-commutation of the isospin $\tau$ matrices and
$\nu_{\sigma} =1/3$ is due to the
non-commutation of the spin $\sigma$ matrices. Each product $\tau_i \tau_i
\tau_j \tau_j$ yields factor 9, whereas
each product $\tau_i \tau_j \tau_i \tau_j$ yields  factor 3. The same
statement is valid for $\sigma$ matrices.
Concluding, one  may retain only the first
diagram (\ref{phi}) in $\Phi$ in this case with mentioned accuracy.

There are several temperature regimes \cite{Voskresensky:2004ux}.
The regime $T\ll {\mbox{min}}\{m_{\rm b}^2 /m_{\rm f} , T_{\rm bl.f}\}$ corresponds to a {\em slightly
heated hadron liquid}, $m_{\rm b}$ is the boson mass, $m_{\rm f}$ is the fermion mass, at $m_{\rm f}\gg m_{\rm b}$.
$T_{\rm bl.f}$ is the temperature, at which
the gap between fermion-antifermion continua becomes completely blurred, see below.
Typically $T_{\rm bl.f}(g)$
is of the order of $\sim m_{\pi}$ for relevant values of the fermion-boson
coupling constant $g$,
$m_{\pi}=140$~MeV is the pion mass.
Bosons are almost free particles in this temperature regime but
fermion distributions  deviate
from the Boltzmann law, obeying the Urbach rule,  due to the
multiple collisions of each fermion on   bosons. At
$m_{b}^{2}/m_f \lsim T\ll \mbox{min} \{T_{\rm bl.f}, m_{b}\}$, if such interval indeed exists,
the fermion mass shell is  already partially blurred
due to multiple re-scatterings of the fermion on bosons. Here we deal with ({\em a warm hadron liquid, partially blurred fermion continuum}).
The quasiparticle approximation for fermions fails, if the fermion-boson
coupling constant $g$ is rather large (e.g., $g\sim 10$
for $\sigma$, $\om$ and $\rho$ meson -- nucleon $N$ and other baryon  interactions).
As the result,
fermion distributions become  essentially enhanced
compared to the ordinary Boltzmann distribution.
For realistic hadron parameters
the regime of {\em{a warm hadron liquid}}
can be realized
only for pions, not for $\sigma$, $\om$ and $\rho$ due to their large masses.
However the enhancement
is not too strong for pions, since $g\sim 1$ (rather than
$\gg 1$) in the latter case.

For $m_f \gg T\gsim T_{\rm bl.f}$ the fermion effective mass significantly decreases and
fermions become essentially relativistic particles.
The hot hadron liquid
comes to the regime of {\em the blurred fermion continuum.}
The fermion sub-system represents then {\em a rather dense packing of
fermion-antifermion pairs.} With further
increase of the
temperature, the boson effective masses substantially decrease.
For  $m_f \gg T\gsim m_b^{*} (T)$ ({\em hot hadron liquid, blurred
boson continuum}), where
$m_b^{*}(T, g)$ is the effective boson mass depending on $g$,
 rapid fermions abundantly
produce effectively
less massive and slower off-shell bosons
and undergo multiple re-scatterings on them, as on
quasi-static impurities.
The fermion  propagator completely
looses the former quasiparticle pole shape it had in a dilute medium.
The fermion-antifermion density,
$n_{f\bar{f}}$ grows exponentially with the temperature
in a wide temperature interval. Bosons rescatter on fermion-antifermion pairs
 and due to that become effectively less massive.
At a temperature $T>T_{\rm cB}>T_{\rm bl,f}$,
the effective   boson mass may vanish
and a   hot Bose condensation may set in. We stress that
the condensate may appear for the temperature
larger than a critical temperature. Similarly,  the condensate of the charged vector field in the form of the lattice may appear at a high magnetic field $H>H_c$, whereas condensation is absent for $H< H_c$, cf. \cite{Voskresensky:2019zcp,Chernodub}.
For realistic values of hadron parameters, the problem of the
determination of
$T_{\rm bl.f}$, $T_{\rm cB}$ and $m^*_b (T)$ is the coupled-channel problem.
Estimations show that
$T_{\rm cB}$ proves to be close to $T_{\rm bl.f}$.

\subsection{Scalar boson -- fermion coupling}
 To simplify derivations consider first
example of the Yukawa interaction of spin $1/2$ non-relativistic
heavy fermion with a light relativistic scalar boson. The
interaction Lagrangian is given by
 \be\label{intLag} L_{\rm
int}=g\bar{\psi}\phi\psi .
 \ee
Examples of different couplings of relativistic fermions and
bosons, including the Yukawa scalar boson-baryon interaction, the vector boson-baryon interaction and   pion-nucleon one, can be found in
\cite{Voskresensky:2004ux}.

\subsubsection{Approximation of a soft thermal loop}

For $m_{\rm f} \gg m_{\rm b}$,  in a broad
temperature range,  boson
occupations are essentially higher than fermion ones. Then, at
such temperatures we may retain in (\ref{selfzf}) only terms
proportional to boson occupations. Using this we find \cite{Voskresensky:2004ux}:
 \be\label{sigm-R0-sim} \Sigma_{\rm f}^{R}(p_{\rm f}) \simeq\int
g^2 \frac{d^3 p_{\rm b}}{(2\pi)^3} \int_{0}^{\infty} \frac{d
\omega_{\rm b}}{2\pi} [G_{\rm f}^{R} (p_{\rm f}+p_{\rm b}) +G_{\rm
f}^{R} (p_{\rm f}-p_{\rm b})] {A}_{\rm b}(p_{\rm b})f_{\rm
b}(\omega_{\rm b} ).
 \ee
 Dropping boson momentum $p_{\rm b}$-dependence of the
fermion Green functions $G_{\rm f}$ in (\ref{sigm-R0-sim}), i.e., in {\em the soft thermal loop} (STL) approximation, Eq.
(\ref{sigm-R0-sim}) is simplified as
 \be\label{sigmJ}
\Sigma_{\rm f}^{R}(p_{\rm f})\simeq {J}\cdot {G}_{\rm
f}^{R}(p_{\rm f}), \quad J=2g^2 \int \frac{d^3 p_{\rm
b}}{(2\pi)^3}\int_{0}^{\infty} \frac{d \omega_{\rm b}}{2\pi} A_{\rm
b}(p_{\rm b}) f_{\rm b}(\omega_{\rm b} ),
 \ee
 $A_{\rm b}$ and $f_{\rm b}$ are the boson spectral and distribution functions.
At $T\gsim m_{\rm b}^{*}(T)$  departure of the fermion energy
from the mass shell, $\delta \omega_{\rm f} \sim \sqrt{J}$, proves to be much
larger than that for bosons, $\delta \omega_{\rm b}\sim \mbox{max}\{
m_{\rm b} -m_{\rm b}^* (T), T\}$, and typical fermion momenta
${p}_{\rm f}\sim \sqrt{2 m_{\rm f}T}$ are much higher than typical
boson momenta ${p}_{\rm b}\sim \mbox{max}\{\sqrt{2 m_{\rm b}^*
(T)T},\,T\}$. Here $m_{\rm b}^* (T)$ is an effective boson mass.
At these conditions the STL approximation should be valid.
Moreover, we assume that $\sqrt{2 m_{\rm f}T}\ll m_{\rm f}$ and
$\sqrt{J}\ll m_{\rm f}$. At these conditions there is a broad region of temperatures, where fermions can be treated as
non-relativistic particles. The latter approximation allows us to
avoid the spin algebra. A general case was considered in \cite{Voskresensky:2004ux}.

As follows from (\ref{sigmJ}), the quantity ${J}$ can be expressed
through the tadpole fluctuation diagram (\ref{ladprtad1}) where now the vertex yields the factor $g^2$.
%% \be\label{ladprtad}
%%J=\includegraphics[width=2cm,clip=true]{tadpole}
%% \ee
 The latter contribution  describes fluctuations of virtual (off-mass shell) bosons, see
%Eqs. (\ref{ladprtad1}),
 Eq. (\ref{phi-beta}). On
the other hand, $J$ can be  interpreted as the density of
quasi-static boson impurities.  Due to multiple repetition of this
diagram in the Dyson series for the fermion, $J$ demonstrates {\em
the intensity of the multiple quasi-elastic scattering} of the
fermion on quasi-static boson impurities. In the kinetical description, with the multiply-repeated diagram (\ref{ladprtad1}) one describes the Landau--Pomeranchuk--Migdal effect, cf. \cite{Knoll:1995gs,Knoll:1995nz}.

The Dyson equation for the retarded fermion Green function is greatly
 simplified in the STL approximation,
  \be
 G_{\rm f}^R =G_{0{\rm f}}^R + G_{0{\rm f}}^R J (G_{\rm f}^R)^2 \,,\label{DysonnonrelFermbl}
 \ee
with a simple analytical solution
 \be\label{tblrel}
 G_{\rm f}^R =\frac{\omega_{\rm f} -\omega_p^{\rm f} \pm
 \sqrt{(\omega_{\rm f} -\omega_p^{\rm f} )^2
 -4J}}{2J}, \quad \omega
 _p^{\rm f} =m_{\rm f} +\frac{p^2_{\rm f}}{2m_{\rm f}}.
  \ee
 In this problem it is convenient to count  fermion energies from the mass shell.

  Only
negative sign solution satisfies the retarded property and should
be retained.
  For $(\omega_{\rm f} -\omega_p^{\rm f}
)^2\gg 4 J$ we recover the quasiparticle (pole-like) solution.
Since then typical energies are $\omega_{\rm f} -\omega
_p^{\rm f} \sim
T$, the quasiparticle approximation   for fermions proves to be valid only
for low temperatures, $J \ll T^2$. Otherwise (for $J\gsim T^2$) fermion Green
function is completely regular. The fermions become  {\em blurred
particles}. From (\ref{tblrel}), using relation $(G^R)^{-1}_{\rm
f}=\omega_{\rm f} -\omega_p^{\rm f}-\mbox{Re}\Sigma^R_{\rm
f}+i\Gamma_{\rm f}/2$, for $4J>(\omega_{\rm f} -\omega_p^{\rm f} )^2$ we
find
 \begin{eqnarray}\label{tblrel1} &\mbox{Re}\Sigma^R_{\rm f}
=\frac{\omega_{\rm f} -\omega_p^{\rm f}}{2},\quad \Gamma_{\rm f}
=\sqrt{4J-(\omega_{\rm f} -\omega_p^{\rm f} )^2 }\,\,\theta
\left(4J-(\omega_{\rm f} -\omega_p^{\rm f} )^2\right),\nonumber\\
&\quad A_{\rm f}=\frac{\sqrt{4J-(\omega_{\rm f} -\omega_p^{\rm f} )^2
}}{J}\,\,\theta \left(4J-(\omega_{\rm f} -\omega_p^{\rm f} )^2\right).
\end{eqnarray}
where the fermion spectral function, $A_{\rm f}$, satisfies the exact sum-rule $\int_{-\infty}^{\infty} A_{\rm f}d\omega/(2\pi)=1$ . Thus, although we used approximations,
their consistency is preserved.

\subsubsection{Intensity of multiple scattering}
\begin{figure*}
\centerline{\includegraphics[clip=true,width=6cm]{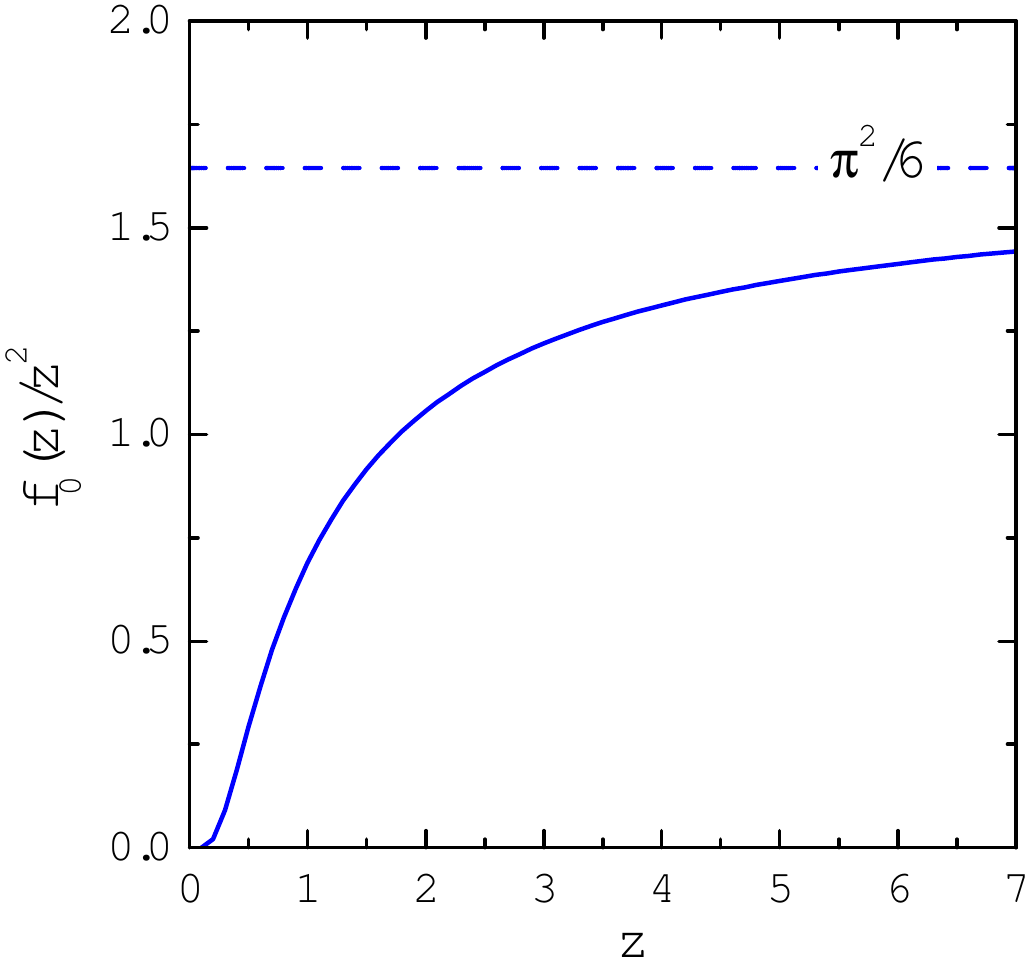}}
\caption{$f_0 (z)/z^2$, cf. Eq. (\ref{f0}). Dash line demonstrates
asymptotic behavior for $z\gg 1$.} \label{fig:f0}
\end{figure*}

Now we are able to evaluate the intensity of multiple scattering
$J$. Assuming that    in the energy-momentum and temperature region
of our interest bosons  are
good quasiparticles we have \cite{Voskresensky:2004ux,Voskresensky:2008ur}
$$A_{\rm b}= 2\pi\delta \left(\om_{\rm b}^2 -{p}_{\rm
b}^{\,\,2}-m_{\rm b}^2 -\mbox{Re} \Sigma_{\rm b}^{R} (\om_{\rm b} ,
{p}_{\rm b})\right)\,,$$   and find
 \begin{eqnarray}\label{Jexp} &&J =
\frac{g^2}{2\pi^2}\int_{0}^{\infty}\frac{ {p}_{\rm b}^{\,\,2} d
{p}_{\rm b}} { \left[ m_{\rm b}^{*2}(T)+ \beta_1{p}_{\rm b}^{\,\,2}+{\beta}_2{p}_{\rm b}^{\,\,4}
\right]^{1/2}}\frac{1}{\mbox{exp}\left[\left( m_{\rm b}^{*2}(T)+
\beta_1{p}_{\rm b}^{\,\,2}+{\beta}_2{p}_{\rm b}^{\,\,4}\right)^{1/2} /T\right]- 1} ,\nonumber
 \end{eqnarray}
where  we adopted a simple form of the
quasiparticle spectrum
  \be\label{branch} \om^2_{\rm b} ({p}_{\rm
b},T) \simeq m_{\rm b}^{*2}(T) +\beta_1 (T){p}_{\rm b}^{\,\,2}
+{\beta}_2{p}_{\rm b}^{\,\,4}+...
 \ee
Then for not too small positive value  $\beta_1$ neglecting term ${\beta}_2$ one finds
 \be\label{f0} J =\frac{g^2 T^2}{12\beta_1^{3/2}}r_s
(z), \quad r_s (z)=\frac{6f_0 (z)}{\pi^2 z^2}, \quad
z=\frac{T}{m_{\rm b}^*}.
 \ee
 Numerical evaluation of the integral
(\ref{f0}) is demonstrated in Fig. \ref{fig:f0}, cf. \cite{Voskresensky:2004ux}.  In the
limiting case of a high temperature  typical values of the boson momenta $\sim T$ and
 \be\label{Jexp-lim2} J \simeq \frac{g^2  T^2}{12\beta_1^{3/2}}
\,,\quad \mbox{for} \quad T\gg m_b^{*}(T) .
 \ee
 Let us focus on  the temperature region $T\gsim m_{\rm
b}^{*}(T)$, when $J$ is not exponentially suppressed.
For fermions in this case the
quasiparticle approximation valid for $T^2 \gg 4J$ would work
 only for weak coupling $g \ll \beta_1^{3/4}\simeq 1$.
Since we are interested in description of strongly interacting
particles, when $g\gsim 1$, we deal with blurred fermions for $T\gsim m_{\rm
b}^{*}(T)$.

The quasiparticle boson density inside the system (compare with Eq. (\ref{npifreevirt})) is given by
 \be\label{boso1} n_{\rm b}=
%N_{\rm b}
\int_{0}^{\infty}\frac{4\pi p^2 d
p}{(2\pi)^3}\int_{0}^{\infty}\frac{ d \om}{2\pi} 2\om A_{\rm b}
f_{\rm b} \simeq \int_{0}^{\infty} \frac{p^2 dp
}{2\pi^2}\frac{1}{e^{\sqrt{m_{\rm b}^{*\,2} +\beta_1 p^2}/T}-1} ,
 \ee
and for $T\gg m_{\rm b}^{*}$ we find
$n_{\rm b}\simeq T^3 \zeta (3)/(\beta_1^{3/2}\pi^2)$, $\zeta
(3)\simeq 1.202$.

\subsubsection{Density of fermion-antifermion pairs}

Let us continue to study the case of the strong interaction, $g \gsim 1$, and assume
$ m_{\rm b}^* (T)\lsim T\ll 5m_{\rm f}\beta_1^{3/4}/g $, see \cite{Voskresensky:2004ux,Voskresensky:2008ur}. Then we can
easily check that, on the one hand, conditions of the STL
approximation are fulfilled and, on the other hand, fermions can
be still treated as non-relativistic particles.

The 3-momentum fermion distribution is as
follows
 \be
{f}_{\rm f}^{(\pm)} (p_{\rm f} )=\int_{0}^{\infty}\frac{d\om_{\rm
f}}{2\pi}A_{\rm f}^{(\pm)}f_{\rm f}(\om_{\rm f}) .
 \ee
Substituting (\ref{tblrel1}) into this expression we find the
3-momentum fermion distribution
 \begin{eqnarray}\label{hdist} &&{f}_{\rm
f}^{(\pm)} (p_{\rm f} )=\int^{2 \sqrt{J}}_ {-2
\sqrt{J}}\frac{d\xi}{\pi} \frac{\sqrt{4J-\xi^2}}{2J}
\frac{1}{\mbox{exp}[ (\xi +\om_p^{\rm f} )/T]+ 1},
 \end{eqnarray}
where the variable $\xi =\om_{\rm f} -\om_p^{\rm f}$
and  $|\xi| <2 \sqrt{J}\ll  m_{\rm f}$. Doing further
replacement $\xi =-2\sqrt{J_s} +Ty$ and using that $m_{\rm f} \gg
T$, we obtain
 \be\label{hdist1} {f}_{\rm f}^{(\pm)} ( p_{\rm f}
)\simeq
 \frac{
T^{3/2}}{\pi J^{3/4}} I \left( \frac{4 \sqrt{J}}{
T}\right)\mbox{exp}\left[-\frac{\om_p^{\rm f} - 2\sqrt{J}
}{T}\right]= F(T) f_{{\rm Bol}}^{(\pm)}(p_{\rm f} ),
 \ee
 where $f_{{\rm Bol}}^{(\pm)}(p_{\rm f} )=e^{-\omega_p^{\rm f}/T}$,
 % \be
$$F(T)=\frac{ T^{3/2}}{\pi J^{3/4}} I \left( \frac{4 \sqrt{J}}{
T}\right)\mbox{exp}\left[\frac{2\sqrt{J} }{T}\right],\quad
 %\ee
%% \be
 %\label{I1}
 I (x)=\int^{x}_{0} dy e^{-y}\sqrt{y-y^2
x^{-1}},\quad x=4\sqrt{J} /  T, $$
%%\ee
%%\begin{eqnarray}\label{I1lim}
%%%I (x)&\simeq& \frac{\pi }{8}x^{3/2}(1-\frac{x}{2}) \,,\quad \mbox{for} \quad x\ll 1,\\
$I (x)\simeq \frac{\sqrt{\pi}}{2}
(1-\frac{3}{4x})\,,\quad \mbox{for} \quad x\gg 1. $
%%\nonumber \end{eqnarray}
As we have mentioned,  for $g \gsim 1$ the condition of the
validity of the quasiparticle approximation for fermions, $x\simeq
4\sqrt{J}/ T \ll 1$, is not fulfilled at temperatures of our
interest here.

For  $z\gsim 1$,  and $x\gg 1$ (i.e. for $g\gg 1$), we find
  \be
F(T)\simeq \frac{2^{1/2}3^{3/4}
\beta_1^{9/8}}{\pi^{1/2}g^{3/2}r_s^{3/4}}\mbox{exp}\left[\frac{gr_s^{1/2}}{3^{1/2}\beta_1^{3/4}}\right]\,.
 \ee

Integrating (\ref{hdist1}) in momenta we obtain the fermion
(antifermion) density:
 \be\label{ratd} n_{\rm f}^{(\pm)}\simeq
F(T) n_{{\rm Bol}}^{(\pm)}, \quad n_{{\rm Bol}}^{(\pm)}\simeq
N_{\rm f} \left(\frac{m_{\rm f}
T}{2\pi}\right)^{3/2}\mbox{exp}\left[ -\frac{m_{\rm
f}}{T}\right],\quad N_{\rm f}=2.
 \ee
We see that with growing  parameter $\sqrt{J_s}/T$, i.e. with
growing temperature, the density of fermion-antifermion pairs
increases significantly compared to the standard Boltzmann value.

The result (\ref{hdist1}) can be interpreted with the help of the
relevant quantity
 \be\label{effermm} m_{\rm f}^{*} (T) =m_{\rm f}
-2\sqrt{J}, \quad 2\sqrt{J}\ll m_{\rm f},
 \ee
which has the meaning of {\em{the effective fermion  (antifermion)
mass.}} However, contrary to the usually introduced effective
mass, the quantity (\ref{effermm}) enters only the exponent in
(\ref{hdist1}). We see that $m_{\rm f}^{* }(T)$
decreases with increase of the intensity of the multiple
scattering $J$, i.e., with growth of the temperature.

Typical (fermion blurring) temperature, when effective fermion mass decreases
significantly, $\sqrt{J}\sim m_{\rm f}$, is as follows
 \be\label{Tcssn} T\sim T_{\rm bl} \sim \beta_1^{3/4} (T_{\rm bl})g^{-1}m_{\rm
f}r_s^{-1/2} (T_{\rm bl}).
 \ee
For $T\gsim T_{\rm bl}$
 the non-relativistic
approximation for fermions, that we  used, fails.  However
exponential increase of the density of fermion-antifermion pairs,
that we have demonstrated, starts already for $T\ll T_{\rm bl}$,
in the region of validity of the non-relativistic approximation. Generalization
to the relativistic case can be found in \cite{Voskresensky:2004ux}.
For $g\gg 1$ and $\beta_1 \simeq 1$ we obtain \be T_{\rm bl}\sim
m_{\rm f}/g\ll m_{\rm f}.\ee Thus already for comparatively low
temperatures the fermion vacuum becomes blurred due to strong
interaction between light boson and heavy fermion sub-systems.

\subsubsection{Boson quasiparticles}
With the help of expression (\ref{tblrel}) we are able to
calculate the boson self-energy (\ref{selfzb}),
 \begin{eqnarray}
&&\mbox{Re}\Sigma_{\rm b}^R (\om_{\rm b}, p_{\rm b})\simeq -2g^2
\mbox{Tr}\int \frac{d^4 p_{\rm f}}{(2\pi)^4}\left[\mbox{Re}G_{\rm
f}^R (p_{\rm f}+p_{\rm b})+\mbox{Re}G_{\rm f}^R (p_{\rm f}-p_{\rm
b})\right]\nonumber\\ &&\times\mbox{Im}G_{\rm f}^R (p_{\rm
f})f_{\rm f} (\om_{\rm f}) ,\label{Resigsc}
  \end{eqnarray}
 \begin{eqnarray}
&&\Gamma_{\rm b}^R (\om_{\rm b}, p_{\rm b})\simeq 4g^2
\mbox{Tr}\int \frac{d^4 p_{\rm f}}{(2\pi)^4}\mbox{Im}G_{\rm f}^R
(p_{\rm f}+p_{\rm b})\mbox{Im}G_{\rm f}^R (p_{\rm f})\nonumber\\
&&\times\left[ f_{\rm f} (\om_{\rm f})-f_{\rm f} (\om_{\rm
f}+\om_{\rm b})\right] .
 \end{eqnarray}
In the limit $x\gg 1$ (i.e., $g\gg 1$):
 \be\label{bm}\mbox{Re}\Sigma_{\rm b}^R
(\om_{\rm b}, p_{\rm b})\simeq -\frac{4g^2 n_{\rm
f}^{(\pm)}}{\sqrt{J}}+\alpha (\om_{\rm b}^2 -\frac{1}{2}p^2_{\rm
b}),\quad \alpha =\frac{4g^2 n_{\rm f}^{(\pm)}}{Jm_{\rm f}},
 \ee
 \be
\Gamma_{\rm b}^R / \mbox{Re}\Sigma_{\rm b}^R =O(T^2 /J)
 \ee
 and thereby $\beta_1 \simeq 1+\alpha/2$ and at $T\sim T_{\rm bl}$ we estimate $\beta_1 \simeq  1+(m_{\rm b}^2/m_{\rm f}^2)$.
 Using
(\ref{bm}) we recover the effective boson mass \be\label{bosc}
m_{\rm b}^{*\,2} \simeq m_{\rm b}^{2}-\frac{4g^2 n_{\rm
f}^{(\pm)}}{\sqrt{J}}.\ee
The contribution $\propto \alpha$ has extra smallness
for $m_{\rm f}\gg m_{\rm b}$ and dependence on it can be
neglected.

We see that for temperatures $T< T_{\rm bl}$ one has
$|\mbox{Im}\Sigma_{\rm b}^R /\mbox{Re}\Sigma_{\rm b}^R|\ll 1$.
Thus, at these temperatures  bosons can be treated within the
quasiparticle approximation. Although the value
$\mbox{Re}\Sigma_{\rm b}^R$ is exponentially
small, we do not set $m^*_{\rm b}= m_{\rm b}$
and $\beta_1 \simeq 1$ because the boson corrections to some thermodynamic
quantities, e.g.,  pressure, can be larger than the corresponding fermion
contributions, cf. \cite{Voskresensky:2008ur}. Moreover, following Ref. \cite{Voskresensky:2004ux}, which treated fermions
in relativistic terms, at a somewhat higher temperature, $T>
T_{\rm bl}$, the value $-\mbox{Re} \Sigma_{\rm b}^R$ sharply increases
and the effective boson mass substantially decreases.

It proves to be that in thermodynamical quantities some interaction boson and fermion terms compensate each other.
The resulting total pressure, energy density and the entropy
are given by \cite{Voskresensky:2008ur},
 \be \label{prtot} P_{\rm
tot}\simeq\frac{\pi^2 T^4}{90}-\frac{m_{\rm b}^2 T^2}{24}+2Tn_{\rm
f}^{(\pm)} ,
 \ee
 \be \label{ertot} E_{\rm tot}\simeq\frac{\pi^2
T^4}{30}-\frac{m_{\rm b}^2 T^2}{24}+2m_{\rm f}n_{\rm f}^{(\pm)}\,,
 \ee
 \be \label{stot} TS_{\rm tot}= E_{\rm tot}+P_{\rm
tot}\simeq\frac{2\pi^2 T^4}{45}-\frac{m_{\rm b}^2 T^2}{12}+2m_{\rm
f}n_{\rm f}^{(\pm)} .
 \ee

Finally we have arrived at the following picture. A hot system of
strongly interacting ($g\gg 1$) light bosons and heavy fermions
 with zero chemical
potentials at temperatures $T_{\rm bl}>T\gsim m_{\rm b}^* (T)$
represents a   gas mixture  of  boson quasiparticles and blurred
fermions. Blurred heavy fermions undergo rapid ($p_{\rm f}\sim
\sqrt{m_{\rm f}T} \gg p_{\rm b}\sim T$) Brownian motion in the
boson quasiparticle gas. The density of blurred fermions is
dramatically increased at $T\sim T_{\rm bl}$ compared to the
standard Bolzmann value. Thermodynamical quantities are such as
for the quasi-ideal gas mixture of quasi-free fermion blurs and
quasi-free bosons. Note however that the fermion distributions and
the energy, pressure and  entropy of the fermion blurs are much
higher (by factor $F(T)$) than the
corresponding  quantities for the ideal Boltzmann gas.

\subsubsection{Hot Bose condensation}

For $T\gsim T_{\rm bl}$ typical fermion momenta $\sqrt{2m_{\rm
f}T}$ continue to remain much smaller than the mass $m_{\rm f}$.
However non-relativistic approximation for fermions, that was for simplicity used above, fails since typical deviation of the fermion energy from the
mass-shell becomes comparable with $m_{\rm f}$.  Nevertheless let
us extrapolate our results to higher temperatures. From
(\ref{bosc}) we see that at $\sqrt{J}(T_{\rm HB})=4g^2 n_{\rm
f}^{(\pm)}/m_{\rm b}^2$  the squared  effective boson mass
reaches zero and it may  become negative for $T>T_{\rm HB}$. In Ref. \cite{Voskresensky:2004ux} the phenomenon was dubbed the  {\em hot Bose condensation}, since
the condensate appears for the temperature
larger than a critical temperature. As the consequence of the strong
boson--fermion-antifermion interaction, the number of fermion degrees of
freedom is dramatically increased that, on the other hand,
results in the increase of the boson
abundance. Bosons  feel
a lack of the
phase space for energies and momenta $\sim T$
and a part of them is forced to
occupy the coherent condensate state, thereby.

 In Ref. \cite{Voskresensky:2004ux} in  relativistic framework  it was found that the value
$T_{\rm HB}$ lies in the vicinity of the value $T_{\rm bl}$. Saturation of the condensate field arises
for $T>T_{\rm HB}$  due to the repulsive boson-boson
interaction $L_{\rm
int}=-\Lambda \phi^4/4$ (with the coupling $\Lambda >0$) as it happens in case of the pion condensation in dense nuclear matter. Further details see in \cite{Voskresensky:2004ux,Voskresensky:2008ur}.

So, one could expect an anomalous enhancement of the boson (e.g. pion
and kaon) production at low momenta ($p_{\rm b}\lsim T$)  and an
anomalous behavior  of fluctuations, e.g. at LHC conditions, as a
signature of the hot Bose condensation for $T>T_{\rm HB}$, if a
similar phenomenon occurred in a realistic problem including all
relevant particle species.

\subsection{Pseudo-vector boson -- fermion coupling}
\subsubsection{Hot pion-nucleon vacuum. Nucleon blurring}
\label{pseudo-vector}
Let us now consider light pseudo-scalar  boson -- spin $\frac{1}{2}$ heavy fermion system
interacting via the pseudo-vector coupling (\ref{f-vertex}).
Let us further speak about $\pi^+ ,\pi^- ,\pi^0$ and nucleons and antinucleons.
To easier demonstrate the key ideas we continue to consider fermions in non-relativistic approximation. For relativistic generalization see \cite{Voskresensky:2004ux}.
Due to the p-wave nature of the interaction, cf. (\ref{f-vertex}), (\ref{f-non-vertex}), one may employ
in Eq. (\ref{branch})  the bare pion mass $m_\pi$ instead of $m^*_\pi$,  but $\beta_1\neq 1$. Moreover, the quantity  $\beta_1 (T)$ decreases with increasing temperature and at  $T>T_{\rm 1,HB}$ it can be that
 $\beta_1 <0$. In the region of sufficiently small $\beta_1$ and of
$\beta_1 <0$ one should keep in (\ref{branch}) the higher order terms, at
least the $\beta_2 k^4$ term. Eqs. (\ref{DysonnonrelFermbl}), (\ref{tblrel}) continue  to hold but now with the intensity of the multiple scattering $J$  given by
\be
J=6f_{\pi NN}^2 \int\frac{\vec{q}^{\,2}d^3 q d\omega}{(2\pi)^4}A_\pi n_\pi (\omega)\,.\label{Jpi}
 \ee
Assuming for a while that   $\beta_1 >0$ and not too small, in the limit case $T\gg m_\pi$ one finds
 \begin{eqnarray}\label{Jexppv-limh}
J(T\gg m_\pi)=J_{\rm lim}=
\frac{\pi^2 f_{\pi NN}^2  T^4}{10\beta_1^{5/2}(T)}
   \,.
\end{eqnarray}
This limiting expression works well already for $T\gsim m_\pi$, since numerical calculation shows that $J(T=m_\pi)\simeq 0.8 J_{\rm lim}$ and $J(T=1.5 m_\pi)\simeq 0.89 J_{\rm lim}$.

The temperature of the blurring of the fermion continuum evaluated within the
non-relativistic approximation for fermions ($T_{\rm bl}^{\rm n.r.}$)
follows from the relation
$J =m_f^2 /4$ (for $\mu_f =0$), from where using
(\ref{Jexppv-limh}) at the condition $\beta_1 >0$ we  estimate,
\be\label{tblps}
T_{\rm bl}^{\rm n.r.}=\frac{5^{1/4}(m_N)^{1/2}\beta_1^{5/8}(T_{\rm bl}^{\rm n.r.})
}{2^{1/4}\pi^{1/2}f_{\pi NN}^{1/2}}\,.
\ee
Taking into account relativistic interactions results in a decrease of the value $T_{\rm bl}$ by the factor $1/(3-\beta_1)^{1/4}$. Then for $\beta_1 = 1$ Ref. \cite{Voskresensky:2004ux} estimated
$T_{\rm bl}\simeq 215$ MeV and for $\beta_1 = 0.5$, $T_{\rm bl}\simeq 132$ MeV. Also, estimates \cite{Voskresensky:2004ux} show that $n_{N\bar{N}}$ reaches $2n_{N\bar{N},{\rm Bol}}$ already for $T\simeq 0.8 m_\pi$.

With the expression similar to (\ref{Resigsc}), but now for pseudovector coupling,  we are able to recover the quantity entering the boson spectrum (\ref{branch}),
\be\label{alpv}
\beta_1 (T) = 1-\frac{4
f^2_{\pi NN} n_{N\bar{N}}}{\sqrt{J}},\quad n_{N\bar{N}}\simeq \frac{m_N^{3/2}T^3 e^{-m_N^*/T}}{2^{3/2}\pi^2 J^{3/4}}\,,
%n_{\rm f\bar{f}}/n_{\rm f\bar{f},Bol}\simeq \frac{3f_{\pi NN}^2 m_NT}{8\pi^2}\,.
\ee
where $m^*_N=m_N-2\sqrt{J}$.
For $\omega \ll T$ Ref. \cite{Voskresensky:2004ux} found
$\mbox{Im} \Sigma_b^{R}\simeq -\beta (k)\omega$, where the quantity
\be
\beta (k)
=\frac{4 f^2_{\pi NN} n_{N\bar{N}}k^{\,2} }{\pi^{1/2}
J^{3/4}T^{1/2}},\quad \mbox{for} \quad 4\sqrt{J}\gg T,
\ee
controls the low-energy part of the spectrum (\ref{branch}).
Note that in these estimates in the boson self-energy for $\beta_1 >0$ we retained
the term $\propto k^2 $ and
dropped terms of the higher order
$\propto k^4$.

For temperatures  $T>T_{{\rm 1,HB}}$ the value $\beta_1 (T)$ may become
negative. In refs \cite{Voskresensky:1982vd,Dyugaev:1982gf,Voskresensky:1989sn,MSTV90,Voskresensky:1993ud} in the problem of the pion behavior in the dense and cold or warm matter the value of the density $n_{c}^{(1)}$,  when there appears minimum in $\tilde{\omega}^2 (k_0)$ at $k_0\neq 0$,
 was called the critical density for the occurrence of
{\em{the liquid/amorphous phase of the pion condensation}}. This phase exists till $n<n_c^\pi$.
Here, for
 $T_{\rm HB}>T>T_{{\rm 1,HB}}$ we may have a similar phenomenon \cite{Voskresensky:2004ux}. There is no yet a long-ranged order in this temperature interval.  However there arise
many virtual boson excitations carrying finite momentum $|\vec{k}_{0}|$.
Actual
values of particle momenta  are near a fixed value  $|\vec{k}_{0}|\neq 0$, but directions of the
momenta are randomly distributed.

At $T$ close to
$T_{{\rm HB}}$ at low $\omega$, the boson spectrum
gets a non-quasiparticle nature.
The dispersion relation
is then given by
\be\label{brim}
i\beta (k)\omega \simeq \omega_0^2 +\beta_0 (k-k_0)^2 ,
\quad k_0\neq 0,\,\omega_0^2 ,\,\beta ,\,\beta_0 >0 ,
\ee
where
%\begin{eqnarray}\label{omze}
$\omega_0^2
%&=&
=\widetilde{\omega}^2 (k_0)
=m_\pi^2 +\mbox{Re} \Sigma_\pi^R (\omega =0, k=k_0 )
%\\&\simeq&
\simeq m_\pi^2 -
\beta_1^2 /(4{\beta}_2).$
%\nonumber
%\end{eqnarray}
The value
\be
k_0 \simeq \sqrt{-\beta_1/(2{\beta}_2)}
\ee
corresponds to the minimum of $\widetilde{\om}^2 (k)$,
$\beta_0 \simeq -2\beta_1>0$.
The form of the spectrum (\ref{brim})
coincides with that we
used above  in the description of the pion behavior in the dense baryon matter.
The boson spectral function has the form
\be\label{vec-nonq}
{A}_\pi\simeq \frac{2\beta \omega}{
[\omega_0^2+\beta_0 (k-k_0)^2]^2 +\beta^2 \omega^2}.
\ee
Replacing (\ref{vec-nonq})  in Eq. (\ref{Jpi})
and using for simplicity the limit $\beta T\gg \om_0^2$,
$\beta \equiv \beta (k_0 )$, we calculate
\be\label{relJk0}
J \simeq \frac{3f_{\pi NN}^2 k_0^{\,4} T}{2\pi
\sqrt{\beta_0}\om_0 }\, .
\ee
 Thus we found an anomalous increase of the intensity of the multiple
scattering for $\omega_0\ll m_\pi$, i.e. the critical opalescence, as the precursor of the first order phase
transition to the state of the hot Bose condensation. In difference with the case of the scalar bosons, here owing to a strong p-wave pion-nucleon attraction the phase transition may occur
to the crystalline-like or liquid crystalline-like state similar to the case of the pion condensation in dense baryon matter studied in Section \ref{Pion-section}.

\subsubsection{$\Delta\bar{\Delta}$ resonance matter}
As we have mentioned, the $\Delta$ isobars play important role in nuclear interactions.
With inclusion of the $\Delta$-isobars  the first diagram  (\ref{phi}) should be replaced by
 \be\label{phiDelta}
\includegraphics[width=7cm,clip=true]{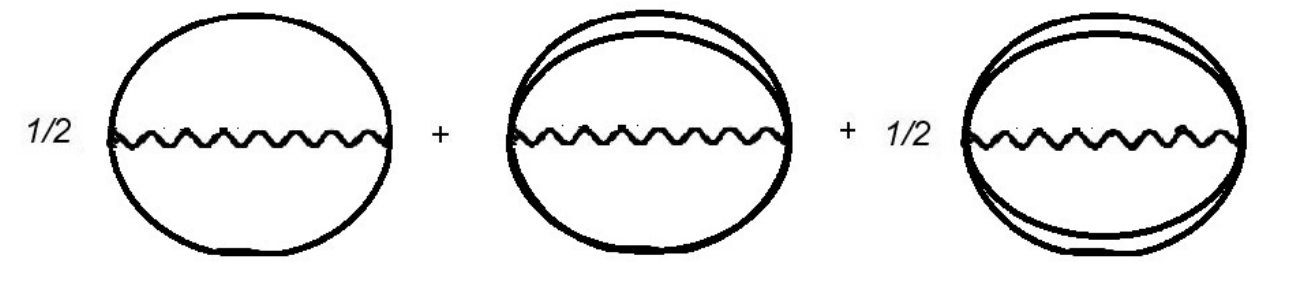}
 \ee
The number of $\Delta\bar{\Delta}$ pairs is estimated as
\be n_{\Delta\bar{\Delta}}\simeq 32 \int_0^{\infty} \frac{d\omega}{2\pi}\int \frac{d^3 p}{(2\pi)^3}A_\Delta e^{-p_0/T}\simeq \frac{16J}{9\omega^2_\Delta} n_{N\bar{N}}\,,
\ee
where $\omega_\Delta =m_\Delta -m_N$, $A_\Delta =-2\Im G^R_\Delta$. Thus we estimate $n_{\Delta\bar{\Delta}}/n_{N\bar{N}}\gsim 1$ in the vicinity of the temperature $T=T_{\rm bl}$. Strange/antistrange baryons also contribute on equal footing with nucleons/antinucleons and $\Delta$ and $\bar{\Delta}$.
Thus the hot vacuum (system at $\mu_N=0$) proves to be  a matter consisting of blurred resonances.

\subsubsection{Towards the description of the state of
hot hadron porridge}\label{Porridge}
We discussed simplified models, like the case of spin $\frac{1}{2}$
fermions (e.g., $N$, or  $N$ and $\Delta$) and their antiparticles coupled with the scalar boson ($\sigma$), or with the
pion. This is, of course,   oversimplification.
In reality nucleons couple with all mesons: $\sigma$,
$\pi$, $\omega$, $\rho$, $K$, etc. Also, higher lying baryon
resonances,
like $\Delta$-isobar, $N^* (1440)$, hyperons, etc., interact with the nucleons, with each other and with mesons.  Moreover, mesons interact with each other.
Reference \cite{Voskresensky:2004ux} dubbed such a  state the {\em hot hadron porridge}, bearing in mind that  the quasiparticle approximation fails
and that for $T\gsim m_{\pi}$ the continuum is likely blurred for all
relevant hadrons.

In realistic situation, also
quark-gluon degrees of freedom are  excited for
$T\gsim m_{\pi}$.  In a sense we deal with a quark-hadron duality, when similar effects follow both from the quark-gluon and the hadron descriptions.
{{The role of quark-gluon fluctuations
rises with increase of the temperature.}}
More likely, the system comes to
{{a strongly
correlated hot hadron--quark-gluon mixed state, representing
a hot hadron--quark-gluon porridge.}}
Taking into account multiple scattering effects,
the number of hadron degrees of freedom is significantly enhanced
simulating at least partially the
same effects, as those from deconfined quarks.  It is not easy to perform  calculations of such a system avoiding the double counting problems.  Deconfinement, in a standard meaning, as the purely quark-gluon state without hadrons is likely not realized.

\section{Non-pionic excitations in  Fermi liquids and possible condensation of scalar quanta}\label{ScalarSection}

\subsection{Particle-hole amplitude and Landau-Migdal parameters}

Let us consider the simplest case of a one-component Fermi liquid of
non-relativistic fermions.
Simplifying, we assume that the system is stable against pairing.
References \cite{Kolomeitsev:2016zid,Kolomeitsev:2017foi} studied low-lying scalar excitation modes
(density-density fluctuations) in  cold normal Fermi liquids
for various values and momentum behavior of the scalar Landau
parameter $f_0$ in the particle-hole channel.  In some density interval  an
interaction in the particle-particle channel is repulsive, $f_0 (n)>0$, and the
system is, therefore, stable against pairing in the s-wave state.
An induced p-wave pairing  possible at the repulsion, cf.  Ref.~\cite{pwave-pairing,pwave-pairing1,pwave-pairing2}, can be
precluded by the assumption that the temperature of the system is
small but higher than the critical temperature for the pairing, $T^{\rm rep}_{{\rm c,p}}$.

The
particle-hole scattering amplitude on the Fermi surface obeys the
equation, cf. \cite{LP1981,Nozieres,GP-FL,M67,Kolomeitsev:2016zid} and Eq. (\ref{TKFS}) above,
\begin{eqnarray}
&\widehat{T}_{\rm ph}(\vec{n}\,',\vec{n};q) =
\widehat{\Gamma}^{\omega}(\vec{n}\,',\vec{n})
 \nonumber\\
&\quad+ \langle \widehat{\Gamma}^{\omega}(\vec{n}\,',\vec{n}'')\,
\mathcal{L}_{\rm ph}(\vec{n}\,'';q)\, \widehat{T}_{\rm
ph}(\vec{n}\,'',\vec{n};q) \rangle_{\vec{n}\,''}\,,
\label{Tph-FL}
\end{eqnarray}
where $\vec{n}$ and $\vec{n}\,'$ are the directions of the fermion momenta
before and after scattering and $q=(\omega,\vec{k})$ is the momentum
transferred in the particle-hole channel. The brackets denote
averaging over the momentum direction $\vec{n}$,
\be
\langle\dots\rangle_{\vec{n}} = \intop\frac{d \Omega_{\vec
n}}{4\,\pi} \big(\dots\big)\,, \label{angl-aver}
\ee
and the particle-hole propagator is
\be
\mathcal{L}_{\rm ph}(\vec{n};q)= \intop_{-\infty}^{+\infty}\!\!
\frac{d \epsilon}{2\pi i}\!\! \intop_{0}^{+\infty}\!\!\frac{
d p\,p^2}{\pi^2} \, {G}_{\rm f}(p_{\rm f+})\,{G}_{\rm f}(p_{\rm f-})\,,
\ee
where  $p_{\rm f
\pm}=(\epsilon\pm\omega/2,p_{\rm F}\,\vec{n}\pm\vec{k}/2)$ and $p_{\rm
F}$ is the Fermi momentum. The quasiparticle contribution
to the full Green function is given by
\be
G_{\rm f}(\epsilon,\vec{p})=\frac{Z_{\rm qp}^{\rm f}}{\epsilon-\xi_{\vec{p}}+i\, 0\, {\rm
sign} \epsilon}\,,\quad\xi_{\vec{p}}=\frac{p^2-p_{\rm F}^2}{2\,m_{\rm
f}^*}\,.\label{Gn-QP}
\ee
Here $m^*_{\rm f}$ is the effective fermion mass, and the value $Z_{\rm qp}^{\rm f}$ determines a quasiparticle
weight in the fermion spectral density, $0<Z_{\rm qp}^{\rm f}\leq 1$, which is expressed through the retarded
fermion self-energy $\Sigma_{ \rm f}^R(\epsilon,p)$ as
$$Z_{\rm qp, f}^{-1}= 1-(\partial \Re\Sigma_{ \rm
f}^R/\partial \epsilon)_0\,.$$ The full Green function contains also a regular background part
$G_{\rm f}^{\rm reg}$, which is encoded in the renormalized particle-hole interaction $\hat{\Gamma}^\omega$ in Eq.~(\ref{Tph-FL}).
The particle-hole (p-h) scattering amplitude on the Fermi surface is determined
by the infinite series of p-h diagrams dressing the local interaction
 separated in scalar and spin channels as (\ref{localint}).
 We neglect  the spin-orbit
interaction, which is suppressed for small transferred momenta
$k\ll p_{{\rm F}}$.

On the Fermi surface $\Gamma_{0,1}$ are functions of the angle $\theta$ between momenta of incoming and outgoing fermions. Amplitudes $\Gamma_{0,1}^{\omega}$ ($F$ and $G$ in (\ref{localint})) are expanded in Legendre polinomials, e.g.,
$$Z_{\rm qp,f}^2 N\,\Gamma_0^{\omega}(\theta)=f(\theta)=\sum_l(2l+1) f_lP_l(\cos\theta)\,.$$
A similar expression exists in the spin channel. Here   $N={\nu m^*_{\rm f}\,p_{\rm F}}/{\pi^2}$ is the density of states at the Fermi surface. The Fermi momentum $p_{\rm F}$ is related to the total fermion density as $n=\nu\,p_{\rm F}^3/(3\pi^2)$ with $\nu =1$ for one type of fermions and $\nu =2$ for two types of fermions, like for the isospin-symmetric nuclear matter. Let us notice that here we use the normalization $N$ different from $C_0^{-1}$ employed in (\ref{localint}).

The Landau parameters $f_l, g_l$  can be calculated or fitted from experiments. For instance, the scalar parameter $f_0$ is related to the incompressibility of the system
$$K = n d^2 E_{\rm f}/d n^2=(1+f_0)p_{\rm F}^2/3m^*_{\rm f}\,,$$
 where $E_{\rm f}$ is the energy density of the fermion system.
Similarly, the square of the first-sound velocity  is expressed as
\be
u^2 =\frac{\partial P}{\partial \rho}=\frac{p_{\rm F}^2}{3m_{\rm f} m^{*}_{\rm f}}
(1+f_{0})\,,
\label{u}
\ee
$P=n\, d E_{\rm f}/d n -E_{\rm f}$ is the pressure, $\rho$ is the mass density.
The positiveness of the incompressibility and of the first-sound velocity squared are assured by
fulfillment of the Pomeranchuk stability condition $f_0>-1$.

Solution of Eq.~(\ref{Tph-FL})
is
\begin{eqnarray}
\widehat{T}_{\rm ph}(\vec{n}\,',\vec{n};q) &=& T_{{\rm ph},0}(q)\,
\sigma'_0\, \sigma_0 + T_{{\rm
ph},1}(q)\,(\vec{\sigma}\,'\vec{\sigma})\,, \nonumber\\ T_{{\rm
ph},0(1)}(q) &=& \frac{1}{1/\Gamma^\omega_{0(1)}- \langle
\mathcal{L}_{\rm ph}(\vec{n};q)\rangle_{\vec{n}}}\,.
\label{Tph-sol}
\end{eqnarray}
The averaged particle-hole propagator is
\begin{eqnarray}
\langle \mathcal{L}_{\rm ph}(\vec{n};q)\rangle_{\vec{n}}=-Z_{\rm qp,f}^2 N
\Phi\big(\frac{\omega}{v_{\rm F}\,k},\frac{k}{p_{\rm F}},T\big)\,,\label{Lindh}
\end{eqnarray}
where for $T\to 0$  the Lindhard function is
\be\Phi(\omega,k)=\frac12 + \sum_{i=\pm} (-i)\frac{z_i^2-1}{4(z_+-z_-)}\ln\frac{z_i + 1}{z_i - 1}
\,,\label{sound-eq}\ee
$z_\pm= \frac{\omega}{kv_{\rm F}} \pm \frac{k}{2p_{\rm F}}\,,
$
 $\omega,k$ are the energy and momentum transferred in the particle-hole channel, $v_{\rm F}=p_{\rm F}/m_{\rm f}^*$ is the Fermi velocity, cf. Eqs. (\ref{PhiT}), (\ref{Pht}), (\ref{rephi-lim-full}).

\subsection{~Scalar excitations and condensation}
\subsubsection{~Spectrum of scalar excitations}
Solution of equation
\be f_0^{-1}=-\Phi(\omega,\vec{k})\label{phiLind}
\ee
 gives the spectrum of excitations in  the scalar channel $\omega(k)$. A similar equation exists in the $g$-channel (with replacement $f_0\to
g_0$). Analytical properties of the solutions  have
been studied in \cite{Sadovnikova,Sadovnikova1,Sadovnikova2}.

Expanding the retarded particle-hole amplitude $T^R_{{\rm
ph},0}(q)$ near the spectrum branch \cite{Kolomeitsev:2016zid},
\begin{eqnarray}
T^R_{{\rm ph},0}(q)\approx \frac{2\omega (k)V^2(k)}{(\omega +i0)^2
-\omega^2 (k)}\,, \quad V^{-2}(k) & =Z_{\rm qp,f}^2 N
\frac{\partial\Phi}{\partial \omega}\Big|_{\omega (k)}\,, \label{Tnr}
\end{eqnarray}
we identify the quantity
\be
D^R(\omega,k)=[(\omega +i0)^2 -\omega^2 (k)]^{-1}\label{TnrD}
\ee
with the near-pole expansion of the retarded propagator of {\em a scalar  boson} with the dispersion relation $\omega =\omega(k)$ and the quantity $V(k)$, as the effective vertex of the fermion-boson interaction. This scalar boson can be associated with a field operator $\hat{\phi}$ in the second quantization scheme.

 For the repulsive interaction,  $f_0 (n)>0$, there exists a real (zero sound) solution of this equation,
%\begin{eqnarray}
\be\omega = k\, v_{\rm F}\,s(k/p_{\rm F})\,,\,\, s(k/p_{\rm F})\approx s_0+s_2 (k/p_{\rm F})^2\,,\,\,k/p_{\rm F}\ll 1\,,\label{zerosound}
\ee
%end{eqnarray}
where $s(x)$ is a function of $x=k/p_{\rm F}$, which can be taken in the form $s(x)\approx s_0 + s_2\,
x^2+s_4\,x^4$, $s_0>1$, see  \cite{Kolomeitsev:2016zid}. The odd powers of $x$ are absent, since $\Re\Phi$ is
an even function of $x$.  The zero-sound mode
exists as a quasi-particle mode for frequencies much larger than the inversed fermion collision time, i.e. $\omega\gg \tau_{\rm col}^{-1}\sim T^2/\epsilon_{\rm F}$, $\epsilon_{\rm F}$ is the Fermi energy, $T$ is the temperature. In the opposite limiting case, $\omega\ll T^2/\epsilon_{\rm F}$, the solution
describes a hydrodynamic (first) sound. For $k> k_{\rm lim}$, $k_{\rm lim} \lsim
p_{\rm F}$, for $s_2 <1/(16(s_0-1))$ the spectrum branch enters the region with $\Im\Phi> 0$,
and the zero sound becomes damped. For $s_2 >1/(16(s_0-1))$ the dissipation appears as result of the decay of the quantum to two quanta of smaller momenta, see \cite{Kolomeitsev:2016zid}.

Above  we assumed that $f$ depends only on $\vec{n}\vec{n}'$. The
results are, however, also valid, if the Landau parameter $f$ is a
very smooth function of $x^2$, e.g., for
\be
\label{fexp}
f_0 (x)\approx f_{00} +f_{02} x^2\,,
\ee
where the parameter $f_{02}$ is determined by the effective range
of the fermion-fermion scattering amplitude and expansion is valid
provided $|f_{00}|\gg |f_{02}|$ for relevant values $k/p_{\rm
F}<1$. According to Ref.~\cite{SaperFayans}, in the case of atomic
nuclei ($n\simeq n_0$), $f_{02}=- f_{00} r_{\rm eff}^2 p_{\rm
F}^2/2$, and  $0.5\lsim r_{\rm eff}\lsim 1$fm, as follows from the
comparison with the Skyrme parametrization of the nucleon-nucleon
interaction and with the experimental data.
 In the point $k=k_{0}$ corresponding to the minimum of
$\omega(k)/k$ the group velocity of the excitation $v_{\rm
gr}=d\omega/dk$ coincides with the phase one $v_{\rm ph}=\omega/k$.
The quantity $\omega (k_0)/k_0$ coincides with the value of the
Landau critical velocity $u_{\rm L}$ for the production of Bose
excitations in the    superfluid moving with the velocity $W>u_{\rm
L}$.

For isospin-symmetric
nuclear matter $f_{00}>0$ for $n> n_0$, $f_{00}<0$ for lower
densities, and in a certain  density interval below $n_0$ one has
$f_{00}<-1$, cf. Refs.~\cite{M67a,SVB,Backman:1984sx,Speth:2014tja,Matsui,Maslov:2015wba}. In the purely neutron
matter one has $-1< f_{00} < 0$ for $n\lsim
n_0$, cf.~\cite{Wambach:1992ik}.

Let us assume $f_{02}=0$. For $-1 < f_{00} < 0$, Eq.~(\ref{sound-eq}) has
only damped solutions. In the density region where $f_0<0$  Eq.~(\ref{sound-eq}) has purely imaginary solution
\begin{eqnarray}
\label{omnegf}
\omega (k) \approx i (2kv_{\rm F}/\pi) \left[z_f - {k^2}/{12p^2_{\rm F}} \right]\,,\quad
z_f=1-1/|f_{0}|\,,
\end{eqnarray}
which is valid for $1-|f_{00}|\ll 1$ and $k\ll p_{\rm F}$.

For $f_0<-1$, $z_f>0$ and in some interval of momenta $\Im\omega (k)>0$. The mode becomes  exponentially growing (Pomeranchuk instability). The function $-i\omega (k)$ has a maximum at $k_m =2p_{\rm F}\sqrt{z_f}$ equal to
\be
\label{maxsound}
-i\omega_m= ({8}/{3\pi}) v_{\rm F}p_{\rm F}\, z_f^{3/2}\,.
\ee
For $T\neq 0$ corrections to these results are as follows: $1+O(T^2/\epsilon_{\rm F}^2)$. Usually this instability is treated as a spinodal instability resulting in the creation of aerosol-like mixture of droplets and bubbles.
Reference \cite{Kolomeitsev:2016zid} suggested a new alternative that at certain conditions  the Pomeranchuk instability may lead to formation of a static Bose condensate of a scalar field.

The first-sound velocity squared follows from (\ref{u}) after the  replacement $f_0\to f_{00}$. For $f_{00}<-1$, as the incompressibility, the first sound velocity $u^2\propto 1+f_{00}$ is negative. Thus, in the region $f_{00} (n)<-1$ the hydrodynamical first sound mode with the frequency $\omega =u k$ for $k\to 0$ proves to be unstable. Note once more that the first sound exists in the hydrodynamical (collisional) regime, i.e. for $\omega\ll \tau_{\rm col}^{-1}\propto T^2$,  which is the opposite limit  to the collision-less regime of the zero sound, i.e. $\omega\gg \tau_{\rm col}^{-1}$. For the  equation of state with the isotherms, $P_T(1/n)$,  having a van der Waals form, the incompressibility and the square of the first-sound velocity prove to be negative in the  spinodal region. After a while the system is separated, as a result of the spinodal instability, in a part,  being in the vapor phase with the density $n_{\rm V}$, and another part, being in the liquid phase with the density $n_{\rm L}$. The corresponding chemical potentials at constant temperature and pressure are equal in the equilibrium state, $\mu_{\rm V}=\mu_{\rm L}$,  forming the  Maxwell line in the $P(1/n)$ dependence, see Fig. \ref{VanderWaals}. The relative fraction of the vapor phase $\chi$ ($0<\chi<1$) is determined by the averaged density $\bar{n} =n_{\rm V} \chi+n_{\rm L} (1-\chi)$.  With a decrease of $T$, isotherms $P_T(n)$ may cross the line $P_T(n)=0$ in two points for $n>0$, one corresponds  to a local maximum of the Landau free energy density and the other one to a local minimum of it. Collapsing to this minimum, the system becomes bound after a radiation of an energy excess, cf. \cite{SVB,Chomaz:2003dz}.

In the simplest case of ideal hydrodynamics the growing first-sound mode has the spectrum, cf. ~\cite{Skokov:2008zp,Skokov:2009yu,Skokov:2010dd},
\begin{eqnarray}\label{hydromode}
 -i\omega =k\sqrt{|u_{\tilde{s}}^2| -ck^2}\,,
\end{eqnarray}
where $u=u_{\tilde{s}}$ is taken at fixed specific entropy, and $c$ is a coefficient related to the surface tension of droplets of one phase in the other one as $\sigma\propto \sqrt{c}$. For a non-zero thermal conductivity the isothermal, $u=u_T$, and adiabatic, $u=u_{\tilde{s}}$, first-sound velocities are different, and $u_T^2$ becomes negative at a higher $T$ than $u_{\tilde{s}}^2$, see Ref.~\cite{Skokov:2010dd} and Fig. \ref{spinT}. After the replacement $u_{\tilde{s}}^2\to u_T^2$, Eq. (\ref{hydromode}) holds also for the case of a large thermal conductivity and small viscosity.
In the limit $T\to 0$ the isothermal and adiabatic first-sound velocities, $u_T$ and $u_{\tilde{s}}$, coincide. Maximum of $\omega$  with respect to $k$ is
\begin{eqnarray}\label{spinod}
-i\omega_m =v_{\rm F}p_{\rm F}(|f_{00}|-1)/(6m_{\rm f}\sqrt{c})\,.
\end{eqnarray}
The growth rate of the spinodal first-sound-like mode decreases with an increase of the surface tension of the droplets, whereas the growth rate of the collision-less zero-sound-like mode does not depend on the surface tension. For
\begin{eqnarray}\label{surftens}
\sqrt{c}>\frac{\pi|f_{00}|^{3/2}}{16 m_{\rm f} (|f_{00}|-1)^{1/2}}
\end{eqnarray}
the zero-sound-like excitations  (\ref{maxsound}) would grow more rapidly than excitations of the  hydrodynamic mode (\ref{spinod}). Note that the presence of the viscosity  may additionally detain formation of the hydrodynamic modes.

As it has been noted, at  first-order phase transitions in  multi-component systems with charged constituents, like neutron stars, the resulting stationary state can be a mixed pasta state, where finite size effects (a surface tension and a charge screening) are very important, cf. Refs.~\cite{Maruyama:2005vb,VYT2003}, contrary to the case of the one-component system, where the stationary state is determined by the Maxwell construction, cf. Ref.~\cite{SVB,Chomaz:2003dz}. For the isospin-symmetric nuclear matter at fixed temperature $T$, $T<T_c\simeq (15\mbox{--}20)$ MeV, there exists a spinodal region in the dependence $P(1/n)$ at nucleon densities below the nuclear saturation density, see Ref.~\cite{SVB,Chomaz:2003dz}. The liquid -- vapor phase transition may occur in heavy-ion collisions. The nuclear fireball prepared in the course of a collision has a rather small size, typically less or of the order of the Debye screening length. Nevertheless,  may be even in this case  some features of the pasta phase could be manifested \cite{Maslov:2019dep}.

\subsubsection{Condensation of scalar field}
In a fermion system with a contact interaction in scalar channel  one can introduce a collective scalar bosonic field by means of the Hubbard--Stratonovich transformation
%~\cite{Kopietz},
or by the formal replacement of the contact interaction to the exchange by a heavy scalar boson ~\cite{Kolomeitsev:2016zid,Kolomeitsev:2017foi}. The effective Lagrangian density of a static  scalar complex condensate field, taken in the simplest form $\phi =\phi_0 e^{-i\omega_c t+i\vec{k}_0\vec{r}}$, can be presented as follows
\begin{eqnarray}
\mathcal{L}_\phi = -{\rm sgn}(f_0)[(\Gamma_0^\omega)^{-1} +
Z_{\rm qp, f}^2 N\Phi(\omega_c,k_0)]|\phi_0|^2 -{\textstyle\frac12}\Lambda(\omega_c, k_0) |\phi_0|^4\,,
\end{eqnarray}
where the self-interacting term is determined by the integral of four fermion Green functions evaluated in~\cite{Brovman,Brovman1}. Energetically preferable is the state $\omega_c =0$, cf. \cite{Kolomeitsev:2016zid}, then
\be
\Lambda(0,k_0)\approx Z_{\rm qp,f}^4 \lambda\, (1+k_0^2/2p_{\rm F}^2)\label{Brovman1}
\ee
for $k_0\ll p_{\rm F}$ with $\lambda = \nu/(\pi^2v_{\rm F}^3)$, cf. general Eq. (\ref{Omoper}) with neglected $h$ term.
Variation of the Lagrangian in  $\phi_0$ yields equation $$-Z_{\rm qp,f}^2 N\tilde{\omega}^2(k_0)\phi_0 -\Lambda(0,k_0)\,|\phi_0|^2\phi_0 = 0\,.$$
The quantity  $\tilde{\omega}^2(k_0)=|f_0(k_0)|^{-1}-\Re\Phi(0,k_0)$ plays now a role of  an effective gap in the excitation spectrum. The  amplitude of the condensate field $|\phi_0|$ and the Bose condensate energy-density term become
\begin{eqnarray}
|\phi_0|^2 = - \frac{N\tilde{\omega}^2(k_0)}{
 Z_{\rm qp,f}^2\lambda}\,\Big(1+\frac{k_0^2}{2p_{\rm F}^2}\Big)^{-1}
\,,\quad
E_{\rm b}=- \frac{N^2\tilde{\omega}^4(k_0)}{ 2
\lambda\,}\Big(1+\frac{k_0^2}{2p_{\rm F}^2}\Big)^{-1}.
\label{EB-MF}
\end{eqnarray}
The inhomogeneous condensation proves to be more energetically favorable compared to the homogeneous one, if
\be
f_{02}\le f_{02}^{\rm cr}=-{f_{00}^2}/{3}-{f_{00}}/{4}\,,
\label{cond3}
\ee
otherwise there appears condensate with $k_0=0$, further details see in \cite{Kolomeitsev:2016zid}.

In the presence of the condensate the incompressibility becomes, $K = K_{\rm f} +K_{\rm b}$, $K_{\rm b}=n\frac{d^2 E_{\rm b}}{d n^2}$,
and, therefore, the scalar Landau parameter changes to
\begin{eqnarray}
f_0\to f_{0}^{\rm tot}=f_{0}+f_{0}^{\rm b}=f_{0}+3m_{\rm f}^*K_{\rm b}[f_{0}^{\rm tot}]/p^2_{\rm F}\,.
\label{diff-ftot}
\end{eqnarray}
Here $f_0$ and $f_0^{\rm b}$ are functions of $k_0$, which should be found from the energy minimization.
In  case of a weak condensate (for $|f_{0} +1|\ll 1$) one can use $E_{\rm b}[f_{0}^{\rm tot}]\approx E_{\rm b}[f_{0}]$. For a developed condensate the perturbative analysis does not work and one should solve Eq.~(\ref{diff-ftot}) self-consistently.

\subsubsection{Possibility of dilute metastable scalar condensate  states}
The spectrum of excitations on top of the condensate is described by the equation
$$Z_{\rm qp,f}^2 N\big[-|f_{0}^{\rm tot}(k)|^{-1} + \Phi(\omega, k)\big] - \delta\Sigma_\phi=0\,,$$
where $\delta\Sigma_\phi= 2\Lambda(\omega,k)\,|\phi_0|^2$ includes the interaction of excitations with the condensate. Making use expression for $|\phi_0|$, one presents the spectrum in the form \cite{Kolomeitsev:2016zid,Kolomeitsev:2017foi}
$${\omega} \approx i\frac{2}{\pi} \tilde{\omega}^2(k_0) \,k v_{\rm F}$$
 for
$-1\ll\tilde{\omega}^2(k_0)<0$\,.
We see that  excitations are damped, i.e., in the presence of the  condensate the Fermi liquid becomes free from the Pomeranchuk instability of the zero-sound-like modes.
\begin{figure}
\centering
%\begin{minipage}{1.0\textwidth}
\includegraphics[width=0.4\textwidth]{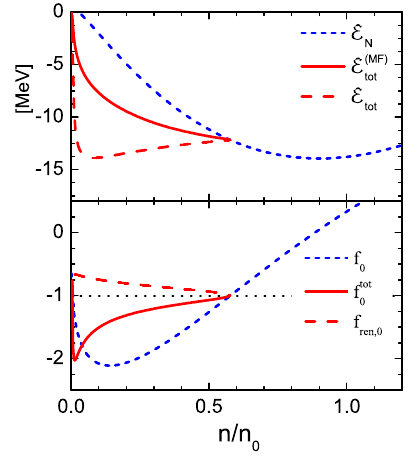}
%\end{minipage}
%\begin{minipage}{0.65\textwidth}
\caption{Energy per particle (upper panel) and the scalar Landau parameter (lower panel) as functions of the nucleon density in a piece of isospin-symmetric nuclear matter with and without the  condensate. The figure is adopted from \cite{Kolomeitsev:2016zid,Kolomeitsev:2017foi}.
\label{fig:ISM-eos}
}
%\end{minipage}
\end{figure}

The particle-hole interaction  also changes in the presence of the condensate.
There appears a new term in equation for the p-h amplitude, which
can be included in the renormalized local interaction as
$$ 1/f_{\rm ren,0}= 1/f_{0}^{\rm tot}(k_0) + 2 \, \tilde{\omega}^2(k_0) \,.$$
For homogenous condensate with $k_0=0$ we have $\tilde{\omega}^2(0)=-z_f=
-1-1/f_{0}^{\rm tot}$ and consequently
$f_{\rm ren,0} = -f_{0}^{\rm tot}/(2f_{0}^{\rm tot}+1)$. Thus,
if originally $f_{0}<-1$ and therefore $f_{0}^{\rm tot}<-1$, the renormalized interaction yields $-1<f_{\rm ren,0} < -1/2$. Hence, {in the Fermi liquid with the condensate the first-sound modes are stable}.  Knowing the value $f_{\rm ren,0}$ one can reconstruct the energy density ${E}_{\rm tot}(n)$ of the Fermi liquid from the differential equation
\begin{eqnarray}
\label{frenE}
d^2 {E}_{\rm tot}(n)/d n^2 = 2\epsilon_{\rm F} (1+f_{\rm ren,0})/(3n)\,,
\end{eqnarray}
which solution should continuously match the original energy density $E_{\rm f}$ at the values of the density, where $f_0=-1$. Note that the energy density includes both mean-field and quadratic-fluctuation contributions. Note also that the results are valid, if on the one hand  $\phi_0$  is rather small and on the other hand  fluctuations on the top of the condensate yield a yet smaller contribution, since the self-interaction of excitations on top of the condensate and  feedback of fluctuations on the mean field were disregarded.

 Let us demonstrate how the scheme works on example of the isospin-symmetric nuclear matter.
References \cite{Kolomeitsev:2016zid,Kolomeitsev:2017foi} considered a system of 125 nucleons, which energy per particle, $\mathcal{E}_N=E_N/n-m_N$, contains the volume and surface parts.
The  volume part satisfies  standard properties of the nuclear saturation: the density $n_0=0.16{\rm fm}^{-3}$, the energy per particle $\mathcal{E}_0=-16$\,MeV and the nuclear incompressibility $9K_{\rm f}=285$\,MeV. The inclusion of the surface term shifts the saturation density to $0.9n_0$ and the energy per particle to $-13.9$\,MeV. Values $\mathcal{E}_N$ and  $f_0$ are shown in Fig.~\ref{fig:ISM-eos} by short-dash lines. We see that $f_0<-1$  in some density interval. With this $f_0$ we solve Eq.~(\ref{diff-ftot}) and obtain $f_{0}^{\rm tot}$ and  $\mathcal{E}_{\rm tot}^{\rm (MF)}=\mathcal{E}_N + E_{\rm b}[f_0^{\rm tot}]/n$
shown by solid lines.
Finally, long-dashed lines show the renormalized Landau parameter $f_{\rm ren,0}$ and  the corresponding energy  per particle ${\mathcal{E}}_{\rm tot}(n) = E_{\rm tot}(n)/n$ with $E_{\rm tot}$ determined by Eq.~(\ref{frenE}).
We see that because of the condensate formation {\em a new metastable dilute nuclear state} may appear (in our example at density $0.081\,n_0$ with the binding energy $13.8$\,MeV).

\section{Condensation of Bose excitations in nonuniform state  in uniformly moving  media}\label{MovingSection}

\subsection{Physical picture and general phenomenological treatment}
\begin{figure}\centering
\includegraphics[width=10.8cm,clip]{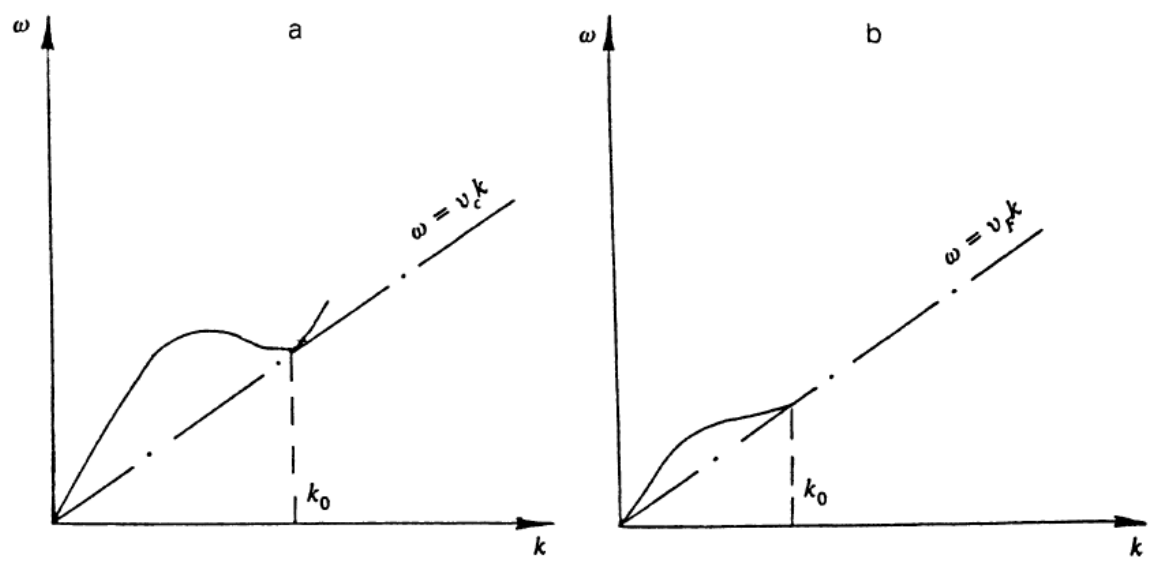}
\caption{(a) Spectrum of Bose phonon-roton excitations  in superfluid helium. (b) Zero sound branch of the spectrum in Fermi liquids.
 }\label{PitaevskiiCond}
\end{figure}

The spectrum of excitations in the $^4$He
is schematically shown in Fig. \ref{PitaevskiiCond} a). During a long time it was thought that the superfluidity is destroyed, if the superfluid moves uniformly with the speed $\vec{W}$ such that $|\vec{W}|$ is larger than the Landau critical velocity $u_{\rm L} =\mbox{min} (\omega (k)/k)$ with respect to a wall. The wall singles out the laboratory frame, with respect to which
the motion is defined. Equivalently one can consider  motion of a massive body with the speed $-\vec{W}$ in the superfluid, cf. \cite{Tilly-Tilly}.
In case of the spectrum shown in Fig. \ref{PitaevskiiCond} a) the local minimum of the frequency is reached at $k=k_0\neq 0$. The criterion
$W<u_{\rm L}$ is usually called the necessary condition for the superfluidity.
However Pitaevskii \cite{Pitaev84}  demonstrated that in case of the superfluid $^4$He moving in the capillary with $W>u_{\rm L}$ (in a nonrelativistic motion) there may appear additional condensate of excitations with
$k=k_0\neq 0$. In \cite{V93jetp} the consideration was generalized for different media including relativistic systems provided the spectrum of excitations has the form shown in Figs. \ref{PitaevskiiCond} a) or b). Both uniformly moving and rotating bodies were considered.  The spectrum of zero-sound excitations in Fermi liquids has the form shown in Fig. \ref{PitaevskiiCond}  b), see discussion in  Section \ref{ScalarSection}.  Possibility of the condensation of the Bose excitations  with $k_0\neq 0$ in the moving Fermi liquids was considered in \cite{Vexp95}. It was shown that Cherenkov-like radiation of strongly interacting excitations may form a Bose condensate with a non-zero wave vector. Explicit results were  presented for pions and for various types of zero sounds in moving  nuclear matter. Then Ref. \cite{BaymPethick2012} studied a possibility of the condensation of excitations  with $k_0\neq 0$ (levons) in cold Bose gases.  Condensation of excitations with $k_0\neq 0$ in moving $^4$He and  superconductors  was studied in \cite{Kolomeitsev:2016isb}.  The results might be applicable to the description of various bosonic systems like superfluid $^4$He, ultra-cold atomic Bose gases, charged pion and kaon condensates in rotating neutron stars, and various superconducting fermionic systems with the pairing, like proton and color-superconducting components in compact stars, metallic superconductors, and neutral fermionic systems with the pairing, like the neutron component in compact stars and ultracold atomic Fermi gases. The photon Cherenkov radiation and shock waves in supersonic fluxes (e.g., shock wave appearing  when an airplane overcomes sound velocity) are related phenomena. However in open systems  produced excitations may run away instead of forming of the condensate.

The key idea of the phenomenon is as follows \cite{Pitaev84,V93jetp}: when a medium  moves with a  velocity as a whole with respect to a laboratory frame (respectively a wall) with a velocity higher than $u_{\rm L}$, it may become energetically favorable to transfer part of its momentum from particles of the moving medium to a Bose condensate of
excitations  with  non-zero momentum $k_0\neq 0$. This would happen, if the spectrum of excitations is soft in some region of momenta. References \cite{Pitaev84,BaymPethick2012,Melnikovsky}  studied the condensation of excitations at $T = 0$ assuming the conservation of the flow velocity. Alternatively, \cite{V93jetp,Vexp95,Kolomeitsev:2016zid,Kolomeitsev:2016isb}  considered  systems under other conditions, namely assuming  conservation of the momentum (or angular momentum for rotating systems), for $T\neq 0$ in \cite{Kolomeitsev:2016isb}.

Let a  system, having spectrum of Bose excitations of the form shown in
Fig. \ref{PitaevskiiCond}  a) or b) moves initially as a whole with the constant velocity $|\vec{W}_{\rm in}|\ll c$ relatively the flat  wall.
In absence of the excitations occupying the state $\omega (k_0)$ the density of the kinetic energy of the moving fluid in the reference frame of the wall is $E^{\rm kin}_{\rm in}=\rho\vec{W}_{\rm in}^{\,2}/2$, where $\rho =M/V_3$, $M$ is the total mass of the system and $V_3$ is its volume. Assume that appearance of the condensate of Bose excitations of the form of the running plane wave,
\be
\phi^{\prime} =\phi^{\prime}_0 e^{-i\omega (\vec{k}_0) t +i\vec{k}_0\vec{r}},\label{runningphirot}
 \ee
can be energetically profitable, where $\phi^{\prime}_0$ is a real quantity. Then, after appearance of the condensate of excitations with $k=k_0$ a part of the momentum $\vec{k_0}\phi^{\prime\,2}_0$ belongs to the condensate of excitations. The conservation of the momentum reads as
\be
\rho \vec{W}_{\rm in}=\rho_M \vec{W}_{\rm fin}+\vec{k}_0 \phi^{\prime 2}_0\,,\label{momentuminfin}
 \ee
where assuming that the momentum of the condensate is $\vec{k}_0 \phi^{\prime 2}_0$ we simplifying consideration disregarded the retardation effects.

The kinetic energy of the system after appearance of the condensate of excitations becomes
\be
E^{\rm kin}_{\rm fin}=\rho\vec{W}_{\rm fin}^{\,2}/2 +\omega (k_0)\phi^{\prime 2}_0 +\Lambda \phi^{\prime 4}_0/2\,,\label{Ekinfin}
 \ee
where the term $\Lambda \phi^{\prime 4}_0/2$ describes self-interaction of the condensate of excitations. Also, simplifying consideration we disregarded interaction of the (``daughter'')
condensate of excitations with the ``mother'' condensate existed already  in the resting system. The latter interaction is disregarded in \cite{Pitaev84,V93jetp,Vexp95,BaymPethick2012} but was then  taken into account in \cite{Kolomeitsev:2016isb}.

After substitution of $\vec{W}_{\rm fin}$ from (\ref{momentuminfin}) to (\ref{Ekinfin})
 the difference  $E^{\rm kin}_{\rm fin}-E^{\rm kin}_{\rm in}$ renders
\be
\delta E^{\rm kin}=E^{\rm kin}_{\rm fin}-E^{\rm kin}_{\rm in}=- [\vec{k}_0\vec{W}_{\rm in}-\omega (k_0)]\phi^{\prime 2}_0+\tilde{\Lambda} \phi^{\prime 4}_0/2\,,\label{deltaEinfin}
 \ee
where $\tilde{\Lambda}=\Lambda +k_0^2/\rho$. It is energetically favorable to take $\vec{k_0}\parallel \vec{W}$ in order the condensate would decrease the intitial velocity of the system. After minimization of (\ref{deltaEinfin}) in the amplitude of the field $\phi^{\prime}_0$
we arrive at the final results
\be
\phi_0^{\prime\,2}=\frac{{k}_0({W}_{\rm in}-u_{\rm L})}{2\tilde{\Lambda}} \theta (W_{\rm in}- u_{\rm L})  \,,\label{deltaphiinfinfin}
 \ee
\be
\delta E^{\rm kin}=- \frac{{k}_0^2({W}_{\rm in}-u_{\rm L})^2}{2\tilde{\Lambda}} \theta (W_{\rm in}- u_{\rm L})  \,.\label{deltaEinfinfin}
 \ee
Thus we see that for $W_{\rm in}>u_{\rm L} = \omega (k_0)/k_0$ formation of the condensate of excitations with $k=k_0$ becomes energetically profitable. A part of the initial momentum of the system moving as a whole is transported to  the inhomogeneous condensate of excitations. As it follows from (\ref{momentuminfin}) and  (\ref{deltaphiinfinfin}), in the limit $\tilde{\Lambda} \to 0$ for arbitrary initial velocity $U_{\rm in}$ (in nonrelativistic system, which was considered) the resulting velocity tends to $u_{\rm L}$.

Note that in absence of the wall or in absence of any interaction, due to the Galilean invariance of the moving  and  resting reference frames excitations are not produced and the condensation cannot occur. In presence of the wall excitations are produced near the walls.

\subsection{Condensation of zero-sound-like excitations with a
non-zero momentum and frequency in a moving Fermi liquid}
\subsubsection{Condensation in rectilinearly moving Fermi liquid}
 Let us apply the constructed scheme to the analysis of
a possibility of the condensation of zero-sound-like excitations with a
non-zero momentum and frequency in a moving Fermi liquid.
As in Refs.~\cite{V93jetp,Vexp95}, we consider a fluid element of the medium
with the mass density $\rho$ moving with a non-relativistic
constant velocity $\vec{W}$. The quasiparticle energy $\omega (k)$ in
the rest frame of the fluid is determined from the dispersion
relation
\be
\label{disp} \Re D_{\phi}^{-1}(\omega , k)=0\,.
\ee
We continue to employ the complex scalar condensate field
described by the simplest running-wave probe function, cf.
Eq.~(\ref{runningphirot}), and the general expression for the Lagrangian density (\ref{Omoper}), but now
for the condensate of excitations, neglecting $h$ term.

The appearance of the condensate with a finite momentum
$\vec{k}_0$, frequency $\omega =\omega (k_0)$ and an amplitude
$\phi_0$ leads to a change of the fluid velocity from $\vec{W}$
to $\vec{W}_{\rm fin}$, as it is required by the momentum
conservation
\be
\label{momentumcons} \rho \vec{W}
=\rho \vec{W}_{\rm fin} +\vec{k}_0 Z^{-1}_{\rm qp,b} |\phi_0^2| \,,
\ee
where $\vec{k}_0 Z^{-1}_{\rm qp,b} |\phi_0^2|$ is the density of the momentum
of the condensate of the boson quasiparticles with the
quasiparticle weight
\be Z^{-1}_{\rm qp,b} (k_0) = \Big[
\frac{\partial}{\partial \omega}\Re D^{-1}_{\phi}(\omega ,
k)\Big]_{\omega(k_0),k_0}>0\,, \label{Zinv}
\ee
at  taking into account the retardation effects.
If in the absence of the condensate of excitations the energy
density of the liquid element was $E_{\rm in}=\rho \vec{W}^{\,2}/2$, then in
the presence of the condensate of excitations, which takes a part
of the momentum, the energy density becomes
\be
\label{Ef} E_{\rm fin}=\frac{1}{2}\rho W_{\rm fin}^2 +\omega (k_0)Z^{-1}_{\rm qp,b}
|\phi_0|^2 +\frac{1}{2} \Lambda (\omega (k_0), k_0) |\phi_0|^4.
\ee
Here the last two terms appear because of the classical field of the condensate of  excitations.
The gain in the energy density due to the condensation, $\delta E=E_{\rm fin} -E_{\rm in}$, is equal to
\be
\delta E = -[\vec{W}\vec{k}_0 -\omega (k_0)]\,Z^{-1}_{\rm qp,b}(k_0)\,
|\phi_0|^2 +\frac{1}{2}\widetilde{\Lambda}|\phi_0|^4\,,
\label{dE}
\ee
where
\be
\label{lambdatilde} \widetilde{\Lambda}=\Lambda (\omega
(k_0),k_0)+(Z^{-1}_{\rm qp,b}(k_0))^2 k_0^2/\rho\,.
\ee
For $\omega =0$, $\Lambda(0, k_0)$ is calculated explicitly, cf.
Eq.~(\ref{Brovman1}). Note that above equations hold also for
$\Lambda =0$.

The condensate of excitations is generated for the velocity of the
medium exceeding the Landau  critical velocity, $W>u_{\rm L} =\omega
(k_0)/k_0$, where the direction of the condensate vector
$\vec{k}_0$ coincides with the direction of the system velocity,
$\vec{k_0}\parallel \vec{W}$, and the magnitude $k_0$ is
determined by the equation $\omega(k_0)/k_0 =d\omega (k_0)/d k$.
The gain in the energy density after the formation of the
classical condensate field with the amplitude $\phi_0$ and the
momentum $k_0$ is then
\be\label{condE}
\delta E =- Z^{-1}_{\rm qp,b}(k_0) [W k_0 -\omega (k_0)]\phi_0^2 + \frac{1}{2}\widetilde{\Lambda} \phi_0^4\,.
\ee
The amplitude of the condensate field is found by minimization of the energy. From (\ref{condE}) one gets
\be
\label{varphi}
\phi_0^2 =Z^{-1}_{\rm qp,b} (k_0)\frac{W k_0 -\omega (k_0)}{\widetilde{\Lambda}}\theta (W-u_{\rm L})\,.
\ee
The resulting velocity of the medium becomes
\be
\label{finvel} W_{\rm fin} =u_{\rm L} +\frac{(W-u_{\rm L})\theta
(W-u_{\rm L})}{1+[Z^{-1}_{\rm qp,b} (k_0)]^2 k_0^2/[\Lambda (\omega
(k_0),k_0)\rho]}\,.
\ee
For a small $\Lambda$, we have $W_{\rm fin} =u_{\rm L} +O(\Lambda)$.

For the repulsive interaction $f_0>0$ there is real zero-sound
branch of excitations $\omega_{\rm s} (k)\approx k v_{\rm{F}} (s_0 +
s_2\, x^2+s_4\,x^4)$, where the parameters $s_i$ depend on the
coupling constants $f_{00}$ and $f_{02}$, cf. Eqs. (\ref{fexp}), (\ref{zerosound}) and
\cite{Kolomeitsev:2016zid}.
%Eqs.~(\ref{s0-coeff}), (\ref{s2-coeff}), and (\ref{s4-coeff}).
The ratio $\omega_{\rm s}
(k)/(k v_{\rm F})$ has a minimum at $k_{0}=p_{\rm F}\sqrt{-s_2/(2s_4)}$
provided $f_{02}$ is smaller than $f_{\rm crit,
02}=-f_{00}^2/[12(s_0^2-1)^2]$.
The Landau critical velocity of the medium is  equal to $u_{\rm
L}/v_{\rm F}=\omega_{\rm s}(k_0)/(v_{\rm F} k_0) \approx s_0-s_2^2/(2s_4)$.
The quasiparticle weight of the zero-sound mode~(\ref{Zinv}) is now
\be
Z_{\rm qp,b}^{-1}(k_0)=\frac{Z_{\rm qp,f}^2
N}{k_0v_{\rm F}}\frac{\partial\Phi(s,x)}{\partial
s}\Big|_{\frac{\omega_s(k_0)}{v_{\rm F} k_0},\frac{k_0}{p_{\rm F}}}\,.
\label{Z-fact}
\ee
The amplitude of the condensate field~(\ref{varphi}) can be written as
\be
\label{phif} \phi^2_0 =Z_{\rm qp,b}^{-1}(k_0) k_0\frac{W -u_{\rm L}
}{\widetilde{\Lambda}} \theta(W-u_{\rm L})\, .
\ee
  The energy density gained
owing to the condensation of the excitations is
\be\label{deltaEex}
\delta E=- k_0^2(Z^{-1}_{\rm qp,b}(k_0))^2\frac{(W -u_{\rm L}
)^2}{2\widetilde{\Lambda}}\theta
 (W-u_{\rm L})\,.
\ee
For a small $\Lambda$, Eqs.~(\ref{phif}), (\ref{deltaEex}) are simplified as
 \begin{eqnarray}
\phi^2_0 &=\rho (W -u_{\rm L}){Z_{\rm qp,b}(k_0) k_0^{-1}}\theta
(W-u_{\rm L}), \nonumber\\ \delta E &=-\frac{\rho}{2}(W-u_{\rm
L})^2 \theta (W-u_{\rm L})\,.
 \end{eqnarray}

Finally we note that the Landau parameter $f_0(k_0)$ should be recalculated with taking into account of the condensate in the moving Fermi liquid, except the vicinity of the critical point, where the corrections are small and can be neglected.

The instability condition for the mode with $\vec{W}\vec{k}_0>\omega (k_0)$, which was found at hand of the energy balance in Eq.~(\ref{dE}), can be obtained differently.  We may transfer the description in the laboratory frame, shifting variables  \cite{Vexp95} $\Phi (\omega, k,n)\to \Phi (\omega -\vec{k}\vec{W}, k,n)$. Then from Eqs.~(\ref{Tph-sol}), (\ref{Lindh}) it follows that the imaginary part of $\Phi$ changes its sign for $\omega <\vec{k}\vec{W}$. Thus, the instability first arises for modes with $\omega <kW$ at $\vec{k}\parallel\vec{W}$, like in case of the Cherenkov radiation. Excitations with the momentum $k=k_0$  start to populate the spectral branch $\omega (k)$, when the minimum $\omega (k)$ touches the value $\vec{k}\vec{W}$ (the condition of the coincidence of the group and phase velocities).

\subsubsection{Condensation in two interpenetrating dilute streams of fermions}
Now let us consider two interpenetrating dilute streams of fermions, which are supposed to interact very weakly. This model may
have a number of physical applications -- as example let us mention a possible manifestation
of pion instabilities and Cherenkov-like radiation in peripheral heavy-ion collisions, cf. \cite{Vexp95,Pirner:1994tt}.

Let us consider a peripheral nucleus-nucleus ($A+A$) collision in the reference frame associated with one of the nuclei (the target frame). For a momentum of the excitation $k$ there exists a minimal
value of the projectile momentum, $p_{\rm lab}$, above which the product of two momentum Fermi distributions, $n_{\rm F}(\vec{p})$, of target and projectile fermions vanishes $n_{\rm F}(\vec{p}\,)n_{\rm F}(\vec{p}+\vec{p}_{\rm lab}+\vec{k}\,)=0$\,. Then excitations from one Fermi sphere cannot overlap with the ground state distribution in the other
Fermi sphere. This condition is satisfied already for the laboratory energy $E_{\rm lab}\gsim 160\,{\rm MeV}/A$. All the results obtained above continue to hold after the replacement
$$f_0(n) \Phi (\omega, k,n)\to f_0(n/2)[\Phi (\omega, k,n/2)+\Phi (\omega -\vec{k}\vec{W}, k,n/2)].$$ For example,  for $\vec{k}\perp\vec{V}$ we have $$f_0(n) \Phi (\omega, k,n)\to 2f_0(n/2) \Phi (\omega, k,n/2)$$ and for $\omega =0$, $k\ll 2 p_{\rm F}$ we arrive at the simple replacement  $f_0(n)\Phi (\omega, k,n)\to 2f_0 (n_0/2)$. For a smooth density dependent value $f_0$ it further reduces to $f_0\Phi (0, k, n)\simeq f_0\to 2f_0$. If so, the Pomeranchuk instability would arise already for the values of $f_0 (n)$ essentially larger than $-1$; e.g., in the just considered simplified case of a smooth $f_0(n)$ function
the instability would occur for $f_0 (n)<-1/2$ rather than for $f_0 (n)<-1$. The instability could provoke a growth of the scalar condensate field $\phi$ with $\vec{k}_0\perp \vec{p}_{\rm lab}$ in the course of peripheral heavy-ion collisions.

Application of the same model of two interpenetrating beams to the  pion modes for $\vec{k}\perp\vec{W}$, cf. \cite{Pirner:1994tt,Voskresensky:2016oee}, leads us to the replacement
$$\Phi (\omega, k,n)\to 2 \Phi (\omega, k,n/2)$$
and for $\omega =0$, $k\ll 2 p_{\rm F}$ we  arrive at the  replacement
$\Phi  (\omega, k,n)\to 2$ for $\vec k \perp \vec p_{\rm lab}$.
 Since the nucleon particle-hole term of the pion polarization function is $\Re\Sigma^R\propto -p_{\rm F}(n)k^2$ for $\omega =0$, $k\ll 2 p_{\rm F}$ and $p_{\rm F}\propto n^{1/3}$, with the above modification
$p_{\rm F}(n)\Phi \to 2p_{\rm F}(n/2)$ we get effectively 4 times larger density and much stronger pion-nucleon attraction compared to the equilibrium case. So, there arises question about a possibility to observe pion condensate with $k_0\neq 0$ in peripheral heavy-ion collisions.

\subsubsection{Vortices}\label{Vortices}

Above we focused our consideration on the cases where either the
vortices are absent (as in a narrow capillary \cite{Pitaev84}) or they leave the system (in  open systems), or the presence of vortices supports a common rigid motion of the normal and superfluid components \cite{Tilly-Tilly} (e.g., as in systems with charged components \cite{Sauls89}, or in rotating systems, like neutron stars \cite{ST83}).

In case of He-II moving in a narrow capillary
vortices do not appear, see~\cite{Pitaev84,Ancilotto05}. For a
rectilinearly moving superfluids, as well as for the rotating superfluids,  there may
appear excitations of the type of vortex rings and other structures, cf. ~\cite{AndreevKagan84,Kolomeitsev:2016isb}.
The energy of the vortex ring is estimated, cf. ~\cite{Ginzburg1976,Khalatnikov}, as $\epsilon^{\rm vort}=2\pi^2
\hbar^2|\psi|^2 R$ $m^{*\,-1} \ln (R/l_0)$, and the momentum is $p^{\rm vort}=2\pi^2\hbar|\psi|^2 R^2$, where $\psi$ is the complex order parameter,  $m^*=1/(2\alpha_3)$ is the  coefficient near the gradient squared term in the Lagrangian density, cf. (\ref{Dexp}), for nonrelativistic particles $m^*$ has the sense of the effective mass of the particle,  $R$ is the radius of the vortex ring and $l_0\sim 1/\sqrt{|T-T_c|}$ is the coherence length. Thus, $u_{c1}=\epsilon^{\rm vort}/p^{\rm vort}= \hbar (R_{\rm tr}\,m^*)^{-1}\ln(R_{\rm tr}/l_0)$ is the Landau critical velocity for the vortex production, where
 in the absence of impurities $R_{\rm tr}$ is  the transverse size of the system. For a system of distributed impurities moving together with the fluid, $R_{\rm tr}$ is a typical distance between the defects. We recovered dependence on $\hbar$ only to remind that vortices have quantum origin.
Vortices are pined to the impurities and move together with them and the superfluid. In an open clean system moving with the velocity  $W>u_{c1}$ the vortex rings are pushed
to infinity by the Magnus and Iordanskii forces. Note that for spatially extended systems the value $u_{c1}$ proves to be lower than the Landau critical velocity $u_{\rm L}$. The flow moving with the
velocity $W$ at $W\geq u_{c1}$ can be considered as metastable,
since the vortex creation probability is hindered by a large
potential barrier, and formation of a vortex takes a long
time~\cite{vortex-creation,vortex-creation1}. The vortex production rate increases, however, strongly when $W$ approaches $u_{\rm L}$.
For a motion in a pipe the vortices are captured by the pipe wall,
forming after a while a stationary sub-system in the frame of the
walls. Periodic solitonic solutions of the Gross-Pitaevskii
equation were studied in~\cite{Tsuzuki}. This situation  might be
rather similar to that of a  condensate moving in a periodic
potential, produced by the spatial variations of the  order
parameter of the condensate of excitations~\cite{BaymPethick2012}. Since in the exterior regions of the
vortices the superfluidity persists, the
consideration of the condensation of excitations for $W>u_{\rm L}$ remains  applicable, cf. \cite{Kolomeitsev:2016isb}.
Note that in He-II under a high external pressure the value $u_{\rm L}$ decreases and at some conditions becomes lower than $u_{c1}$, see~\cite{McClintock}, and in the interval $u_{\rm L}<W<u_{c1}$ there are no vortices. Therefore for $u_{\rm L}<W<u_{c1}$ the condensate of excitations with $k_0\neq 0$  is not influenced by the vortices related to the mother condensate, cf.  \cite{Kolomeitsev:2016isb}.

Above we considered vortices appearing on the ground of the uniform mother condensate (at $k_0=0$). If the mother condensate is characterized by the wave vector  $\vec{k}_0\neq 0$, the form of the vortices is different. Instead of vortex filaments there may appear vortex sheets, see \cite{MSTV90}. The same relates to the daughter condensate of excitations.

Finally, let us notice that in superconducting systems vortices, if formed, are involved in a common motion with the superconducting sub-system due to the appearance of a tiny London field~\cite{Sauls89} distributed throughout the medium, that supports the condition $W=0$. In rotating superfluids vortices appear at rotation frequency
$\Omega >\Omega_{c1}= \frac{\hbar}{m^*\,R^2}\ln(R/l_0)$, where for the spherical system $R$ is the size of the system (transversal size for the cylindrical system), and their number grows with an increase of $\Omega$.
When the density of vortices becomes sufficiently large, they form the Abrikosov  lattice, cf.~\cite{Tilly-Tilly}, forcing, thereby, the superfluid and normal components to move as a rigid body, i.e. with $W\to 0$.

\section{Quantum and finite size effects in nonequilibrium particle
distributions}\label{NonequilibriumSection}
\subsection{Bose--Einstein condensation of relativistic bosons at  a dynamically fixed particle number}

The heavy-ion collision experiments at SPS, RHIC and LHC energies demonstrated that  at midrapidity a  baryon-poor
medium is formed  \cite{Afanasiev,Alt,Nayak,Abelev,Adamczyk} with pion number exceeding the baryon/antibaryon one more than by the order. To describe such a matter authors of the hadron resonance gas model, e.g., cf. Ref. \cite{Stachel}
and references therein, assume that hadrons  are produced at the hadronization temperature
$T_{\rm had}$  after cooling of an expanding quark-gluon fireball.
At the temperature of the chemical freeze-out, $T_{\rm chem}$, all inelastic processes cease and for lower temperatures only elastic processes are  possible.  In this model the  chemical potentials of the conserving baryonic, electric  and strangeness charges  are fitting parameters. The  hadron resonance gas state is assumed to be a mixture of ideal gases of all stable hadrons and resonances.
At the sudden break up of the system at the same temperature  $T_{\rm chem}$  mesons, e.g.  pions, are assumed to be in the ideal pion gas state at the thermal equilibrium at $\mu =0$. Fitting the ratios of the measured particle yields at  collision  energies $\sqrt{s}_{NN}>20$ the authors  found values   $T\simeq T_{\rm had}\simeq T_{\rm chem}\simeq (155-160)$ MeV. Sometimes,  repulsive interactions are modelled with an ``excluded volume'' prescription, which is however inherently a low density approach.

Authors of a chemical nonequilibrium approach  \cite{Letessier,Petran}
performed the analysis of top SPS and LHC data on mean
particle multiplicities assuming a sudden decay of the quark-gluon
medium into hadrons streaming then towards detector
without a hadron re-scattering phase, i.e. keeping information on the value $T\simeq T_{\rm had}\simeq T_{\rm chem}\simeq T_{\rm kin}$, where  $T=T_{\rm kin}$ is the temperature of the kinetic (thermal) freeze-out, when the system breaks up. For the best fits the authors  used all particle distributions, including those for pions, with non-zero fugacities determined by the values of the quark fugacities. They extracted smaller values  $T_{\rm chem}\simeq (140-145)$ MeV than \cite{Stachel} and values of the pion chemical potentials $\mu_\pi$ to be rather close to $m_\pi$. Note that following  \cite{Petran} at $T=T_{\rm chem}$ about $80\%$ of pions are
hidden in hadron resonances producing “secondary pions” at
the freeze-out. The later ones contribute to the total number of pions and to the extracted values $\mu_\pi$ together with primary
pions from the ideal gas state. The lattice QCD calculations \cite{Ding2019} found
that the temperature of a hypothetical critical point of a chiral phase transition should
not exceed a value of $132 +3-6$ MeV.

We considered  the behavior of the baryon-poor equilibrium hot hadronic medium in Section \ref{Blurring}, where  an important role of  strong interaction effects was demonstrated. Strongly interacting baryons prove to be  blurred at temperatures $T\sim T_{\rm bl}\lsim m_\pi$ and we discussed a possibility of the hot Bose condensation for $T>T_{\rm HB}>T_{\rm bl}$. In Section \ref{wherepions-sect} we demonstrated  influence of the in-medium effects on the  particle distributions at assumption of the sudden freeze out, further details see in \cite{Senatorov:1989cg,Voskresensky:1989sn,MSTV90,Voskresensky:1993ud,
Voskresensky:1995tx}.

Spectra of  pions produced in experiments at SPS, RHIC and LHC proved
to be approximately exponential at intermediate transverse
momenta, $2m_\pi \lsim p_T \lsim 7 m_\pi$, but show a significant  enhancement at
low transverse momenta $p_T \lsim m_\pi$. Already first
attempts  to fit the pion $p_T$  distributions in heavy-ion
collisions at 200 AGeV by ideal gas expressions required usage of the
pion chemical potential $\mu_\pi\simeq (120-130)$ MeV, cf. \cite{Kataja1990,Mishustin1992}.  Subsequent  analyzes of the SPS
data \cite{Ferenc1999,Tomasik2002}, employing the method proposed in \cite{Bertsch} of extraction
of the pion freeze-out density from the mid-rapidity particle
densities and the femtoscopic radii, supported statement
about significant enhancement of pion distributions at small
transverse momenta. Using the results of the first pion femtoscopy experiments, the value of the density of the pion system at the kinetic freeze-out was estimated as $n_\pi \sim (1-6)n_0$. Subsequent estimates, cf. \cite{Teaney,BFR2014}, yielded values  $n_\pi\sim (0.8-2.5)n_0$  for the pion production at LHC energies.
Analysis \cite{Stachel} extracted the fireball volume  $V_3=5280\pm 410$ fm$^{-3}$ that corresponds to the pion density $n_\pi \sim 2.5 n_0$.

Estimates \cite{Goity,Gerber} showed that at temperatures $T\simeq  (130-140)$ MeV the rate of the pion absorption becomes smaller than the rate of the re-scattering.
During subsequent pion fireball expansion from the chemical freeze-out state, when  $T = T_{\rm chem}$, till  a  kinetic
freeze-out state, at which $T$ reaches the value  $T_{\rm kin}<T_{\rm chem}$, the total pion number can be considered as approximately fixed,   see \cite{Goity,Gerber,Hung}.  Estimates  \cite{Teaney,Pratt,Melo,Prorok} gave the value  $T_{\rm kin}\simeq (100-120)$ MeV.

Recently,  the ALICE Collaboration observed a significant suppression of three and four pion Bose– Einstein correlations in Pb-Pb collisions at
$\sqrt{s_{NN}} =2.76$ TeV at the LHC \cite{Abelev2014,Adam2016}. This can be interpreted as there is a considerable degree of coherent pion
emission in relativistic heavy-ion collisions \cite{Akkelin2002,Wong2007}. Analysis \cite{Begun2015} indicated that about $5\%$ of pions could stem from the Bose--Einstein condensate. Also, a discussion of a possibility of the  Bose--Einstein condensation  in heavy-ion collisions
at LHC energies can be found in the review \cite{Shuryak2017}.

As it was mentioned, Section \ref{pseudo-vector}  considered description of the equilibrium strongly interacting pion enriched matter at a rather high temperature. Now we assume that such a matter exists for $T_c>T > T_{\rm chem}$, where $T_c$ is the pseudo-critical temperature of the deconfinement crossover transition. Now we will focus on the description of the expanding pion fireball  in the regime $T_{\rm kin}<T(t)<T_{\rm chem}$.

Actually, the problem is a more general.
Above we considered various examples of the Bose condensation in equilibrium systems. In $^4$He at $T=0$ all particles are in the condensate state and their number is fixed. At $T\neq 0$ a part of particles is in the condensate and a part is in ``normal'' excitations. In this case  only the total number of the condensate-  and over-condensate- particles is fixed. Similar situation occurs for the ordinary metallic  superconductors, where at $T\neq 0$ only a part of fermions is paired. The pairs form the Bose condensate with not fixed particle number. In case of the p-wave neutral pion, $\pi^0$, condensation in baryonic matter there appears condensate of virtual particles,  the number of neutral relativistic bosons in the equilibrium system is not fixed. The same relates to the $\pi^{\pm , 0}$ condensate in the isospin-symmetrical finite-size systems. In  these cases  there arise static $\pi^0$ and $\pi^{\pm , 0}$ condensate fields, respectively. In the electrically neutral neutron star matter it is energetically profitable to produce the $\pi^{\pm}$ condensate with a net negative electric charge. An excess of the charged  $\pi^-$ condensate particles together with electrons and $\mu^-$ muons  compensate the positive charge of the protons \cite{MSTV90}. These problems have been considered in the given review.

In case of the Bose--Einstein condensation  of nonrelativistic bosons in equilibrium matter one deals with fixed averaged number of bosons determined by the value of their chemical potential, cf. \cite{LL5}.  However the Bose--Einstein condensation may also develop in nonequilibrium systems at some conditions.  For instance, let  a system of Bose particles is rapidly cooled down below the value $T_c$  during the typical time $\tau_{\rm cool}\ll \tau_{\rm abs}$,  where $T_c$ is the critical temperature of the Bose--Einstein condensation and $\tau_{\rm abs}$ is the time characterizing the absorption of bosons.     We  assume that thermalization occurs in elastic processes at typical time scale $\tau_T\lsim \tau_{\rm cool}$. In this case the system on a time scale $t\ll \tau_{\rm abs}$ is characterized by the dynamically fixed particle number.
Such a kind of  the Bose--Einstein condensation  was studied  in  \cite{Voskresensky1994,Kolomeitsev1995,Voskresensky:1995tx,KKV1996,
Voskresensky1996,BFR2014,Kolomeitsev2018,Nazarova2019,Blaschke:2020afk,
Kolomeitsev:2019bju} and   other works on example of
pions  produced in  ultra-relativistic heavy-ion collisions.  On the time scale $t\ll \tau_{\rm abs}$ the dynamical behavior of the pion Bose--Einstein condensate is described by the equations of the two-fluid hydrodynamics similarly to the description of  superfluids at $T\neq 0$.
Note that descriptions of the hydrodynamical evolution of the Bose--Einstein condensate in approximation of the ideal gas and the self-interacting gas are essentially different, cf.  \cite{Putterman1974,LL06}. Both cases were considered in \cite{Voskresensky1994}. A number of subsequent works continued to study the  Bose-Einstein condensation of the ideal pion gas, e.g., cf. \cite{Begun2008,BFR2014}.
Behavior of the interacting pion gas was studied in the $\lambda\phi^4$ and  the $\sigma$ models for $\lambda>0$, cf. \cite{Voskresensky1994,Kolomeitsev1995,Voskresensky:1995tx,
Voskresensky1996,Kolomeitsev2018,Nazarova2019,Blaschke:2020afk,Kolomeitsev:2019bju}, and in the model with the Weinberg interaction \cite{KKV1996}.

The condensate component and the normal component do not interact provided  the velocity of the expansion of the pion fireball $W$ is low, $W<u_{c1}\sim [l\ln (R/l_0)]/R$, where $R$ is the typical size of the fireball (transversal size in case of one-dimensional Bjorken flux) and $l_0\ll R$ is the correlation (coherence) length associated with the pion condensate field, estimated in \cite{Voskresensky:1995tx,Voskresensky1996}. For $W >u_{c1}$ the vortex filaments and rings are produced.
Reference \cite{Voskresensky1994} suggested to seek such a pion soliton-like inhomogeneities in heavy-ion collisions. In presence of rotation in the interval of rotation frequences $\Omega_{c1}<\Omega <\Omega_{c2}$, $\Omega_{c1}\sim l_0\ln (R/l_0)/R^2$, $\Omega_{c2} \sim 1/l_0\sim (10^{21}-10^{22})$Hz,  there may appear the Abrikosov lattice of vortices.
Recall however that although this picture is similar to that occurs in case of   superfluids and superconductors, in case of pions  there exist  dissipative processes and the pion Bose-Einstein condensation may occur only for $t\ll \tau_{\rm abs}$.

 The pion Bose--Einstein condensation  in the Hartree approximation was studied in \cite{KKV1996,Kolomeitsev2018,Kolomeitsev:2019bju}. Pions acquire effective mass $m_{\pi}^* (T)>m_{\pi}$. The critical temperature for the self-interacting pion gas, at which  $\mu_\pi (T_{\rm BEC})=m^* (T_{\rm BEC})$, is smaller  than that would be for the ideal gas. At $T_{\rm BEC}$ there occurs the second-order phase transition. However already for $T=T^{\rm ind}_{\rm BEC}>T_{\rm BEC}$, at which
$\mu_\pi (T^{\rm ind}_{\rm BEC})$ reaches the free pion mass $m_\pi$, there may occur the first-order phase transition to the state of the ``induced Bose--Einstein condensation.'' A role
of inelastic reactions $\pi^0\pi^0\leftrightarrow \pi^-\pi^+$, which however conserve the net pion  number, was also discussed. The Lagrangian of the pion system describing not only elastically interacting pions but also permitting processes  $\pi^0\pi^0\leftrightarrow \pi^-\pi^+$ but not permitting other inelastic reactions was studied in \cite{Kolomeitsev2018,Kolomeitsev:2019bju}.

The kinetics of the Bose--Einstein condensation of photons in totally ionized plasma and shock waves of photons were  considered in \cite{Zeldovich} with the help of the kinetic equation for photons in electron gas derived in \cite{Kompaneets}. Then kinetics of the Bose--Einstein condensation in
nonrelativistic systems was studied in a number of works, cf. \cite{Kagan1992,Tkachev1995,Tkachev1997}. In relativistic systems kinetics of the Bose--Einstein condensation of pions was considered in \cite{Voskresensky:1995tx,
Voskresensky1996,Blaschke:2020afk}. Reference \cite{Voskresensky1996} employed the    Boltzmann equation (previously generalized to deal with high boson occupations $f(p)>1$), Ref.
\cite{Voskresensky:1995tx} used framework of the quantum kinetic Kadanoff--Baym equation and Ref. \cite{Blaschke:2020afk} considered the
 theoretical background provided by the Zubarev formalism of the nonequilibrium statistical operator \cite{Zubarev1971,Zubarev1996}.
Following \cite{Voskresensky1996}, if the initial state is appropriately overpopulated by pions undergoing predominantly  elastic collisions, then the self-interacting pions at low energies, $10^{-2}m^*\lsim |\omega -m^*|\ll m^*$,  enter   the region of a nonlinear Kolmogorov turbulence. For  nonrelativistic bosons the stationary state can be realized, characterized by the distribution function $f(\epsilon)\propto \epsilon^{-7/6}$ at constant flux of particles to the low energies, where  $\epsilon \ll m$ is the nonrelativistic particle energy, cf. \cite{Kagan1992}.
With a time passage, for still lower particle energies a self-similar solution is formed $f(\epsilon)\propto \epsilon^{-1.24}$, which is then  destroyed at the  formation of the Bose--Einstein condensate  for $t>\tau_{\rm BEC}$, cf.  \cite{Tkachev1995,Tkachev1997}. A similar picture holds in case of relativistic self-interacting pions forming the pion Bose--Einstein condensate \cite{Voskresensky1996}. For $t>\tau_{\rm BEC}$ the solution of the corresponding Boltzmann equation for the pion distribution should be supported by the solution of additional equation for the condensate field. This problem can be adequately considered, e.g., within Zubarev formalism of the nonequilibrium statistical operator, cf.  \cite{Blaschke:2020afk}.

The higher is the pion multiplicity the more probable is
to observe effects related to the pion Bose--Einstein condensation \cite{Voskresensky1994,Voskresensky1996}.  The fluctuation
effects  are increased in the vicinity of the critical
point of any phase transition, cf. discussion in Sects. \ref{FlhomSect} and \ref{flnonzero-sect}. In particular, second-order
phase transitions are accompanied by fluctuations of the order
parameter observed in various critical opalescence phenomena
in equilibrium systems \cite{LL8}.   If the system crosses the
spinodal instability border at a first order phase transition,
fluctuations begin to grow exponentially, cf. \cite{Skokov:2008zp,Skokov:2009yu,Skokov:2010dd,Maslov:2019dep} and Section  \ref{unstable}.

In case of the nonequilibrium matter the normalized variance of the density is expressed in terms of the structure factor \cite{RVKB2018},
\be S(\vec{q}\to 0, t, \vec{r})=(\langle \hat{n}^2\rangle -n^2)/n=\int_{-\infty}^{\infty}dq_0[-i\Sigma^{00}_{-+}(q_0,\vec{q}\to 0, t, \vec{r})]/(2\pi)\,,
\ee
where $\mu,\nu =0,1,2,3$,  $-i\Sigma^{\mu\nu}_{-+}(q,t, \vec{r})$ is the Wigner transform of the 4-current--4-current auto-correlation function expressed in terms of the nonequilibrium  diagram technique \cite{Knoll:1995gs,Knoll:1995nz}, $\hat{n}(t, \vec{r})$ is the density operator, the local density is $\langle\hat{n}(t, \vec{r})\rangle$. In the thermal equilibrium $S (\vec{q}\to 0)=T/(\partial\mu/\partial n)_T$.

In the ideal gas approximation   the normalized
variance of the number of produced pions diverges at the critical
point of the Bose--Einstein condensation, cf. \cite{Begun2008}. In experiments at SPS energies a growth of the normalized variance for the pion number
with an increase of the collision energy and the number of
produced pions was reported in Ref. \cite{Anticic}. Also an enhancement of the normalized variance
with an increase of the pion multiplicity was observed
for $pp$ collisions \cite{Kokoulina,Ryadovikov} in the energy range (50-70) GeV. However, a care should be taken comparing
expectations for the thermal fluctuation characteristics
with results of actual measurements, which incorporate
background contributions, the dependence on the center-of-mass
energy, other dynamical effects, collision centrality, kinematic
cuts, etc.

The most simple and still relevant
description of fluctuations in a quasi-equilibrium system
formed in heavy-ion collisions can be performed employing
the grand-canonical ensemble formulation, since usually
only a part of the system, typically around mid-rapidity, is
considered. Thus energy and conserved quantum numbers
may be exchanged with the rest of the system, which serves
as a heat bath. Although in the time interval between chemical and
kinetic freeze-outs the total number of pions remains fixed, an
exchange of particles between pion species continues owing
to the $2\leftrightarrow 2$ reactions. Thus, if pions are measured in
experiments with incomplete geometry and/or in a restricted
momentum range, then the elastic pion-pion reactions and
processes of the type $\pi^0\pi^0\leftrightarrow \pi^+\pi^-$
change populations
of pions of different isospin species and in different momentum
bins. Therefore, there exists a kind of thermodynamic
reservoir for the sub-system of pions, which reach the detector
later, and the grand-canonical formulation can be relevant in
such a situation. If one measures correlations between pions
emitted at different angles and in various momentum bins,
one may get an information about the state of the pion fireball
at the kinetic freeze-out. Self-consistent account for a pion-pion interaction in the
Hartree approximation demonstrated that variance of the pion
number in the system with an equal averaged number of pion
species remains finite at the critical temperature \cite{Kolomeitsev2018} as well
as the skewness and kurtosis \cite{Borisov}. A suppression of fluctuation
effects occurs also due to the finiteness of the system, cf. \cite{Begun2008,Kolomeitsev:2019bju}. Nevertheless, they remain to be enhanced near the critical
point of the Bose--Einstein condensation. In \cite{Kolomeitsev:2019bju} it was  shown that in the
case of the system with equal averaged numbers of isospin
species, the variance of the charge, $Q=N_{+}-N_{-}$, diverges
at $T\to T_{\rm BEC}$, whereas variances of the total particle
number, $N=N_{+}-N_{-}$, and of a relative abundance of
charged and neutral pions, $G=(N_{+}-N_{-})/2 N_0$, remain
finite in the critical point.  Thus, the appearance of a significant
increase of particle number fluctuations could be considered
as a signal that the pion system formed in heavy-ion collisions
is approaching the critical point of the Bose--Einstein condensation.
However to avoid a possible misunderstanding let us notice that the  cumulants of the net baryon/charge
distributions are increased in the vicinity of the critical point in case of other second-order phase transitions. Also recall about   the dynamical slowing-down effect for the near-critical fluctuations in the dynamical systems, cf. \cite{Berdnikov,Skokov:2009yu}.

\subsection{Breaking of  potential box filled by particles}

Comparison with the  data obtained at RHIC and  LHC for
proton-proton, proton-nucleus, and nucleus-nucleus collisions demonstrates that nucleon and pion  distribution functions (for $p_\perp>1$ GeV) follow the power laws, although at intermediate transverse momenta ($2m_\pi\lsim p_\perp\lsim 1$ GeV) these particles follow thermal  equilibrium  distributions
and for $p_\perp\lsim m_\pi$ the pion distributions are enhanced.
  The parton
model for scattering of point particles yields
\be
\frac{d\sigma}{dyd^2p_\perp}\propto p^{-n}_\perp
 \ee
with   $n=4$, as it follows from the dimensional analysis, cf. \cite{Kapusta2021}. However, scale breaking and realistic parton distribution functions in the projectile and
target increase $n$ significantly for hadrons, whereas  $n \simeq 4.5$ to $5.5$ for jets. In \cite{Wong2013,Wong2015} these features were attributed to the Tsalis distributions. Let us demonstrate that    strongly nonequilibrium particle distributions can be also characterized by power-law high-momentum tails.

Let us discuss a toy quantum mechanical
model, when one can easily find nucleon and pion distributions. Assume that nucleons of one species are placed into the box with infinite walls,
\begin{eqnarray}\label{U}
U=\begin{array}{cc}
\infty ,&r<R\\
0,&r>R\end{array}
\end{eqnarray}
and  occupy stationary states, e.g., with $l=0,m=0,n=1,...,N$, where $N$ is
the total number of nucleons of fixed spin and species. These
distributions  essentially differ from the thermal equilibrium distributions. If the
nucleon gas is so rare that collisions can be neglected, the initial
distributions can be considered as not changed at the time scale of our interest.

\subsection{Nucleon distributions}
Now let us assume that the walls  are suddenly removed. Let
us find the nucleon
distributions in the momentum space. For this aim we need the nucleon
$\psi$ function in the momentum representation
\begin{equation}\label{psipn}
\psi_{nlm}(\vec p )=\int_0^R \psi_{nlm}(\vec r )
\psi_{\vec{p}}^{\ast}(\vec r )\, r^2drd\Omega,
\end{equation}
where $\psi_{\vec{p}}(\vec r)=\mbox{exp}(i\vec{p}\vec{r})
/(2\pi)^{3/2}$ is the eigen function of the $\hat{\vec{p}}$
operator. From Eq. (\ref{psipn}) it follows that
\begin{equation}\label{psin-norm}
\int \mid \psi_{nlm} (\vec p )\mid^2 d^3 p =
\int \mid \psi_{nlm}(\vec r )\mid^2 d^3 x ,\,\,\,\,
dN /\left[ d^3 p /(2\pi )^3 \right] =(2\pi )^3 \sum_{nlm}\mid
\psi_{nlm} (\vec p )\mid^2 ,
\end{equation}
$\sum_{nlm} 1 =N$, where the sum is taken over the occupied states,
i.e. $l=m=0$ in our case. Replacing
\begin{equation}\label{psin}
\psi_{n00}(\vec r )=
A_{n0}\frac{\mbox{sin}(k_nr)}{r}Y_{00},\,\,k_n=\pi n/R,\,\,
A_{n0}=(2/R)^{1/2},\,\,Y_{00}=1/(4\pi)^{1/2}
\end{equation}
in Eq. (\ref{psipn}) one obtains
\begin{equation}\label{psixi}
\psi_{n00}(\vec p )=
\frac{(-1)^{(n+1)}R^{1/2}n\mbox{sin}(pR)}{\pi^2p(n^2-\xi^2)},\,\,
\xi=pR/\pi.
\end{equation}
The momentum distribution of the nucleons (of fixed spin) is as
follows
\begin{equation}\label{nJ}
n^N_p=dN /\left[ V_3~d^3 p /(2\pi )^3 \right] =
\frac{\mbox{sin}^2(pR)J}{(pR)^2},
\end{equation}
where
\begin{equation}\label{Jsum}
J=\frac{6}{\pi^2}\sum_{n=1}^N \frac{n^2}{(n^2-\xi^2)^2},
\end{equation}
and $V_3$ is the volume of the box.
In the limit case $\xi\ll 1$, $N\gg 1$, we have
\begin{equation}\label{nR}
n^N_p\simeq \frac{\mbox{sin}^2(pR)}{(pR)^2}\simeq
1-\frac{(pR)^2}{3}.
\end{equation}
Thus, for $pR\ll 1$ we get $n_{p}^{N}\simeq 1$. In the limit
$\xi\gg N\gg 1$ we obtain an oscillating solution
\begin{equation}\label{nRop}
n^N_p\simeq \frac{2\pi^2 N^3\mbox{sin}^2(pR)}{ (pR)^6}.
\end{equation}
One can see from (\ref{nR}) and (\ref{nRop}) that
nucleon distributions  behave
in this nonequilibrium quantum model
quite differently compared to the quasi-equilibrium thermal
distribution. Quantum effects are important for $p \lsim  \pi
N/R\lsim N^{2/3}m_{\pi}$, where the system size
is $R\sim N^{1/3}/m_{\pi}$. Also, Eq. (\ref{nRop}) demonstrates a
power-law tail at very large momenta $p \gg \pi N/R\sim
N^{2/3}m_{\pi}$, whereas thermal distributions show exponential
behaviour $\sim
\mbox{exp}(-\epsilon_p /T)$, cf. \cite{Senatorov:1989cg,Voskresensky:1989sp,MSTV90}. Another peculiarity   of the distribution (\ref{nRop}) is  the presence of oscillations  (factor $\mbox{sin}^2(pR)$).

\subsection{Pion distributions}
 Now with obtained nucleon distributions let us
calculate the pion
distributions. Supposing that
the number of pions is much smaller than the
number of nucleons, i.e. $N_{\pi}\ll N$, we may neglect the back
reaction of the light pion sub-system on the heavy nucleon
sub-system. The pion distribution yields
\begin{equation}  \label{prefact}
n_k^{\pi^-}=\frac{-i\Sigma^{- +}_{\pi^-} (t, \omega,
\vec{k})}{A_{\pi^-} (t, \omega,
\vec{k})}=\frac{\int_{\epsilon_0}^{\infty}
d\epsilon_pn^N_{\epsilon_p+\omega}
(1-n^N_{\epsilon_p})}{\int_{\epsilon_0}^{\infty}
d\epsilon_p(n^N_{\epsilon_p}- n^N_{\epsilon_p+\omega})}\,,
\end{equation}
where $\Sigma^{- +}_{\pi^-}$ is the ${\pi^-}$ self-energy expressed in terms of the nonequilibrium diagram technique, cf. \cite{Ivanov:1999tj}, and we assumed that the pion self-energy is determined by the diagram (\ref{selfzb}).
With the help of Eq. (\ref{nRop}) we get
\begin{equation}\label{impioutr}
A_\pi =-2\Im\Sigma^{R}_\pi\simeq \frac{\alpha N^3 \pi^2}
{4  m^{3}_N R^6}
\left(\frac{1}{\epsilon_0^2}-\frac{1}{(\epsilon_0+\omega)^2}\right),
\end{equation}
and
\begin{equation}\label{impiout-+}
-i\Sigma^{- +}_\pi\simeq \frac{\alpha N^3 \pi^2}{4 m^{3}_N R^6}
\frac{1}{(\epsilon_0+\omega)^2}\,,
\end{equation}
with
%\begin{equation} \label{alph}
$\alpha =-f^{2}_{\pi NN}q^2
m_N^{\ast 2}/\pi k,$
%\end{equation}
and
\begin{eqnarray}\label{betaeps-a}
\epsilon_0
%= \frac{[\omega -((m_N^{\ast})^2+k^2)^{1/2}+m_N^{\ast}]^{2}}
%{4[((m_N^{\ast})^2+k^2)^{1/2}-m_N^{\ast}]}
\simeq(\omega +q^2/2m_N^{\ast})^2
m_N^{\ast}/ 2k^2 .
\end{eqnarray}
The condition $\xi \gg N$  is satisfied, if $\epsilon_0\gg
N^2\pi^2/m_NR^2$. Thus
we obtain
\begin{equation}\label{pi-q}
n^{\pi^-}_k (t , \omega , k)\simeq
\frac{\epsilon_0^2 (\omega )}{2\epsilon_0 (\omega )
\omega+\omega^2}
\end{equation}
for the virtual pion distribution characterized by disconnected values $\omega, k$. Setting $\omega =\omega_k$ we find ${\pi^-}$ distribution at infinity,
\begin{equation}\label{nepres}
n_k^{\pi -}(\rm free) \simeq \frac{\epsilon_0^2 (\omega_k
)}{2\epsilon_0 (\omega_k )
\omega_k+\omega^2_k},\,\, \quad \omega_k = (m_\pi^2+k^2)^{1/2}\,.
\end{equation}
Both Eqs. (\ref{pi-q}) and (\ref{nepres}) are quite different from
the  thermal equilibrium distributions  (\ref{freepiondistr}), (\ref{piondistrbr}), (\ref{boso1}).  Instead of exponentially suppressed thermal distributions we obtained  enhanced pion distributions.

From Eq. (\ref{nepres}) we see a pronounced enhancement
of the pion distribution at small momenta,
\begin{equation}\label{smk}
n_k^{\pi-}({\rm free})\rightarrow m_\pi m^{\ast}_N/4k^2\,,
\end{equation}
for $k\rightarrow 0$, whereas for thermal distribution one would
rather  have $n_k^{\pi
-}({\rm free})\rightarrow const$ for $k\rightarrow 0$.
Introducing the value $T_c m_N /8 \simeq 0.8 m_{\pi}$, we present
Eq. (\ref{smk}) in the form $n_k^{\pi-}({\rm free})=2m_{\pi}T_c /k^2$.
Such a behaviour is typical for the thermal Bose distribution at
small momenta characterized by the pion chemical potential $\mu_{\pi}
=m_{\pi}$ and the temperature $T_c$, corresponding to the critical point of the Bose--Einstein
condensation. Please note that  enhancement of
the soft pion production is required in order to describe experimental pion
differential cross sections in a broad interval of heavy-ion collision energies, from GSI to LHC energies, cf. \cite{Voskresensky1996,Kolomeitsev2018,Kolomeitsev:2019bju}.

For the momenta ($m_N \gg k\gg m_\pi$) we have $n_k^{\pi
-}({\rm free})\simeq (m^{\ast}_N)^2 /4k^2$, whereas in
ultrarelativistic limit ($k \gg m_N $) we get $n_k^{\pi
-}({\rm free})\simeq (m^{\ast}_N)^4/4k^{4}$ and the pion number ($N^{\pi^-}$) is converged.
Thus  we see that the pion
momentum distribution is characterized by
three different slopes at $k\lsim m_\pi$, $m_N \gsim k\gg m_\pi$ and $k \gg m_N $.
Inclusion of the $\Delta$ isobars into consideration can be done with
the help of the replacement $\omega \rightarrow
\pm \omega -\omega_{\Delta}$. It does not bring about new peculiarities
to the problem.

\section{Conclusion}
Description of the phase diagram of the strongly interacting matter with  possible phase transitions between various phases is the great challenge for the researchers during many years. In nuclear physics   microscopic   description of the transition from the quark-gluon degrees of freedom to the hadron ones is absent due to absence of the solution of the confinement problem. In spite of intensive experimental studies of nonequilibrium nuclear matter have been performed  on accelerators  new  and more trick measurements are required.
A hope to get new exciting results is connected with  commissioning  of the NICA and FAIR facilities in the nearest future and with astrophysical studies. In this situation it is reasonable to employ  methods borrowed from the condensed matter physics with its much wider experimental facilities.

Phenomenological methods like the Ginzburg--Landau theory are widely used together with various semi-phenomenological and microscopic descriptions, such as the theory of finite Fermi systems, chiral perturbation theory, etc. Methods for description of inhomogeneous phases and phase transitions in nuclear systems to the states characterized by the  non-zero wave vectors are  less developed. Dynamics of the phase transitions and the structure formation are still less studied.

This manuscript deals with a broad range of problems associated with  phase transitions in various systems characterized by the strong interaction between particles and with formation of structures,  focusing on hadron systems. In Section \ref{General-sect} we started with a general phenomenological mean-field model constructed for the description of  phase transitions of the first and the second order to the homogeneous,  $k_0=0$, and inhomogeneous, $\vec{k}_0\neq 0$, states,  the latter transition may occur even in case, when the interaction is translation-invariant. First we studied the phase transitions within the mean-field approximation and then focus was made on the role of long-range fluctuations of the order parameter. These fluctuations are especially strong in case of the phase transition to the  state $\vec{k}_0\neq 0$
due to their large phase-space volume. Their inclusion results in that
the phase transition to the  state $\vec{k}_0\neq 0$ necessarily becomes  the  transition of the first order. Especially strong are fluctuations at $T\neq 0$.

Various specific features   of the phase transitions to the  state $\vec{k}_0\neq 0$ are then considered such as the anisotropic spectrum of excitations, a possibility of formation of various structures including running and standing waves, three-axis structures, the chiral waves, pasta states, etc. The Higgs effect is specific for $\vec{k}_0\neq 0$ due to the anisotropy. The Goldstone mode remains even in case of the charged running wave condensate. In case of the pasta mixed phase attention is focused on the charge screening effects resulting in that the hadron--quark and kaon condensate equations of state are closer to those described by the Maxwell construction than by the homogeneous Gibbs conditions. Next, a formal transition to hydrodynamical variables is performed. Derived equation for the phase of the order parameter is formally  similar to Navier-Stokes equation of non-ideal hydrodynamics. Such an analogy can be helpful, since one may use well developed methods of hydrodynamics.

In Section  \ref{First-order-transitions}  focus is made on description of the dynamics of the order parameter at the phase transitions to the states with $\vec{k}_0= 0$ and $\vec{k}_0\neq 0$. In case of the first-order phase transition, seeds of the form of slabs of any size of the
stable phase inside the metastable one grow to the stable state,
whereas rods and droplets of under-critical size are melt and overcritical seeds grow. In case of the phase transition from homogeneous state to the homogeneous state the seeds of some specific strongly non-spherical form  may grow to the new phase at any their initial size, similar to the slabs. In case of the phase transition from the inhomogeneous state to the inhomogeneous state the dynamics has specific features.  For example, initially spherical  seeds change their form during the time evolution.
Also, we discussed specificity of the transitions between homogeneous and inhomogeneous phases.

In Section \ref{hydro} the non-ideal hydrodynamical description of the phase transitions of the liquid--vapor type  was studied. In nuclear physics the phase transition of the liquid--gas type occurs in heavy-ion collision reactions with isospin-symmetric nuclei at very low collision energies ($\lsim 100$  GeV$/ A$). Following \cite{Skokov:2008zp,Skokov:2009yu,Skokov:2010dd,Steinheimer:2013gla,
Steinheimer:2016bet} it was also conjectured that the hadron--quark phase transition might be  of the liquid-gas type.
It was argued that the ordinary Ginzburg--Landau model is not applicable for description of an initial inertial stage of the seeds. Dynamics is determined by the inertial/viscous parameter $\beta$ proportional to the surface tension and inversely proportional to the viscosity. Stages of the nuclear rain and fog were discussed.
 Then we focused on the dynamics of seeds in the spinodal region.
At this stage quasi-periodic  structures are developed, see Fig. \ref{band}.
Crucial role played by the transport coefficients was emphasized. Linear and nonlinear dynamical stages   were considered   as for transitions to the homogeneous liquid state as to inhomogeneous state.
Effect of the cooling of the system, aging of materials and  sticking of domains were discussed.

The specific example of the pion condensation phase transition to the $\vec{k}_0\neq 0$ state in dense, cold and warm nuclear matter  was considered in Section \ref{Pion-section}. First, the Fermi-liquid description of the nuclear systems with explicit separation of the soft pion mode  was performed according \cite{Migdal78,MSTV90}. Then we focused on peculiarities of the $\pi$ condensate phase transition. Actually, due to strong  fluctuations of the pion field with $k_0\neq 0$, cf. \cite{Voskresensky:1981zd,Voskresensky:1982vd,Dyugaev:1982gf,Dyg1,Schulz:1984cb},
already for $n>n_{c}^{(1)}\simeq (0.5-0.8)n_0$ there appears, so called liquid or amorphous phase of the pion condensate, cf. \cite{Dyg1,Voskresensky:1989sn,Voskresensky:1993ud}. For $n>n_c>n_{c}^{(1)}$ by the first-order phase transition there appears the p-wave pion condensate with the liquid-crystalline-like or solid-like structure.
We focused on the discussion of the fluctuations with $k_0\neq 0$ at $T\neq 0$ and their contribution to thermodynamical quantities and observable pion distributions.
The pion mass-width effect is very essential, since it drives the phase transition. Following \cite{Ivanov:2000ma} we considered dynamics of the pion condensate transition.

Then in Section \ref{Blurring}
the high temperature -- small baryon chemical potential system was studied, when baryons  become completely  blurred in the sense that their Green function looses information about the quasiparticle pole due to multiple re-scatterings on virtual boson impurities. In dynamics this is similar to the Landau-Pomeranchuk-Migdal effect.
For $T>T_{\rm bl}$, with $T_{\rm bl}\lsim m_\pi$, the number of nucleon-antinucleon pairs increases appreciably compared to the standard Boltzmann result.
Light bosons, e.g., pions, may condense either in $\vec{k}_0= 0$ or $\vec{k}_0\neq 0$ states. This phenomenon can be called the hot Bose condensation, since it may occur for $T>T_{\rm HB}$ slightly exceeding the value $T_{\rm bl}$. Number of bosons is significantly increased due to the increase of the number of nucleon-antinucleon pairs for $T\sim T_{\rm bl}$ contributing to the nucleon-antinucleon loops. Then we focused on the example of the pion-nucleon sub-system, then  included $\Delta$ degrees of freedom and demonstrated that probably $\Delta$ contribute even stronger than nucleons. We argued that at $T\sim m_\pi$ one may speak about the state of a hot hadron porridge, when many hadron species contribute on equal footing.
The resulting state is rather dense and hot, so quarks and gluons may also contribute significantly.

In Section \ref{ScalarSection}   the phenomena of the  Pomeranchuk instability and the condensation of the scalar collective modes were studied. The latter phenomenon may result in appearance of a metastable nuclear state in dilute nuclear matter, as was conjectured  in \cite{Kolomeitsev:2016zid}.

In Section \ref{MovingSection} we considered   condensation of the Bose collective excitations in the $\vec{k}_0\neq 0$ state in rectilinearly moving media with the speed larger than the critical Landau velocity. The condensation of scalar quanta and the pion condensation with $\vec{k}_0\neq 0$ in peripheral heavy ion collisions can be considered as one of possible observable effects, cf. \cite{Pirner:1994tt}.

In Section \ref{NonequilibriumSection} we started with a possibility of the Bose-Einstein condensation of bosons  characterized by the  dynamically fixed particle number, i.e., when inelastic processes can be neglected at the typical time of the evolution of the system.
The consideration was then applied to the description of the pion production in heavy-ion collisions at ultrarelativistic energies.
 Then, on the example of the sudden break up of the box filled by nucleons  we discussed  a specificity of the purely nonequilibrium effects in production of nucleons and pions and oscillations in the nucleon distributions.

Finally note that the structure formations at the rotation and in magnetic fields, $\alpha$ Bose condensation,   as well as many other relevant problems, were not considered in the given review. These problems should be discussed elsewhere.

{\bf{Acknowledgments.}} Fruitful discussions with E. E. Kolomeitsev are acknowledged.

\end{document}